\documentclass[aps,noshowpacs,nofootinbib,superscriptaddress,notitlepage]{revtex4-1}

\usepackage{color}

\usepackage{epsf} 
\usepackage{graphicx} 
\usepackage{amsmath}
\usepackage{amssymb}
\usepackage{slashed}
\usepackage{hyperref}

\usepackage{epsf} 
\usepackage{graphicx} 
\usepackage{amsmath}
\usepackage{amssymb}
\usepackage{slashed}
\usepackage{hyperref}

\usepackage{microtype}

\def\be{\begin{equation}}
\def\ee{\end{equation}}
\def\bea{\begin{eqnarray}}
\def\eea{\end{eqnarray}}
\def\bear{\begin{array}}
\def\ear{\end{array}}
\def\bfig{\begin{figure}}
\def\efig{\end{figure}}
\def\bcen{\begin{center}}
\def\ecen{\end{center}}
\def\bi{\begin{itemize}}
\def\ei{\end{itemize}}

\def\raw{\rightarrow}

\def\la{\label}
\def\chic{\scriptscriptstyle}

\def\D{\displaystyle}

\begin{document}

\title{Progress and open questions in the physics of neutrino cross sections\\ at intermediate energies}

\author{L. Alvarez-Ruso}
\affiliation{Instituto de F\'\i sica Corpuscular (IFIC), Centro Mixto CSIC-Universidad de Valencia, E-46071 Valencia, Spain} 
\author{Y. Hayato}
\affiliation{University of Tokyo, Institute for Cosmic Ray Research, Kamioka Observatory, Kamioka, Japan}
\author{J. Nieves} 
\affiliation{Instituto de F\'\i sica Corpuscular (IFIC), Centro Mixto CSIC-Universidad de Valencia, E-46071 Valencia, Spain} 

\begin{abstract}
New and more precise measurements of neutrino cross sections have
renewed the interest in a better understanding of electroweak
interactions on nucleons and nuclei. This effort is crucial to achieve
the precision goals of the neutrino oscillation program, making new
discoveries, like the CP violation in the leptonic sector,
possible. We review the recent progress in the physics of neutrino
cross sections, putting emphasis on the open questions that arise in
the comparison with new experimental data. Following an overview 
of recent neutrino experiments and future plans, we present some details about
the theoretical development in the description of
(anti)neutrino-induced quasielastic scattering and the role of
multi-nucleon quasielastic-like mechanisms. We cover not only pion
production in nucleons and nuclei but also other inelastic channels
including strangeness production and photon emission. Coherent
reaction channels on nuclear targets are also discussed. Finally, we
briefly describe some of the Monte Carlo event generators, which are at the
core of all neutrino oscillation and cross section measurements.
\end{abstract}

\maketitle
\tableofcontents
\newpage

\section{Introduction}
Recent years have witnessed an intense experimental and theoretical
activity aimed at a better understanding of neutrino interactions with
nucleons and nuclei. Although this activity has been stimulated mostly
by the needs of neutrino oscillation experiments in their quest for a
precise determination of neutrino properties, the relevance of
neutrino interactions with matter extends over a large variety of
topics in astrophysics, physics beyond the Standard Model, 
hadronic and nuclear physics.

{\bf Oscillation experiments:} At present, the main motivation for
neutrino cross section studies comes from oscillation
experiments. They aim at a precise determination of mass-squared
differences and mixing angles in $\nu_\mu$ disappearance
and $\nu_e$ appearance measurements. The ability to reconstruct the
neutrino energy is crucial for this program. Indeed, oscillation
probabilities, such as
\begin{equation}
P(\nu_\mu \rightarrow \nu_\tau) = \sin^2{2 \theta_{23}} \, \sin^2{\frac{\Delta m^2_{32} L}{4 E_\nu}}
\end{equation}
depend on the neutrino energy $E_\nu$, which is not known for broad
fluxes. A reliable determination of the neutrino energies in nuclear
targets requires a good understanding of the reaction mechanisms and a
precise simulation of final state interactions. There are also
irreducible backgrounds, for example from neutral current $\pi^0$ or
$\gamma$ production when these particle produced showers that are
misidentified as electrons from $\nu_e \, n \rightarrow e^- \, p$.

{\bf Astrophysics:} Neutrinos play an important role in astrophysical
phenomena and carry information about the emitting sources. In
particular, the dynamics of core-collapse supernovae is controlled by
neutrino interactions. The neutron rich environment of supernovae is a
candidate site for r-process nucleosynthesis because radiated
neutrinos convert neutrons into protons. To address these questions a
good knowledge of low energy neutrino production and detection cross
sections is required~\cite{Bertulani:2009mf,Volpe:2013kxa}.

{\bf Physics beyond the Standard Model:}  Non-standard neutrino interactions leading, for example, 
 to deviations from universality in the weak couplings or flavor violation in neutral current 
processes could affect neutrino production, propagation, and detection processes as subleading effects (see Ref.~\cite{Ohlsson:2012kf} for a recent review). Long and short baseline experiments allow to set bounds on these interactions.

{\bf Hadronic physics:} Neutrino cross section measurements allow to
investigate the axial structure of the nucleon and baryon resonances,
enlarging our views of hadron structure beyond what is presently known
from experiments with hadronic and electromagnetic probes, not
forgetting about lattice QCD. Another fundamental and open question is
the strangeness content of the nucleon spin which can be best
unraveled in $\nu \, p(n) \rightarrow \nu \, p(n)$ studies.

{\bf Nuclear physics:} Modern neutrino experiments are performed with
nuclear targets.  For nuclear physics this represents a challenge and
an opportunity. A challenge because the precise knowledge of neutrino
and baryon properties can only be achieved if nuclear effects are
under control. An opportunity because neutrino cross sections
incorporate richer information than electron-scattering ones,
providing an excellent testing ground for nuclear structure, many-body
mechanisms and reaction models.

We discuss the progress in the physics of neutrino interactions with
nucleons and nuclei at intermediate energies, highlighting some of the
open questions and challenges standing ahead. We start with an
overview of recent neutrino experiments and future prospects. The
theory of neutrino interactions on nucleons is introduced making
emphasis on the symmetries, the sources of experimental information
and major uncertainties. Pion production is covered, but also other no
less interesting inelastic reactions with smaller cross sections. We
present different descriptions of neutrino-nucleus scattering from a
common perspective, indicating the different approximations and
assumptions adopted. The role of multinucleon mechanisms is stressed
as well as their impact in the neutrino energy reconstruction. Nuclear
medium effects and final state interactions in particle production off
nuclear targets are also addressed, paying special attention to the
coherent reaction channels and their different theoretical approaches.
The last part is devoted to the Monte Carlo models, which are applied 
to the simulation and analysis of neutrino experiments.

The present review complements other excellent articles of this kind
that have recently
appeared~\cite{Gallagher:2011zza,Morfin:2012kn,Kopeliovich:2012kw,Formaggio:2013kya}. In
Ref.~\cite{Formaggio:2013kya} a very broad range of energies from the
eV to EeV is covered, providing an impressive account of the existing
experimental data. As we focus on the few-GeV region, we do not
discuss the deep inelastic regime, nor questions like shadowing or
duality, all of which can be found in
Refs.~\cite{Morfin:2012kn,Kopeliovich:2012kw}. Instead, we address
topics like superscaling, quasielastic and inelastic strangeness
production, photon emission or Monte Carlo generators,
which are at most mentioned in previous reviews.

\section{Recent and future oscillation and cross section experiments}

 The discovery of atmospheric neutrino oscillations~\cite{Fukuda:1998mi}
was made through the observed distribution of the outgoing charged
leptons, produced by the interactions of atmospheric neutrinos
with water in the Super-Kamiokande (SK) detector.
The signature of the $\nu_\mu$ disappearance was clearly
seen as an up-down asymmetry in the zenith-angle distribution 
of muons, which is expected to be almost symmetric in the absence of 
neutrino oscillations. This result was quite robust against the uncertainties 
in the neutrino interactions considering the observed number
of events at that time. By now, this experiment has accumulated a significant amount of atmospheric neutrino data which allowed more detailed analyses, establishing that these data are well explained by $\nu_\mu \leftrightarrow \nu_\tau$ oscillations and determining accurately the oscillation parameters~\cite{Ashie:2005ik,Abe:2012jj}. The generalization to the three flavor scenario, including matter effects has also been performed~\cite{Wendell:2010md}.  

In the mean time, several new accelerator-based long-baseline neutrino 
oscillation experiments, like K2K~\cite{Ahn:2001cq},
MINOS~\cite{Adamson:2007gu}, OPERA~\cite{Agafonova:2010dc}, 
and T2K~\cite{Abe:2011ks} have started taking data to confirm the
SK results and to measure more precisely the oscillation parameters.
All these experiments use high purity $\nu_\mu$ and/or $\bar\nu_\mu$
beams and search for the $\nu_\mu$ ($\bar\nu_\mu$) oscillation 
into the other neutrino flavors.

\begin{figure}[htbp]
\begin{center}
  \includegraphics[width=0.7\textwidth]{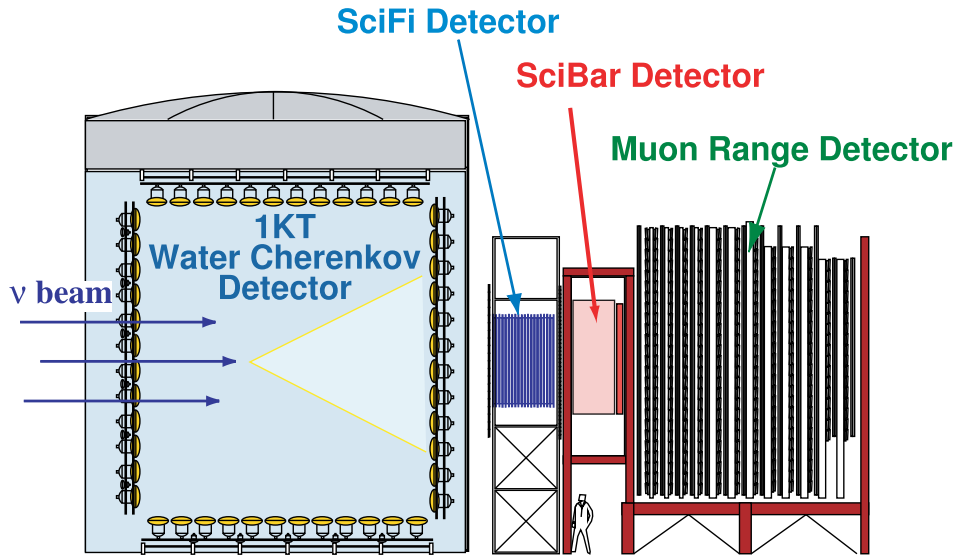}
  \caption{Near detector complex of the K2K experiment.}
  \label{Exp:K2K}
\end{center}
\end{figure}

 The K2K experiment started operating in 1999 as the first
 accelerator-based long baseline neutrino experiment.  The K2K
 experiment intended to confirm the neutrino oscillation phenomena
 observed in atmospheric neutrinos. It also aimed at the search for
 $\nu_\mu \to \nu_e$ appearance signals.  The accelerator and the
 near detectors were located in KEK, Tsukuba, Japan. The average neutrino
 energy was 1.3 GeV, and the SK detector, situated 250 km away from KEK,
 was used as the far detector. K2K reported a deficit in the number of
  $\mu$-like events observed at SK, and confirmed the $\nu_\mu$
 oscillation phenomenon.  However, no indication of $\nu_e$ appearance
 events was observed. The near detector of the K2K experiment
 consisted of several components: a 1 kt water Cherenkov detector, a
 water target scintillator fiber tracker (SciFi), an iron target muon
 range and a lead glass (LG) detectors. In the later stage of the 
 experiment, the LG dectector was replaced with a fully active scintillator 
 tracking detector (SciBar) (see Fig.~\ref{Exp:K2K}).
With these detectors, various
 neutrino interactions were studied. One of the interesting
 observations was a strong suppression of muons in the forward
 direction~\cite{Gran:2006jn} and an enhancement of events for large scattering
 angles. This suppression was observed not only in SciFi but also in the water Cherenkov and the SciBar
 detectors.  These measurements could be well described by increasing
 the axial vector masses ($M_A$) entering in the used models for
 the charged current quasi-elastic (CCQE) scattering and the charged
 current single pion production (CC1$\pi$).  Moreover, to achieve a good
 agreement with data, corrections on the parton distribution functions
 (PDF's) ~\cite{Bodek:2003wc} were also necessary.  The CCQE interaction
 was extensively studied with the SciFi detector and as result  an effective
 value of $M_A = 1.20 \pm 0.12$  GeV for the nucleon
 axial vector mass was extracted from the data~\cite{Gran:2006jn}.
The model dependence in the determination of $M_A$ in experiments with nuclear targets, 
in the light of new theoretical developments, is discussed in Sec.~\ref{sec:CCQE}.  

 The neutral current $\pi^0$ production, which is one of the
 background sources in the  search for  $\nu_e$ appearance signatures, was also studied using the
 1kt water Cherenkov detector.  The predicted number of interactions
 from the simulation  agreed quite well with the data,  but the
 $\pi^0$ momentum  distribution turned out to be slightly
 shifted~\cite{Nakayama:2004dp}.  Another interesting result from the
 K2K experiment was the  measurement of the charged current
 coherent pion production cross-section using the SciBar detector.
 The reported  cross-section was almost  consistent with zero and much smaller than
 the predictions from simple PCAC (partial conservation of the axial
 current)  models~\cite{Hasegawa:2005td}, which
 however successfully described  old high energy experiments. This finding stimulated the 
theoretical work on coherent reactions induced by (anti)neutrinos described in Sec.~\ref{sec:coh}.

 The MINOS experiment is also devoted to the study of $\nu_\mu$ and
 $\bar\nu_\mu$ oscillations. The neutrino beam is produced at the
 Fermi National Laboratory (FNAL) in Illinois, US using 120 GeV
 protons from the main injector.  The mean energy of the neutrino beam
 is adjustable by changing the configurations of the target and the
 magnetic horns.  Because the neutrino mass difference  $\Delta m^2_{32}$
 turned out to be quite small, the
 nominal energy of the beam was set to 3 GeV. The produced neutrinos
 are studied in a near detector  at FNAL and also in the far detector
 located in the SOUDAN mine, 750 km away from the target.  Both the near
 and far detectors are steel-scintillator sampling calorimeters with
 tracking and energy measurement capabilities.  These detectors have
 a magnet field; thus, it is possible to measure the particle
 momentum and identify its charge.  Owing to the intense neutrino
 beam, the most stringent limit was set on the squared mass difference
 $\Delta m^2_{32}$ and the consistency of the oscillation parameters
 between $\nu_\mu$ and $\bar\nu_\mu$ was also
 discussed~\cite{Adamson:2013whj}.  They also studied CCQE scattering in the
 near detector and obtained a  value for the effective axial nucleon
 mass of  around $1.2$ GeV~\cite{Dorman:2009zz},
 which is consistent with the result from K2K.

 The OPERA experiment is slightly different from these 
two previous ones. It aimed to study the $\nu_\mu
\to \nu_\tau$ oscillation in the appearance channel.
The neutrino beam was produced using the 400 GeV SPS proton line
in CERN. The far detector is located in the Gran Sasso
Laboratory (LNGS) in Italy, which is 732 km away from the target.
It consists of an emulsion detector  and
a high precision tracker, which are used to identify $\tau$ leptons 
by their decay products. The experiment identified
3 $\tau$ candidates and confirmed the appearance of
$\tau$'s from $\nu_\mu$ oscillation~\cite{Agafonova:2013dtp}.

The MiniBooNE experiment~\cite{AguilarArevalo:2010zc} was constructed to study
the neutrino oscillation pattern observed by the LSND
experiment~\cite{Aguilar:2001ty}, which can not be explained
with the standard three neutrino scenario. MiniBooNE 
used a line of 8 GeV kinetic-energy protons taken from the Booster at
FNAL. The averaged energy of the produced neutrino beam was around 800 MeV
and the detector is located 541 m away from the target.
The MiniBooNE tank contains 800 tons of mineral oil which is used to detect
Cherenkov and scintillation lights. Because the beam intensity
was high and the mass of the detector was large, this experiment
accumulated more than 100 thousands of neutrino and anti-neutrino
events from 2002 to 2012. MiniBooNE measured  CCQE
\cite{AguilarArevalo:2013hm, Katori:2009zz},  
CC1$\pi$ differential cross-sections~\cite{AguilarArevalo:2010xt, AguilarArevalo:2010bm},  
and also the corresponding neutral current (NC) channels~\cite{AguilarArevalo:2010cx,AguilarArevalo:2009ww}.
Their results  corroborated  the muon forward suppression observed in K2K and MINOS, and 
a number of interactions significantly larger than expected
from the theoretical predictions with the nominal theoretical parameters,
i.e. $M_A = 1.0$ GeV, was also obtained.
Besides, the 
pion momentum distribution  of the charged current single $\pi^+$
events is broader and larger than Monte Carlo predictions, which might indicate that
the effects of $\pi^+$ rescattering  in the nucleus are much 
smaller than expected from the theoretical simulations. On the 
other hand, the momentum distribution of  the NC$\pi^0$ production sample ç
is more consistent with the theoretical predictions. 
More details in this respect can be found in Sec.~\ref{sec:pion}. 
MiniBooNE also observed a larger number of $\nu_e-$like events
 than theoretically expected. This could not be explained
by the known backgrounds. Various new models including  possible 
``non-standard'' oscillation scenarios  
have been proposed,  but no definitive conclusion has yet 
been reached~\cite{AguilarArevalo:2008rc, Aguilar-Arevalo:2013pmq}.

Another short baseline experiment, NOMAD~\cite{Altegoer:1997gv}, searched for the appearance of $\nu_\tau$ neutrinos in the CERN SPS wideband neutrino beam with a 24~GeV neutrino energy. This study was motivated by the conjecture that $\nu_\tau$ could have a mass of 1~eV or more. The NOMAD detector has a 2.7 tons active target made of drift chambers in a 0.4~T magnetic field. The target is followed by a transition radiation detector for electron identification, a preshower detector, a lead-glass electromagnetic calorimeter, a hadron calorimeter and two stations of drift chambers for muon detection~\cite{Altegoer:1997gv}. The analysis found no evidence of $\nu_\tau$ appearance~\cite{Astier:2001yj} but the good quality of event reconstruction and the large data set collected made detailed neutrino interaction studies possible. Such measurements at NOMAD include  muon (anti)neutrino CCQE cross section~\cite{Lyubushkin:2008pe}, NC coherent pion~\cite{Kullenberg:2009pu} and CC coherent $\rho$ production~\cite{fortheNOMAD:2013gba}, strangeness and charm production yields~\cite{Astier:2001vi,Naumov:2004wa,Astier:2001ri}.

 In 2007 and 2008, another experiment called SciBooNE took data in 
 the Booster neutrino beamline, which was also used by MiniBooNE.
 This detector was located at a distance of about 100 m from the neutrino production
 target. The detector complex consists of a fully active scintillator
 tracking detector, called SciBar, and a muon tracker.  The SciBar
 detector was used in KEK for the K2K experiment and transported to
 FNAL to study low energy neutrino and anti-neutrino interactions with
 much higher statistics. This detector was capable to find low
 momentum protons and thus, it had power to select CCQE events
 exclusively~\cite{Aunion:2010zz}.  The measured CCQE total cross section 
 was found to be $\sim$10\% smaller than in MiniBooNE~\cite{Katori:2013nca}. 
 The charged
 current coherent pion production reaction was also studied and consistent
 results with the K2K findings were
 obtained~\cite{Hiraide:2008eu}. This time, the neutral current
 coherent $\pi^0$ production cross-section was also measured and found
 to be consistent with the predictions from PCAC based
 models~\cite{Kurimoto:2010rc}.  The inclusive CC cross-section was
 also measured in SciBooNE. In the energy region of 0.5-1.5 GeV, it turned out
 to be significantly higher than NEUT and NUANCE 
 Monte Carlo predictions~\cite{Nakajima:2010fp}.

 The ArgoNeuT experiment studies neutrino interactions using a
 liquid argon (LAr) time projection chamber
 (TPC)~\cite{Anderson:2012vc}. This LAr TPC has the capability of tracking
 low energy charged particles, which is difficult for the other detectors. 
 For this reason,  ArgoNeuT is expected to provide more precise information on various
 neutrino cross sections.  The detector is located just in front of the
 MINOS near detector, which is used as a muon spectrometer. Despite the small 
 accumulated statistics, because of the
 limited run period,  the inclusive
 $\nu_\mu$ charged current differential cross sections on
 argon has been measured~\cite{Anderson:2011ce}. Recently, ArgoNeuT succeeded
 in tracking a large number of protons after the neutrino interaction,
 and reported on the observed proton multiplicity. In particular,  events with a final charged lepton and
 two back-to-back protons are quite interesting.  These back-to-back
 protons might have been originated via neutrino interactions with  two
 bound nucleons or from the  absorption of a produced pion.  This kind of data sets will be
 useful in the future to improve the performance of the event generators used in neutrino experiments and 
as a new complement in the study of nuclear structure and the nucleon properties inside the nuclear medium.

 The MINER$\nu$A experiment placed the detector after ArgoNeuT
 finished taking data. The setup consists of a fully active
 scintillator tracking detector, an electromagnetic and a hadron
 calorimeters, together with various nuclear targets, including helium,
 carbon, water, iron and lead.  This experiment also uses the MINOS
 near detector as a muon spectrometer.  MINER$\nu$A can measure cross
 sections for several nuclei using the same neutrino beam, which will
 allow to study the nuclear dependence of the different neutrino
 interactions with a minimized impact of systematic uncertainties. 
 Charged current $\nu_\mu$ and $\bar\nu_\mu$
 quasi-elastic differential cross-sections in carbon have been
 recently published. The interaction rates are consistent with a value
 of nucleon axial mass $\rm{M_A}$ $\sim 1.0$ GeV~\cite{Fields:2013zhk,Fiorentini:2013ezn}, in agreement
 with previous results from NOMAD~\cite{Lyubushkin:2008pe}, and in
 contradiction to the other recent experiments. On the other hand,
 deviations are found  between the measured
 $q^2$ differential cross section  and the expectations from a model of
 independent nucleons in a relativistic Fermi gas  with a 
small $M_A$ value~\cite{Fiorentini:2013ezn}.

\begin{figure}[htbp]
\begin{center}
  \includegraphics[width=0.9\textwidth]{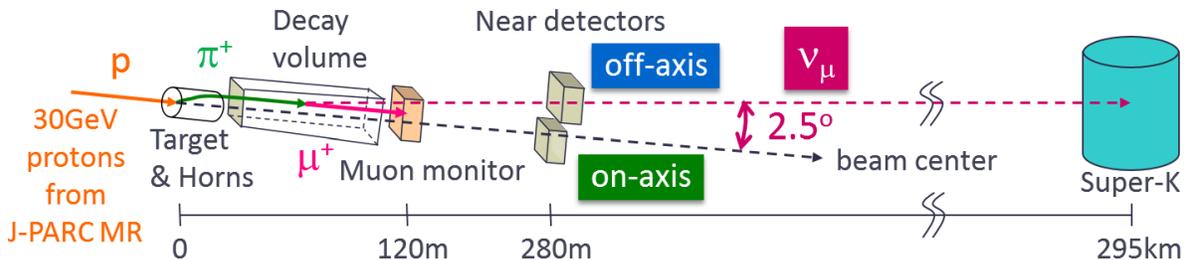}
  \caption{Schematic diagram of the T2K experiment.}
  \label{Exp:T2K}
\end{center}
\end{figure}

 After the K2K experiment was successfully finished in 2005, the T2K
 experiment started in 2009 as its natural upgrade. T2K aims to
 search for $\nu_\mu \to \nu_e$ oscillations and measure or set a
 limit for the $\theta_{13}$ mixing angle.  This experiment also tries
 to improve by one order of magnitude the precision on the
 determination of $\Delta m^2_{32}$ and $\theta_{23}$. These
 parameters are essential to disentangle the pattern of CP violation
 in the lepton sector.  The T2K experiment reported the first indication
 of a non-zero value for $\theta_{13}$ in 2011 at 2.5$\sigma$
 level~\cite{Abe:2011sj} and the significance reached 7.5$\sigma$
 in 2013~\cite{Abe:2013hdq}.  After the first result from T2K,
 $\theta_{13}$ was also measured in nuclear reactor experiments in the
 anti-electron neutrino disappearance channel, and the precisions are
 being continuously improved~\cite{An:2013zwz,Ahn:2012nd,Abe:2013sxa}.  
 The neutrino beam for the T2K experiment is produced by a 30 GeV proton line from the J-PARC
 proton synchrotron, located in Tokai, Ibaraki, Japan.  The direction
 of the resulting neutrino beam is 2.5 degree shifted from the direction of the
 far detector, SK located at 295~km from the target as shown in Fig.~\ref{Exp:T2K}. In this manner, the
 neutrino spectrum becomes narrower and peaks at lower energies. 
This configuration is
 called {\it off-axis}. With this method, the average energy of the
 neutrino beam is adjusted to peak around 700~MeV to match the oscillation maximum.  
There are several neutrino detectors, located at 280 meters from the
 target, where the neutrino interactions before oscillations are measured.  There are two
 sets of near detectors. The first one is aligned with the neutrino beam
 direction and consists of a steel-scintillator sampling tracker that
 allows to monitor the direction and stability of the neutrino
 beam.  The second set is aligned approximately in the direction to
 SK. This off-axis detector complex has full active scintillator
 tracking detectors, TPC's together with calorimeters. These detectors
 are located in a magnet that allows to identify the charge and 
 momentum of the produced particles.  Various
 studies on neutrino interactions are been carried out using both
 on-axis and off-axis detectors.  The inclusive CC and the CCQE
 cross-sections have already been measured. Regarding CCQE, the distributions
 seem to favor a large $M_A$ value. Total CC cross-sections have
 been measured in carbon and iron, which turn out to be similar,
 once divided by the number of active nucleons, with small differences
 at the level of 3\%.

 In 2013, the NO$\nu$A experiment has started  taking data. This
experiment utilizes a fully active liquid scintillator tracking
detector to accurately measure $\theta_{13}$, study  CP violation 
and the neutrino mass hierarchy. The neutrino beam 
is produced at FNAL, and is formally used also for MINOS, 
but upgraded for this new experiment. The far detector is located
in the Ash river, 810 km away from the target and 14 mrad off-axis
to adjust the maximum neutrino energy. The construction of the 
far detector is still in process and shall be completed in 2014.

 Until now, three mixing angles and two squared mass differences 
have been measured, but there is no information on whether CP is violated 
or not in the lepton sector. The mass hierarchy pattern is also unknown. 
In order to answer these questions, several high precision experiments with much
higher statistics and precisions have been proposed. 
In Japan, a gigantic water Cherenkov detector,
called Hyper-Kamiokande (HK) has been proposed. The fiducial volume of
HK  is more than 20 times larger than SK.
It is planned to be located in the Kamioka mine
and thus, it will be also possible to study atmospheric neutrinos with much
higher precision. Furthermore, the existing neutrino beam line in J-PARC
is expected to be upgraded.  
In the US, a project called LBNE has been launched. The proposed location 
of the far detector is the Sanford Laboratory, located at 1280 km from  FNAL. 
Two LAr TPC's, with a total fiducial
volume of 34 kton, have been proposed as far detectors. This type
of detector allows to measure low momentum particles; it is also expected to have a large electron--photon discriminating power. 
In spite of its relatively small mass compared to HK, the LBNE baseline is rather long, 
and thus this experiment would be quite sensitive
to both the CP $\delta$ phase and the neutrino mass hierarchy. 
In Europe, the LBNO/Laguna project is being discussed. The proposal includes
three types of the detectors: a 100 kton LAr TPC called 
GLACIER, a 500 kton water Cherenkov tank (MEMPHYS) and 
a 50 kton liquid scintillator detector (LENA).
Several locations are under consideration including several detector
configurations exploiting different advantages. Further study is underway.

This next generation of experiments would reduce the uncertainties of
the neutrino interactions down to the few percent level from around
the 10\% affecting the current experiments. More precise theoretical
models will be needed to understand and describe the future wealth of
neutrino-nucleus data.

\section{Quasielastic and quasielastic-like scattering}
\la{sec:CCQE}

\subsection{Quasielastic scattering on the nucleon}
\la{subsec:CCQE-Nucleon}

Let us first consider the processes 
\bea
\nu_l(k) \, n(p) &\raw&  l^-(k') \, p(p') \,, \qquad (\nu\mathrm{CCQE}) \la{eq:nuCCQE} \\
\nu_l(k) \, N(p) &\raw& \nu_l(k') \, N(p') \,, \quad N=n,p  \quad (\nu\mathrm{NCE}) \la{eq:nuNCE}
\eea
standing for charge-current (CC)  quasielastic (CCQE) and neutral current (NC) elastic (NCE) scattering on nucleons induced by neutrinos. The corresponding reactions with antineutrinos are
\bea 
\bar \nu_l(k) \, p(p) &\raw&  l^+(k') \, n(p') \,, \qquad (\bar\nu\mathrm{CCQE}) \la{eq:anuCCQE} \\
\bar\nu_l(k) \, N(p) &\raw& \bar\nu_l(k') \, N(p') \,,\quad N=n,p  \quad (\bar\nu\mathrm{NCE}) \la{eq:anuNCE}
\eea
At energies low enough that the 4-momentum squared  transferred to the nucleon is much smaller than the intermediate vector boson mass squared [$q^2 = (k - k')^2 \ll M^2_{W,Z}$], their cross sections can be cast as
\be
\la{eq:dsdq2}
\frac{d \sigma}{d q^2} = \frac{1}{32 \pi} \frac{1}{M^2 E_\nu^2} G^2 c_{\chic \mathrm{EW}}^2 L_{\alpha \beta} H^{\beta \alpha} \,, \quad \mathrm{EW = CC,\,NC}\,
\ee
where $E_\nu \equiv k^0$ is the (anti)neutrino energy in the Laboratory frame and $M$ the nucleon mass (isospin symmetry is assumed); $G$ is the Fermi constant while $c_{\chic \mathrm{CC}} = \cos{\theta_C}$, in terms of the Cabibbo angle, and $c_{\chic \mathrm{NC}} = 1/4$. The leptonic tensor is 
\be
\la{eq:leptensor}
L_{\alpha \beta} = k_\alpha k'_\beta  + k'_\alpha k_\beta - g_{\alpha \beta} k\cdot k' \pm i \epsilon_{\alpha \beta \sigma \delta} k'^\sigma k^\delta \,
\ee 
where the $+$($-$) sign is valid for $\nu$($\bar\nu$) interactions ($\epsilon_{0123} =+1$). The hadronic part is contained in the hadronic tensor $H$ which, in the case of CCQE and NCE processes, takes the form
\be
\la{eq:hadtensor}
H^{\alpha \beta} =\mathrm{Tr}\left[ \left( \slashed{p} + M \right) \gamma^0 \left( \Gamma^\alpha \right)^\dagger \gamma^0 \left( \slashed{p}' + M \right) \Gamma^\beta  \right] \,.
\ee
The amputated amplitudes $\Gamma^\alpha$ enter the weak charged and neutral currents. The latter ones can be written in the most general way consistent with the symmetries of the Standard Model, in terms of form factors that contain the information about nucleon properties
\be
\la{eq:QEcurrent}
J^\alpha = \bar u(p') \Gamma^\alpha u(p) = V^\alpha - A^\alpha \,, 
\ee
with the vector and axial currents given by
\be
V^\alpha = \bar u(p') \left[ \gamma^\alpha F_1(q^2)+\frac{i}{2M}\sigma^{\alpha\beta}q_\beta F_2(q^2) \right] u(p)
\ee
and
\be
 A^\alpha =  \bar u(p') \left[\gamma^\alpha \gamma_5 F_A(q^2) + \gamma_5 \frac{q^\alpha}{M} F_P(q^2)  \right]  u(p) \,.
\ee
Vector-current conservation $q_\alpha V^\alpha =0$ and isospin symmetry imply that vector form factors $F_{1,2}$ are given in terms of the electromagnetic form factors of protons and neutrons. On the other hand, the PCAC $q_\alpha A^\alpha = i (m_u + m_d) \bar q_u \gamma_5  q_d \raw 0$ in the chiral limit of QCD  ($m_q \raw 0$), together with the approximation of the pion-pole dominance of the pseudoscalar form factor $F_P$, allows to relate $F_P$ to the axial form factor $F_A$. Explicit expressions for these relations among form factors can be found, for instance, in Ref.~\cite{Nieves:2004wx} for the CC case, and in Ref.~\cite{Nieves:2005rq} for the NC one. 

With these ingredients, a compact expression of the cross section in Eq.~(\ref{eq:dsdq2}) as a function of the form factors was obtained~\cite{LlewellynSmith:1971zm} and widely utilized~\cite{Formaggio:2013kya}. Alternatively, it is instructive to write this cross section as an expansion in small variables\footnote{Close to threshold ($E_\nu \sim m_l$), and for CCQE with $\tau$-neutrinos (due to the large $m_\tau$ value) the counting is different.} $q^2,m_l^2 \ll M^2,E^2_\nu$. One has that 
\be
\la{eq:dsdq2expand}
\frac{d \sigma}{d q^2} = \frac{1}{2 \pi} G^2 c_{\chic \mathrm{EW}}^2 \left[ R -\frac{m_l^2}{4 E_\nu^2} S + \frac{q^2}{4 E_\nu^2} T \right]  
+ \mathcal{O}(q^4,m_l^4,m_l^2 q^2) \,.  
\ee
For CC reactions, Eqs.~(\ref{eq:nuCCQE}) and (\ref{eq:anuCCQE})
\bea
\la{la:RST_CC}
R_{\chic \mathrm{CC}} &=& 1 + g_A^2 \,, \\ [0.2cm]
S_{\chic \mathrm{CC}} &=& \frac{2 E_\nu + M}{M} + g_A^2 \frac{2 E_\nu - M}{M} \,, \\ [0.2cm]
T_{\chic \mathrm{CC}} &=& 1 - g_A^2 + 2 \frac{E_\nu}{M} \left( 1 \mp  g_A \right)^2 \mp 4 \frac{E_\nu}{M} g_A \kappa^{\chic \mathrm{V}} 
- \left(\frac{E_\nu}{M} \kappa^{\chic \mathrm{V}} \right)^2 \nonumber \\ 
&+& 4 E_\nu^2 \left[ \frac{1}{3} \left( \langle r_p^2 \rangle 
- \langle r_n^2 \rangle + g_A^2 \langle r_A^2 \rangle \right) 
- \frac{1}{2 M^2} \kappa^{\chic \mathrm{V}} \right] \,,
\eea 
where $\kappa^{\chic \mathrm{V}} = \mu_p - \mu_n - 1$. The upper (lower) sign stands for $\nu$($\bar \nu$)CCQE. For NC reactions, Eqs.~(\ref{eq:nuNCE}) and (\ref{eq:anuNCE}), Eq.~(\ref{eq:dsdq2expand}) remains valid when $m_l \raw 0$ and the $R$, $T$ functions are given by 
\bea
\la{la:RST_NCp}
R^{(p)}_{\chic \mathrm{NC}} &=& \alpha_{\chic \mathrm{V}}^2  + (g_A - \Delta s)^2 \,, \\ [0.2cm]
T^{(p)}_{\chic \mathrm{NC}} &=& \alpha_{\chic \mathrm{V}}^2  - (g_A - \Delta s)^2  
+ 2 \frac{E_\nu}{M} \left[ \alpha_{\chic \mathrm{V}}  \mp  (g_A - \Delta s) \right]^2 
\mp 4 \frac{E_\nu}{M} (g_A- \Delta s) \kappa_{\chic \mathrm{NC}}^{(p)} 
-  \left(\frac{E_\nu}{M} \kappa_{\chic \mathrm{NC}}^{(p)} \right)^2 \nonumber \\
&+& 4 E_\nu^2 \left\{ \alpha_{\chic \mathrm{V}} \left[ \frac{1}{3} \left( \alpha_{\chic \mathrm{V}}  \langle r_p^2 \rangle 
- \langle r_n^2 \rangle  -  \langle r_s^2 \rangle \right) - \frac{1}{2 M^2} \kappa_{\chic \mathrm{NC}}^{(p)} \right] + 
\frac{1}{3} \left(g_A - \Delta s\right) \left(g_A \langle r_A^2 \rangle - \Delta s \langle r_{As}^2 \rangle  \right) \right\}  \,,
\eea
on proton targets, with $\kappa_{\chic \mathrm{NC}}^{(p)} = \alpha_{\chic \mathrm{V}} (\mu_p  -1) - \mu_n - \mu_s$ and the (very small) quantity 
$\alpha_{\chic \mathrm{V}} = 1 - 4 \sin^2{\theta_W}$, where $\theta_W$ is the weak angle. On neutrons
\bea
\la{la:RST_NCn}
R^{(n)}_{\chic \mathrm{NC}} &=& 1  + (g_A + \Delta s)^2 \,, \\ [0.2cm]
T^{(n)}_{\chic \mathrm{NC}} &=& 1 - (g_A + \Delta s)^2  
+ 2 \frac{E_\nu}{M} \left[ 1  \mp  (g_A + \Delta s) \right]^2 
\pm 4 \frac{E_\nu}{M} (g_A + \Delta s) \kappa_{\chic \mathrm{NC}}^{(n)} 
-  \left(\frac{E_\nu}{M} \kappa_{\chic \mathrm{NC}}^{(n)} \right)^2 \nonumber \\
&+& 4 E_\nu^2 \left\{ -\frac{1}{3} \left( \alpha_{\chic \mathrm{V}}  \langle r_n^2 \rangle 
- \langle r_p^2 \rangle  -  \langle r_s^2 \rangle \right) + \frac{1}{2 M^2} \kappa_{\chic \mathrm{NC}}^{(n)}  + 
\frac{1}{3} \left(g_A + \Delta s\right) \left(g_A \langle r_A^2 \rangle + \Delta s \langle r_{As}^2 \rangle  \right) \right\}  \,,
\eea
with $\kappa_{\chic \mathrm{NC}}^{(n)} = 1 - \mu_p + \alpha_{\chic \mathrm{V}} \mu_n - \mu_s$.

In the $q^2,m_l^2 \ll M^2,E^2_\nu$ limit, which is valid for most of the integrated cross section in the few-GeV region, the CCQE cross section is determined by a few experimentally well measured electromagnetic properties of the nucleon: charge, magnetic moments $\mu_{p,n}$  and mean squared charge radii 
\be
\la{eq:radii}
\langle r_p^2 \rangle = \left.\frac{6}{G^{(p)}_{\mathrm{E}}(0)} \frac{dG^{(p)}_{\mathrm{E}}(q^2)}{dq^2}\right|_{q^2=0} \,, \qquad
\langle r_n^2 \rangle = 6 \left.\frac{dG^{(n)}_{\mathrm{E}}(q^2)}{dq^2}\right|_{q^2=0}  \,,    
\ee
where $G^{(p,n)}_{\mathrm{E}}$ are the electric form factors, and two axial quantities: coupling $g_A$, also well known from neutron $\beta$ decay, and radius
\be
\la{eq:axradius}
\langle r_A^2 \rangle = \left.\frac{6}{F_A(0)} \frac{dF_A(q^2)}{dq^2}\right|_{q^2=0} \,. 
\ee
It is remarkable that $\langle r_A^2 \rangle$ can be extracted from single pion electroproduction data independently of neutrino experiments which could be distrusted due to low statistics, inaccuracies in the neutrino flux determinations or the influence of nuclear effects. It has been shown~\cite{Bernard:1992ys,Bernard:2001rs} that, up to $\mathcal{O}(p^3)$ in a chiral expansion in small momenta and quark masses (Chiral Perturbation Theory) 
\be
\la{eq:E0+}
6 \left. \frac{d E_{0+}^{(-)}}{d q^2} \right|_{q^2=0} = \langle r_A^2 \rangle + \frac{3}{M} \left( \kappa^{\chic \mathrm{V}} + \frac{1}{2} \right) + \frac{3}{64 f_{\pi}^2} \left( 1 - \frac{12}{\pi^2} \right) \,,
\ee
where $E_{0+}^{(-)}$ is an s-wave electric dipole amplitude in a specific isospin combination (see Ref.~\cite{Bernard:2001rs} and references therein for more details). The derivative is taken over the virtual photon 4-momentum squared, and $f_\pi$ is the pion decay constant. Using an effective Lagrangian model to extrapolate pion electroproduction data to $q^2 = 0$~\cite{Liesenfeld:1999mv} and taking into account the hadronic corrections of Eq.~(\ref{eq:E0+})~\cite{Bernard:1992ys} one finds that 
\be
\la{eq:axradius2}
\langle r_A^2 \rangle = 0.455 \pm 0.012\, \mathrm{fm}^2 \,.
\ee

At higher $q^2$, the relation between pion electroproduction amplitudes and the axial form factor becomes more uncertain and model dependent. Therefore, for information on the $q^2$ dependence of $F_A$ one has to rely chiefly on (anti)neutrino experiments off hydrogen and deuterium targets. Although the contribution of the term proportional to $F_A$  to the parity-violating asymmetry in electron-proton elastic scattering with polarized beams is typically orders of magnitude smaller than the dominant (magnetic) one, a detailed study at backward angles might also help constraining this form factor~\cite{GonzalezJimenez:2011fq}. 

The axial form factor is usually parametrized with a dipole ansatz
\be
\la{eq:FA}
F_A (q^2) = g_A \left( 1 - \frac{q^2}{M_A^2} \right)^{-2} \,
\ee
which corresponds to an exponential shape for the axial charge-density distribution. For such a one parameter function, the so-called axial mass $M_A$ is directly related to the axial radius
\be
\la{eq:MA}
\langle r_A^2 \rangle = \frac{12}{M_A^2}\,.
\ee
The value of $M_A$ extracted from early CCQE measurements on deuterium and, to lesser extent, hydrogen targets is 
$M_A = 1.016 \pm 0.026$~GeV~\cite{Bodek:2007ym}. It is in excellent agreement with the pion electroproduction result, $M_A = 1.014 \pm 0.016$~GeV, obtained from the axial radius, Eq.~(\ref{eq:axradius2}), using the relation of Eq.~(\ref{eq:MA}). In spite of the fact that deviations from the dipole form have not been observed so far, it is worth stressing that the dipole parametrization is not well justified from a theoretical point of view. In the case of the electromagnetic form factor, the dipole behavior arises from cancellations between monopole terms that appear naturally in the vector meson dominance picture~\cite{Crawford:2010gv}. The situation is more uncertain in the axial sector but a similar scenario might be in place from the interplay of two or more axial mesons~\cite{Masjuan:2012sk}. The lack of knowledge about the $q^2$ dependence of $F_A$ may result in large uncertainties in the CCQE cross section, specially at large energies as shown in Ref.~\cite{Amaro:2013yna}, although the constrains in the axial radius from pion electroproduction would make the bands in Fig.~5 of that paper narrower. 

In addition to the nucleon electromagnetic and axial properties that define the CCQE hadronic tensor at $q^2,m_l^2 \ll M^2,E^2_\nu$, the NCE one depends on the strangeness content of the nucleon via the strange mean squared radius $\langle r_s^2 \rangle$, magnetic moment $\mu_s$, the strange axial coupling $\Delta s$, which is the strange quark contribution to the nucleon spin, and the corresponding axial radius $\langle r_{As}^2 \rangle$. The impact of $\langle r_s^2 \rangle$, $\mu_s$, $\langle r_{As}^2 \rangle$, which are not only small (see Ref.~\cite{Pate:2013wra} for a recent global fit) but also appear in the subleading $T^{(p,n)}_{\chic \mathrm{NC}}$ functions, is insignificant. This is not the case of  $\Delta s$ which, if different from zero, could change the NCE cross section appreciably. This becomes evident by looking at the ratio
\be
\la{eq:NCratio}
\left. \frac{d \sigma^{(p)}_{\chic \mathrm{NC}}/d q^2}{d \sigma^{(n)}_{\chic \mathrm{NC}}/d q^2} \right|_{q^2=0} = \frac{\alpha_{\chic \mathrm{V}}^2  + (g_A - \Delta s)^2}{1  + (g_A + \Delta s)^2} \approx \frac{(g_A - \Delta s)^2}{1  + (g_A + \Delta s)^2}  \approx \left\{ 
\begin{array}{l} 0.62 \qquad \mathrm{if}\,\, \Delta s =0 \\ 1.27 \qquad \mathrm{if}\,\, \Delta s =-0.3   \end{array} \right. \,.
\ee
The MiniBooNE experiment performed a detailed study of NC nucleon knock-out on mineral oil (CH$_2$) but their measurement turned out to be rather insensitive to $\Delta s$ because of difficulties to distinguish between protons and neutrons~\cite{AguilarArevalo:2010cx,GonzalezJimenez:2012bz}. The recently proposed MiniBooNE+ experiment with improved sensitivity to neutrons might allow a better determination of this important nucleon property~\cite{Dharmapalan:2013zcy}. The same is true about MicroBooNE~\cite{Chen:2007ae}, where a reliable identification of the low energy protons knocked out of argon should be possible. As a result, the error in the determination of $\Delta s$ would be drastically reduced~\cite{Pate:2013wra}. It should be recalled that for both  MiniBooNE+ and MicroBooNE, running with nuclear targets, the presence of multinucleon contributions (discussed in Sec.~\ref{sec:2p2h}) and other inelastic mechanisms, together with final state interactions~\cite{Nieves:2005rq,Leitner:2006sp} will certainly affect the extraction of $\Delta s$ and should be carefully studied. In fact, we think that dedicated (anti)neutrino-nucleon experiments are needed in order to understand the axial structure of the nucleon in depth.

\subsection{Quasielastic scattering on nuclei}
\la{subsec:CCQE-Nucleus}

\begin{figure}[htb!]
\begin{center}
\includegraphics[width=\textwidth]{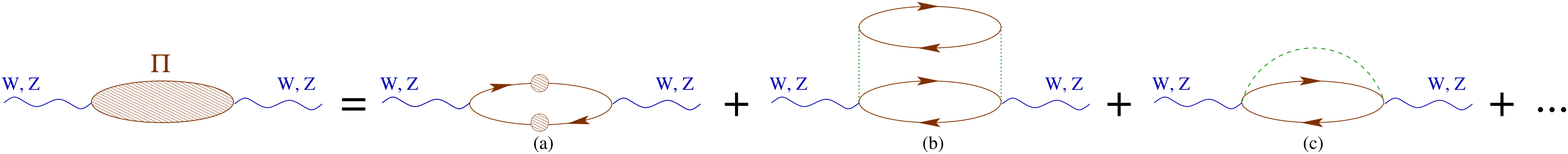}
\caption{\label{fig:polar}
  Diagrammatic representation of many-body contributions to the polarization propagator. Solid (dashed) lines correspond to
free nucleon (pion) propagators; dotted lines stand for effective nucleon-nucleon interactions. The solid lines with a blob represent full (dressed) nucleon propagators. For nucleons, the lines pointing to the right (left) denote particle (hole) states.}
\end{center}
\end{figure}

For (anti)neutrinos interacting with nuclear targets, 
\bea
\nu_l,\bar\nu_l(k) \, A_Z &\raw& l^{\mp}(k') \, X \\
\nu_l,\bar\nu_l(k) \,  A_Z &\raw& \nu_l,\bar\nu_l(k')  \, X
\eea
the inclusive cross section per unit volume, which is the proper quantity for an extended system, is given by
\be
\la{eq:inclcs}
\frac{d}{d^3r} \left(\frac{d \sigma}{d \Omega(k') dk'^0} \right) = \frac{G^2 c_{\chic \mathrm{EW}}^2}{4 \pi^2} \frac{| \vec{k}' |}{| \vec{k} |} \, L_{\alpha \beta} W^{\alpha \beta} \,.
\ee
This formula stands for both CC and NC processes, with  $c_{\chic \mathrm{CC}} = \cos{\theta_C}$ and $c_{\chic \mathrm{NC}} = 1/4$. The lepton tensor is defined in Eq.~(\ref{eq:leptensor}) as in the nucleon case. By construction, the hadronic tensor can be decomposed as 
\be
\la{eq:had_tensor}
W^{\alpha \beta} = W_s^{\alpha \beta} + i W_a^{\alpha \beta}
\ee 
with $W_s$ ($W_a$) being a real symmetric (antisymmetric) tensor. Furthermore, it can be expressed in terms of the polarization propagator (or tensor)
\be
\la{eq:polar}
W^{\alpha \beta}_{(s,a)} = -\frac{1}{\pi} \, \mathrm{Im}\, \Pi^{\alpha \beta}_{(s,a)} \,.
\ee
This is a classic result, known for the response of many-body systems to external probes~\cite{fetterwalecka}, that also holds for neutrino interactions. Some examples of different contributions to the polarization propagator are diagrammatically represented in Fig~\ref{fig:polar}. Diagram (c) contributes mostly to pion production\footnote{Mostly, because real pions produced inside the nuclei can be absorbed by two or more nucleons.} discussed in Sec.~\ref{sec:pion}. Diagram (b) is an example of a meson exchange current (MEC) contribution to the interaction mechanisms involving two nucleons from the nucleus, also known as two-particle-two-hole (2p2h) terms. They play a very important role in neutrino scattering in the few-GeV region and will be covered in Sec.~\ref{sec:2p2h}. 

Let us now consider diagram (a) of Fig~\ref{fig:polar}. This piece of the polarization tensor contributes mainly to the quasielastic (QE) peak, where the interaction takes place on a single nucleon which is knocked out; these are one-particle-one-hole (1p1h) terms. Nevertheless, through the correlations in the initial or final state, this diagram also contains 2p2h, other multinucleon and even particle emission 
contributions. It can be shown that for the polarization propagator of diagram (a) in Fig~\ref{fig:polar}  
\be
\la{eq:polar(a)}
\mathrm{Im}\, \Pi^{\alpha \beta}_{(s,a)} = - 2 \pi^2 \int \frac{d^4 p}{(2 \pi)^4} H^{\beta \alpha}_{(s,a)} \mathcal{A}_p(p+q) \, \mathcal{A}_h(p)
\ee
where $H^{\alpha \beta}_{(s,a)}$ are the symmetric and antisymmetric components of the QE hadronic tensor for nucleons introduced in Eq.~(\ref{eq:hadtensor}). In this way, one disregards off-shell effects on the nucleon current and adopts the same form factors as for the free nucleon.  
\be
\la{eq:spectral}
\mathcal{A}_{p,h}(p) = \mp \frac{1}{\pi} \frac{\mathrm{Im}  \Sigma(p)}{[p^2-M^2-\mathrm{Re} \Sigma(p)]^2 + [\mathrm{Im}  \Sigma(p)]^2}  
\ee
are particle and hole spectral functions related to the scalar (averaged over spins) nucleon selfenergy $\Sigma$ present in the full (dressed) in-medium nucleon propagator as a result of nucleon-nucleon ($NN$) interactions. Here $p^0 \geq \mu$ and $p^0 \leq \mu$, with $\mu$ being the chemical potential, for $\mathcal{A}_{p}$ and $\mathcal{A}_{h}$, respectively. In presence of the nuclear medium, nucleons have a modified dispersion relation and, in addition, become broad states.

Practically all approximations employed to calculate neutrino, but also electron, QE scattering with nuclei can be obtained from Eq.~(\ref{eq:polar(a)}). The simplest description, present in most event generators used in the analysis of neutrino experiments, is the relativistic global Fermi gas (RgFG)~\cite{Smith:1972xh}. In this model, the nuclear ground state is a Fermi gas of non-interacting nucleons characterized by a global Fermi momentum $p_F$ and a constant binding energy $E_B$. Then, the hole spectral function is 
\be
\la{eq:sfhole}
\mathcal{A}_{h}(p) = \frac{1}{2 E(\vec{p})} \delta\left(p^0-E(\vec{p})\right) \,\theta\left(p_F - |\vec{p}|\right)\,,
\ee
where $E(\vec{p}) = \sqrt{M^2 + \vec{p}^{\,2}} - E_B$. The step function $\theta\left(p_F - |\vec{p}|\right)$,
with $p_F$ a constant Fermi momentum, accounts for the Fermi motion of the nucleons in the nucleus. For the particle
\be
\la{eq:sfparticle}
\mathcal{A}_{p}(p) = \frac{1}{2 E(\vec{p})} \delta\left(p^0-E(\vec{p})\right) \,\left[1-\theta\left(p_F - |\vec{p}|\right)\right] \,,
\ee
with $E(\vec{p})$ the on-shell energy. The factor $1-\theta\left(p_F - |\vec{p}|\right)$ takes Pauli blocking into account. Such a simple picture with only  two parameters $(p_F ,E_B)$ explains qualitatively 
inclusive QE electron scattering data but fails in the details. A better description requires a more realistic treatment of nuclear dynamics. 

An improvement over the RgFG is the so called relativistic local
Fermi Gas (RlFG) where the Fermi momentum is fixed according to the 
local density of protons and neutrons $\rho_{p,n}(r)$
\be
\la{eq:localFermi}
p_F^{p,n}(r) = \left[ 3 \pi^2 \rho_{p,n}(r) \right]^{1/3} \,.
\ee
Equations (\ref{eq:sfhole}) and (\ref{eq:sfparticle}) also hold after replacing the global Fermi momentum by the local one. The binding energy is often neglected but a minimal excitation energy required for the transition to the ground state of the final nucleus has been taken into account in the CCQE model of Ref.~\cite{Nieves:2004wx}.

Nuclear spectral functions like the one of Ref.~\cite{Benhar:1994hw}, obtained using empirical single particle wave functions and realistic calculations in nuclear matter [adapted for finite nuclei employing the local-density approximation (LDA)], from the convolution model~\cite{CiofidegliAtti:1995qe} and from other (semi)phenomenological models~\cite{FernandezdeCordoba:1991wf,Kulagin:2004ie} have been applied to electron and neutrino scattering in the QE region~\cite{CiofidegliAtti:1990vn,Gil:1997bm,Nieves:2004wx,Benhar:2005dj,Benhar:2006nr,Ankowski:2007uy,Butkevich:2012zr}. Sometimes, realistic spectral functions for holes are combined with the use of plane waves for nucleons in the final state, eventually including Pauli blocking with a global Fermi momentum like in Eq.~(\ref{eq:sfparticle}). Examples of such an approach, known as plane wave impulse approximation (PWIA), for CCQE scattering can be found in Refs.~\cite{Ankowski:2005wi,Benhar:2005dj,Benhar:2006nr,Butkevich:2012zr}. 

\begin{figure}[htb!]
\begin{center}
\includegraphics[width=0.5\textwidth]{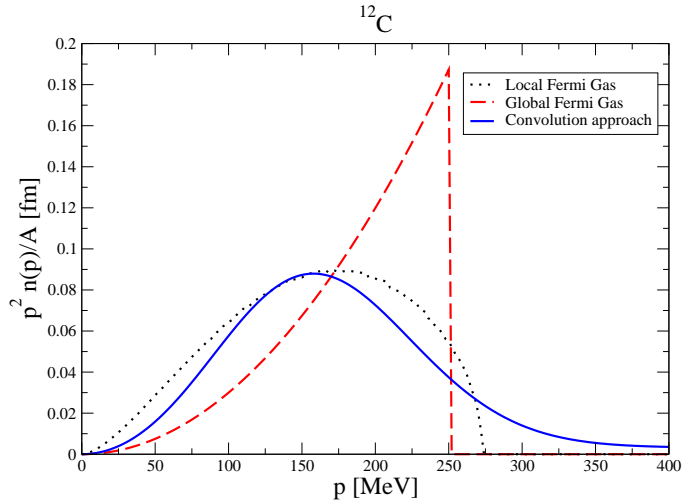}
\caption{\label{fig:n(p)}
Comparison of the nucleon momentum distributions on $^{12}$C for RlFG, RgFG and the convolution model of Ref.~\cite{CiofidegliAtti:1995qe}.}
\end{center}
\end{figure}

To put in perspective the RgFG, RlFG and realistic spectral function descriptions of the nuclear ground state, we consider the nucleon momentum distribution in nuclei, related to the hole spectral function by 
\be
\la{eq:momdistRgFG}
n^{\mathrm{RgFG}}(|\vec{p}|) = \frac{4}{(2 \pi)^3} \mathcal{V} \int dp_0 (2 p_0) \mathcal{A}_{h}(p)
\ee     
where $\mathcal{V}$ is the nuclear volume, and 
\be
\la{eq:momdistLDA}
n^{\mathrm{LDA}}(|\vec{p}|) = 4 \int \frac{d^3r}{(2 \pi)^3} dp_0 (2 p_0) \mathcal{A}_{h}(p)
\ee 
within the LDA approximation. The momentum distribution is normalized as 
\be
\la{eq:norm}
(4 \pi) \int d|\vec{p}| \vec{p}^{\,2} n(|\vec{p}|) = A \,,
\ee
where $A$ is the nuclear mass number. For simplicity, the same momentum distribution for protons and neutrons, as in symmetric nuclei, has been assumed. In Fig.~\ref{fig:n(p)}, $\vec{p}^{\,2} n(|\vec{p}|)$ on $^{12}$C is plotted for RgFG (neglecting the binding energy), the RlFG (with density profiles taken from Ref.~\cite{Nieves:1993ev}) and the convolution approach of Ref.~\cite{CiofidegliAtti:1995qe} that provides convenient parametrizations of these distributions. 
The plot shows the tail at high momenta of the realistic distribution due to short-range correlations, which is absent in the Fermi gas distributions. These correlations are investigated in two-nucleon knockout electron scattering experiments at specific kinematics where they are bound to be relevant~\cite{Shneor:2007tu}. In spite of its simplicity, the RlFG model introduces space-momentum correlations that translate into a considerably more realistic description of the momentum distribution than the RgFG.   

Besides the short range correlations, the description of the nuclear ground state can be improved with more realistic treatments of the single particle content beyond the non-interacting Fermi-type picture. In the GiBUU model~\cite{Buss:2011mx}, the hole spectral function is modified with respect to the one of the RlFG by the presence of a real position- and momentum-dependent mean-field potential generated by the spectator particles. The hole states remain narrow but the dispersion relation is modified: $p^2 = [M^*(\vec{r},\vec{p})]^2$, where $M^*(\vec{r},\vec{p})$ is the effective mass of the bound nucleon. The initial nucleons have also been treated as shell-model bound states with wave functions obtained as solutions of the Dirac equation in a $\sigma-\omega$ potential~\cite{Alberico:1997vh,Maieron:2003df,Butkevich:2007gm}, or from a phenomenological energy and target dependent optical potential (OP)~\cite{Meucci:2004ip}. This is known as the relativistic mean field (RMF) approach. 

Although the details of the particle spectral functions, and the final state interactions (FSI) in general, do not affect the inclusive integrated cross section, they are important for the differential ones, particularly in the region of low-energy transfer, and are crucial to achieve a realistic description of the reaction final state. It is precisely through the detection of (some of the) final state particles that neutrinos can be detected and its energy, a priori not known in broad beams, identified. The role of final-state correlations is particularly important for NC processes because the outgoing neutrino cannot be detected. In an inclusive CC process of the type of diagram (a) in Fig.~\ref{fig:polar}, the chief effects of the medium in the particle spectral function are an energy shift of the cross section caused by the mean field and a redistribution of the strength caused by $NN$ interactions~\cite{Benhar:2005dj}. Particle spectral functions have been obtained in a variety of approaches: generalizing Glauber theory of multiple scattering~\cite{Benhar:1991af}, also using phenomenological potentials for the real part of the nucleon selfenergy, and the low density limit  to relate $\mathrm{Im}\Sigma$ to the $NN$ cross section~\cite{Ankowski:2007uy,Leitner:2008ue}. In Ref.~\cite{FernandezdeCordoba:1991wf},  $\mathrm{Im}\Sigma$ is also expressed in terms of the free $NN$ cross section in a LDA framework and taking polarization effects into account; $\mathrm{Re}\Sigma$ is obtained from the imaginary part using dispersion relations. All of them have been employed to study neutrino-induced QE reactions~\cite{Nieves:2004wx,Benhar:2005dj,Ankowski:2007uy,Leitner:2008ue}.   

The wave functions of the outgoing nucleons can be distorted with complex OP in an approach known as distorted-wave impulse approximation (DWIA)~\cite{Alberico:1997vh,Maieron:2003df,Meucci:2004ip,Martinez:2005xe,Butkevich:2007gm,Meucci:2011ce,Butkevich:2012zr} or with the Glauber multiple scattering approximation for knocked out nucleons with more than 1~GeV kinetic energy~\cite{Martinez:2005xe}.  DWIA models are successful in describing a large amount of exclusive proton knockout $(e,e'p)$ data and are valid when the residual nucleus is left in a given state~\cite{Boffi:1993gs,Udias:1993xy,Giusti:2011it} but are not appropriate for inclusive scattering. Indeed, the imaginary part of the OP produces an absorption and a reduction of the cross section  which accounts for the flux lost towards other channels. This is not correct for an inclusive reaction where all elastic and inelastic channels contribute and the total flux must be conserved. DWIA calculations using the same real potential (RMF) used to describe the initial state, or the real part of a phenomenological OP~\cite{Maieron:2003df} are more suitable for inclusive processes although one should recall that optical potentials have to be complex owing to the presence of inelastic channels~\cite{Giusti:2009ym}. 
An alternative is the relativistic Green function (RGF) approach~\cite{Meucci:2003cv,Meucci:2011vd,Meucci:2012yq,Meucci:2014pka} which also starts from a phenomenological complex OP 
that describes proton-nucleus scattering data, but recovers the flux lost into nonelastic channels in the inclusive case. This in principle renders the model more appropriate for the study of CCQE scattering when nucleons are not detected in the final state, as in the MiniBooNE measurement (see the related discussion in Sec~\ref{sec:2p2h}).  NCQE\footnote{We prefer to call the $\nu,\bar\nu \, A \raw \nu,\bar\nu \, N \, X$ reactions NCQE (or NCQE-like, depending on the primary mechanism) rather than NCE, which we keep for the corresponding reactions on nucleons, Eqs.~(\ref{eq:nuNCE}) and (\ref{eq:anuNCE}).} processes, instead, are identified by the detection of the knocked-out nucleon with no lepton in the final state. The RGF model may include channels which are not present in the measurement but, on the other hand, it contains contributions that are not present in DWIA calculations. DWIA, RMF and RGF approaches have been recently compared for different NCQE kinematics~\cite{Gonzalez-Jimenez:2013xpa}. It is argued that RGF may provide an upper limit to the NCQE cross section while DWIA gives the lower one but one should keep in mind that other mechanisms like MEC not accounted in these approaches can also contribute.   

A detailed description of the final state interactions can be achieved using semiclassical Monte-Carlo methods. Moreover, at $E_\nu \gtrsim 500$~MeV, this is the only viable way to simulate the final hadronic state. These techniques allow to take into account account rescattering causing energy losses, charge exchange and multiple nucleon emissions~\cite{Nieves:2005rq,Leitner:2006ww,Leitner:2006sp}. Furthermore, there are QE-like processes in which a pion produced in the primary interaction and then absorbed, leads only to nucleons in the final state. These mechanisms, which can be very hard, if at all, to disentangle from those originated in an elastic or QE interaction on the nucleon, can be naturally incorporated to the framework. Other approaches as those discussed above have to rely on the subtraction of QE-like events performed by the different experiments, using event generators that do not necessarily incorporate state-of-the-art physics input.  

\subsubsection{Electron scattering and the superscaling approach}
\la{subsubsec:susa}

Electron scattering  is a major source of information for neutrino interactions studies, providing  not only the vector form factors of nucleons but also a testing ground for nuclear scattering models, which can be confronted with a large set of good quality data.  As mentioned above, the RgFG model describes the main features of the inclusive $(e,e')$ cross section at the QE peak. However, this model fails to reproduce simultaneously the longitudinal and transverse responses. In particular, the longitudinal response is overestimated (see for instance Ref.~\cite{Meziani:1984is}). With the RGF approach, the longitudinal response is  well reproduced, while the transverse is underestimated~\cite{Meucci:2003uy}. This disagreement can be attributed to the lack of MEC and resonance excitation in the model, which are much more important in the transverse than in the longitudinal channel. In microscopic calculations with realistic spectral functions, the longitudinal response, quenched with respect to the Fermi gas estimate, agrees with data, as well as the transverse one at energies low enough for inelastic processes to be negligible (see Fig.~32 of Ref.~\cite{Benhar:2006wy}, showing the comparison of the theoretical results of Ref.~\cite{Fabrocini:1989nw} with data). The nonrelativistic model of Ref.~\cite{Gil:1997bm} incorporates a semiphenomenological particle spectral function, MEC and $\Delta(1232)$ degrees of freedom in a many-body framework for nuclear matter applied to final nuclei using the LDA. A good description of both responses is obtained as can be appreciated in Figs.~42 and 43 of Ref.~\cite{Gil:1997bm}. Further details about different theoretical approaches to electron scattering and their comparison to data can be found in the review articles of Refs.~\cite{Boffi:1993gs,Benhar:2006wy}.

\begin{figure}[h!]
\begin{center}
\includegraphics[width=0.5\textwidth]{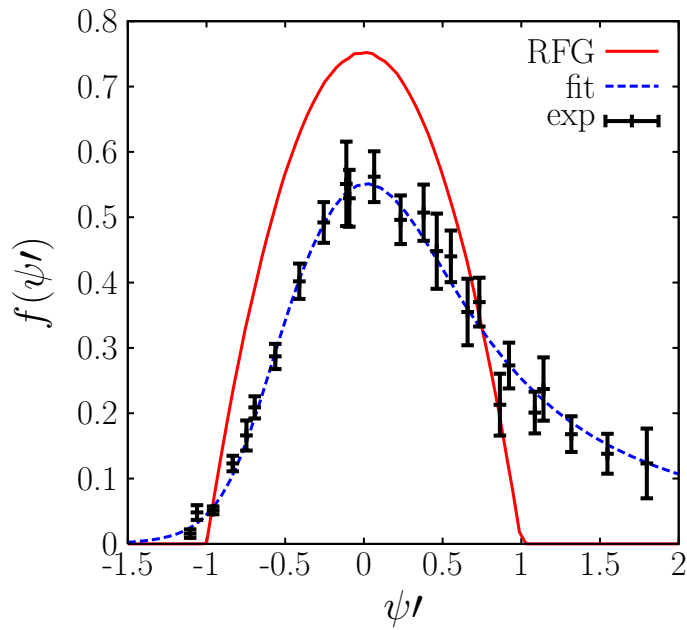}
\caption{\label{fig:susa} 
The QE scaling function obtained as a fit to the experimental data in the longitudinal channel~\cite{Amaro:2004bs} compared to the RgFG result. Data are from Ref.~\cite{Jourdan:1996np}.}
\end{center}
\end{figure}

Inclusive electron scattering data exhibit interesting systematics that can be used to predict (anti)neutrino-nucleus cross sections. When the  experimental $(e,e')$ differential cross sections are divided by the corresponding single nucleon ones and multiplied by a global Fermi momentum, the resulting function 
\be
\la{eq:scaling}
f= p_F \, \frac{\frac{\D d\sigma}{\D d\Omega' dk'^0}}{\D Z \sigma_{ep} + N \sigma_{en}}
\ee
is found to depend on energy and 3-momentum transfers $(q^0, |\vec{q}|)$ through a specific combination, the scaling variable $\psi'$, and to be largely independent of the specific nucleus. This property is known as superscaling~\cite{Donnelly:1998xg}. 
Scaling violations reside mainly in the transverse channel~\cite{Donnelly:1999sw} and have their origin in the excitation of resonances and meson production in general, 2p2h mechanisms and even the tail of deep inelastic scattering. Therefore, an experimental scaling function $f(\psi')$ could be reliably extracted by fitting the data for the longitudinal response~\cite{Amaro:2004bs}. The experimental $f(\psi')$ has an asymmetric shape with a tail at positive $\psi'$ (large $q^0$) as can be seen in Fig~\ref{fig:susa}. 
The requirement of a correct description of the scaling function is a constraint for nuclear models. The RgFG model fulfills superscaling exactly and has a very simple scaling function
\be
\la{eq:scalingRgFG}
f^{\mathrm{RgFG}} = \frac{3}{4} \left( 1 - \psi'^2 \right) \theta \left( 1 - \psi'^2 \right) \,,
\ee
but with the wrong symmetric shape. It has been observed that models based on the impulse approximation are consistent with the superscaling. However, while PWIA and DWIA approaches with complex OP fail to reproduce the asymmetric shape and the tail of the scaling function, a very good description of it is achieved within the RMF model~\cite{Caballero:2005sj}. In addition, the RMF model predicts a larger scaling function in the transverse mode than in the longitudinal one ($f_T > f_L$)~\cite{Caballero:2007tz}, which seems to be supported by data. Extensive studies with a wide class of models reveals the importance of a proper description of the interaction of knocked-out nucleons with the residual nucleus~\cite{Antonov:2011bi} to obtain the tail of the scaling function.

With the superscaling approximation (SuSA) a good representation of the nuclear response can be obtained by embedding nuclear effects in the scaling function: the observables can be calculated with the simple RgFG model followed by the replacement $f_{\mathrm {RgFG}} \raw f_{exp}$. The same strategy can be used to predict neutrino QE cross section, minimizing in this way the model dependence of the results. It has been found that the SuSA approach predicts a 15~\% smaller integrated CCQE cross section compared to the RgFG, close to the result obtained with the RMF model (Fig.~3 of Ref.~\cite{Amaro:2006tf}). Nevertheless, it should be remembered that scaling fails at $\omega < 40$~MeV and $|\vec{q}| < 400$~MeV due to collective effects.  

\subsubsection{Long-range RPA correlations}
\la{subsubsec:rpa}

The theoretical models discussed so far assume the validity of the impulse approximation according to which the (anti)neutrino interacts with a single nucleon in the nucleus. The influence of the spectator nucleons is only present in the hole and particle spectral functions or in FSI. However, when the 3-momentum transferred to the target $|\vec{q}|$ is small and $1/|\vec{q}|$ becomes comparable with the internucleon distance, one should not expect this approximation to hold. The comparison with inclusive electron scattering data shows that at $|\vec{q}| \lesssim 350-400$~MeV, systematic discrepancies associated with the breakdown of the impulse approximation start to show up~\cite{Ankowski:2007uy,Leitner:2008ue}. 

\begin{figure}[h!]
\begin{center}
\includegraphics[width=0.8\textwidth]{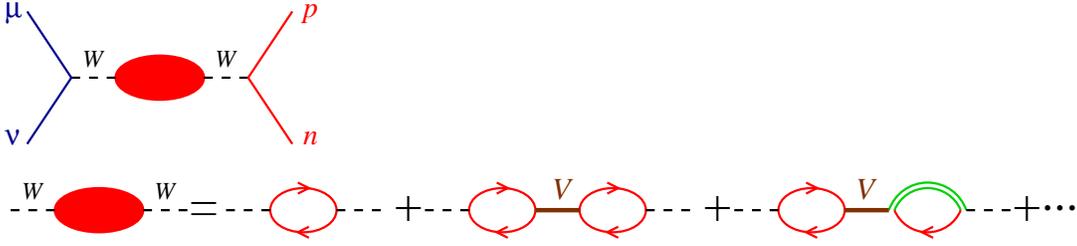}
\caption{\label{fig:rpadiag} 
RPA correlations in CCQE scattering. Solid lines pointing to the right (left) denote particle (hole) states. The double line stands for the $\Delta(1232)$.}
\end{center}
\end{figure}

Collective effects can be handled within the RPA (random phase) ring approximation using the bare polarization propagators as input. For CCQE scattering this is illustrated in Fig.~\ref{fig:rpadiag}. 
These RPA correlations renormalize the different components of the hadronic tensor. One expects them to be relevant for neutrino interactions on the base of the well established quenching of $g_A$ in nuclear Gamow-Teller $\beta$ decay. For CCQE scattering, these long-range correlations were taken into account in Ref.~\cite{Singh:1992dc} using a RlFG model and an effective $NN$ interaction in the nuclear medium (denoted $V$ in Fig.~\ref{fig:rpadiag}) consisting of pion and rho-meson exchange plus a short range part effectively included in the phenomenological constant $g'$, taken to be the same in the longitudinal and transverse channels. The contributions of the RPA sums to the tensor are expressed in an analytic form in terms of the 1p1h and $\Delta$h polarization propagators.
Such a model has been subsequently applied to obtain inclusive and semi-inclusive cross sections in different experimental situations~\cite{Kosmas:1996fh,Singh:1998md,SajjadAthar:2005ke,Athar:2005hu}. This model was improved in Refs.~\cite{Nieves:2004wx,Nieves:2005rq} by a more rigorous resummation of the RPA series in the vector-isovector channel, and the introduction of scalar-isoscalar, scalar-isovector and vector-isoscalar contact density dependent terms of the effective interaction (see Eqs.~(32-36) of Ref.~\cite{Nieves:2004wx} for details). The approach set up in Refs~\cite{Marteau:1999kt,Marteau:1999jp,Martini:2009uj} and the one of Ref.~\cite{Kim:1994zea} are quite similar. In this case the RPA equations are solved numerically. An algebraic solution of these equations was developed in Ref~\cite{Graczyk:2003ru} and applied to the study of $\tau$ polarizations in CC $\nu_\tau,\bar\nu_\tau$-nucleus scattering~\cite{Graczyk:2004uy}.

\begin{figure}[h!]
\begin{center}
\includegraphics[width=0.5\textwidth]{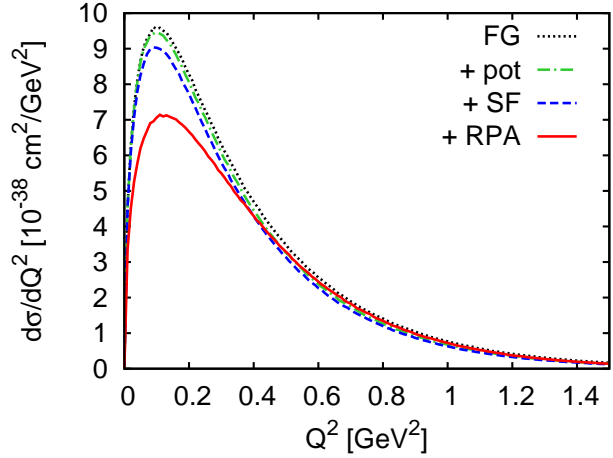}
\caption{\label{fig:rpaplot} 
Differential cross sections for the CCQE reaction on $^{12}$C averaged over the MiniBooNE flux~\cite{AguilarArevalo:2008yp} as a function of the 4-momentum transfer squared $Q^2 \equiv -q^2$. Dotted lines represent the RlFG model with Fermi motion and Pauli blocking. In the dash-dotted lines the nucleons are exposed to the mean field potential while the dashed ones also incorporate spectral functions for the outgoing nucleons~\cite{Leitner:2006ww,Leitner:2008ue}. The full model with long range (RPA) correlations is denoted by solid lines~\cite{AlvarezRuso:2009ad}.}
\end{center}
\end{figure}

The impact of RPA correlations at MiniBooNE energies $\langle E_\nu \rangle \sim 750$~MeV is quite significant. This is  shown in Fig.~\ref{fig:rpaplot}, where they have been calculated for CCQE on $^{12}$C  following Ref.~\cite{Nieves:2004wx}. At low $Q^2 \equiv -q^2 < 0.3$~GeV$^2$ these collective effects cause a sizable reduction of the cross section which, as will be discussed in the next section, is crucial to understand CCQE MiniBooNE data.  

In the models outlined in this section, the effective $NN$ interaction contains contact and energy independent interactions with strengths determined from low-energy collective excitations~\cite{Speth:1980kw}. For this reason, the validity of $V$, like some of the non-relativistic approximations present in the models~\cite{Nieves:2004wx,Martini:2009uj}, can be questioned at the rather high energy transferred of a few hundreds of MeV that can be encountered at MiniBooNE and other neutrino experiments. However, it should be remembered, that the inclusion of dynamical pion and rho propagators in the longitudinal and transverse interactions, as well as the presence of $\Delta$h excitations in addition to 1p1h ones , extend the validity of the framework towards higher energies. Furthermore, once $q^2 < 0$, high $q^0$ correspond to even higher $|\vec{q}|$; one then enters the realm of the impulse approximation, where collective RPA corrections are bound to be negligible. Therefore, even large corrections to the model should have a small impact on the cross sections. A similar situation takes place in inclusive electron scattering on nuclear targets. 

Another point of concern is the validity of the RlFG description in the kinematic region of $q^0 \lesssim 50$~MeV, which can account for a large part of the cross section even at high neutrino energies~\cite{Amaro:2013yna}. It is indeed true that at such low $q^0$, details of the nuclear structure that are beyond reach for a non-interacting Fermi gas model become important, and continuum RPA~\cite{Kolbe:1992xu,Volpe:2000zn,Jachowicz:2002rr,Pandey:2013cca} or even shell-model calculations~\cite{Alberico:1997vh,Butkevich:2007gm} should be more reliable. Nevertheless, there are indications~\cite{Amaro:2004cm} that the RlFG models with RPA corrections lead to realistic predictions for integrated quantities, for which the details of the excitation spectrum are not so relevant. The success of Ref.~\cite{Nieves:2004wx} describing simultaneously inclusive
muon capture on $^{12}$C and the low-energy LSND inclusive CCQE measurements reinforces this conclusion.

\subsection{The role of 2p2h excitations}
\la{sec:2p2h}

In May 2009, at NuInt09, the MiniBooNE Collaboration presented a new CCQE cross section measurement using a high-statistics sample of $\nu_\mu$ interactions on $^{12}$C~\cite{Katori:2009du}. The subsequent publication~\cite{AguilarArevalo:2010zc}, describing the first measurement of the double differential CCQE cross section $d^2\sigma/(dk'^0 d\cos{\theta'})$ as a function of the energy and angle of the outgoing muon, reaffirmed the surprising result reported earlier: a cross section per nucleon $\sim  20$~\% higher than expected from bubble chamber low-energy measurements.  The size of the cross section was found to be well described with the RgFG model using the dipole parametrization of the axial form factor given in Eq.~(\ref{eq:FA}) with an axial mass of  $M_A=1.35 \pm 0.17$~GeV obtained with a shape-only fit to data~\cite{AguilarArevalo:2010zc}. Such a high value of $M_A$ is in contradiction with the previous determinations discussed in Sec.~\ref{subsec:CCQE-Nucleon}. 

The origin of this {\it CCQE puzzle} has been extensively debated. It was conjectured that nuclear effects on $^{12}$C were influencing the determination of $M_A$, which should be understood as {\it effective}. Such a pragmatic attitude that could bring short term benefits in the analysis of neutrino oscillations was dangerous in the long run because the underlying physics was not properly understood. The RgFG model used in the analysis of MiniBooNE data was too simple, but it turned out that more realistic models of the kind reviewed in the previous section also underestimated the data. This situation is illustrated in Fig.~\ref{CCQE} for some of the calculations collected in Ref.~\cite{Boyd:2009zz}. Theoretical predictions with $M_A \sim 1$~GeV lie on a rather narrow band (narrower than the experimental errorbars) clearly below the data.  
\bfig[h!]
\includegraphics[width=0.55\textwidth]{CCQE_C_3.eps}
\caption{(color online) Summary of CCQE integrated cross sections as a function of the neutrino energy. Solid lines denote the models from Refs~\cite{Ankowski:2007uy}, \cite{SajjadAthar:2009rd}, \cite{Benhar:2006nr}, \cite{Leitner:2006ww}, \cite{Maieron:2003df}, \cite{Martini:2009uj} and \cite{Nieves:2004wx} in this order, as reported in Ref.~\cite{Boyd:2009zz}. The dash-dotted and dotted lines are RgFG calculations with $p_F = 220$~MeV, $E_B =34$~MeV and $M_A = 1$ and 1.35~GeV respectively. The dashed and the dashed double-dotted lines are the result of Refs.~\cite{Martini:2009uj} and \cite{Nieves:2011yp} after adding the 2p2h contributions. The data points are from MiniBooNE~\cite{AguilarArevalo:2010zc} as a function of the reconstructed neutrino energy.}
\label{CCQE}
\efig

In a CCQE measurement on nuclear targets, there are CCQE-like events for which CC pion production is followed by pion absorption. The subtraction of this background relies partially on the Monte Carlo  simulation. In the case of the MiniBooNE CCQE measurement, the event generator has been adjusted to the CC$\pi^+$ data in order to reduce the uncertainty. In a rather light nucleus like $^{12}$C absorption is not prominent and one should expect a rather small number of these CCQE-like events: it is unlikely that they were so badly underestimated to be responsible for the large CCQE cross section. Martini et al.~\cite{Martini:2009uj} pointed out that another so far unaccounted source of CCQE-like cross section arises from the contributions of two (or more) interacting nucleons (2p2h excitations), because ejected low energy nucleons are not detected at MiniBooNE.
  
One should recall that certain 2p2h mechanisms [diagrams (1) and (2) of Fig.~\ref{2p2h}] had been already taken into account in former CCQE calculations as part of the more general nucleon spectral functions for particles~\cite{Nieves:2004wx,Benhar:2005dj,Ankowski:2007uy,Leitner:2008ue} and holes~\cite{Benhar:2005dj,Ankowski:2007uy} and have a small effect on the integrated cross section, although they are known to play an important role in the description of the QE peak in inclusive electron scattering on nuclei~\cite{Gil:1997bm,Benhar:2005dj}. Diagram (7) in Fig.~\ref{2p2h} has also been considered in calculations of the inclusive neutrino-nucleus cross sections as part of the in-medium $\Delta(1232)$ spectral function~\cite{Benhar:2005dj,Ankowski:2007uy,Leitner:2008ue}.
\bfig[h!]
\includegraphics[width=0.65\textwidth]{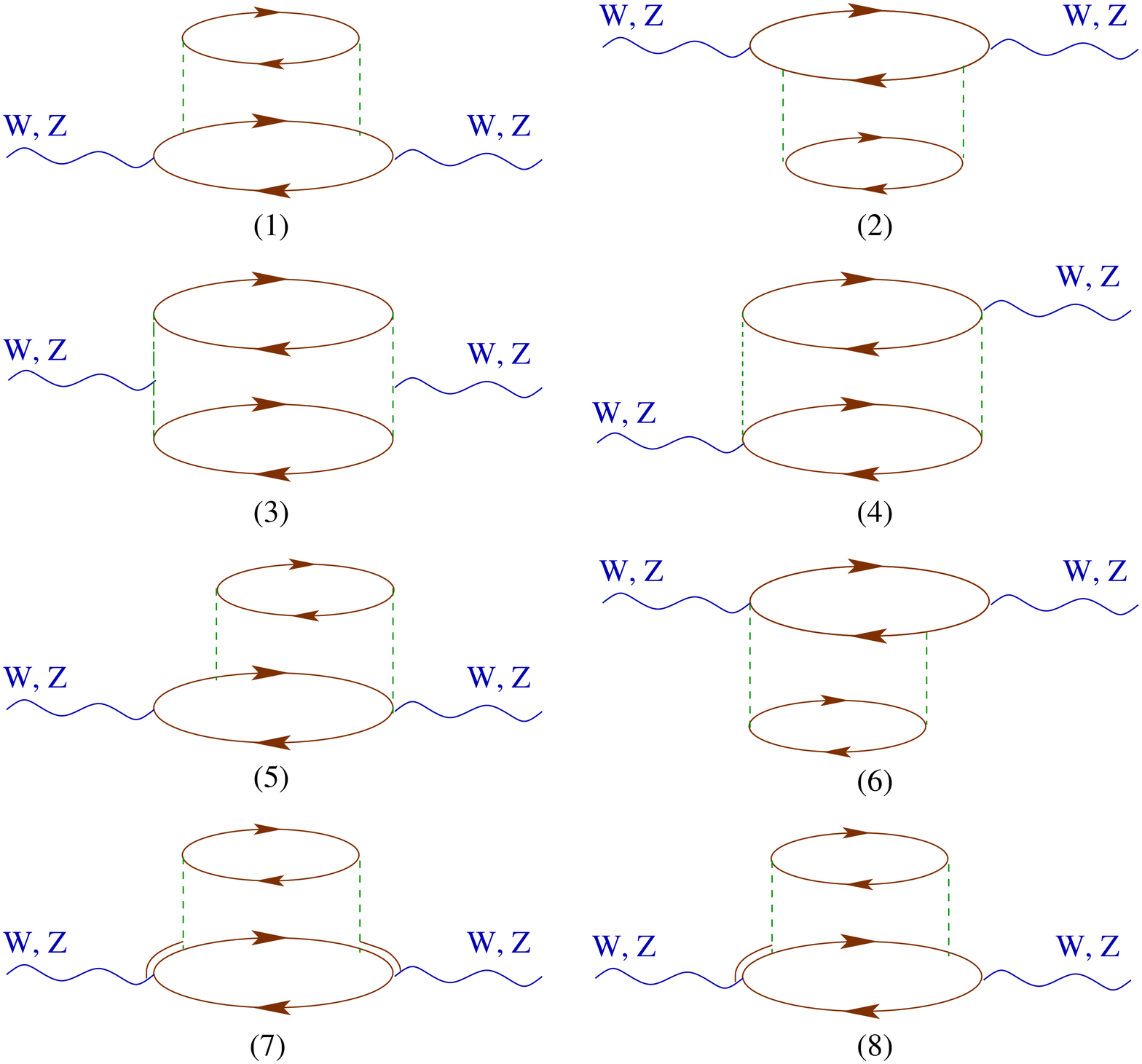}
\caption{Some 2p2h contributions to the polarization propagators. Solid (dashed) lines denote nucleon (pion) propagators. Double lines represent $\Delta(1232)$ propagators. Solid lines pointing to the right (left) denote particle (hole) states.}
\label{2p2h}
\efig        
On the other hand, there are many 2p2h terms [for example MEC diagrams (3), (4) and interference diagram (8) in Fig.~\ref{2p2h}] that are not reduced to particle, hole or resonance spectral functions. They are required for a satisfactory description of the dip region between the QE and the $\Delta(1232)$ peaks in inclusive electron scattering~\cite{Gil:1997bm}. Some of these 2p2h contributions have been taken approximately into account in Ref.~\cite{Martini:2009uj} using two different parametrizations of the multinucleon terms, from pion absorption~\cite{Shimizu:1980kb} and from electron scattering~\cite{Alberico:1983zg}, extrapolated to the kinematic region of neutrino interactions. Once they were added to the {\it true} CCQE cross section, a very good agreement with the MiniBooNE data was obtained (see the dashed line in Fig.~\ref{CCQE}). A good description of the data, quite similar to the one obtained with the RgFG and $M_A = 1.35$~GeV, is also obtained with the microscopic model for 2p2h excitations developed in Ref.~\cite{Nieves:2011pp} (dashed double-dotted line in Fig.~\ref{CCQE}).

The importance of MEC has been further stressed in the recent {\it ab initio} calculation of the sum rules of the weak NC response functions on $^{12}$C~\cite{Lovato:2014eva}. A significant enhancement ($\sim 30$\%) of the weak response is due to two-nucleon currents. This approach implements non-relativistic currents and treats the $\Delta$ in the static limit but provides a state-of-the-art description of the nuclear ground state and the nuclear correlations. Therefore, it represents a benchmark for more phenomenological methods. Based on a recent calculation of sum rules for the electromagnetic response obtained in Ref.~\cite{Lovato:2013cua} within the {\it ab initio} Green Function Monte Carlo framework, it has been suggested~\cite{Benhar:2013bba} that 2p2h terms arising from the interference between one-body and two-body currents play a significant role in neutrino scattering. Indeed, the transverse sum rule on $^{12}$C has a sizable contribution of this kind, as can be seen in Fig.~1 of Ref.~\cite{Benhar:2013bba}. Diagrams (5) and (6) of Fig.~\ref{2p2h} are examples of these pieces in the polarization propagator: in the nomenclature of Ref.~\cite{Benhar:2013bba}, diagram (5)[(6)] accounts for a MEC-final (initial) state correlation interference contribution to the 2p2h hadronic tensor. It should be stressed that, with the caveats discussed above about the kinematic extrapolations, the calculation of Martini et al.~\cite{Martini:2009uj} incorporates such interference terms, as can be seen in Sec.~3.4 of Ref.~\cite{Shimizu:1980kb} and in Sec.~4 of Ref.~\cite{Alberico:1983zg}. In the model of Ref.~\cite{Nieves:2011pp}, the final state correlation terms, and the interferences with MEC mechanisms, are included as part of the generic 2p2h diagram of Figs.~4 and 9 of that reference, while the initial state ones are neglected because they are of higher order in an expansion in powers of the nuclear density [as diagram (2) of Fig.~\ref{2p2h} is]. Although the transverse sum rule is indicative, a systematic study of observable quantities at the kinematics encountered in neutrino experiments is required to establish the importance of these mechanisms for the QE-like cross section.

In neutrino experiments, the incident energy is not known a priori and its determination involves some model dependence (see the discussion in Sec.~\ref{subsec:erec}). For this reason, theoretical models should be directly confronted with the double differential cross section of Ref.~\cite{AguilarArevalo:2010zc} for neutrinos and Ref.~\cite{AguilarArevalo:2013hm} for antineutrinos. Such a comparison has been performed in Refs.~\cite{Amaro:2010sd,Nieves:2011yp,Martini:2011wp} for neutrinos, and in Refs.~\cite{Amaro:2011aa,Nieves:2013fr,Martini:2013sha} for antineutrinos. In Refs.~\cite{Amaro:2010sd,Amaro:2011aa} the addition of the vector part of the MEC mechanisms to the result of the SuSa approach led to a better agreement with data. Within the model of Ref.~\cite{Nieves:2011pp} a very good agreement with the $\nu$-$^{12}$C double differential cross section has been achieved with $M_A = 1.077 \pm 0.027$~GeV (see Table I of Ref.~\cite{Nieves:2011yp}). This value is certainly much lower and closer to the determinations from CCQE on deuterium and pion electroproduction discussed in Sec.~\ref{subsec:CCQE-Nucleon} than the one obtained by the RgFG model.  

The role of the different reaction mechanisms according to the model of Refs.~\cite{Nieves:2011pp,Nieves:2011yp} can be appreciated in Fig.~\ref{fig:2diff} where $d^2\sigma/(dk'^0 d\cos{\theta'})$ is shown for a single muon-angle bin. Not only multinucleon mechanisms, but also RPA corrections are essential to understand the data. Indeed, RPA strongly decreases the cross section at low muon energies, there where multinucleon contributions accumulate.  Therefore, the final picture arises from a delicate balance between the dominant single nucleon scattering, corrected by collective effects, and mechanisms that involve two or more nucleons.
\bfig[hb!]
\bcen
\includegraphics[width=0.5\linewidth]{new85.NPJ.eps}
\caption{$\nu$-$^{12}$C double differential cross section  averaged over the MiniBooNE flux~\cite{AguilarArevalo:2008yp}
 as a function of the muon kinetic energy and for the  $0.80 < \cos\theta_\mu < 0.90$ angular bin~\cite{Nieves:2011yp}. The thick solid line stands for the full model (RlFG+RPA+2p2h). The dashed, dotted and dash-dotted lines show partial results for only RlFG, RlFG+RPA and only 2p2h, respectively. All these curves are obtained with $M_A = 1.049$~GeV while the thin solid line is calculated with the RlFG and $M_A = 1.32$~GeV. The data of Ref.~\cite{AguilarArevalo:2010zc} have been rescaled by a factor 0.9 (compatible with flux uncertainties).
\label{fig:2diff}}
\ecen
\efig   

A good description of MiniBooNE data for both $d^2\sigma/(dk'^0 d\cos{\theta'})$ and $\sigma(E_\nu)$ with an $M_A = 1.03$~GeV
has also been found~\cite{Meucci:2011vd} with the RGF model with empirical OP briefly covered in the previous section. This has been achieved with a model that takes into account those multinucleon contributions that can be ascribed to the particle spectral function (like, for example, diagram (1) in Fig.~\ref{2p2h}) but does not contain MEC mechanisms or in-medium $\Delta$ modifications. This finding, at odds with the picture outlined above, should be interpreted with care. First of all, the results depend rather strongly on the choice of OP: compare green and red lines in Figs.~1-3 of Ref.~\cite{Meucci:2011vd}. Second and most importantly, the imaginary part of the OP, which adds to the total cross section in the RGF model, is due to inelastic channels. These inelastic channels include pion emission and absorption, and have already been subtracted in the MiniBooNE analysis. Therefore, it would be very interesting to confront the RGF results  with fully inclusive CC data.

Ultimately, it would be important to find a more direct experimental signature for the multinucleon processes. Possible observables have been considered in Refs.~\cite{Lalakulich:2012ac,Sobczyk:2012ms}. It is found that 2p2h primary interactions do lead to an increase of multinucleon events. Such an enhancement may indeed be revealed in measurements by looking, for example, at proton pairs in the final state, or the total visible energy which would contain contributions from protons below reconstruction threshold. However, the primary distributions will be heavily distorted by FSI and, therefore, model discrimination would require a high precision and a considerable improvements in the Monte Carlo simulations.

\subsubsection{Multinucleon mechanisms and neutrino energy reconstruction}
\la{subsec:erec}

Neutrino beams are not monochromatic so that the energy of an interaction event is a priori not known. As the oscillation probability is energy dependent, the neutrino energy determination is important for oscillation analyses and also, needless to say, to measure the energy dependence of different cross sections, like the CCQE one from MiniBooNE~\cite{AguilarArevalo:2010zc} shown in Fig.~\ref{CCQE}. There are different strategies to reconstruct $E_\nu$. In high energy experiments such as MINOS ($1 \lesssim E_\nu \lesssim 50$~GeV) it is reconstructed as the sum of the muon energy and the hadronic shower energy. As the detection of the final particles is never perfect, the procedure partially relies on the theoretical models contained in the simulation program~\cite{Kordosky:2006gt}. 

At lower energies it is common to rely on a kinematic energy reconstruction based on the event identification as CCQE. In this case, the neutrino energy can be obtained from the measured angle and energy of the outgoing lepton using two-body kinematics
\be
\la{eq:Enu}
E^{\mathrm{rec}}_\nu = \frac{2 M_n k'^0 - m_l^2 - M_n^2 + M_p^2}{2 \left( M_n - k'^0 + \sqrt{(k'^0)^2 -m_l^2} \cos{\theta'} \right)}  \,.
\ee
This formula, sometimes modified to incorporate the constant binding energy of the RgFG model, is only valid for a target neutron at rest. The Fermi motion of the nucleons in the nucleus causes a smearing of the reconstructed energy around the true value but the procedure remains accurate enough for oscillation analyses, and the energy dependence of the cross section is not affected. On the contrary, CCQE-like events from absorbed pions produce a systematic error in the neutrino energy determination, which is too large to be neglected. This was known and taken into account in the MiniBooNE analysis, by treating these events as a background that must be subtracted~\cite{AguilarArevalo:2010zc}. However, the same is true for multinucleon contributions. Once they are sizable, the effect on the $E_\nu$ determination is significant~\cite{Martini:2012fa,Nieves:2012yz,Martini:2012uc}.

\bfig[h!]
\bcen
\includegraphics[width=0.48\linewidth]{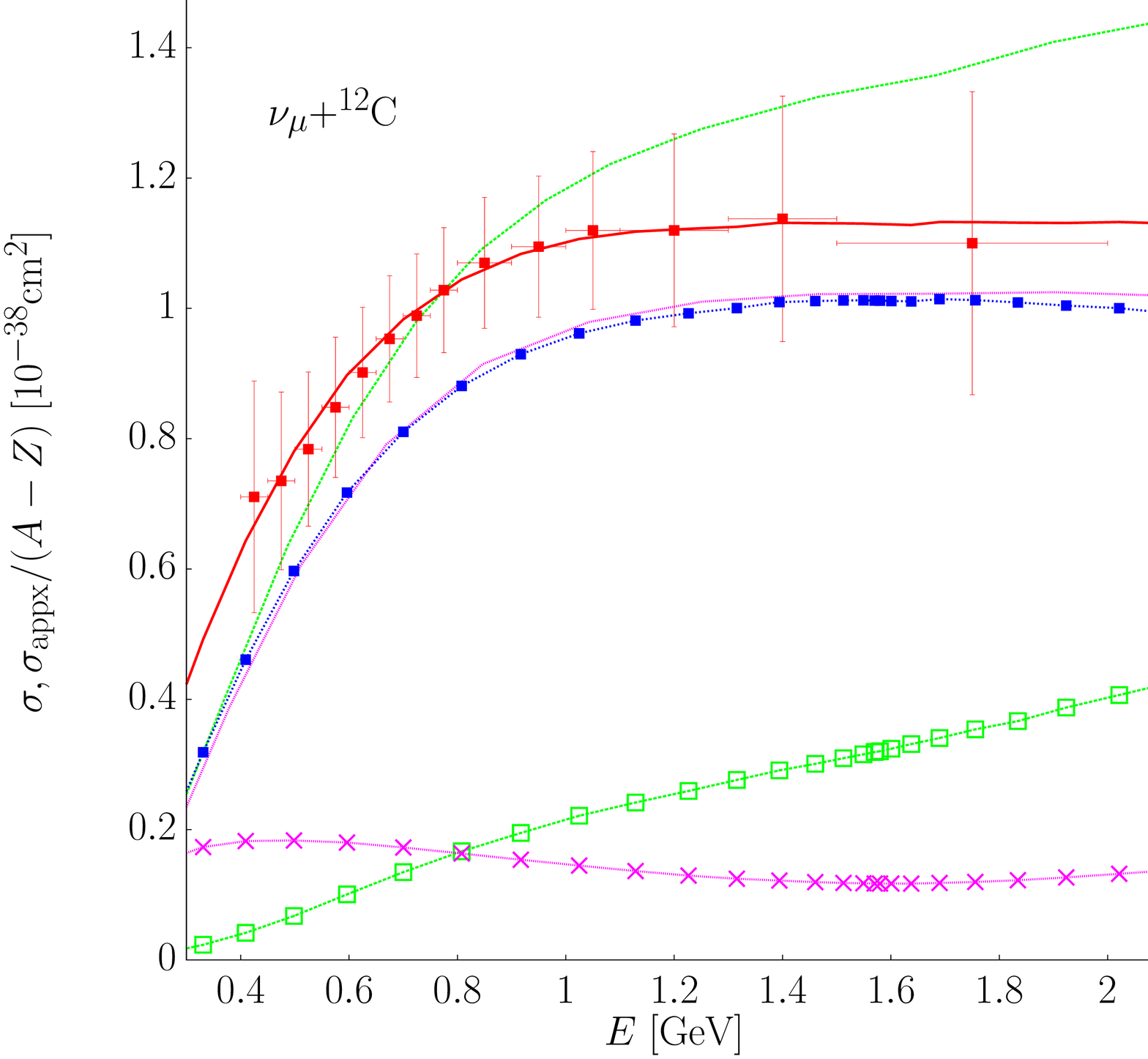}
\caption{CCQE-like cross section as a function of the true and the reconstructed neutrino energies~\cite{Nieves:2012yz}. The RlFG+RPA cross section, given by the blue filled squares plotted as a function of $E^{\mathrm{rec}}_\nu$,  is practically unaffected by the energy reconstruction as can be seen from the comparison with the solid magenta line. The 2p2h parts as a function of $E_\nu$ and  $E^{\mathrm{rec}}_\nu$ are given by the open green squares and the magenta crosses respectively. The corresponding curves for the total RlFG+RPA+2p2h cross section are the green dotted and the red ones. MiniBooNE data~\cite{AguilarArevalo:2010zc} have been rescaled by a factor 0.9.
\label{fig:erec}}
\ecen
\efig   
Figure~\ref{fig:erec} demonstrates the effect of the neutrino-energy reconstruction on the $E_\nu$ dependence of the CCQE-like (RlFG+RPA+2p2h) cross section according to the model of Refs.~\cite{Nieves:2011pp,Nieves:2011yp}. The {\it true} CCQE (RlFG+RPA) cross section is unaffected by the energy reconstruction. The 2p2h contribution instead is a very different function of the true neutrino energy (green open squares) than as a function of $E^{\mathrm{rec}}_\nu$ (magenta crosses). In other words,  $E^{\mathrm{rec}}_\nu$  is a poor estimate of the actual energy for multinucleon mechanisms. As a consequence, the total theoretical RlFG+RPA+2p2h cross section, when plotted in terms of $E^{\mathrm{rec}}_\nu$ shows a remarkably good agreement with the  MiniBooNE data of Ref.~\cite{AguilarArevalo:2010zc} rescaled by a factor 0.9, which is consistent with the experimental normalization error of 10.7\%. In conclusion, the actual energy dependence of the  CCQE-like cross section is not given by the MiniBooNE data but would be steeper and closer to the dotted (green) line in Fig.~\ref{fig:erec}. A similar finding has been made with the model of Refs.~\cite{Martini:2009uj} as can be seen in Fig.~14 of Ref.~\cite{Martini:2012uc}.

The misreconstruction of QE events resulting from many-body dynamics is bound to have an impact on the oscillation analyses of experiments like MiniBooNE, T2K and LBNE~\cite{Lalakulich:2012hs,Martini:2012uc,Mosel:2013fxa}. The bias in the determination of oscillation parameters may remain even after the near detector has been taken into account~\cite{Coloma:2013rqa,Coloma:2013tba}. 

\subsubsection{The high $E_\nu > 1$~GeV region}

As the neutrino energy increases, so does the range of possible energies that can be transferred to the target. Large energy transfers make possible the excitation of baryon resonances heavier than the $\Delta(1232)$ not taken into account in the 2p2h models that have been developed so far. Based on the experience with weak resonance excitation and pion production (see Section~\ref{subsec:pion-nucleon}) one could expect the $D_{13}(1520)$ to play a role.
The effective $NN$ interaction used to compute RPA correlations are not realistic at high energies as discussed in Sec.~\ref{subsubsec:rpa} although collective effects for this kinematics should be small. In addition, the MEC vertices present in the models come from effective low-energy interactions. Nevertheless, the model of Refs.~\cite{Nieves:2011pp,Nieves:2011yp} has been applied to neutrino energies of up to 10~GeV, but limiting the 2p2h contribution to $|\vec{q}| < 1.2$~GeV~\cite{Gran:2013kda}. The results obtained in this way, averaged over the neutrino and antineutrino fluxes at MINERvA, have been confronted with data~\cite{Gran:2013kda} for the reconstructed $q^2$ distribution obtained using  $E^{\mathrm{rec}}_\nu$. As can be seen in Fig.~\ref{fig:Minerva}, the agreement is quite good, with a slight overestimation of the data. The impact of the reconstruction procedure in the case of MINERvA flux is small.
\bfig[h!]
\includegraphics[width=0.48\linewidth]{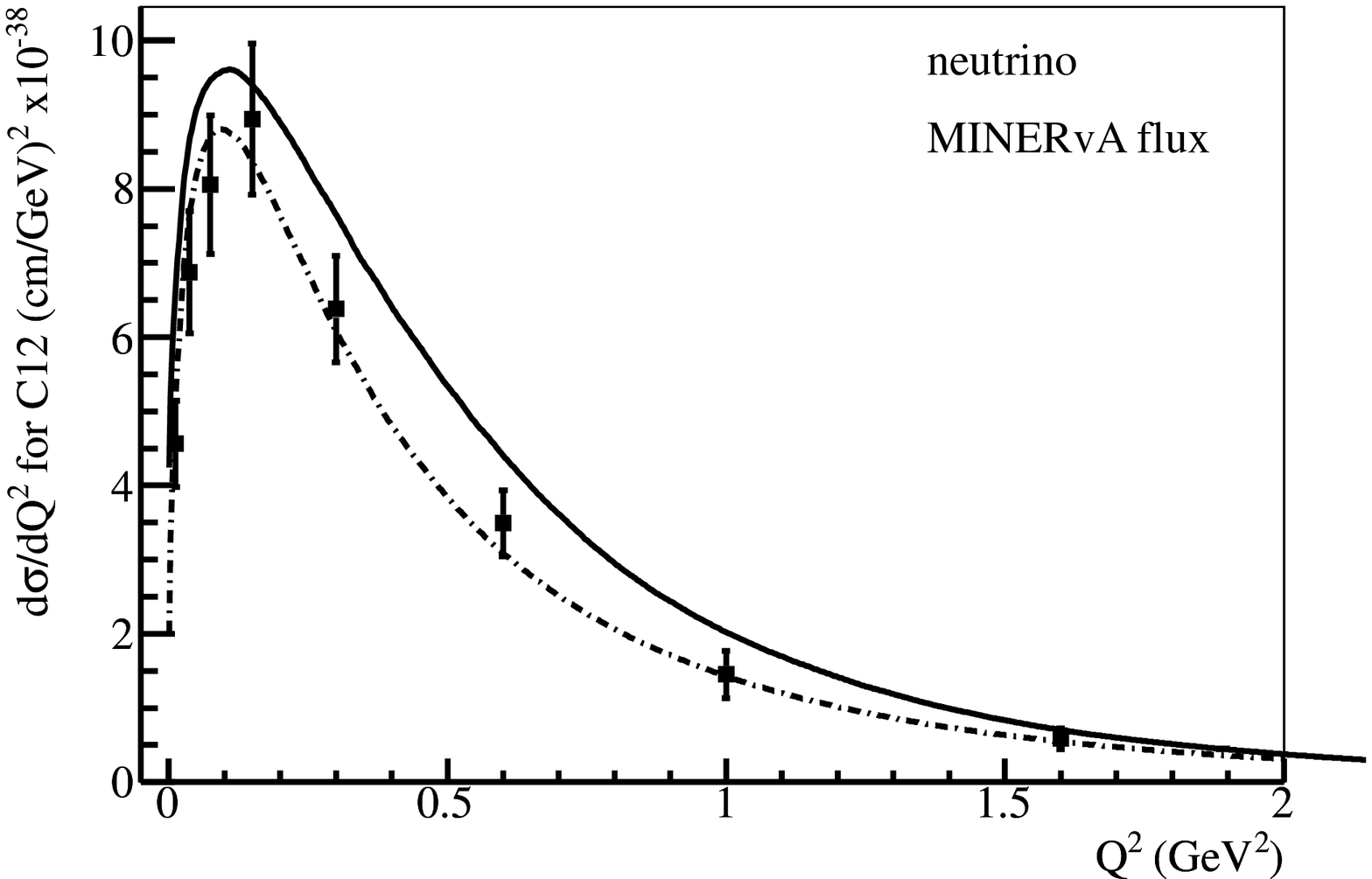}
\includegraphics[width=0.48\linewidth]{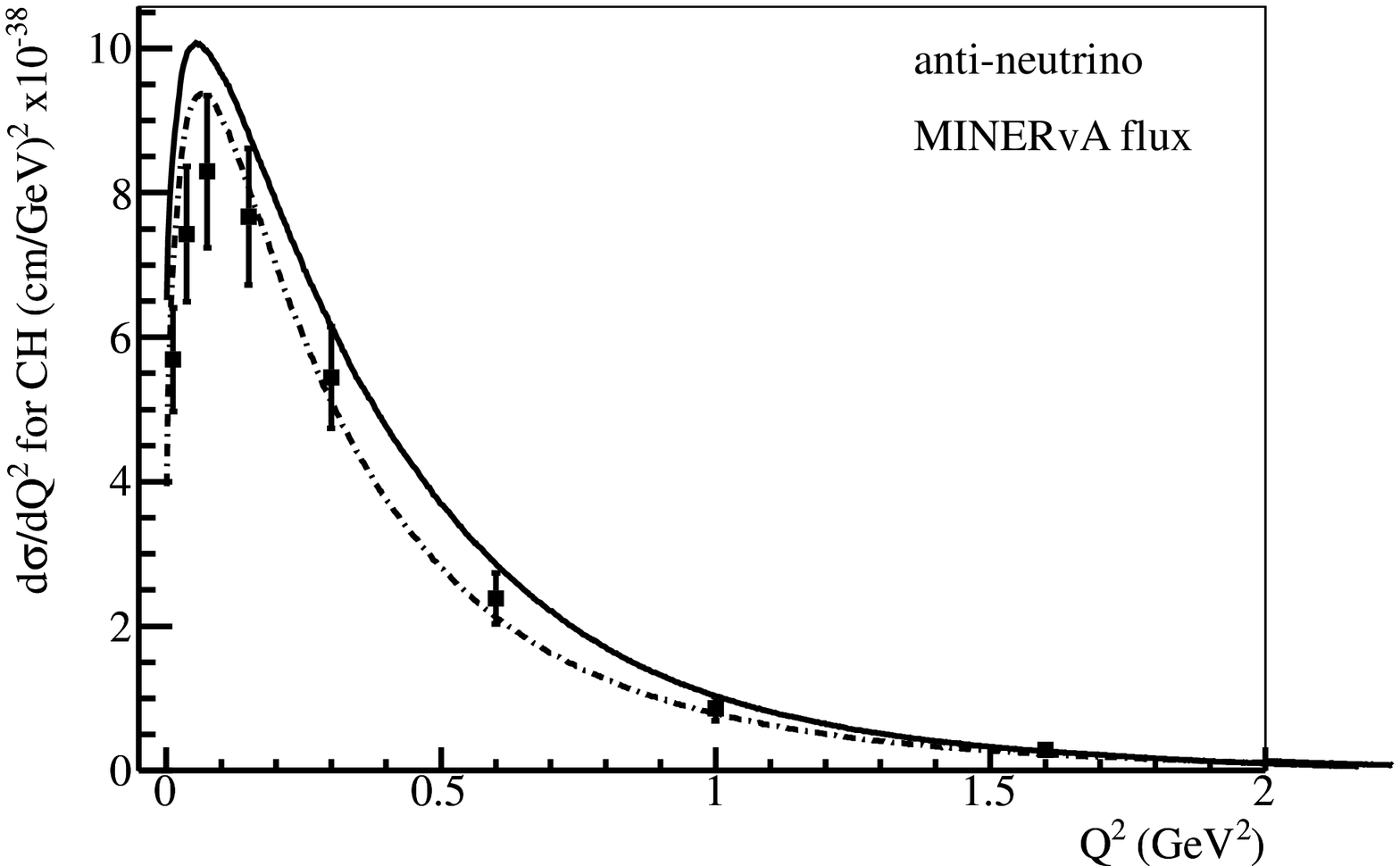}
\caption{Differential $Q^2$ distribution averaged over the MINERvA
  $\nu_\mu$ and $\bar\nu_\mu$
  fluxes~\cite{Fiorentini:2013ezn,Fields:2013zhk} as a function of the
  reconstructed $Q^2$ for the RlFG (dash-dotted lines) and the RlFG+RPA+2p2h (solid lines) models. 
Data are from Refs.~\cite{Fiorentini:2013ezn,Fields:2013zhk}.
\label{fig:Minerva}}
\efig 

According to Ref.~\cite{Gran:2013kda}, as the neutrino energy increases, up to 10~GeV, the 2p2h contribution saturates to $\sim 30$\% of the QE cross section. A priori, there is no reason for this trend to change drastically at even higher energies. This brings us to a question that remains open: the compatibility of MiniBooNE results with the NOMAD one of $M_A = 1.05 \pm 0.02(stat) \pm 0.06(syst)$~GeV~\cite{Lyubushkin:2008pe}. The answer is not obvious and requires further investigations. The NOMAD measurement includes events with a muon track and one or no knocked out proton track. In principle the events without nucleon tracks should contain at least some 2p2h contributions producing  unobservable low energy protons or neutrons but it is possible that, due to the high excitation energies involved, a fraction of the 2p2h events are observed as multi-track ones and removed from the CCQE sample.

\subsection{Quasielastic production of hyperons}
\la{subec:hyper}

The existence of flavor changing CC converting $u$ quarks into $s$ quarks make the QE production of hyperons induced by antineutrinos possible. One has the following $\Delta S = 1$, Cabibbo suppressed, reactions on nucleons
\bea 
\bar \nu_l(k) \, p(p) &\raw&  l^+(k') \, Y^0(p') \,, \quad Y^0=\Lambda,\Sigma^0 \la{eq:anuY1} \\
\bar\nu_l(k) \, n(p) &\raw& l^+(k')  \, \Sigma^-(p') \,, \la{eq:anuY2}
\eea
which are related to the semileptonic decays of hyperons~\cite{Cabibbo:2003cu}. The theoretical study of these reactions has been undertaken in Refs.~\cite{Singh:2006xp,Mintz:2007zz,Kuzmin:2008zz}. The framework outlined in Sec.~\ref{subsec:CCQE-Nucleon} for QE scattering on nucleons remains valid, now with $c_{\chic \mathrm{EW}} = \sin{\theta_C}$ in Eq.~(\ref{eq:dsdq2}),
\be
\la{eq:hadtensorY}
H^{\alpha \beta} =\mathrm{Tr}\left[ \left( \slashed{p} + M \right) \gamma^0 \left( \Gamma^\alpha \right)^\dagger \gamma^0 \left( \slashed{p}' + M_Y \right) \Gamma^\beta  \right] \,,
\ee
where $M_Y$ is the hyperon mass and
\be
\la{eq:QEcurrentY}
J^\alpha = \bar u_Y(p') \Gamma^\alpha u(p) = V^\alpha - A^\alpha \,, 
\ee
with the vector and axial currents given by
\be
V^\alpha = \bar u_Y(p') \left[ \gamma^\alpha f_1(q^2)+ i \sigma^{\alpha\beta} \frac{q_\beta}{M+M_Y} f_2(q^2) + \frac{q^\alpha}{M_Y} f_3(q^2) \right] u(p)
\ee
and
\be
 A^\alpha =  \bar u_Y(p') \left[\gamma^\alpha g_1(q^2) +   i \sigma^{\alpha\beta} \frac{q_\beta}{M+M_Y} g_2(q^2)  + \frac{q^\alpha}{M_Y} g_3(q^2)  \right] \gamma_5  \,u(p) \,.
\ee 
Assuming SU(3) symmetry, the form factors can be related to the electromagnetic and axial form factors of nucleons (see for example Table II of Ref.~\cite{Singh:2006xp}). In this limit, $f_3 = g_2 =0$. SU(3) breaking corrections, which can be systematically studied using chiral perturbation theory~\cite{Zhu:2000zf}, are small for the accuracy presently achievable in neutrino experiments. 

Weak hyperon emission off nuclear targets has been addressed in Ref.~\cite{Singh:2006xp}. Apart from the Fermi motion of the initial nucleon and the mean field potential felt by the hyperons, estimated to be negligible~\cite{Singh:2006xp}, there are important FSI effects. The hyperons produced in the reactions of Eqs.~(\ref{eq:anuY1}) and (\ref{eq:anuY2}) undergo elastic and charge exchange scattering; $\Sigma^0$ can be converted into $\Lambda$ via radiative decay $\Sigma^0 \raw \Lambda \, \gamma$. These processes, which alter the composition and momentum distributions of the emitted hyperons, have been modeled in Ref.~\cite{Singh:2006xp} by a Monte Carlo cascade simulation using experimental information on hyperon-nucleon cross sections as input. Another consequence of FSI is that $\Sigma^+$ hyperons, not produced in the primary reactions, can emerge due to processes like $\Lambda  \, p \raw \Sigma^+ \, n$ or $\Sigma^0  \, p \raw \Sigma^+ \, n$ although at a small rate (compare Fig.~9 of Ref.~\cite{Singh:2006xp} with Figs~4-6 of the same article).

An important issue brought up in Ref.~\cite{Singh:2006xp} and elaborated further in Ref.~\cite{Alam:2013cra} is that, owing to their weak decays $Y \raw N \, \pi$, hyperons become a source of pions in experiments with antineutrino beams. As can be seen from Fig.~\ref{fig:PifromY}, at low incident energies (550~MeV for $\pi^-$ and 650 for $\pi^0$) the cross section for pion production from hyperons becomes larger than the one from $\Delta(1232)$ excitation, which is the dominant mechanism at higher energies (see Sec.~\ref{sec:pion} for more details about weak pion production). Unlike $\Delta$ resonances, hyperons have a large mean life and decay predominantly outside the nucleus. Therefore, the resulting pions are not absorbed in the nucleus. This partially compensates the Cabibbo suppression, particularly for heavy nuclei where absorption is strong. 
\bfig[ht!]
\includegraphics[width=0.48\linewidth]{PifromY.eps}
\caption{Integrated cross section for $\pi$ production on $^{16}$O induced by antineutrinos as a function of the antineutrino energy in the laboratory frame. Results for pions produced from $\Delta$ and hyperon (Y) decay are compared. Adapted from Ref.~\cite{Singh:2006xp}.
\label{fig:PifromY}}
\efig 
In Ref.~\cite{Alam:2013cra} it has been shown that for atmospheric and MiniBooNE $\bar \nu$ fluxes, a significant fraction of the $\pi^-$ and $\pi^0$ originate indeed from hyperon decays.

\section{Weak pion production and other inelastic channels}
\label{sec:pion}

\subsection{Introduction}

Pion production cross section becomes quite relevant for neutrino
energies above 400 or 500 MeV, and plays a central role for neutrino
energies in the 1 GeV region, of the greatest importance for neutrino
oscillation experiments
as MiniBooNE or T2K. Recently, the MiniBooNE Collaboration has
published one pion production cross sections on mineral oil by
$\nu_\mu$ and $\bar\nu_\mu$ neutrinos with energies below 2 GeV. The
data include $\nu_\mu$ and $\bar\nu_\mu$ NC single $\pi^0$
production~\cite{AguilarArevalo:2009ww}, as well as $\nu_\mu$ induced
CC charged and neutral pion production~\cite{AguilarArevalo:2010bm,
  AguilarArevalo:2010xt}.  These are the first\footnote{There exist
  other two other 
recent measurements of NC $\pi^0$ production
(K2K~\cite{Nakayama:2004dp} and SciBooNE~\cite{Kurimoto:2009wq}),
which however  do not provide  absolutely normalized cross-sections, 
but report only the ratios $\sigma(NC1\pi^0)/\sigma(CC)$, where
$\sigma(CC)$ is the total CC cross section. } pion production cross
sections to be measured since the old deuterium bubble chamber experiments
carried out at Argonne National Laboratory (ANL) 
\cite{Campbell:1973wg,Radecky:1981fn} and Brookhaven
National Laboratory (BNL)~\cite{Kitagaki:1986ct}. 

These new data show interesting deviations from the
predictions of present theoretical models that we
will briefly discuss in what follows.

\subsection{Pion production off nucleons}
\label{subsec:pion-nucleon}

Pion production in weak interactions are a window to the poorly known
axial properties of baryon resonances. In addition, the first
requirement to put neutrino induced pion production on nuclear targets
on a firm ground is to have a realistic model at the nucleon level.
There have been several theoretical studies of the weak pion
production off the nucleon at intermediate
energies~\cite{LlewellynSmith:1971zm, Schreiner:1973ka, Fogli:1979cz,
  Fogli:1979qj, Rein:1980wg, AlvarezRuso:1997jr, AlvarezRuso:1998hi,
  Sato:2003rq, Paschos:2003qr, Lalakulich:2005cs, Lalakulich:2006sw,
  Leitner:2008ue, Hernandez:2007qq, Hernandez:2010bx,
  Hernandez:2013jka, Graczyk:2007bc, Graczyk:2009qm, Barbero:2008zza,
  Barbero:2013eqa, Serot:2012rd}. Most of them describe the pion
production process by means of the weak
excitation of the $\Delta(1232)$ resonance and its subsequent decay
into $N\pi$, and do not incorporate any background terms. The models
of Ref.~\cite{Fogli:1979cz, Fogli:1979qj, Rein:1980wg, Lalakulich:2005cs, Lalakulich:2006sw,
  Leitner:2008ue} include also the weak excitation of several
resonance contributions as intermediate states. In these schemes, the
vector form factors were fixed from helicity amplitudes extracted in
the analysis of pion electroproduction data, while the axial couplings
were obtained from PCAC. The most complete model in this respect is the one of
Ref.~\cite{Leitner:2008ue}, where all 4-star resonances below 1.8~GeV
have been included, with vector form factors taken directly over from
the MAID analysis (\cite{MAID} and \cite{Tiator:2011pw}). The vector
part of the background and its interference with the vector part of
the resonant contributions was fixed using the empirical pion
electroproduction amplitudes extracted in the MAID
analysis~\cite{MAID}.  The axial background part (including the
vector-axial interference) was taken to be proportional to the vector
one. The proportionality constant was adjusted to the old bubble
chamber ANL and BNL data, neglecting deuteron effects.  According to
\cite{Leitner:2008ue}, the $D_{13}(1520)$ resonance, besides the
$\Delta(1232)$, is the only one playing a significant role for
neutrino energies below 1--1.5 GeV. However, at $E_\nu > 1$ GeV
(MINER$\nu$A) higher $N^*$ in general will become important.

\begin{figure}[htb]
\begin{center}
\makebox[0pt]{\includegraphics[width=0.35\textwidth]{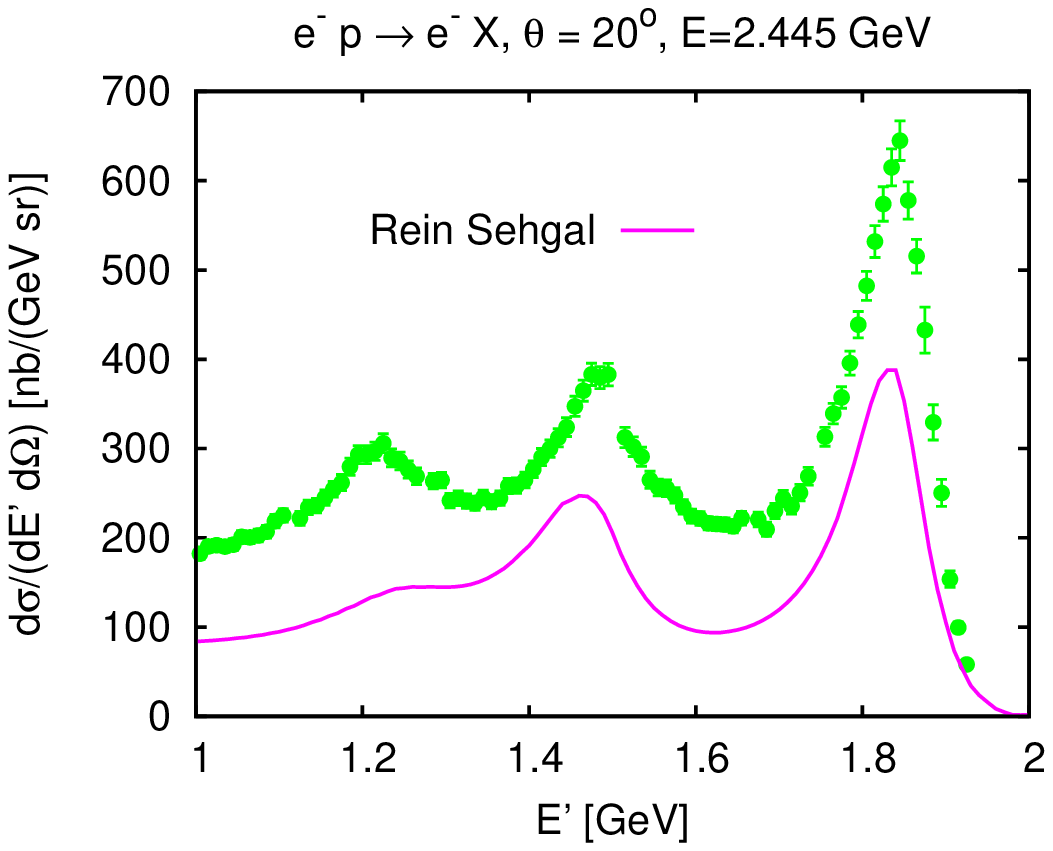}\hspace{0.5cm}\includegraphics[width=0.5\textwidth]{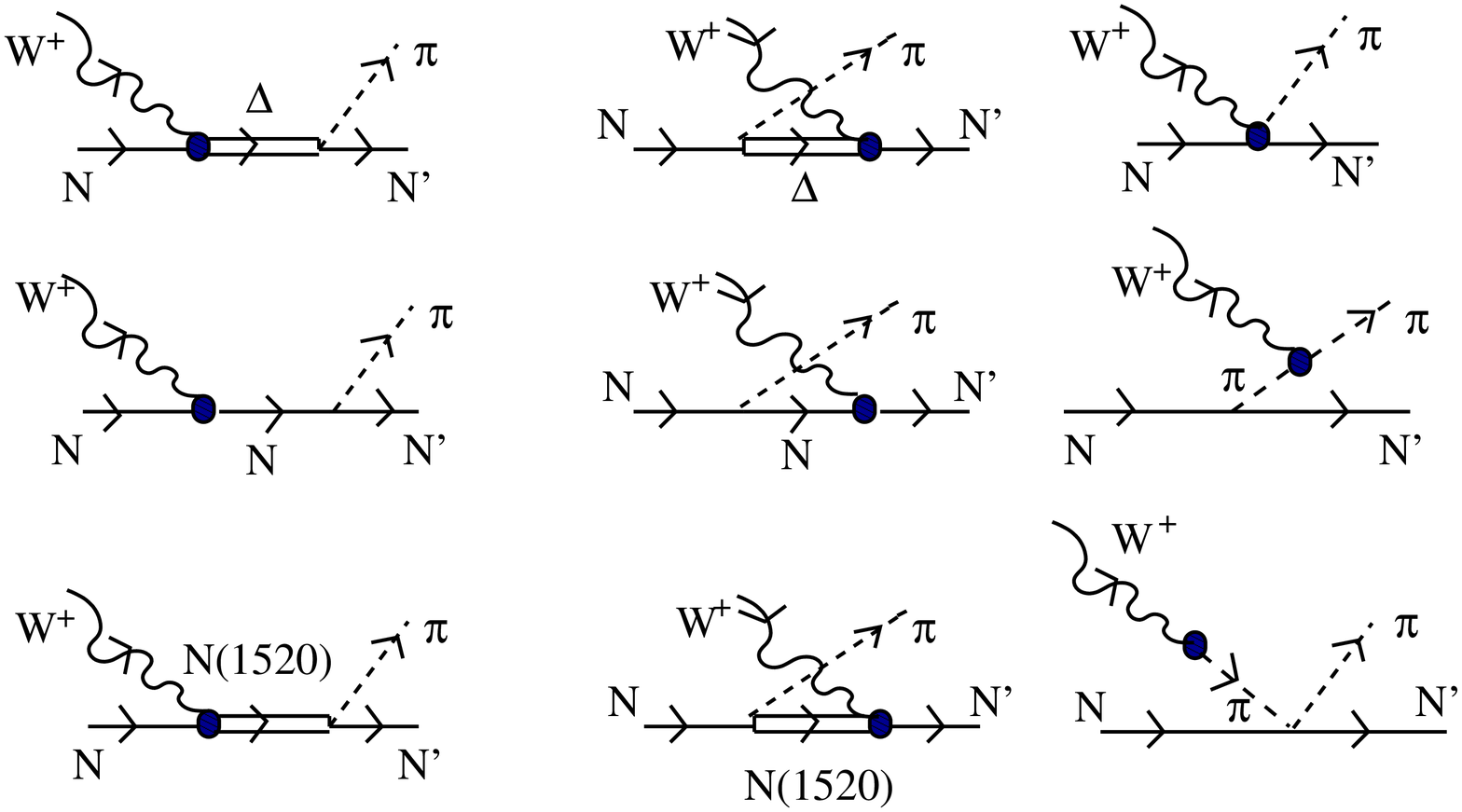}}
\end{center}
\caption{Left: Double differential cross sections for scattering of
  electrons off protons. The predictions of the model of
 Ref.~\cite{Rein:1980wg}  are confronted to data from JLAB~\cite{JLab}
 (see \cite{Leitner:2008fg} for details). Right: Model of
  Refs.~\cite{Hernandez:2007qq,Hernandez:2010bx,Hernandez:2013jka} for
  the $W^+N \to N' \pi$ amplitude. It consists of direct and crossed
  $\Delta(1232)$, $N(1520)$ and nucleon pole terms, contact and pion
  pole contribution, and the pion-in-flight term. }
\label{fig:coh0}
\end{figure}

We should pay a special attention to the Rein-Sehgal
model~\cite{Rein:1980wg}, because it is used by almost all Monte Carlo
(MC) generators. It was an attempt to describe all data available in
1980 on neutrino production of single pions in the resonance region up
to $\pi N$ invariant masses $W_{N\pi}$ of around 2 GeV. The basic
assumption is that single pion production is mediated by all
interfering resonances below 2 GeV, supplemented with a simple
non-interfering, non-resonant phenomenological background of isospin
1/2. The needed transition matrix elements are calculated using the
relativistic quark model of
Feynman-Kislinger-Ravndal~\cite{Feynman:1971wr} (formulated in 1970)
with SU(6) spin-flavor symmetry, and a total of 18 baryon resonances
considered. The original work of Ref.~\cite{Rein:1980wg} assumes
massless leptons. Subsequently, the model for the CC reaction was
extended in \cite{Berger:2007rq,Graczyk:2007xk} to include finite
lepton mass effects in a manner consistent with PCAC. However, we
should stress the Rein-Sehgal model provides a really poor description
of the pion electroproduction data on protons~\cite{Leitner:2008fg,
  Graczyk:2007bc}. Actually, it underestimates significantly the
electron data, as can be appreciated in the left panel of
Fig.~\ref{fig:coh0}, and the more accurate predictions from the model
of Ref.~\cite{Leitner:2008ue}. The model of Rein and Sehgal also
reveals itself unsatisfactory in the axial sector at $q^2 =0$, where
the divergence of axial current can be related to the $\pi N$
amplitude by PCAC~\cite{Paschos:2011ye}.  This connexion was exploited
in Ref.~\cite{Kamano:2012id} to obtain the forward neutrino cross
sections using a dynamical coupled channel model that successfully
fits a large set of $\pi N$ and $\gamma N$ data. The comparison with
the Rein-Sehgal model, given in Fig.~3 of
Ref.~\cite{Nakamura:2013zaa}, shows a clear disagreement with the more
realistic description, underestimating the $\Delta$ peak but
overestimating the higher $W_{\pi N}$ region.

Background non-resonant terms, required and totally fixed at threshold
by chiral symmetry, were evaluated in
\cite{Hernandez:2007qq}\footnote{Some background terms were also
  considered in Refs.~\cite{Fogli:1979cz, Fogli:1979qj} and
  \cite{Sato:2003rq}. In the latter reference, the chiral counting was
  broken to account explicitly for $\rho$ and $\omega$ exchanges in
  the $t-$channel, while the first two works are not fully consistent
  with the chiral counting either, since contact terms were not
  included, and use a rather small axial mass ($\sim $ 650 MeV) for
  the $N\Delta$ transition form factor.}. The background terms in
\cite{Hernandez:2007qq} are the leading contributions of a SU(2)
chiral Lagrangian supplemented with well known form factors in a way
that respects both conservation of vector current (CVC) and PCAC.  The
interference between the $\Delta$ and the background terms produces
parity-violating contributions to the pion angular differential cross
section, which are intimately linked to $T-$odd
correlations\footnote{However, these correlations do not imply a
  genuine violation of time-reversal invariance because of the
  existence of strong final re-scattering effects.} which would be
interesting to measure. The model of Refs.~\cite{Hernandez:2007qq,
  Hernandez:2013jka} is diagrammatically shown in the right panel of
Fig.~\ref{fig:coh0} (note that the $D_{13}$ contributions were not
considered originally in \cite{Hernandez:2007qq}).  Vector form
factors are taken from fits to empirical helicity
amplitudes~\cite{Lalakulich:2005cs}. The axial $N\Delta$ transition is
parametrized in terms of four form factors $C_{3,4,5,6}^A(q^2)$, as in
\cite{LlewellynSmith:1971zm}. Among the axial form factors the most
important contribution comes from $C^A_5$, whose
  numerical value is related to the pseudoscalar form factor $C^A_6$
  by PCAC.  
Moreover, in the massless lepton
  limit, the direct $\Delta$ pole term gives
\begin{equation}
\frac{d\sigma}{dq^2} \propto \left \{ [C_5^A(0)]^2 + q^2 a(q^2) \right \} \,,
\end{equation}
with $C_{3,4}^A(0)$ contributing to $a(q^2)$, i.e. to ${\cal
O}(q^2)$, which also gets contributions from vector form factors and
terms proportional to $dC_5^A/dq^2\big|_{q^2=0}$.
For the sub-leading $C_{3,4}^A$ form factors, the Adler's
  parametrizations~\cite{Adler:1968tw, Bijtebier:1970ku} were adopted
  ($C_3^A=0, C_4^A=-C_5^A/5$), while the dominant axial ($C^A_5
  (q^2)$) form-factor was fitted to the flux averaged $\nu_\mu p \to
  \mu^- p\pi^+$ ANL $q^2-$differential cross section
  data~\cite{Radecky:1981fn}, finding a correction of the order of
  30\% to the off diagonal Goldberger-Treiman relation (GTR)
  prediction $C_5^A(0) \sim 1.15$. Considering electroproduction
  experiments as benchmark, the model provides an accurate description 
  of the data up to pion-nucleon invariant masses of the order of 
   $W_{\pi N} < 1.4$ GeV~\cite{Lalakulich:2010ss}.

Deuterium effects and BNL data were not taken into account in the
analysis carried out in \cite{Hernandez:2007qq}. It is well known that
there exists some tension between ANL and BNL $p\pi^+$ data samples.  It has 
become a relevant issue 
 that the more than 30-year old ANL and BNL low
statistics deuterium pion production data are  still nowdays the best
source of information about the $N\Delta$ transition matrix
element.  The authors of \cite{Graczyk:2009qm} made a simultaneous
fit, considering only the $\Delta-$mechanism, to both ANL and BNL data
including separate overall flux normalization uncertainties for each
experiment.  The main conclusion was that ANL and BNL data are in fact
consistent only when these systematic uncertainties are taken into
account. This strategy was followed in \cite{Hernandez:2010bx}, where
the $N\Delta$ axial transition form-factors within the model derived
in \cite{Hernandez:2007qq} were simultaneously fitted to both bubble
chamber data sets. Deuterium effects were also taken into account. As
a result of this improved analysis, a value of $1.00\pm 0.11$ for
$C_5^A(0)$, 2$\sigma$ away from the GTR estimate\footnote{The approach
  does not satisfy Watson theorem; the inclusion of constraints
  derived from this latter requirement could diminish the
  discrepancies with the GTR prediction.}, was
found~\cite{Hernandez:2010bx}. This model, which includes a
non-resonant chiral background, was further refined in
\cite{Hernandez:2013jka} by adding a new resonance ($D_{13}(1520)$)
aiming to extend the model to higher energies above the $\Delta$
resonance region for which it was originally developed. 
Thus, this scheme emerges as one of  the most adequate ones, 
from a  theoretical perspective, to analyze pion
production data and neutrino energies up to 1 or 1.2 GeV.

Another theoretical description is based on a dynamical
model of photo-, electro- and weak pion production~\cite{Sato:2003rq}. Starting
from an effective Hamiltonian with $N\Delta$  couplings obtained
with the constituent quark model (30 \% below
the measured ones), the $T$  matrix is obtained by solving
the Lippmann-Schwinger equation in coupled channels.
In this way the bare couplings get renormalized by meson
clouds. The predicted cross sections  are in good agreement with
data (Figs. 5-8 of Ref.~\cite{Sato:2003rq}).

More recently, a Lorentz-covariant effective field theory scheme that
contains nucleons, pions, $\Delta$, isoscalar scalar ($\sigma$) and
vector ($\omega$) fields, and isovector vector ($\rho$) fields
consistent with chiral symmetry has been also employed to study the
neutrino-production of pions from nucleons~\cite{Serot:2012rd}. At low
neutrino energies, below 500 MeV, the convergence of the
power-counting scheme used in \cite{Serot:2012rd} is fast and
next-to-leading-order tree-level corrections are found to be small.
To go beyond this energy regime, the authors of \cite{Serot:2012rd} use
phenomenological form factors. Nevertheless, they  are
mostly concerned with the $E_\nu < 0.5$ GeV region, where a
satisfactory agreement with ANL data is obtained.

Finally, we just mention the approach of Ref.~\cite{Paschos:2011ye} entirely based on
PCAC and valid only at low energies and in  the small $q^2$ region.

\subsection{Pion cross sections in nuclei and the MiniBooNE
puzzle}
\label{sec:pion-inela}

The main contribution to MiniBooNE data comes from $^{12}$C and this
poses an extra problem to theoretical calculations because the
in-medium modifications of the production mechanisms, and the FSI
effects on the produced pions are important. 
As in QE scattering, pion production in nuclei is affected by the
description of the initial nucleus. Although the most common
approximation adopted for resonance production, and inelastic
scattering in general, on nuclear targets is the Fermi gas in its
global~\cite{Kim:1996bt} and local~\cite{Singh:1998ha} versions, more
precise descriptions based on realistic spectral
functions~\cite{Benhar:2005dj} or bound-sate wave
functions~\cite{Praet:2008yn} have been developed. The integrated cross sections
obtained with Fermi gas models are very similar to those from
sophisticated approaches (see for instance Fig.~7 of
Ref.~\cite{Praet:2008yn}). This reflects the fact that at the higher
energy transfers present in inelastic processes, the details of nuclear structure
are less relevant. Given the prevalent role of the
$\Delta(1232)$ excitation in pion production, it is not surprising
that the in-medium modification of its properties represents the most
important nuclear effect, as already stressed in the early work of
Refs.~\cite{Kim:1996bt,Singh:1998ha} and illustrated below.

On the other hand, FSI takes into account that pions can be absorbed
in their way out of the nucleus, and can also suffer different
quasielastic collisions that modify their energy, angle, and charge
when they come off the nucleus.  For instance, in the case of NC
$\pi^0$ production, signal events originate mostly from a NC1$\pi^0$
primary interaction with a $\pi^0$ not being affected by FSI, but also
from a NC1$\pi^+$ primary interaction with the $\pi^+$ being
transformed into $\pi^0$ in a charge exchange FSI reaction. In this
particular case, an additional difficulty in interpreting the
NC$\pi^0$ production comes from the presence of a coherent
contribution. FSI definitely alters the signature of the event and
thus the correct simulation of pion production requires a model not
only able to describe the elementary reactions (discussed in the
previous subsection), but also the final state interactions.

\begin{figure}[ht]
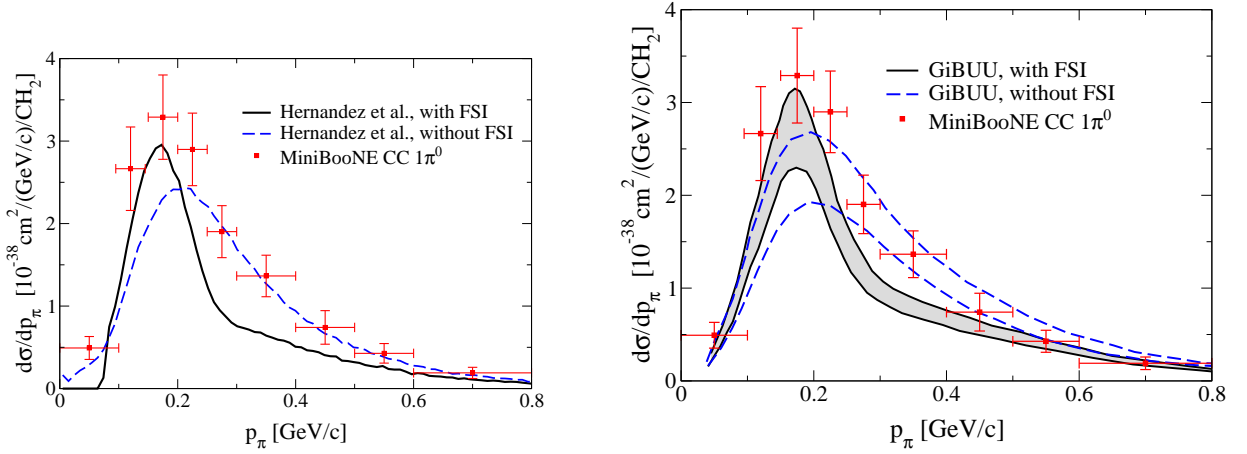

\begin{center}
\makebox[0pt]{\includegraphics[width=0.4\textwidth]{1ccpi0mom.eps}\hspace{1.cm}\includegraphics[width=0.45\textwidth]{1ccpi0mom.mosel.eps}}
\end{center}
\caption{MiniBooNE flux-folded differential $d\sigma/dp_\pi$ cross
  section for CC 1$\pi^0$ production by $\nu_\mu$ in mineral oil. Data
  are from Ref.~\cite{AguilarArevalo:2010xt}. Left: Predictions from the
  cascade approach of Ref.~\cite{Hernandez:2013jka}. The solid
  curve corresponds to the full model and the dashed one stands for
  the results obtained neglecting FSI effects. Right: Predictions from
  the GiBUU transport model of Ref.~\cite{Lalakulich:2012cj}. The
  dashed curves give the results before FSI, the solid curves those
  with all FSI effects included. Two different form factors
  $C_5^A(q^2)$, tuned to the ANL and BNL data-sets have been employed
  and give rise to the systematic uncertainty bands displayed in the
  figure.  }
\label{fig:coh0bis}
\end{figure}

MiniBooNE presented their results as measurements of final states with
only one pion, with the appropriate charge, and no other mesons. A
variety of flux integrated differential cross sections were reported
in final pion momentum or scattering angle. For the case of CC
reactions, total cross sections as a function of the reconstructed
neutrino energy were provided as well. MiniBooNE data show substantial
 discrepancies (enhancement) with respect to the  Rein-Sehgal
interaction~\cite{Rein:1980wg} prediction with FSI effects as implemented in
the NUANCE event generator~\cite{Casper:2002sd}. Some
important discrepancies still remain when these measurements are
compared with the most comprehensive
approaches~\cite{Hernandez:2013jka, Lalakulich:2012cj} available in
the literature until now, as highlighted in the case of CC 1$\pi^0$ in
Fig.~\ref{fig:coh0bis}. Both theoretical approaches start from a quite
complete microscopic description of the pion production on the nucleon
at MiniBooNE energies (\cite{Hernandez:2007qq, Hernandez:2010bx,
  Hernandez:2013jka} and \cite{Leitner:2008ue}, respectively) and
incorporate a number of standard nuclear medium effects in the initial
interaction model (Pauli Blocking, $\Delta(1232)-$spreading potential,
etc..). 

The work of Ref.~\cite{Lalakulich:2012cj} uses the Giessen
Boltzmann-Uehling-Uhlenbeck (GiBUU) model to account for FSI
effects. It is a transport model where FSI are implemented 
by solving the semi-classical Boltzmann-Uehling-Uhlenbeck equation. 
It describes the dynamical evolution of the phase space density for 
each particle species under the influence of the mean field 
potential, introduced in the description of the initial nucleus state. 
Equations for various particle species are coupled through this mean 
field and also through the collision term. GiBUU  provides a unified framework 
for nucleon--, nucleus--, pion--, electron-- and neutrino interactions with nuclei, 
from around a hundred MeV to tens of GeV.  It has been extensively and successfully
used~\cite{Buss:2011mx} in the last years, 
with special attention to pion- and photo-nuclear  reactions. 
The study of \cite{Lalakulich:2012cj} is limited only to the
incoherent part of the CC induced reaction. The comparison with
the MiniBooNE NC$\pi$ data was presented and discussed at the NUFACT
and NUINT conferences ~\cite{Leitner:2009de,Leitner:2009ec}. Previous
work of the group could be found in Refs.~\cite{Leitner:2008ue, Leitner:2008wx}.

Coherent contributions, when relevant, are included in the work of
Ref.~\cite{Hernandez:2013jka} from the microscopic approach of
Refs.~\cite{Amaro:2008hd, Hernandez:2010jf}. This is based on the same
pion production model on
nucleons~\cite{Hernandez:2007qq,Hernandez:2010bx} 
employed in ~\cite{Hernandez:2013jka}. On the other hand
\cite{Hernandez:2013jka} makes use of the LDA to evaluate the
incoherent production on finite nuclei.  This approximation, also part of
 the GiBUU model, turns out to be quite convenient when using a
cascade algorithm, as we will see. The work of
Ref.~\cite{Hernandez:2013jka} is based on the many body scheme set up
in \cite{Nieves:2004wx, Nieves:2011pp} to study inclusive
neutrino--nucleus reactions. It establishes a systematic many body
expansion of the gauge boson absorption modes that includes one, two
and even three body mechanisms, as well as the excitation of $\Delta$
resonances and pion production.

To compute the incoherent pion production on a nucleus, one should sum
the nucleon cross section over all nucleons in the nucleus.  For a
neutrino CC process (for antineutrino or NC induced reactions the
discussion is similar) within the LDA one gets for initial pion
production (prior to any pion FSI)~\cite{Hernandez:2013jka}
\begin{equation}
 \frac{d\sigma}{d|\vec{k}\,|4\pi  r^2\,dr\,d\cos\theta_\pi\, dE_\pi}=\Phi(|\vec k |)\,
 \hspace{-.15cm}
  \sum_{N=n,p}
 2\int \frac{d^3p_N}{(2\pi)^3}\
\theta(E_F^N(r)-E_N) \,\theta(E_N+q^0-E_\pi-E_F^{N'}(r))
\frac{d\hat\sigma(\nu N\to l^-N'\pi)}{d\cos\theta_\pi dE_\pi}
\label{eq:eq1}
\end{equation}
with $E_F^N(r)=\sqrt{M^2+(p_F^N(r))^2}$, given in terms of the local
Fermi momentum $p_F^N(r)$ defined in Eq.~(\ref{eq:localFermi}). The
step functions implement Fermi motion and Pauli blocking. Besides, $\Phi(|\vec k|)$ is
the neutrino flux as a function of the incoming-neutrino energy $E_\nu
\equiv |\vec k|$; $\hat\sigma(\nu N\to l^-N'\pi)$ is the cross
section at the nucleon level modified by medium effects, which in this
case consist in the modification of the $\Delta(1232)$ spectral
function. The $\Delta$ properties are strongly modified in the nuclear
medium~\cite{Oset:1987re}, and since the direct $\Delta$-mechanism is
dominant, a correct treatment is needed for $\pi$ production inside a
nucleus. This is accounted for, both in the works of
Refs.~\cite{Hernandez:2013jka} and \cite{Lalakulich:2012cj}, using a
realistic spreading potential ($\Delta-$selfenergy). In the nuclear
medium, on one hand, the width is reduced due to Pauli blocking, but
on the other hand, it is increased by the collisions inside the
nucleus.  For example, via the processes $\Delta N \to NN$ and $\Delta
NN \to NNN$, the $\Delta$ can disappear without producing a
pion. Secondary pion production is also possible, namely via the
process $\Delta N \to \pi NN$. These processes contribute to the in
medium $\Delta$ width that generally becomes larger than in the free
space. 

The in medium differential cross section of Eq.~(\ref{eq:eq1})
is used in a simulation code to generate, at a given point $\vec r$
inside the nucleus and by neutrinos of a given energy, on-shell pions
with a certain momentum. These pions are followed through their path
across the nucleus. To evaluate these FSI effects, the authors of
~\cite{Hernandez:2013jka} follow the approach of
Ref.~\cite{Salcedo:1987md}, where a computer simulation code was
developed to describe inclusive pion nucleus reactions (quasielastic,
single charge exchange, double charge exchange, and absorption). The
$\pi N$ interaction is dominated by the $\Delta$ resonance excitation,
modified in the nuclear medium in the same way as it was modified in
$\hat\sigma(\nu N\to l^-N'\pi)$. The different contributions to the
imaginary part of its self-energy account for pion, two- and
three-nucleon absorption and quasielastic processes. The probabilities
for the different processes are evaluated in nuclear matter as a
function of the density, then the LDA prescription is used to
obtain results in finite nuclei. After a quasielastic event, pions
change momentum and may change its electric charge. The probability
for charge exchange and the final momentum distribution after a
quasielastic interaction were computed in
Ref.~\cite{Salcedo:1987md}. That information is used in the simulation
program to generate the pion resulting from such a collision.

The model of Ref.~\cite{Hernandez:2013jka} provides an overall
acceptable description of MiniBooNE data, better for NC than for CC
channels, although the theoretical predictions are systematically
below data. Differential cross sections, folded with the full neutrino
flux, show that most of the missing pions lie in the forward direction
and at high energies. An example of this is shown in
Fig.~\ref{fig:coh0bis} for CC$1\pi^0$. FSI effects are clearly visible in the
distribution. Because of the FSI some pions are absorbed, but other
ones are scattered and loose to nucleons part of their energy. FSI is
essential to fill the low momentum part of the distribution. On the
other hand, the combined effect of quasielastic scattering and pion
absorption through $\Delta$ excitation, depletes the $p_\pi= 250 \sim
450$ MeV region producing a distortion of the differential cross
section shape that significantly worsens the description of the data. The
artificial exclusion of the FSI effects leads to a better description of the high
momentum tail of the $d\sigma/dp_\pi$ distribution.  

These findings are fully supported by the results obtained within the
GiBUU scheme of Ref.~\cite{Lalakulich:2012cj}. Actually both
approaches produce quite similar results, as can be seen in
Fig.~\ref{fig:coh0bis} for the particular case of CC$1\pi^0$. The
authors of \cite{Lalakulich:2012cj} also pointed out that MiniBooNE seems to
suggest that the higher elementary BNL data for pion production are
correct and that the ANL data underestimate the elementary production
cross section.  Nevertheless, the discrepancy with theory is not only
due to an overall normalization factor, but as stressed above, the
experimental shape of the pion momentum distributions considerably
differ from those predicted by both approaches. However the theoretical
model calculations~\cite{Lalakulich:2012cj, Gil:1997jg} turn out to be
in agreement with experimental results for the photo-production of pions on
nuclei, predicting a suppression in the pion spectra around the
$\Delta$ resonance region which is not seen in the MiniBooNE neutrino data.

FSI effects could be reduced by considering the so called “formation
zone”, that among other effects includes the propagation of the
$\Delta$ before decaying into a $\pi N$ pair. The NuWro MC event
generator~\cite{Juszczak:2005zs, Juszczak:2009qa} includes the concept
of formation zone, and its predictions for NC$1\pi^0$ have been
compared in \cite{Golan:2012wx} to MiniBooNE
data~\cite{AguilarArevalo:2009ww}. The overall agreement is quite
satisfactory\footnote{To describe the pion production on the nucleon,
  a simple theoretical model, that does not consider background terms,
  is used in \cite{Golan:2012wx}. Moreover, the coherent pion
  production is calculated using the Rein-Sehgal model
  \cite{Rein:1982pf}, which is not appropriate at low energies (this
  will be discussed in detail in Sect.~\ref{sec:cohpi}).}, better
for neutrino induced reactions than for antineutrino ones.  Shapes of
the distributions of final state $\pi^0$'s are affected by an
interplay between pion FSI such as absorption and formation time
effects, and turn out to agree significantly better with the data than those found
in Refs.~\cite{Hernandez:2013jka,Lalakulich:2012cj, Leitner:2009de,Leitner:2009ec}.

Of course, the “formation zone” could be adjusted to reproduce
data. However, these kind of modifications of the FSI could be difficult to
justify, as they might be in conflict with much other
phenomenology; it might  hide our ignorance on the 
relevant dynamics. Even when they could help
reproducing some observables, if we lack of a correct understanding of 
the physical mechanisms responsible for them,
they might lead to wrong predictions for other observables sensible 
to other kinematics, dynamical mechanisms or nuclear corrections.

In this respect, the DUET experiment at TRIUMF, that uses the PIA$\nu$O
detector~\cite{piano1}, will provide quite valuable information in
the future. Pion FSI models were tuned to
 available pion--nucleus data measured in the 1980's and affected by large
uncertainties.  The objective of the experiment  is to measure $\pi$
absorption and $\pi$ charge exchange cross section with $\sim$ 10\%
and $\sim$ 20\% accuracy, respectively.

Another useful MiniBooNE measurement was the ratio of CC1$\pi^+$-like
(one pion in the final state) to CCQE-like (no pions in the final
state) cross-sections on CH$_2$, as a function of the neutrino
energy. This ratio was reported in \cite{AguilarArevalo:2009eb} with
an accuracy of around 10\% (The K2K Collaboration has also measured
this ratio~\cite{Rodriguez:2008aa}, but the reported errors are much
larger than those in the MiniBooNE data). This ratio becomes quite
interesting because of the apparent data/MC normalization discrepancy
for the CC$1\pi^+$ production channel at
MiniBooNE~\cite{AguilarArevalo:2010bm}, since it would be free from
the overall flux normalization uncertainty. However, it is not a
directly observable quantity, because in the experimental analysis it
is necessary to reconstruct the neutrino energy. This measurement puts
constraints on the theoretical models which include QE, $\Delta$
excitation and MEC/2p2h dynamics. In addition, the theoretical
approaches need to include FSI effects as well.  There exist three
theoretical estimates of this ratio: i) The GiBUU group found a
significant discrepancy between the model and the MiniBooNE data,
being its prediction smaller than the experimental
ratio~\cite{Leitner:2008wx}, ii) NuWro MC
results~\cite{Graczyk:2009qm} are slightly below the MiniBooNE data
for larger neutrino energies and iii) the Aligarh
group~\cite{SajjadAthar:2009rc} predictions\footnote{In what concerns
  the initial interaction, this approach
  treats the $\Delta$  inside of the nucleus in a way that has many
  resemblances to the scheme of Ref.~\cite{Hernandez:2013jka} and also
  accounts for FSI effects by means of a cascade algorithm. However,
  the cascade  uses free space $\pi N\to \pi' N'$ cross sections instead of cross
  sections appropriately modified in the nuclear medium. Moreover, the
  model did not include contributions from the non-resonant background
  and from higher resonances, and include an unexpectedly large $\pi^+$
  coherent production cross section.}  agree with MiniBooNE
measurement for $E_\nu < 1$ GeV and are below MiniBooNE data for
larger neutrino energies.

However, one cannot extract any robust conclusion from these
comparisons, since none of the theoretical estimates for the CCQE-like
cross section include multinucleon mechanisms (2p2h)  effects or properly
compute RPA corrections. Moreover, as discussed in
Sec~\ref{subsec:erec}, because of the 2p2h, 
the algorithm used to reconstruct the neutrino energy is not
adequate when dealing with quasielastic-like events, and a distortion
of the total flux-unfolded cross-section shape is produced
(redistribution of strength from high to
low energies, which gives rise to a sizable excess [deficit] of low
[high] energy neutrinos in QE-like distributions)~\cite{Nieves:2012yz}.

Forthcoming T2K and MINER$\nu$A pion production data will hopefully
shed light into the existing puzzle originated by the large
discrepancy between MiniBooNE pion production measurements and
theoretical model predictions. It is worth noting that the GiBUU group
has already studied pion production at the T2K and MINER$\nu$A
experiments within its model~\cite{Lalakulich:2013iaa,
  Mosel:2014lja}. It is found that pion absorption is less pronounced
at the MINER$\nu$A energies than for MiniBooNE/T2K experiments. This
is attributed in \cite{Mosel:2014lja} to a minimum in the $\pi N$
cross section at around 0.7 GeV pion kinetic energy.

\subsection{Other inelastic processes}
\la{sec:OtherInel}

\subsubsection{Weak $K$ and $\bar K$ production}
\la{subsec:kaonprod}

Although in the few-GeV region, the attention has been focused on pion production because it is the inelastic process with the largest cross section, (anti)kaon, and strangeness production in general, are also relevant. In particular, kaon production induced by atmospheric neutrinos is a potential background for the proton decay mode with a kaon in the final state  ($p \raw \bar{\nu} \, K^+$), which has large branching ratios in different theories beyond the Standard Model. Several neutrino oscillation experiments like MINOS, NO$\nu$A, T2K or LBNE have or will have fluxes that extend to energies where strange particles can be produced. Therefore, a better understanding of these reactions is important to reduce systematic errors, which will be the dominant ones in the  era of precise neutrino oscillation measurements.  Such a progress is even more urging for experiments running in the  $\bar{\nu}$ mode as no measurements of strange-particle production cross sections exist with $\bar{\nu}$ fluxes.  

The main reaction channels at these energies are single $K$ production ($\Delta S =1$)
\bea
\nu_l \, p &\raw&  l^- \, K^+ \, p \,, \la{singleK_1} \\
\nu_l \, n &\raw&  l^- \, K^+ \, n \,, \la{singleK_2} \\
\nu_l \, n &\raw&  l^- \, K^0 \, p \,, \la{singleK_3} 
\eea
single $\bar K$ production ($\Delta S =-1$)
\bea
\bar \nu_l \, p &\raw&  l^+ \, K^- \, p \,, \la{singleaK_1} \\
\bar \nu_l \, n &\raw&  l^+ \, K^- \, n \,, \la{singleaK_2} \\
\bar \nu_l \, p &\raw&  l^+ \, \bar K^0 \, n \,, \la{singleaK_3} 
\eea
and associated strangeness production ($\Delta S =0$)
\bea
\nu_l \, n &\raw&  l^- \, K^+ \, (\Lambda,\Sigma^0) \,, \la{aso_1} \\
\nu_l \, n &\raw&  l^- \, K^0 \, \Sigma^+ \,, \la{aso_2} \\
\nu_l \, p &\raw&  l^- \, K^+ \, \Sigma^+ \,, \la{aso_3} \\
\bar \nu_l \, p &\raw&  l^+ \, K^+ \, \Sigma^- \,, \la{aso_4} \\
\bar \nu_l \, p &\raw&  l^+ \, K^0 \, (\Lambda,\Sigma^0) \,, \la{aso_5} \\
\bar \nu_l \, n &\raw&  l^+ \, K^0 \, \Sigma^- \,, \la{aso_6} \\
(\nu_l, \bar \nu_l ) \, p &\raw&  (\nu_l, \bar \nu_l)  \, K^+ \, (\Lambda,\Sigma^0) \,, \la{aso_7} \\
(\nu_l, \bar \nu_l)  \, n &\raw&  (\nu_l, \bar \nu_l)  \, K^+ \, \Sigma^- \,, \la{aso_8} \\
(\nu_l, \bar \nu_l)  n \,  &\raw&  (\nu_l, \bar \nu_l)  \, K^0 \, (\Lambda,\Sigma^0) \,, \la{aso_9} \\
(\nu_l, \bar \nu_l)  \, p &\raw&  (\nu_l, \bar \nu_l)  \, K^0 \, \Sigma^+ \,. \la{aso_10}
\eea
Associated strangeness production is the dominant process but it has a high threshold  because both a kaon and a hyperon are produced. This leaves single kaon production as the main source of kaons at low energies, in spite of being Cabibbo suppressed. Owing to the absence of flavor-changing NC in the Standard Model, $\Delta S =\pm 1$ reactions proceed via CC interactions. Channels (\ref{singleK_1}), (\ref{singleK_2}), (\ref{singleaK_1}) and (\ref{singleaK_2}) make coherent $K$ ($\bar K$) production in nuclei induced by $\nu$($\bar \nu$) possible. These are discussed in Sec.~\ref{subsec:CohK}. 

\begin{figure}[htb!]
\begin{center}
\begin{minipage}{0.53\textwidth}
\includegraphics[width=\textwidth]{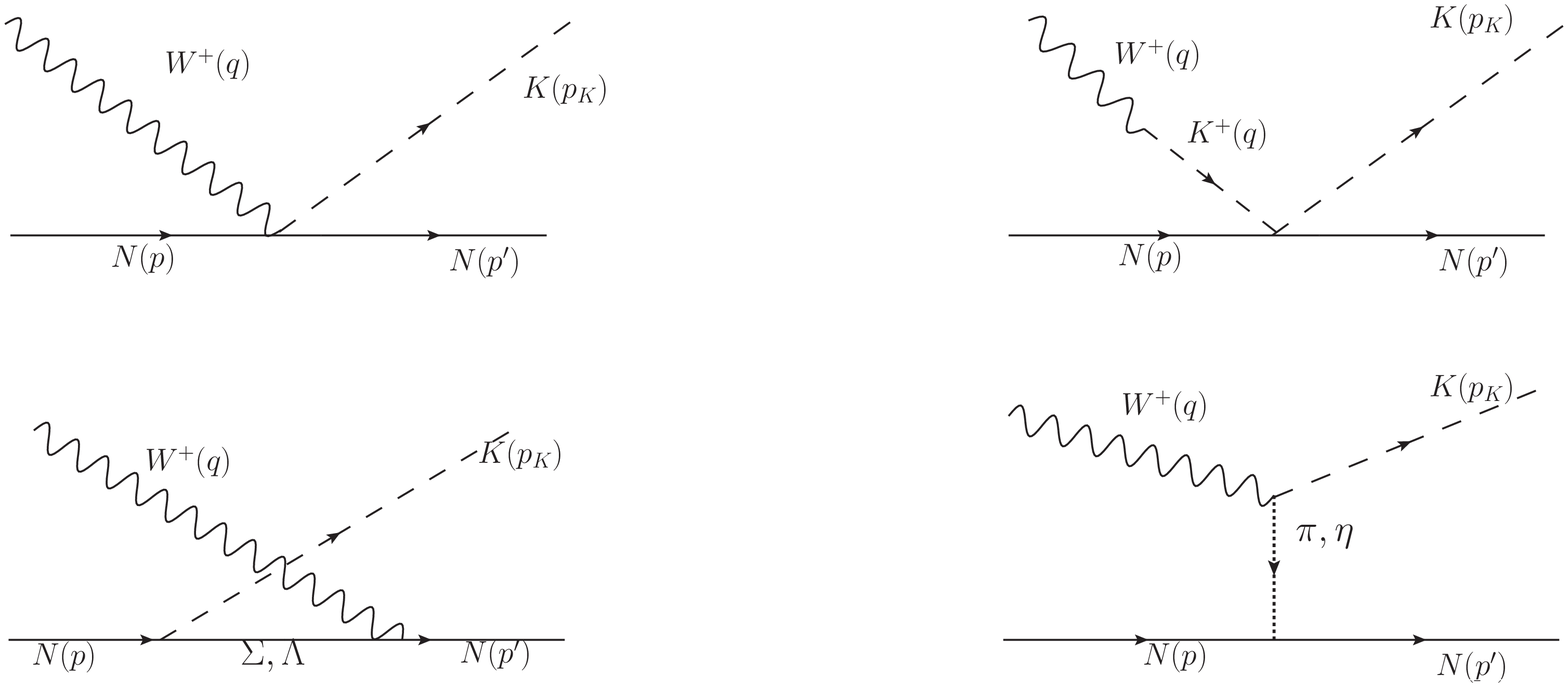}
\end{minipage}
\hfill
\begin{minipage}{0.45\textwidth}
\includegraphics[width=\textwidth]{kaoncs}
\end{minipage}
\caption{\label{fig:diags}
 Left panel: Feynman diagrams for $W^+ N\to  N K$~\cite{RafiAlam:2010kf}. From the
  upper left corner in clockwise order: contact term (CT), Kaon pole (KP),
  $\pi$ and $\eta$ in flight ($\pi$P, $\eta$P) and u-channel hyperon exchange (Cr$\Sigma$, Cr$\Lambda$) terms. Right panel: Integrated cross section for the reaction of Eq.~(\ref{singleK_1}) with $l=\mu$ and the contribution of different terms singled out. The band corresponds to a 10\%  error in $M_F$~\cite{RafiAlam:2010kf}.}
\end{center}
\end{figure}

After the early work of Ref.~\cite{Dewan:1981ab}, new theoretical developments on $\Delta S =1$ reactions (\ref{singleK_1}-\ref{singleK_3}) have become available only recently~\cite{RafiAlam:2010kf}. In the description of Ref.~\cite{RafiAlam:2010kf}, the reaction mechanisms are derived from a Lagrangian that implements the QCD chiral symmetry breaking pattern. Although the vertices are SU(3) symmetric, this flavor symmetry is broken in the amplitudes by the physical hadron masses. The resulting set of diagrams for the hadronic currents are shown in Fig.~\ref{fig:diags} and are refereed as contact (CT),  kaon pole (KP), u-channel crossed $\Sigma$ (Cr$\Sigma$) and $\Lambda$ (Cr$\Lambda$), pion in flight ($\pi$P) and eta in flight ($\eta$P) terms. The absence of $S=1$ baryons implies that there are no s-channel amplitudes.  The structure of these currents and the corresponding cross sections at threshold are fully determined by chiral symmetry, with couplings fixed from nucleon and hyperon semileptonic decays and pion decay. Some corrections of next order, whose couplings are constrained by measured values of nucleon magnetic moments, have also been included. PCAC is implemented for the axial part of the currents.  As the dependence of the different terms of the hadronic current on the momentum transferred to the nucleon is poorly known, if at all, the authors of Ref.~\cite{RafiAlam:2010kf} adopt a global dipole form factor 
\be
\la{eq:kaonFF}
F(q^2)=\left(1-\frac{q^2}{M_F^2} \right)^{-2} \,,
\ee
assuming $M_F=1$~GeV. 
 
The results obtained in Ref.~\cite{RafiAlam:2010kf} for the integrated cross sections are shown in Fig.~\ref{fig:diags} for the reaction of Eq.~(\ref{singleK_1}). In the validity region assumed for the model ($E_\nu \leq 2$~GeV), the CT amplitude, not included in Ref.~\cite{Dewan:1981ab}, is dominant and interferes destructively with the rest. The KP term is negligible, while the Cr$\Sigma$ contribution is much smaller than the Cr$\Lambda$ one because of the much smaller coupling in the strong vertex. The cross section of the reaction of Eq.~(\ref{singleK_3}) has similar size and exhibits much the same features. Instead, the reaction of Eq.~(\ref{singleK_2}) has a 4-5 times smaller cross section almost completely determined by the CT.   

\begin{figure}[htb!]
\begin{center}
\begin{minipage}{0.53\textwidth}
\includegraphics[width=\textwidth]{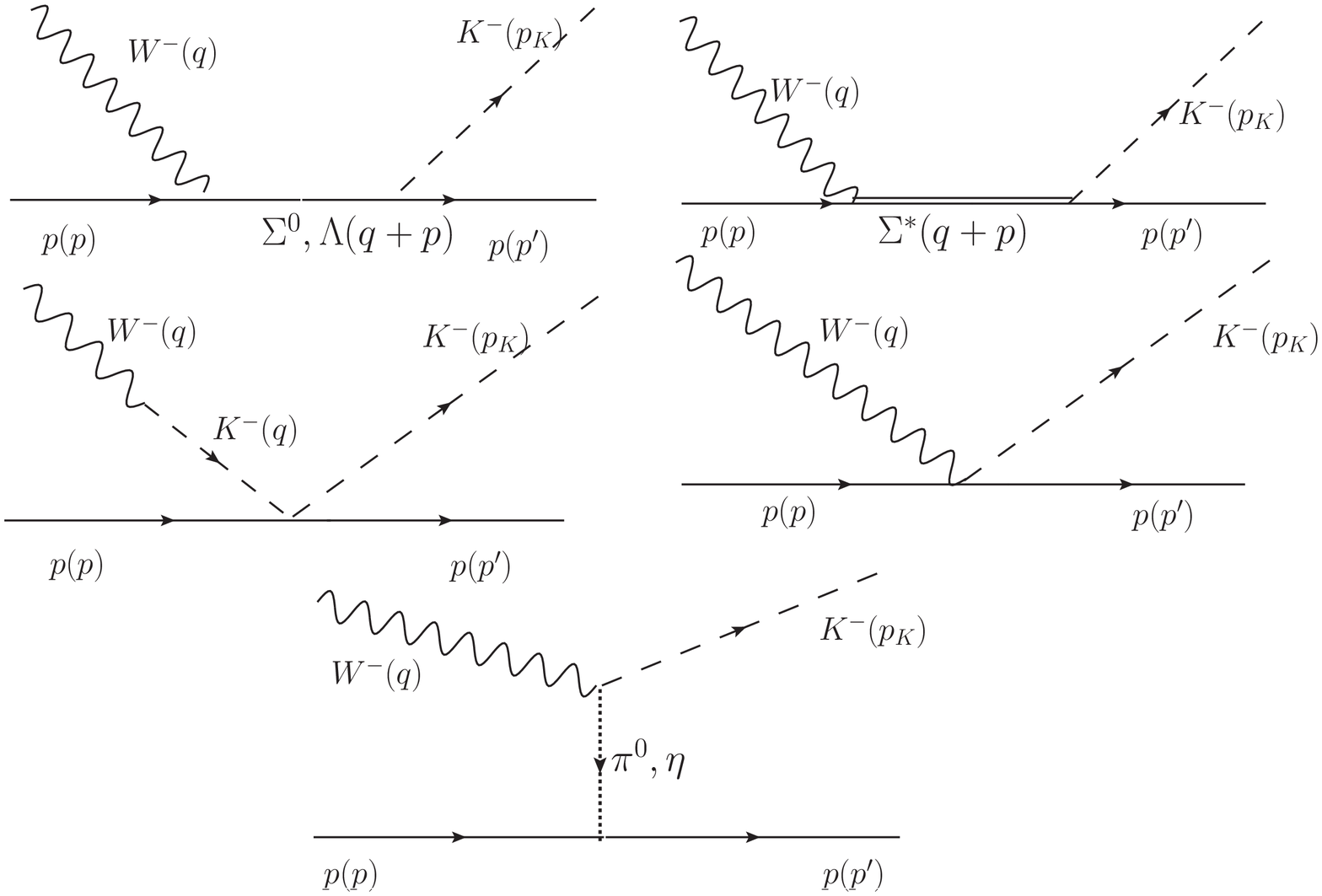}
\end{minipage}
\hfill
\begin{minipage}{0.45\textwidth}
\includegraphics[width=\textwidth]{akaoncs}
\end{minipage}
\caption{\label{fig:adiags}
 In the left panel the Feynman diagrams for $W^- N \to  N \bar{K}$~\cite{Alam:2011xq} are shown. First row:  s-channel $\Sigma, \Lambda$ and $\Sigma^*$ exchange terms; second row: contact (CT) and kaon pole (KP) terms; last row: $\pi$ and $\eta$ in flight ($\pi$P, $\eta$P) terms. The right panel shows the cross section for the reaction of Eq.~(\ref{singleaK_1}) with $l=\mu$ as a function of the neutrino energy together with the isolated contributions from different mechanisms. The band corresponds to a 10\%  error around $M_F=1$~GeV~\cite{Alam:2011xq}.}
\end{center}
\end{figure}

For the $\Delta S =-1$ reactions (\ref{singleaK_1}-\ref{singleaK_3}) close to threshold, the relevant mechanisms can also be obtained  from chiral SU(3) Lagrangians (see Fig~\ref{fig:adiags}). The CT, KP, $\pi$P and $\eta$P contributions to the hadronic current are present but now the $\Lambda$ and $\Sigma$ hyperons appear in the s-channel.  As for $K^+$ production, the structure of these amplitudes close to threshold is fully defined by chiral symmetry, with the couplings determined from semileptonic decays. In Ref.~\cite{Alam:2011xq}, the $q^2$ dependence of Eq.~(\ref{eq:kaonFF}) was also adopted. In pion production reactions, the excitation of the spin-3/2 $\Delta(1232)$ plays a dominant role at relatively low excitation energies ($\sim 200$~MeV). Therefore, the corresponding state of the baryon decuplet $\Sigma^*(1385)$ that couples to $N\bar{K}$ should be considered here. The vector and axial $N-\Sigma^*$ form factors, which are not known, were related to the better known $N-\Delta(1232)$ ones using SU(3) rotations~\cite{Alam:2011xq}.

As can be seen in Fig.~\ref{fig:adiags}, the CT provides the largest contribution to the cross section. The small impact of the $\Sigma^*$ resonance, contrasting with the dominance of the $\Delta$ in the pion case, can be explained by the fact that the $\Sigma^*$ is below the kaon production threshold~\cite{Alam:2011xq}. The other two channels (\ref{singleaK_2}) and (\ref{singleaK_3}) have similar cross sections~\cite{Alam:2011xq}. It should be mentioned that while the weak $K$ production model described above represents a theoretically solid prediction at threshold, the situation is different for the $\bar K$ channels. The presence of a baryon resonance, $\Lambda (1405)$, just below the $\bar K N$ threshold might have a non negligible influence on the weak $\bar K$ production cross sections that needs to be investigated. 

The same formalism has been applied to the study of CC associated strangeness production reactions, Eqs.~(\ref{aso_1}-\ref{aso_6}), at low (anti)neutrino energies. Again, it is found that the CT vertices, present in the leading order chiral Lagrangian and ignored in previous calculations~\cite{Shrock:1975an,Adera:2010zz} are responsible for most of the cross section~\cite{Alam:2013woa,Alam:2013vwa}. Channels (\ref{aso_2}) and (\ref{aso_4}) where CTs are not allowed (at leading order) show relatively smaller cross sections. However, we must point out there may be important resonant contributions to the $\Delta S =0$ processes, such as the $N^*(1535)$ below the $K \Lambda$ threshold, which have not been considered. This is known to be the case for associated strangeness photoproduction. Recently, a model that describes associated strangeness production induced by pions and photons has been used to predict the corresponding reactions induced by neutrinos in the forward direction by applying PCAC~\cite{Kamano:2012id}. 

The dynamics of strange-particle production in nuclear targets is considerably more involved. The interaction of kaons with the nuclear medium is not so strong due to the absence of baryon resonances but quasielastic and charge-exchange scattering are present. Furthermore, as the energy increases, inelastic processes like secondary kaon production $K \, N \raw K' \, N' \pi$ become sizable. Instead, the $\bar K$ interaction with the nucleons is strong starting from very low energies due to the presence of the $\Lambda (1405)$; $\bar K$ can disappear leading to hyperons via $\bar K \, N \raw \pi \, Y$, $\eta \, Y$ with $Y = \Lambda$, $\Sigma$. Due to FSI, (anti)kaons can be produced in secondary collisions such as $\pi \, N \raw Y \, K$, $K \, \bar K \, N'$ and $N \, N \raw N \, Y \,  K$. Indeed, the number of low-energy $K$ produced in neutrino-nucleus collisions is actually enhanced by FSI; in the case of $\bar K$, secondary interactions tend to compensate their absorption~\cite{Lalakulich:2012gm}.  For this reason, from the ongoing exclusive strangeness production measurements on nuclear targets at MINER$\nu$A, it will be very hard to extract the corresponding reactions on the nucleon, unless a very reliable modeling of the nuclear dynamics is implemented~\cite{Lalakulich:2012gm}.

\subsubsection{NC photon emission}
\la{subsec:NCgamma}

One of the possible inelastic reaction channels is photon emission induced by NC interactions (NC$\gamma$), which can take place on single nucleons
\be
\label{eq:reac_nucleon}
\nu (\bar{\nu})\, N \rightarrow \nu (\bar{\nu})\, \gamma \, N \,,
\ee
or on nuclear targets
\bea
\label{eq:reac_incoh}
\nu (\bar{\nu})\, A &\rightarrow& \nu (\bar{\nu})\, \gamma \, X  \, \\
\label{eq:reac_coh}
\nu (\bar{\nu})\, A &\rightarrow& \nu (\bar{\nu})\, \gamma \, A \,\\
\nu (\bar{\nu})\, A &\rightarrow& \nu (\bar{\nu})\, A'^* \, N \label{eq:reac_2step}
\raw \nu (\bar{\nu}) \, \gamma \, A'  \, N\,,
\eea
via incoherent [Eq.~(\ref{eq:reac_incoh})] or coherent  [Eq.~(\ref{eq:reac_coh})] scattering. It is also possible that, after nucleon knockout, the residual excited nucleus decays emitting $\gamma$ rays  [Eq.~(\ref{eq:reac_2step})]. This mechanism has been identified as an important source of low ($\sim 10$~MeV) photons for neutrinos of intermediate energies, whose main reaction mechanism is QE scattering~\cite{Ankowski:2011ei}.

Weak photon emission has a small cross section compared, for example, with pion production. Indeed, while pion production involves predominantly two weak vertices followed by a strong (resonant) decay, in NC$\gamma$ one has a much weaker electromagnetic vertex instead of the strong one. In spite of this, NC$\gamma$ turns out to be one of the largest backgrounds in appearance $\nu_\mu \rightarrow \nu_e$($\bar{\nu}_\mu \rightarrow \bar{\nu}_e$) experiments when photons with hundreds of MeV energies are misidentified as $e^\mp$ from CCQE scattering of $\nu_e (\bar{\nu}_e)$ in the detector. This was the case at MiniBooNE, where the photon background is estimated from the measured NC$\pi^0$ rate assuming that it comes form radiative decay of weakly produced resonances, mainly $\Delta \raw N \, \gamma$. The experiment has found an excess of events with respect to the predicted background in both $\nu $ and $\bar{\nu}$ modes. In the $\bar{\nu}$ mode, the data are consistent with $\bar{\nu}_\mu \raw \bar{\nu}_e$ oscillations and have some overlap with a previous  LSND result~\cite{Aguilar-Arevalo:2013pmq}. In contrast, the reconstructed neutrino-energy distribution ($E_\nu^{QE}$)  of $e$-like events in the $\nu$ mode is only marginally compatible with a two-neutrino oscillation model, showing an unexplained excess of events for $200 < E_\nu^{QE} < 475$~MeV~\cite{Aguilar-Arevalo:2013pmq,AguilarArevalo:2007it}. This puzzle triggered a theoretical interest in the NC$\gamma$ processes as an important background to the MiniBooNE measurement, which was not directly constrained by data.

\bfig[htb!]
\includegraphics[width=0.2\textwidth]{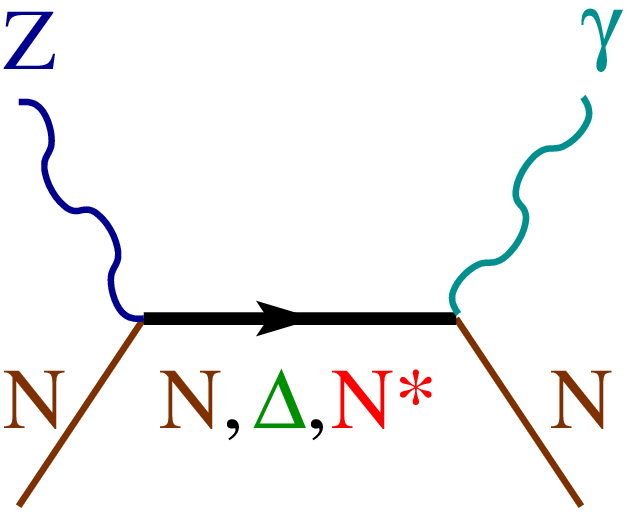}
\hspace{.05\textwidth} 
\includegraphics[width=0.2\textwidth]{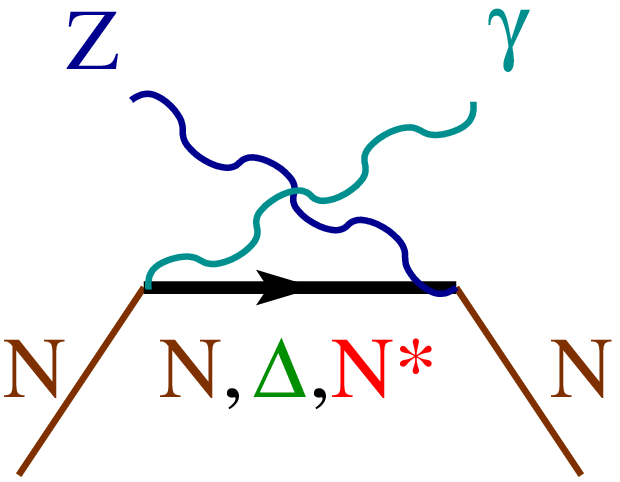}
\hspace{.05\textwidth} 
\includegraphics[width=0.2\textwidth]{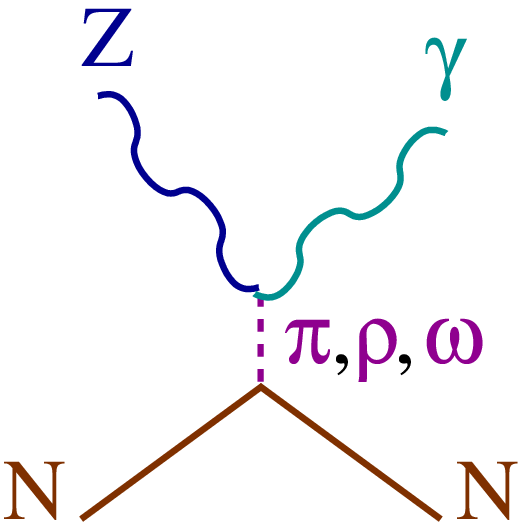}
\caption{\label{fig:NCgamma_diags}  Feynman diagrams for NC photon emission considered in the literature. The first two diagrams stand for direct and crossed baryon pole terms with nucleons and resonances in the intermediate state: $BP$ and $CBP$ with $B=N$, $\Delta(1232)$, $N^*(1440)$, $N^*(1520)$, $N^*(1535)$. The third diagram represents $t$-channel meson exchange contributions $mEx$ with $m=\pi$, $\rho$, $\omega$.}
\efig
Theoretical models for the reaction of Eq.~(\ref{eq:reac_nucleon}) in the few-GeV region have been developed in Refs.~\cite{Hill:2009ek,Zhang:2012xn,Wang:2013wva}. These calculations incorporate $s$- and $u$-channel amplitudes with nucleons and $\Delta(1232)$ in the intermediate state (see Fig.~\ref{fig:NCgamma_diags}). The structure of nucleon pole terms, $NP$ and $CNP$, at threshold is fully determined by the symmetries of the Standard Model. They are infrared divergent when the photon energy $E_\gamma \raw 0$ but this becomes irrelevant when the experimental detection threshold ($E_\gamma > 140$~MeV in the case of MiniBooNE) is taken into account. The extension towards higher energy transfers required to make predictions for the neutrino cross sections is performed by the introduction of phenomenological parametrization of the weak and electromagnetic form factors. The same strategy has been followed for the $\Delta P$ and $C \Delta P$ terms. As can be seen in Fig.~\ref{fig:NCgamma_nucleon}, where the NC$\gamma$ cross sections for the different mechanisms according to the model of Ref.~\cite{Wang:2013wva} are displayed, the $\Delta(1232)$ excitation followed by its radiative decay is the dominant mechanism. Heavier $N^*$ resonances, $P_{11}(1440)$, $D_{13}(1520)$ and $S_{11}(1535)$, were included as intermediate states in Ref.~\cite{Wang:2013wva}. The contribution from the $D_{13}(1520)$ on proton targets is sizable above $E_\nu \sim 1.5$~GeV. Instead, the other two $N^*$ are negligible (see Fig.~\ref{fig:NCgamma_nucleon}).   
\bfig[htb!]
\includegraphics[width=0.45\textwidth]{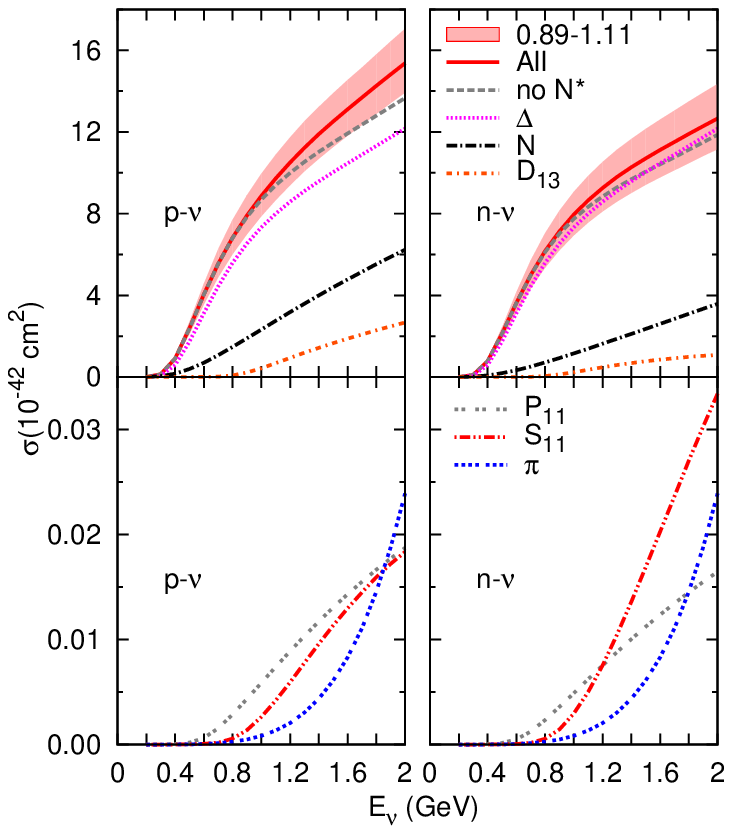}
\includegraphics[width=0.45\textwidth]{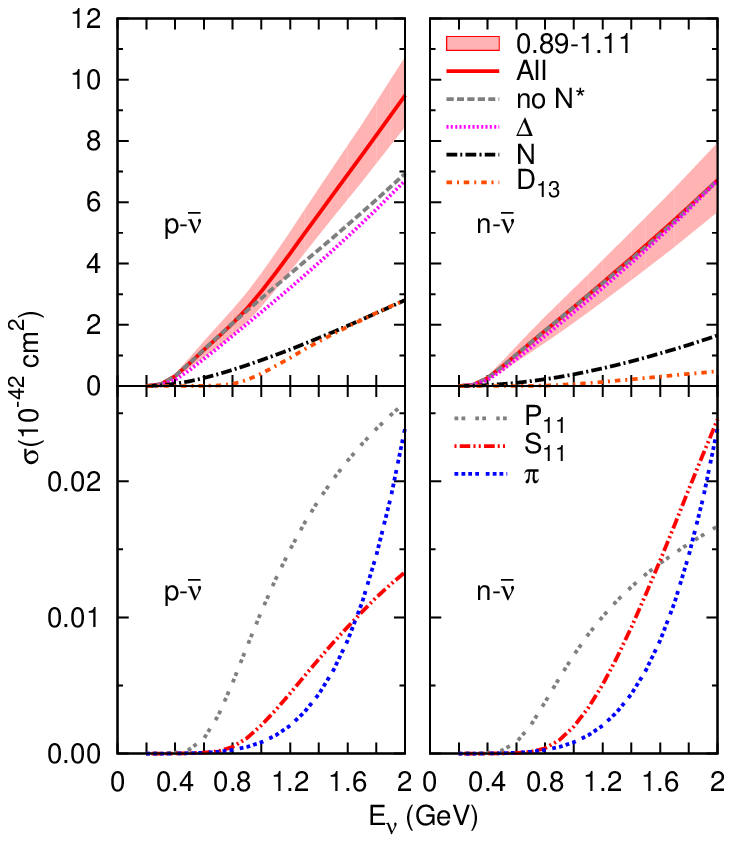}
\caption{\label{fig:NCgamma_nucleon} $\nu N \to \nu N \gamma$ (left) and $\bar\nu N
  \to \bar\nu N \gamma$ (right) cross sections on protons and neutrons as a function of the
  (anti)neutrino energy, obtained with the model of Ref.~\cite{Wang:2013wva}. 
A cut of $E_\gamma \geq 140$~MeV in the phase space integrals has been applied.
The error bands in the full-model results (solid lines) represent the uncertainty in the  
axial $N\Delta$ coupling $C^A_5(0)=1.00 \pm 0.11$ 
according to the determination of Ref.~\cite{Hernandez:2010bx}. The curves labeled $N$, 
$\Delta$, $D_{13}$, $P_{11}$ and $S_{11}$ stand for the partial contributions of the $BP$ and $CBP$ mechanisms 
of Fig~\ref{fig:NCgamma_diags}; the label $\pi$ corresponds to the $\pi Ex$ one.
The lines labeled as ``no $N^*$'' display the predictions without the $N^*$ contributions.}
\efig

 The pion pole ($\pi Ex$) mechanism originates from the $Z^0\gamma\pi$ vertex, which is fixed by the axial anomaly of QCD. It is nominally of higher order~\cite{Serot:2012rd} and, indeed, gives a very small contribution to the cross section as shown in the lower panels of Fig.~\ref{fig:NCgamma_nucleon}. Other $t$-channel mechanisms from the exchange of vector ($\rho Ex$) and pseudoscalar ($\omega Ex$) mesons~\cite{Hill:2009ek} arise from the anomaly-mediated $Z^0\gamma \rho$ and $Z^0\gamma \omega$ interactions~\cite{Harvey:2007rd}. Among them, the $\omega Ex$ contribution is favored by the size of the couplings. In addition, the isoscalar nature of the $\omega$ meson makes the $\omega Ex$ mechanism potentially interesting for the coherent scattering reaction of Eq.~(\ref{eq:reac_coh}). This is discussed in Sec.~\ref{subsec:CohGamma}. The $Z^0\gamma \omega$ vertex has been revisited using a framework which incorporates vector mesons as composite gauge bosons of the spontaneously broken hidden local symmetry~\cite{Harada:2011xx}. It is shown that this  vertex arises from the homogeneous part of the general solution to the anomaly equation and is not fully determined by the anomaly; the corresponding free parameters are related to the $\omega \raw \pi^0 \, \gamma$ decay.  Reference~\cite{Serot:2012rd} assumes that the $\rho Ex$ and $\omega Ex$ mechanisms, taken from Ref.~\cite{Hill:2009ek}, saturate the low-energy constants in the contact terms, although it is emphasized that other sources are possible. The contribution of these contact terms, and of the $\omega Ex$ in particular, to the NC$\gamma$ cross section on the nucleon is very small at $E_\nu \leq 550$~MeV~\cite{Hill:2009ek,Serot:2012rd}, as expected from power counting arguments. The extension to higher energies requires the introduction of poorly understood form factors~\cite{Hill:2009ek,Zhang:2012xn}. The cross section from these mechanisms increases fast with energy. This rapid growth might be a concern for experiments at higher energies, or with a high-energy tail in the neutrino flux (like T2K), as a source of NC$\gamma$ events and, therefore, unconstrained background. However, one should recall that this trend will be limited by unitarity bounds: in a realistic framework, these amplitudes will be modified by loop contributions and partially canceled by contact terms of even higher orders.

The incoherent NC$\gamma$ reaction, Eq.~(\ref{eq:reac_incoh}), on nuclear targets has been studied in Refs.~\cite{Zhang:2012aka,Wang:2013wva} using the relativistic local Fermi gas approximation. The broadening of the $\Delta$ resonance in the medium has also been  incorporated to the models using a spreading potential in Refs.~\cite{Zhang:2012aka,Zhang:2012xn}, while Ref.~\cite{Wang:2013wva} uses the parametrization of the imaginary part of the in-medium $\Delta$ selfenergy as a function of the local nuclear density derived in Ref.~\cite{Oset:1987re}. In the left panel of Fig.~\ref{fig:NCgamma_nucleus}, the cross sections on $^{12}$C obtained with the model of  Ref.~\cite{Wang:2013wva} are shown. It is clear that the  neglect of nuclear medium corrections is a poor approximation.  
By taking into account Fermi motion and Pauli blocking, the cross section already goes down by more than 10\%. With the full model the reduction is  of the order of 30\%. A similar net effect is obtained in Ref.~\cite{Zhang:2012xn}. However, the reduction quoted for the direct $\Delta$ mechanism is substantially larger for neutrinos ($\sim 50\%$)  but not for antineutrinos, which is hard to understand (see the comparison in Fig.~(9) of Ref.~\cite{Wang:2013wva} and the related discussion). The cross section shows an approximate scaling with the target mass (A), as can be seen in the right panel of Fig.~\ref{fig:NCgamma_nucleus}. Nevertheless, the cross section is smaller for heavier nuclei, particularly $^{208}$Pb.
\bfig[htb!]
\includegraphics[width=0.42\textwidth]{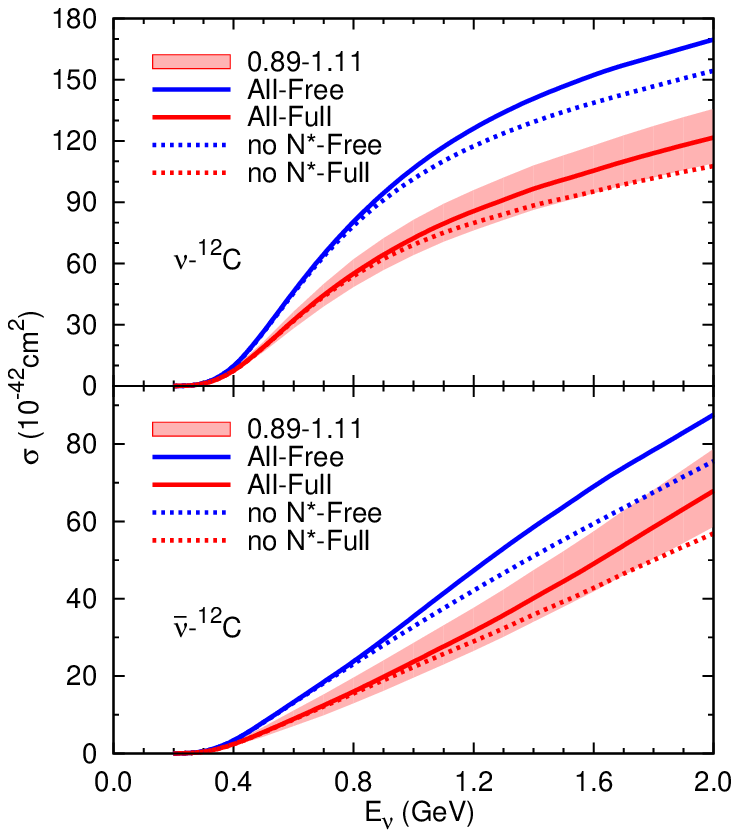}
\includegraphics[width=0.42\textwidth]{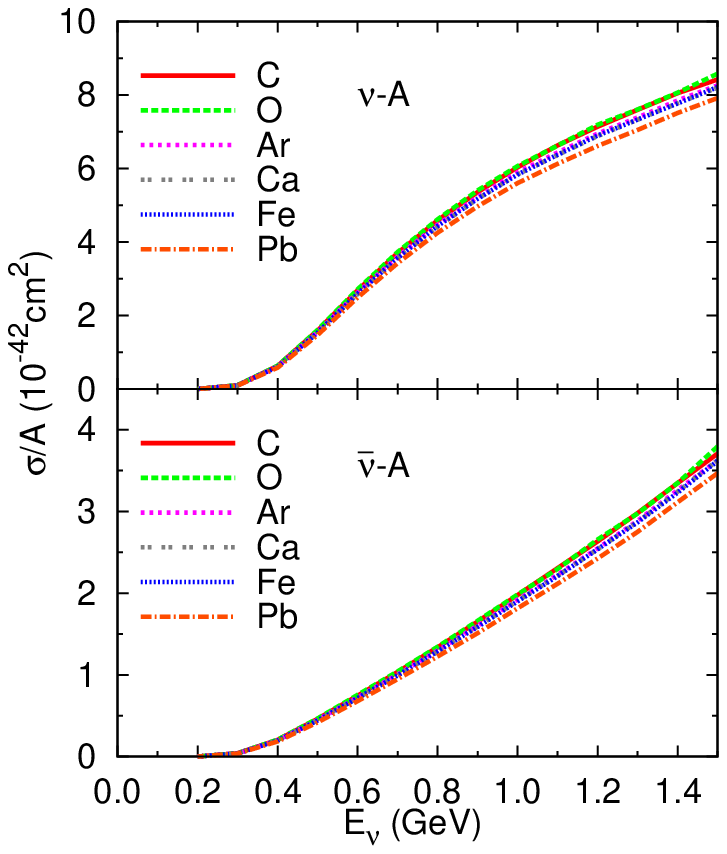}
\caption{\label{fig:NCgamma_nucleus} Left panel: Neutrino (top) and antineutrino (bottom)
 incoherent NC$\gamma$ cross sections on $^{12}$C according to the model of Ref.~\cite{Wang:2013wva}. 
All curves have been obtained with an $E_\gamma \geq 140$~MeV cut in the phase space. Solid lines stand for
results from the complete model at the nucleon level,  while the dotted lines display the predicted cross sections 
without the $N^*$ contributions. Curves denoted as ``Free'' (upper blue curves) do
  not include any nuclear correction: ($\sigma_A=Z\sigma_p+N\sigma_n)$. 
 Curves labeled as ``Full'' (lower red curves) take into account Pauli
  blocking, Fermi motion and the in medium $\Delta$ resonance
  broadening. The error bands on the full model result shows the uncertainty from the the axial $N\Delta$ coupling 
($C^A_5(0) = 1.00 \pm 0.11$). Right panel: Results for  different nuclei ($^{12}$C,$^{16}$O,$^{40}$Ar, $^{40}$Ca,$^{56}$Fe
  and $^{208}$Pb) divided by the number of nucleons~\cite{Wang:2013wva}.}
\efig

The theoretical models outlined above have been used to calculate the NC$\gamma$ events at MiniBooNE (including the coherent contribution described in Sec~\ref{subsec:CohGamma}). With the model of Ref.~\cite{Hill:2009ek} the number of these events were calculated to be twice as many as expected from the MiniBooNE {\it in situ} estimate~\cite{Hill:2010zy}. The conclusion was that NC$\gamma$ events give a significant contribution to the low-energy excess of $e$-like events. However, in Ref.~\cite{Hill:2010zy} the nuclear target ($^{12}$C)  was treated as an ensemble of nucleons, neglecting the important nuclear-medium corrections. Furthermore, an energy independent and rather high efficiency correction compared with the presently available figures~\cite{MiniBooNEweb} was assumed in the analysis. In contrast, the predictions based on the models of Refs~\cite{Zhang:2012xn,Wang:2013wva} are compatible with the MiniBooNE determination in spite of the quantitative differences in these approaches~\cite{Zhang:2012xn,Alvarez-Ruso:2013ica,WangMB}. One would then conclude that the NC$\gamma$ reactions cannot explain the excess of $e$-like events at low $E_\nu^{QE}$ observed at MiniBooNE, which remains an open question.

\section{Weak coherent processes at intermediate energies}
\la{sec:coh}

\subsection{Introduction}

As discussed in Sec.~\ref{sec:pion}, 
neutrino-induced pion, photon and (anti)kaon production off nucleons
and nuclei in the intermediate energy region are not only important for 
neutrino oscillation experiments but are also sources of relevant
data on the structure of hadrons and specially of their axial
properties.  At intermediate energies, pions, antikaons
or photons are mainly produced through resonance excitation and these
reactions can be used to extract information on nucleon-to-resonance
axial transition form factors. In reactions on nuclei, these outgoing
particles can be produced incoherently or coherently. In the latter
case the nucleus is left in its ground state in
contrast with the incoherent case where it is either broken or
left in some excited state. For instance, the
charge current (CC) coherent  pion production (COH$\pi$) reaction reads
\begin{equation}
  \nu_l (k) +\, A_Z|_{gs}(p_A)  \to l^- (k^\prime) +
  A_Z|_{gs}(p^\prime_A) +\, \pi^+(k_\pi) 
\label{eq:reac}
\end{equation}
The same is true for the rest of neutrino induced coherent channels 
reviewed in this section.

 A proper understanding of the coherent processes is very important in
 the analysis of neutrino oscillation experiments. For instance,
 coherent $\pi^0$ or $\gamma$ production by NC are
 among the most important $\nu_\mu$-induced backgrounds to experiments
 that measure $\nu_\mu\to\nu_e$ oscillations in the neutrino energy
 range around $1$~GeV~\cite{AguilarArevalo:2007it}. This is because NC
 $\gamma$ or $\pi^0$ events can mimic $\nu_e$ signal events, since the
 electromagnetic showers instigated by electrons or positrons and
 photons are not distinguishable in the large Cherenkov tanks used as
 far detectors\footnote{In the case of $\pi^0$ production, the
 misidentification can occur when one of the
two photons from the $\pi^0\to \gamma\gamma$ decay is not
detected. This might happen when the photon exits the detector before
showering or does not have enough energy to initiate a shower.}.
 Similarly, coherent CC $\pi^+$ production  is a background in $\nu_\mu$ disappearance
 searches~\cite{Hiraide:2006zq}.

Coherent reactions have smaller cross sections and are clearly 
more forward peaked than incoherent ones. 
Indeed, large momentum transfers to
the nucleus are suppressed by the nuclear form factor (Fourier
transform of the nuclear density), which takes its largest values when
the lepton transferred ($\vec{q}\,$) and produced particle
($\vec{k}_{\rm coh}=\vec{k}_\pi, \vec{k}_\gamma, \vec{k}_K \cdots$) momenta are
similar. Because the  kinetic energy of the final nucleus is
negligible, the energy of the outgoing particle, $k^0_{\rm
  coh}$, coincides, in a very good approximation, with the lepton
transferred energy $q^0$. The limiting case $\vec{q}=\vec{k}_{\rm
  coh}$ would correspond to $q^2 = m^2$, being $m$ the mass of the
produced particle (photon, pion, kaon,...),  
which is not kinematically accessible (except for photon emission).   
Instead, $q^2=0$ can be reached for massless leptons in
the strictly forward kinematics. In this case, the lepton tensor is such that
\begin{equation}
L_{\mu\sigma}^{(\nu,\bar\nu)} \sim q_\mu q_\sigma,
\end{equation}
and, because of CVC, its contraction with the vector part of the hadron tensor 
vanishes. This is the reason why the COH$\pi$ part of  electron and photon induced
reactions turned out to be a quite small
fraction of the total inclusive nuclear absorption cross section~\cite{Carrasco:1991we,Hirenzaki:1993jc}. 
The largest
differential cross sections in coherent particle production with electromagnetic probes 
arise in kinematics that optimize the product of the amplitude squared of the
elementary process times the nuclear form factor. However, in weak processes, 
there is an axial part which is not suppressed for kinematics where $\vec{q}$ and $\vec{k}_{\rm
  coh}$ are almost equal (the exception being coherent photon emission, as discussed in Sec~\ref{subsec:CohGamma}). 
Thus, the reduction induced by the
nuclear form factor is less significant, and the relative contribution of the coherent production channel to the 
total cross section, larger. A similar scenario is encountered in some hadronic reactions such as 
$(^3\mathrm{He},^3\mathrm{H} \,\pi^+)$  in nuclei~\cite{FernandezdeCordoba:1992ky}.

It is worth stressing the important role played by nuclear effects. 
The $\Delta(1232)$ resonance, which is the dominant intermediate state in coherent 
photon and pion production, is strongly modified in the nuclear medium.  In addition, 
the pion and (anti)kaon outgoing wave functions are distorted inside the nuclei. 
This distortion is particularly strong in the  case of pions, owing to the presence 
of the $\Delta(1232)$ resonance in the pion-nucleus optical potential and 
rather mild for kaons due to the absence of $K N$ resonances. 
Thus, these processes are quite sensitive to the pion or (anti-)kaon dynamics in nuclei. 
We should also mention the nonlocality in the $\Delta$ propagation which is often
neglected in the microscopic models. The effects
of this approximation were discussed in Ref.~\cite{Leitner:2009ph} in the
context of COH$\pi$ reactions, without considering the modification of $\Delta$
properties in the nucleus and the distortion of the outgoing pion. Sizable effects
(reductions as large as a factor of two at $E_\nu \sim 500$ MeV) were
claimed in \cite{Leitner:2009ph}, which might question the validity of
the local $\Delta$ propagation.  
In the more realistic description of Ref.~\cite{Nakamura:2009iq}, 
the nonlocality is preserved for the 
$\Delta$ kinetic term in a linearized version of the $\Delta$ propagator 
but, at the same time, a local approximation for vertices and $\Delta$ selfenergy 
have been adopted. Nevertheless, the mismatch  between the non-local recoil effects 
and the local approximation is likely to be minimized by the fact that the 
parameters in the $\Delta$ selfenergy are adjusted to describe pion-nucleus 
scattering data with the same model.  The problem of nonlocality  
should be further investigated, performing a
realistic calculation including full nuclear effects, and
disentangling up to what extent some of the non-local effects are  
effectively accounted for in the empirical $\Delta-$nucleus
optical-potential, usually tested/fitted in pion-nucleus reactions,
employed to  describe the dynamics of the $\Delta$ resonance in the
nuclear environment.

\subsection{CC and NC coherent $\pi$ production reactions}
\label{sec:cohpi}

Models for COH$\pi$ can be classified as PCAC or microscopic. 
The dominance of the axial contributions at $q^2=0$ 
has been extensively
exploited, through PCAC, 
to relate\footnote{At $q^2=0$ only
the axial current survives, being its contribution proportional to its
divergence $(q_\mu A^\mu )$. The relation with the pion-nucleus
elastic differential cross section follows from PCAC 
\begin{equation}
\partial_\mu A^\mu (x) \sim f_\pi m_\pi^2 \pi(x)
\end{equation}
with $f_\pi$ the pion decay constant and $\pi(x)$ a pion field. In
addition, one should assume dominance of the $\vec{q}=\vec{k}_\pi$
kinematics (thanks to the nuclear form factor) and neglect some
off-shell effects since $q^2=0\ne m_\pi^2$. } the neutrino COH$\pi$
cross sections $\left[\sigma_{\rm COH} (\nu + A_Z|_{gs} \to \nu +
  A_Z|_{gs} + \pi^0), \sigma_{\rm COH}(\nu + A_Z|_{gs} \to \ell^- +
  A_Z|_{gs} + \pi^+)\right]$ with the pion-nucleus elastic ones
$\left[\sigma (\pi^0 + A_Z|_{gs} \to \pi^0 + A_Z|_{gs}), \sigma (\pi^+
  + A_Z|_{gs} \to \pi^+ +
  A_Z|_{gs})\right]$~\cite{Rein:1982pf,Kopeliovich:1992ym,Rein:2006di,Paschos:2005km,Berger:2007rq},
and similarly for antineutrino induced reactions. On the other hand,
microscopic approaches~\cite{Singh:2006bm, AlvarezRuso:2007tt,
  AlvarezRuso:2007it, Amaro:2008hd, Martini:2009uj, Nakamura:2009iq,
  Hernandez:2010jf, Zhang:2012xi} start from a model for weak pion
production on the nucleon, and perform a coherent sum over all nucleon
contributions, taking into account modifications of the elementary
amplitudes in the nuclear medium. Since the nucleus remains in its
ground state, a quantum treatment of pion distortion becomes
possible. In microscopic COH$\pi$ models, the hadronic and nuclear
physics input is the same employed to describe the related incoherent
pion production process. The main drawbacks of this kind of approaches
are: i) the available descriptions are restricted to the kinematic
region where pion production is dominated by the excitation of the
$\Delta(1232)$ resonance and cannot be easily extended to higher
energies, and ii) these models are also technically more involved than
PCAC ones and difficult to implement in MC simulations.

  High  neutrino
energy  ($E_\nu \ge 2$  GeV) COH$\pi$  production data  (including the
recent    NOMAD    measurement   of    the    NC   COH$\pi^0$    cross
section~\cite{Kullenberg:2009pu}) were successfully explained with the
PCAC   based  model   of  Ref.~\cite{Rein:1982pf}.   The  experimental
investigation at $E_\nu \sim 1$ GeV started only recently. Contrary to
PCAC based models expectations, the K2K Collaboration obtained only an
upper  bound for  CC  COH$\pi$  at $\langle  E_\nu\rangle$  = 1.2  GeV
\cite{Hasegawa:2005td}.  This unexpected  result  triggered a  renewed
theoretical  interest  in   this  process.  The  present  experimental
situation   is  puzzling   because  the   upper  limits   obtained  by
K2K~\cite{Hasegawa:2005td}  and SciBooNE~\cite{Hiraide:2008eu} coexist
with          measurements          of          NC          COH$\pi^0$
(MiniBooNE~\cite{AguilarArevalo:2009ww}                             and
SciBooNE~\cite{Kurimoto:2010rc})   with  a  presumably   larger  cross
section. Indeed, SciBooNE
reported~\cite{Kurimoto:2010rc} a value
\begin{equation}
\frac{\sigma_{{\rm
    CC-COH}\pi^+}}{\sigma_{\rm NC-COH\pi^0}}= 0.14^{+0.30}_{-0.28}
\end{equation}
for carbon and an average neutrino energy in the 0.8 GeV region.
This result is difficult to  accommodate with the relation $\sigma_{{\rm
    CC-COH}\pi^+}/\sigma_{\rm NC-COH\pi^0}=2$ which, up to
kinematic corrections,  follows from PCAC and  isospin invariance in
the case of isoscalar nuclei. There exists a general consensus within the theory community,
that though this ratio in carbon at these energies could deviate from
2, it is in any case expected to be close to 1.5. This is  more than 4$\sigma$ far
from the SciBooNE measurement.

\begin{figure}[htb!]
\begin{center}
\begin{minipage}{0.49\textwidth}
\includegraphics[width=0.9\textwidth]{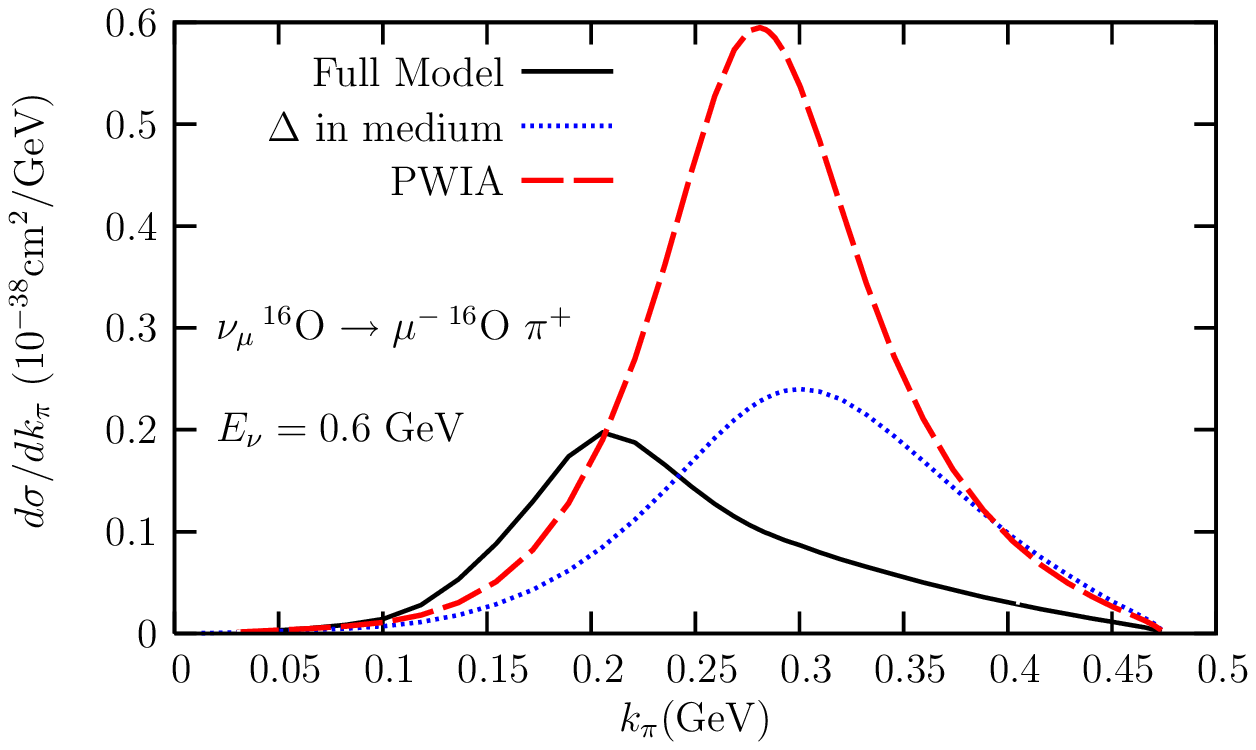}
\end{minipage}
\hfill
\begin{minipage}{0.49\textwidth}
\includegraphics[width=0.9\textwidth]{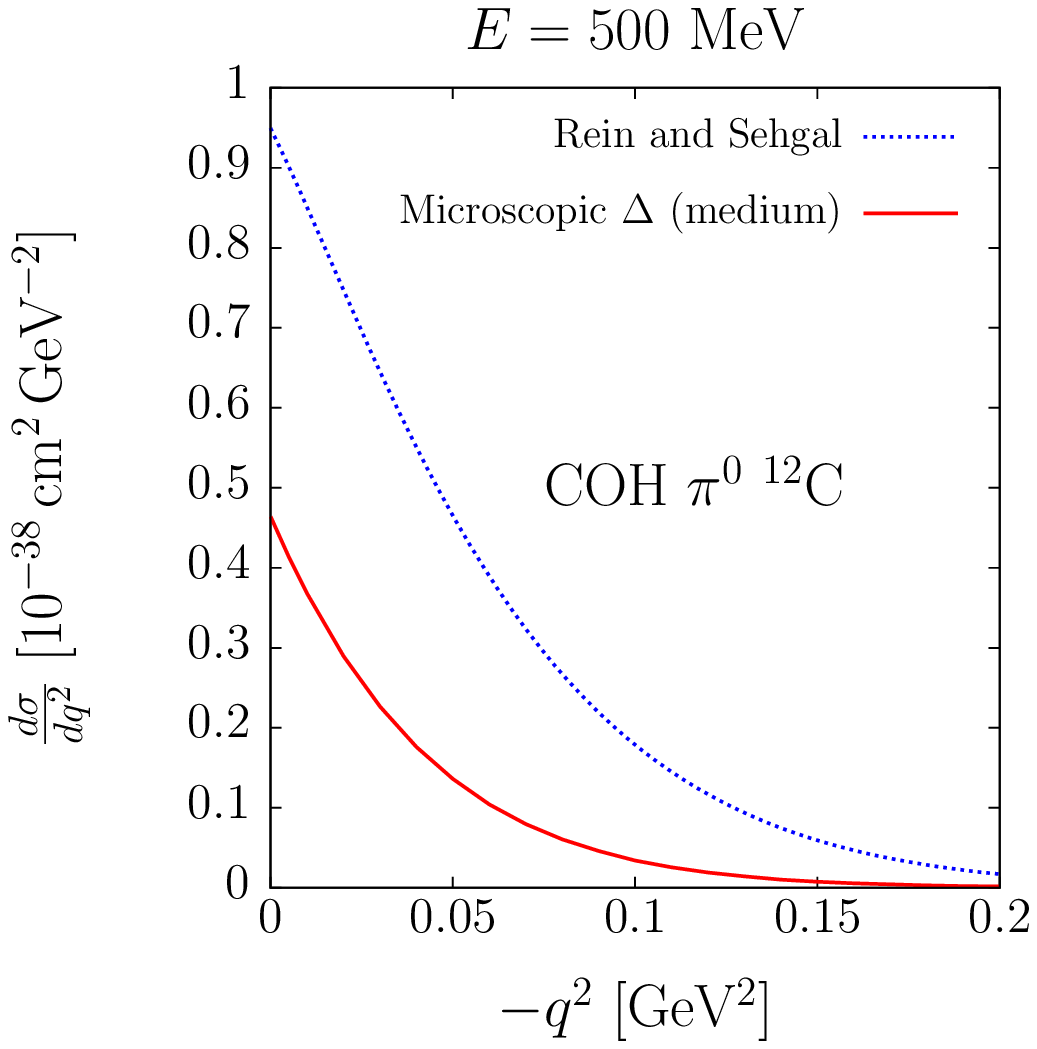}
\end{minipage}
\\ \vspace{0.6cm}
\begin{minipage}{0.49\textwidth}
\includegraphics[width=0.9\textwidth]{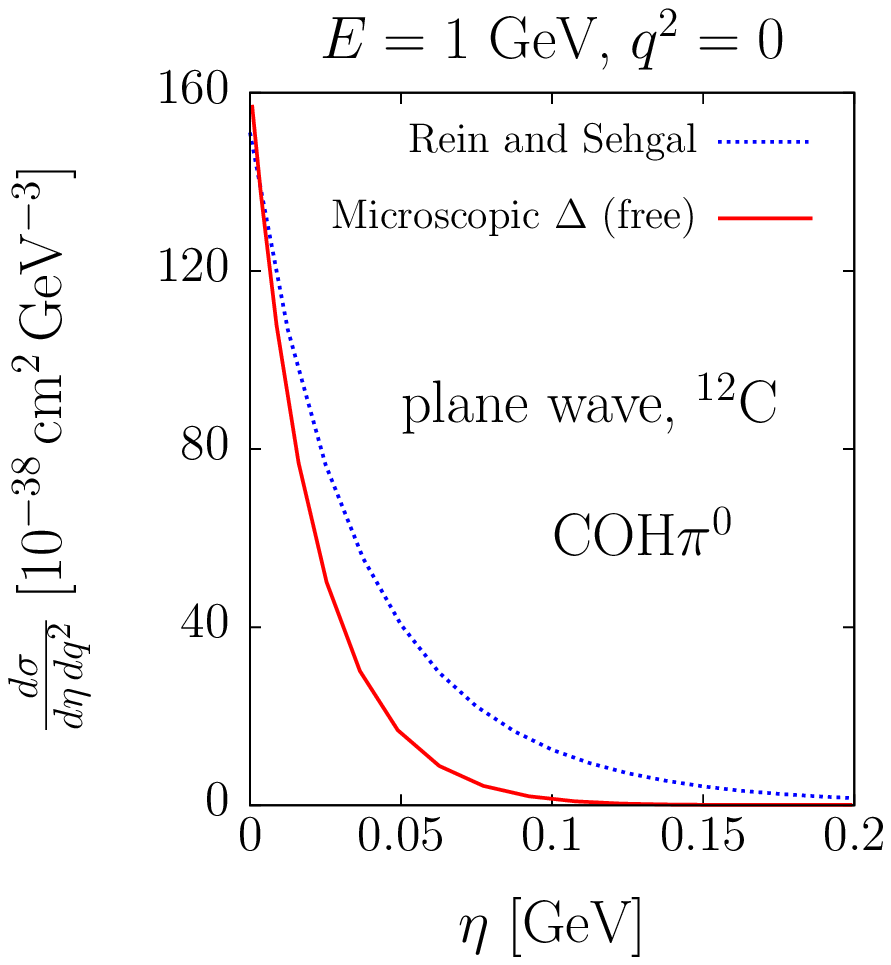}
\end{minipage}
\hfill
\begin{minipage}{0.49\textwidth}
\includegraphics[width=0.9\textwidth]{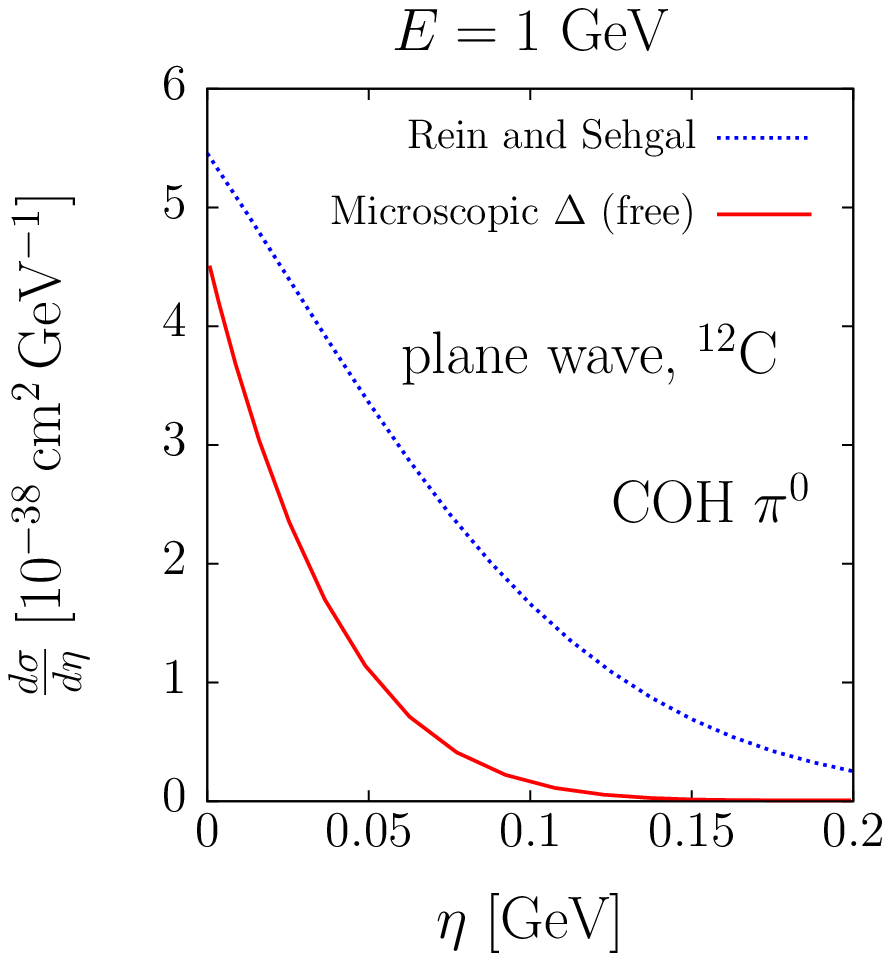}
\end{minipage}
\caption{Laboratory frame COH$\pi$ differential cross section results 
 from the microscopic $\Delta$
  and the PCAC Rein and  Sehgal  models of Refs.~\cite{Amaro:2008hd}
  and \cite{Rein:1982pf}, respectively. The variable $\eta$ is defined
  as  $E_\pi\left(1-\cos\theta_\pi\right)$, with the pion angle defined with
  respect the incoming neutrino momentum. In the right top panel and in the
  bottom panels, the microscopic $\Delta$
  contribution is calculated following \cite{Amaro:2008hd}, but using $C_5^A(0)=1.2$. }
\label{fig:coh1}
\end{center}
\end{figure}

PCAC was used by Rein and Sehgal~\cite{Rein:1982pf} to study the
COH$\pi^0$ reaction, extending it to $q^2\ne 0$ by means of a
phenomenological form factor. Subsequently, the model for the CC
reaction, has been upgraded \cite{Berger:2007rq} to include lepton
mass effects important for low $E_\nu$ studies.  This simple model is
a reference in the field of neutrino interactions and has been adopted
by the MC simulations employed in neutrino experiments\footnote{An
  important improvement~\cite{Hernandez:2009vm,Berger:2008xs,Paschos:2009ag} 
in these models is the use of a
  better input for the elastic $\pi + A_Z|_{gs} \to \pi + A_Z|_{gs}$
  angular differential cross section than that employed in the Rein
  and Sehgal model~\cite{Rein:1982pf}. There, this distribution was
  approximated by the forward cross section modulated by the nuclear
  form factor and an angular-independent attenuation factor that
  should take into account effects of the outgoing pion absorption in
  the nucleus. As a result in \cite{Berger:2008xs,Paschos:2009ag}, and
  using experimental information on the angular dependence of the
  elastic pion-nucleus cross section, the predicted COH$\pi$ cross
  section became reduced by a factor of 2-3 for neutrino energies
  around 1 GeV and for light nuclei in the carbon region. Apart from
  the obvious limitation coming from the lack of experimental data for
  many pion energies and nuclei, it could be also argued that because
  of the strong distortion of the incoming pion in the on-shell
  elastic pion-nucleus process, one cannot directly relate the 
  amplitude of the latter reaction to that of pion production induced by a weak current (more
details can be found in Ref.~\cite{Hernandez:2009vm}).}. However in the $E_\nu < 2$
GeV region, and for light nuclei, there are drawbacks that render the
model inaccurate~\cite{Amaro:2008hd, Hernandez:2009vm}. Far from the
$q^2=0$ kinematic point, PCAC  models cannot be safely used to
determine the angular distribution of the outgoing pions.  Terms that
vanish at $q^2=0$, and that are not considered in PCAC based models,
provide much more forward peaked pion angular
distributions~\cite{Hernandez:2009vm}. (We will discuss this in
certain detail below,  Fig.~\ref{fig:coh1}). For neutrino energies above 2 GeV, the involved
momenta, in most of the available phase space, are sufficiently large to
guaranty that only kinematics close to the $\vec{q} =\vec{k}_{\rm
  coh}$ configuration lead to non totally negligible values of the
nuclear form factor, making the $q^2=0$ approximation quite
appropriate. Hence, PCAC  models provide accurate predictions of
neutrino COH$\pi$ cross sections for high neutrino energies. However,
for lower neutrino energies, these models overestimate both the CC and
NC total coherent cross sections as compared to the results obtained in  
more realistic microscopical models.

Within microscopic models there are still, in principle, various
approaches e.g due to differences in the treatment of the non-resonant
background. A common assumption is that the reaction proceeds solely
via the $\Delta$ excitation \cite{Singh:2006bm, AlvarezRuso:2007tt, Martini:2009uj}. 
Non-resonant contributions
required by chiral symmetry at threshold have been also included in
Refs.~\cite{AlvarezRuso:2007it,Amaro:2008hd,Hernandez:2010jf} using the model for the
weak pion production off the nucleon of
Refs.~\cite{Hernandez:2007qq,Hernandez:2010bx}, though they turn out
to be very small because of large cancellations, some of them exact
for isospin symmetric nuclei. The absolute normalization of the
predicted cross-section depends on the adopted value of the dominant
axial $N\to \Delta$ form factor $C^A_5 (0)$, as the process is
dominated in large extent by the axial part of the weak current
(central values for the total cross sections may suffer by some
20\%–30\% uncertainty), while differential cross sections do not
appreciably change their shape~\cite{Hernandez:2010jf}. 

In  Fig.~\ref{fig:coh1}, we show some results at low and intermediate neutrino energies from the
microscopic $\Delta-$hole and the PCAC Rein and Sehgal models of
Refs.~\cite{Amaro:2008hd} and \cite{Rein:1982pf}, respectively. The scheme of
Refs.~\cite{AlvarezRuso:2007tt, AlvarezRuso:2007it} is quite similar
to that employed in \cite{Amaro:2008hd} and leads to qualitative
similar results, while the approach of \cite{Singh:2006bm}, although 
less sophisticated in the treatment of the distortion of the outgoing pion
waves, makes use of the same model to account for the $\Delta$
properties in the nuclear medium.  The
importance of the nuclear medium effects is firstly stressed in the
top left panel of Fig.~\ref{fig:coh1}. To this end,
different predictions for the pion momentum
differential cross section are displayed~\cite{Amaro:2008hd}. The long-dashed line (in
red) has been calculated using planes waves for the outgoing pion and
without including any in-medium correction for the $\Delta$. Results
with $\Delta$ nuclear medium effects are shown by the dotted line (in
blue). The full model calculation of Ref.~\cite{Amaro:2008hd}
including medium effects on the $\Delta$ and the distortion of the
outgoing pion wave function\footnote{The outgoing pion wave function
  is obtained in \cite{Amaro:2008hd}, as in \cite{AlvarezRuso:2007tt,
    AlvarezRuso:2007it,Hernandez:2010jf}, by solving the Klein-Gordon
  equation with a non-local pion-nucleus optical potential, based on
  the $\Delta-$hole model plus some other low energy terms, that
  successfully describes the interaction of pions with nuclei at low
  and intermediate energies. This is a improvement with respect to the
  simpler eikonal approximation used for example in Ref.~\cite{Singh:2006bm}.} is shown by the solid line (in
black). We can see the in-medium modifications of the $\Delta$ properties produce a
strong reduction and broadening of the peak. Pion distortion further
decreases the cross section and moves the maximum to lower energies.

Pion angular and $q^2$ distributions are displayed in the other three
panels of Fig.~\ref{fig:coh1}, where the results from a microscopic calculation following \cite{Amaro:2008hd},
but neglecting non-resonant background terms and using $C_5^A(0)=1.2$,
is compared with the predictions of the PCAC Rein--Sehgal
model~\cite{Rein:1982pf}. Outgoing pion distortion is only included in
the right top panel, where the modifications of the $\Delta$ properties in the nuclear
medium are also taken into account within the microscopical approach. In
the bottom panels, plane waves for the outgoing pion and a free
$\Delta$ resonance were employed.  From the figure, we can conclude:
\begin{enumerate}
\item Rein-Sehgal model predicts much larger and wider $q^2$ differential
  cross sections. As mentioned above, this is mainly due to the poor
  approximation assumed in this model for the the elastic $\pi +
  A_Z|_{gs} \to \pi + A_Z|_{gs}$ angular differential cross
  section. The effects of this approximation decrease with the
  neutrino energy and atomic number, but they are still important for
  $E_\nu\sim 1.5-2$ GeV in a medium sized nucleus like
  calcium~\cite{Hernandez:2009vm}.  
\item Regarding the outgoing pion distributions, the
  microscopic calculation is much more peaked around
  $\eta=E_\pi\left(1-\cos\theta_\pi\right)=0$ than the Rein-Sehgal
  model results, specially far from $q^2=0$. Terms that vanish at
  $q^2=0$, and that are not considered in PCAC based models, provide
  much more forward peaked pion angular distributions. This is also
  true for results obtained with pion distortion~\cite{Hernandez:2009vm}.
\end{enumerate}

Finally we briefly review other approaches. The model of
\cite{Zhang:2012xi} starts from Lorentz-covariant effective field
theories with nucleon, pion, $\Delta(1232)$ but also scalar ($\sigma$)
and vector ($\rho$, $\omega$) mesons as the relevant degrees of
freedom, and exhibit a nonlinear realization of (approximate) $SU(2)_L
\otimes SU(2)_R$ chiral symmetry. This is the same scheme as that
employed in \cite{Serot:2012rd} for the neutrino-production of pions
from nucleons. A special attention is paid to the power counting,
which is shown to be valid only for quite low neutrino energies below
550 MeV.  On the other hand, the approach of
Ref.~\cite{Martini:2009uj} shares many of the ingredients of previous
models (in-medium $\Delta$ modification, effective $\Delta-$hole
interactions) but there, the coherent cross section is related to the
coherent part of the nuclear response (mostly longitudinal). This
response is obtained within the RPA approximation when the
intermediate pion is placed on the mass shell and thus, some RPA
corrections driven by short distance dynamics in the spin-isospin
longitudinal channel are neglected. In the forward direction, the
authors of Ref.~\cite{Martini:2009uj} explicitly relate their forward
neutrino coherent cross section to the elastic cross section of
physical pions, as predicted by Adler's theorem~\cite{Adler:1964yx},
thus making contact with the PCAC based models mentioned above
(actually the results of \cite{Martini:2009uj} agree quite
well, see ~\cite{Boyd:2009zz}, with those found in
\cite{Berger:2008xs,Paschos:2009ag} that use the elastic differential
pion-carbon data).  This infinite
nuclear matter approach derived in Ref.~\cite{Martini:2009uj} uses 
plain (undistorted) pion wave-functions.  In spite
of this, the obtained COH$\pi$ cross sections do not differ much~\cite{Boyd:2009zz} 
from those obtained with the microscopic models addressed above.

A different microscopic COH$\pi$ model is derived in
\cite{Nakamura:2009iq} starting from a dynamical model in coupled
channels, where the bare $N\Delta$ transition from a constituent quark
model is renormalized by meson clouds. Pion distortion and $\Delta$
spreading potential effects are also taken into account. The free
parameters in the spreading potential and pion-nucleus optical
potential are fitted to pion-nucleus elastic scattering data. The
results of this approach reasonably agree with those found in
\cite{Amaro:2008hd, Hernandez:2010jf} (see Figs. 8 and 9 of
Ref.~\cite{Boyd:2009zz}).

To conclude, we stress once more that all microscopical theoretical models, though
lead to different cross sections, however, predict a value for the
ratio $\sigma_{{\rm CC-COH}\pi^+}/\sigma_{\rm NC-COH\pi^0}$ in carbon always
around 1.5 for $E_\nu \sim 0.8$ GeV, which is in clear contradiction
with the SciBooNE measurement ($0.14^{+0.30}_{-0.28}$). This
becomes an open problem and a challenge for both, theory and experimental communities.

\subsection{CC coherent kaon and antikaon production reactions}
\label{subsec:CohK}

Here we briefly discuss the coherent production of charged
kaons. Namely, we consider the $\Delta S = \pm 1$, Cabibbo suppressed,
weak strangeness production reactions
\begin{equation}
  \nu_l (k) +\, A_Z|_{gs}(p_A)  \to l^- (k^\prime) +
  A_Z|_{gs}(p^\prime_A) +\, K^+(k_K), \qquad
  \bar\nu_l (k) +\, A_Z|_{gs}(p_A)  \to l^+ (k^\prime) +
  A_Z|_{gs}(p^\prime_A) +\, K^-(k_K) \,.
\label{eq:reac2}
\end{equation}
These reactions have been theoretically studied in the recent work of
Ref.~\cite{AlvarezRuso:2012fc} using a microscopic approach along the
lines of the COH$\pi$ study of Ref.~\cite{Amaro:2008hd} described
above. At the nucleon level, the model of
Ref.~\cite{RafiAlam:2010kf,Alam:2012zz}, presented in
Sec.~\ref{subsec:kaonprod}, was implemented. The distortion of the
outgoing kaons is treated in a quantum-mechanical way by solving the
Klein-Gordon equation with realistic optical potentials accounting for
the (very different) $K$ and $\bar{K}$ interactions in the nuclear
medium.

\begin{figure}[htb]
\begin{center}
\makebox[0pt]{\rotatebox{270}{\includegraphics[width=0.3\textwidth]{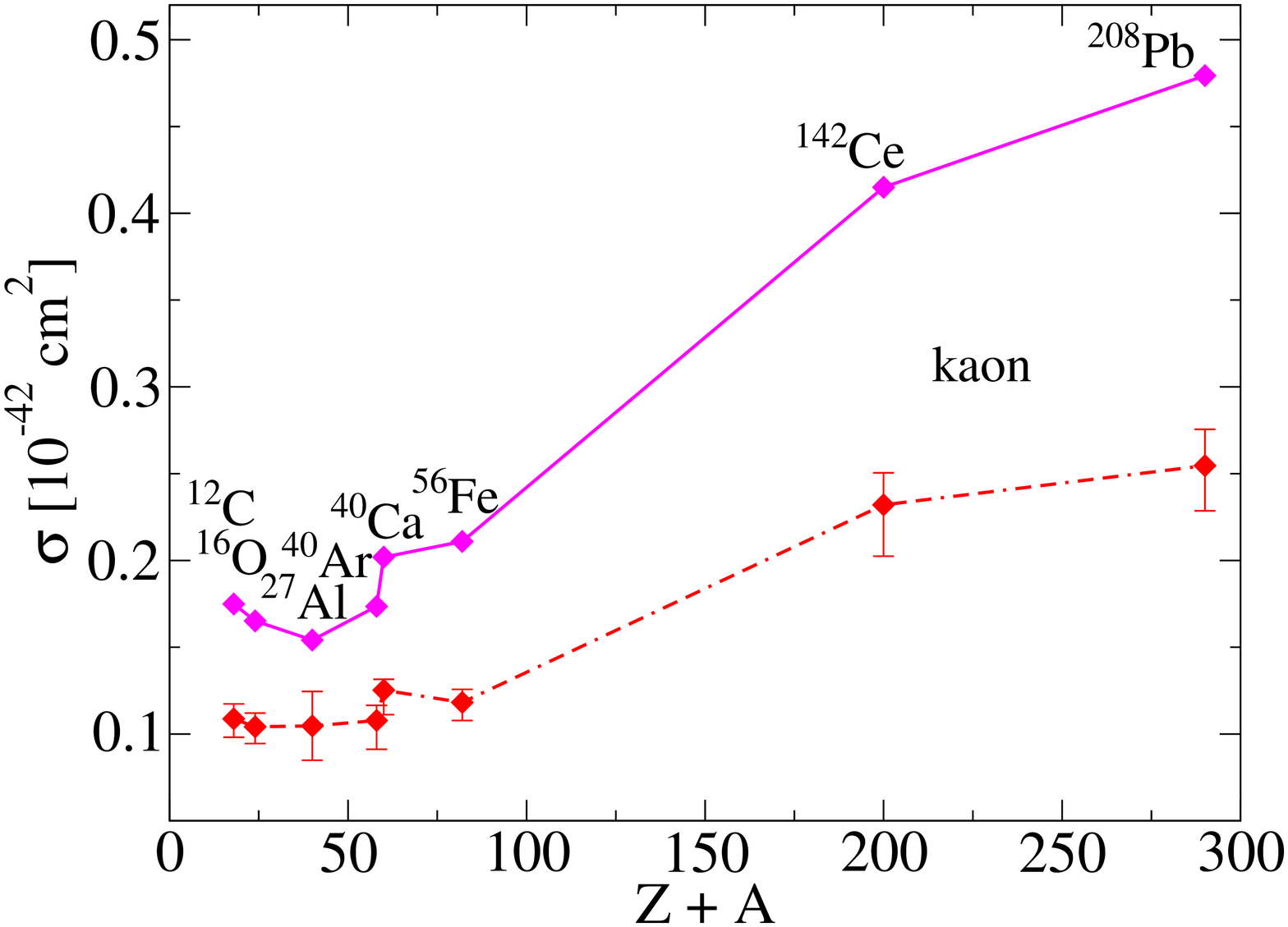}}\rotatebox{270}{\includegraphics[width=0.3\textwidth]{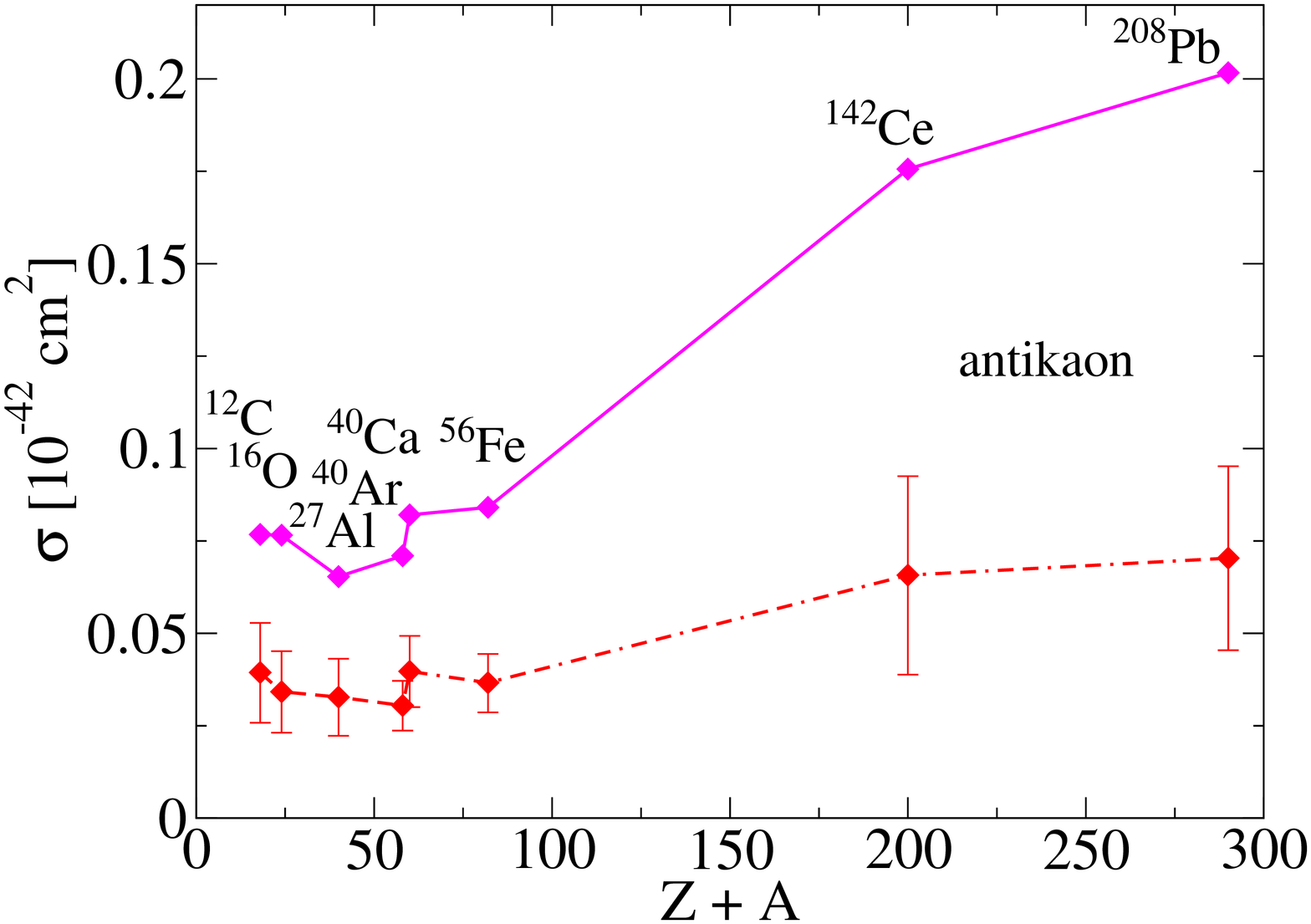}}}
\end{center}
\caption{ Predictions from the model of Ref.~\cite{AlvarezRuso:2012fc}
  for the coherent $K^{\pm}$ production cross sections in several
  targets at an incident (anti)neutrino energy of 1 GeV. The solid
  (dashed) lines are obtained without (with) kaon distortion.}
\label{fig:coh3}
\end{figure}
The resulting cross sections for incident muon neutrinos of 1-2~GeV
are small, with cross sections per nucleon much smaller than the
corresponding ones on free nucleons. This can be explained by the
rather large momentum transferred to the nucleus (due to the large
value of the kaon mass compared to the typical kaon momenta) which
reduces drastically the nuclear form factors. At these large momentum
transfers, the nuclear form factors depend strongly on the details of
the proton and neutron density distributions. Angular kaon and lepton
momentum distributions are forward peaked, as it is normally the case
in coherent processes. The cross section dependence on the atomic (Z)
and mass numbers (A) of the target nuclei is shown in
Fig.~\ref{fig:coh3} for 1~GeV (anti)neutrinos. The isospin factors of
the dominant CT mechanisms (see Figs~\ref{fig:diags} and
~\ref{fig:adiags}) suggest a quadratic dependence of the cross section
with (A+Z) but no significant enhancement for heavy nuclei is
observed, even if the distortion is neglected.  To understand this, one
should recall that heavier nuclei have narrower form factors, which
causes a larger suppression at high momentum transfers. The error-bars
in the full-model results represent the uncertainties in the
model. The errors in the proton and neutron density distributions, as
well as a 10~\% one in $M_F$ (Eq.~(\ref{eq:kaonFF})), accounting for the uncertainty in the
elementary production model, have been propagated to the final
results. In the case of $\bar K$ production, the uncertainty in the
$K^-$ distortion, which turns out to be the major error source, has
also been estimated~\cite{AlvarezRuso:2012fc}.

At higher energies, where the present model is not directly
applicable, larger kaon momenta are present so that the suppressing
role of the kaon mass is less important. A fast increase of the cross
section is therefore expected. In view of this, measuring this
reaction at MINER$\nu$A would be quite interesting.

\subsection{NC coherent $\gamma$ production reactions}
\label{subsec:CohGamma}

The coherent contribution  (NC COH$\gamma$), Eq.~(\ref{eq:reac_coh}), is an important ingredient in a realistic description of NC photon emission on nuclear targets. At high energies, it was firstly studied in the early eighties~\cite{Gershtein:1980wu,Rein:1981ys}. A discussion about these works can be found in Section V.E of Ref.~\cite{Hill:2009ek}. Here, we focus on 
the intermediate energy region, of relevance for the MiniBooNE and T2K experiments.  At these energies, there exist three recent
theoretical calculations of this reaction channel. The model of Ref~\cite{Hill:2009ek} mostly ignores nuclear corrections. The nucleus is treated as a scalar particle, including a nuclear form factor to ensure that the coherence is restricted to low-momentum transfers. 
The microscopical approaches of Refs.~\cite{Zhang:2012xi} and \cite{Wang:2013wva} are
more robust and rely on the same formalism and approximations used to study COH$\pi$  processes~\cite{Amaro:2008hd,Zhang:2012xi}. 
In  Ref.~\cite{Wang:2013wva}, the NC$\gamma$ model on the nucleon is directly applied by summing the different amplitudes coherently. The total cross section is clearly dominated by the $\Delta(1232)$, with small corrections from $D_{13}(1520)$ excitation (see Fig.~11 of Ref.~\cite{Wang:2013wva}). Nucleon-pole contributions are negligible because the coherent kinematics favors a strong cancellation between the direct and crossed terms. The $\pi Ex$ terms vanish exactly for isospin symmetric nuclei because amplitudes for protons and neutrons cancel with each other. The results from the $\Delta P+ C\Delta P+NP+CNP$ part of the model of Refs.~\cite{Zhang:2012xi,Zhang:2012xn} are within the uncertainty band of those in Ref.~\cite{Wang:2013wva} up to (anti)neutrino energies of 1.4--1.5 GeV.

Unlike $\pi$ and $\rho$ $t$-channel terms,  the coherent contribution of the $\omega Ex$ does not vanish for symmetric nuclei because amplitudes on protons and neutrons add up rather than cancel. In Ref.~\cite{Hill:2009ek} it was found that the COH $\omega Ex$ mechanism plays a sub-dominant role at $E_\nu \sim 1$~GeV, compared to naive estimates, being suppressed by form factors and recoil. On the other hand, because of the strong energy dependence of the contact terms in Ref.~\cite{Zhang:2012xn}, with couplings derived from the $\rho Ex$ and $\omega Ex$ amplitudes of Ref.~\cite{Hill:2009ek}, the NC COH$\gamma$ cross section above $E_{\nu,\bar\nu}=0.65$ GeV is dominated by these contact terms and not by the $\Delta$. However, as discussed in Sec.~\ref{subsec:NCgamma} in the nucleon case, the results are not only highly sensitive to unknown form factors but should also be constrained by unitarity.

\begin{figure}[htb]
\begin{center}
\makebox[0pt]{\includegraphics[width=0.4\textwidth]{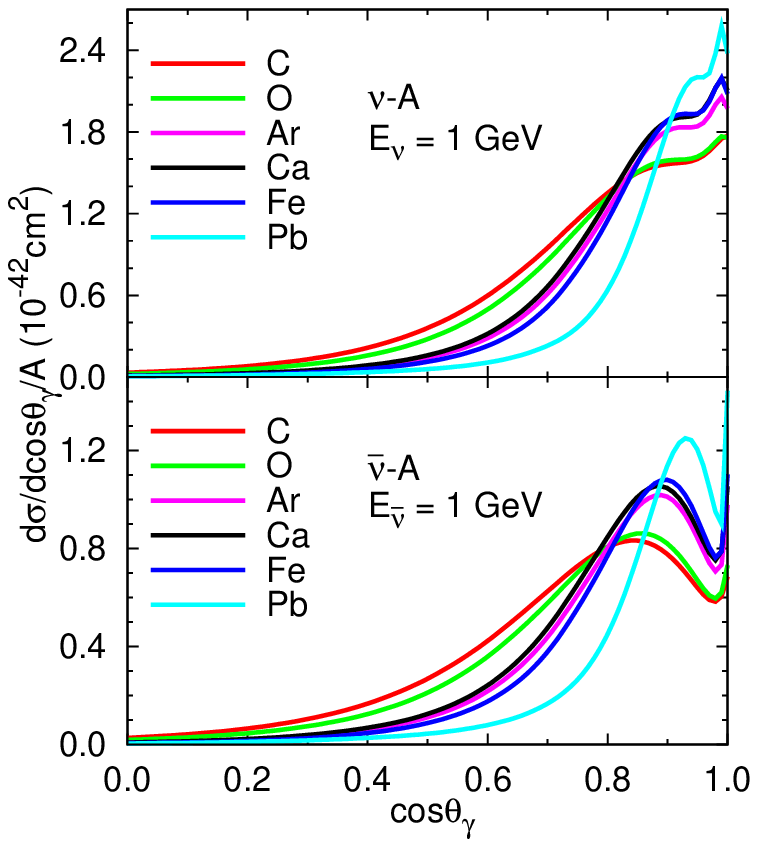}\includegraphics[width=0.4\textwidth]{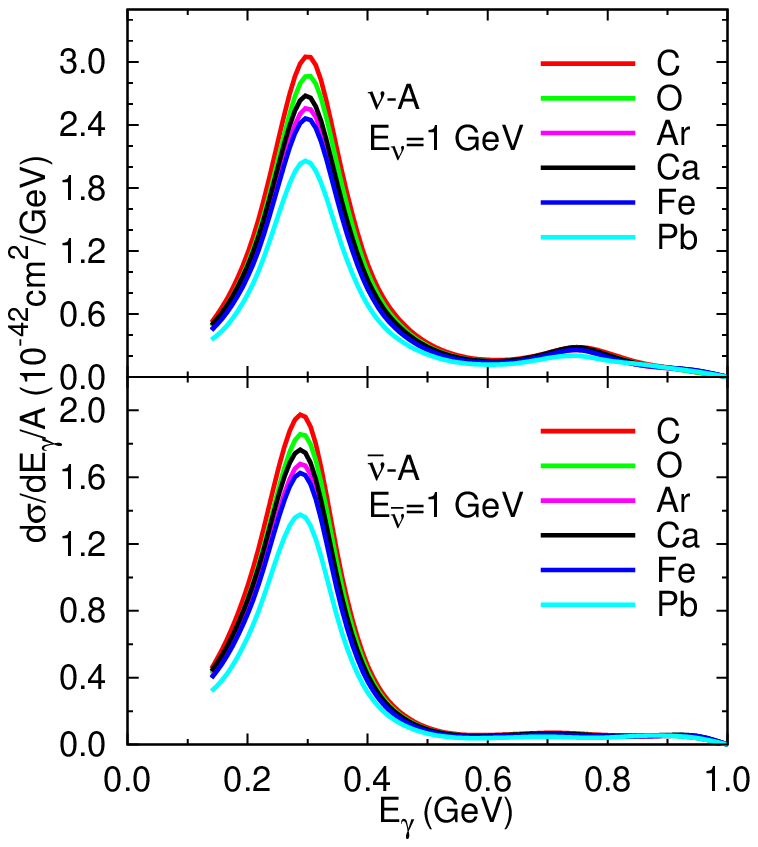}}
\end{center}
\caption{ NC COH$\gamma$ neutrino (top) and antineutrino (bottom)
  photon angular and energy differential distributions for various
  nuclei according to the model of
  Ref.~\cite{Wang:2013wva}. }
\label{fig:coh2}
\end{figure}
There are two features that make the dynamics of NC COH$\gamma$
processes substantially different to that governing the COH$\pi$
ones. First, the outgoing particle ($\gamma$) does not suffer from
strong distortion effects during its way out of the nucleus, and, 
second,  the axial contribution turns out to be purely transverse $\sim
(\vec{k}_\gamma \times \vec{q}\,)$ and also vanishes when $\vec{q} =
\vec{k}_\gamma$ (optimal configuration for the nuclear form factor),
which corresponds to $q^2 = 0$. Therefore, the largest differential cross
sections arise in kinematics that optimize the product of the
amplitude squared of the elementary process times the nuclear form
factor, as in pion coherent production reactions induced by electrons
and photons~\cite{Carrasco:1991we,Hirenzaki:1993jc}. The photon
angular dependence exhibited in the left panels of
Fig.~\ref{fig:coh2}, obtained with the model of
Ref.~\cite{Wang:2013wva}, should be understood from this perspective. 
Notice  that $\theta_\gamma$ is the photon angle with
respect to the direction of the incoming (anti)neutrino beam and not
the angle formed by $\vec{q}$ and $\vec{k}_\gamma$, which is not
observable. Indeed, for each $\theta_\gamma$, and integration over all
possible $\vec{q}$ is performed. The interference pattern  
between the direct dominant $\Delta$ mechanism and the cross $\Delta$
and the $N(1520)-$pole (weak excitation of this resonance and its subsequent
decay into $N\gamma$) terms, also included in the elementary model
employed in \cite{Wang:2013wva}, strongly influences the angular
distributions shown in Fig.~\ref{fig:coh2}.   Neutrino
(antineutrino) COH$\gamma$ cross sections are about a factor 15 (10)
smaller than the incoherent ones~\cite{Wang:2013wva}. These 
 proportions are similar to those found for pion production
reactions~\cite{Hernandez:2010jf,Hernandez:2013jka} in spite of the suppression of 
the axial current for forward kinematics in COH$\gamma$ reactions. 
This suppression is however partially compensated by the reduction
of about a factor of two due to the strong distortion of the outgoing pion, which is 
not present in the photon case. 

Predictions~\cite{Wang:2013wva}  for the outgoing
photon energy distributions in various nuclei are displayed in the right panels of
Fig.~\ref{fig:coh2}. The pronounced peak is produced by the dominant
$\Delta$ resonance. The peak position does not appreciably change 
from nucleus to nucleus, but it gets wider as $A$ increases. The
second, smaller and broader peak that can be seen for neutrinos
but not for antineutrinos corresponds to the excitation of the
$D_{13}(1520)$ resonance.  For the integrated cross sections it is found~~\cite{Wang:2013wva}
that these neither scale with $A$,
like the incoherent one approximately does, nor with $A^2$ as one
would expect from the coherence of the dominant $\Delta$ mechanism.

\section{Montecarlo generators}

Neutrino beams are not mono-energetic and
usually contain several flavors.  In the experiments, it is necessary
to identify the flavor of the interacting neutrino and  reconstruct
its direction and  energy from the  particles observed in the detectors. 
Therefore, a detailed knowledge of the final state after a neutrino interaction 
is essential to analyze  and interpret the data. 
 In order to evaluate the efficiency, resolution and
purity of the selected event samples, or to establish the method to
analyze the data, simulation program libraries, developed 
specifically for neutrino interactions, are extensively used.
These libraries, which provide all the information on the produced
particles for the different neutrino interaction mechanisms,  
are called Monte-Carlo (MC) neutrino event generators.  

There are several generators available, such as ANIS \cite{Gazizov:2004va},
GENIE \cite{Andreopoulos:2009rq}, GiBUU \cite{Buss:2011mx}, NEGN
\cite{Autiero:2005ve}, NEUT \cite{Hayato:2009zz}, NUANCE
\cite{Casper:2002sd}, FLUKA \cite{Battistoni:2009zzb} or NuWRO
\cite{Juszczak:2005zs}.  Some of them have been developed
by different experimental groups to be used in their analysis.  There
are several reasons why the different collaborations developed their own
generators.  The first reason is that the neutrino
beam energy spectrum is specific  for each experiment and thus, the
dominant/relevant interaction modes could be quite different. Unfortunately, there
is no {\it unified} framework able to describe all neutrino
interactions in a broad energy range, so that each generator needs to combine
various  models.  Sometimes  rather simplified schemes
are used  to describe some interactions, which are not too relevant for a given experiment.  
The second reason is that the target nuclei and detectors used in each experiment are 
different as well. Furthermore, the detection
efficiencies for each type of particle are detector dependent and
can become quite different. A widespread strategy to deal with these challenges is 
to choose the underlying models according to the specific needs of the experiment 
and tune them from the data measured at the (near) detectors. 
These specialized generators are useful within a single 
experiment but, sometimes, their diversity makes the comparison of results 
from different experiments difficult. This also obscures the physical interpretation of the 
cross section data and slows down the development of more precise approaches. 
In order to avoid this kind of problems, the GENIE collaboration undertook the development of a new
MC aiming  to cover all the relevant neutrino energy ranges.  On the other hand, some theory groups have also developed generators to
interpret the  experimental data sets that have become available in the last years.

In general, an event generator is required to provide, for any neutrino flavor and energy, and target nucleus:
\begin{itemize}
\item{the total cross section,}
\item{the partial contributions to the cross section from each interaction mode,}
 \item{the energy and direction of all the produced particles, simulating 
particle re-interactions inside the target nucleus when necessary.}
\end{itemize}

Usually, the procedure employed  to simulate an event
in the actual detector can be separated into several steps:
\begin{enumerate}

  \item The target nucleus and the neutrino energy are selected,
    taking into account  the composition of the detector, the total cross section
    for each material and the spectrum of the neutrino beam.

 \item The four-momentum of the target nucleon and the
    position inside the nucleus where the neutrino interaction takes place are chosen.

  \item The type of interaction (QE, pion production, etc) is selected according to 
the corresponding probabilities for the neutrino energy picked up in the first step.

  \item The neutrino-nucleon interaction is simulated, and the four-momenta
    of the outgoing lepton, nucleon(s) and other particles are fixed
    using the differential cross section of the particular type
    of interaction considered.
 
  \item The produced nucleons and  mesons are followed along their path through 
the nucleus; their subsequent interactions are also simulated.

\end{enumerate}

 As mentioned, the total cross section is used in the first step. 
Generally, the generator creates a table of interaction probabilities, which are calculated as 
$ {\rm flux}(E_\nu) \times \sum_{{\rm nuclear\, targets}\, (A)} \sigma_A(E_\nu)\times r_A$, where $r_A$ is the 
relative abundance (weight) of type $A$ nuclei.  This energy-dependent interaction probability is
 used to determine the energy of the incoming neutrino that should be
 simulated. A similar strategy is adopted to select the target
 nucleus and the specific neutrino interaction considered for each
 event. Once the energy and target nucleus are fixed, the actual event
 simulation procedure starts, which is different for each generator
 depending on the choice of models. The
 initial nucleon momentum is selected using a probability density
 profile. The simplest form is a step function, deduced
 from a global Fermi-gas model.  This probability density function
 could be corrected or re-weighted to take into account 
 kinematical bounds such as Paul-blocking in the final state. The
 position within the nucleus where the first neutrino interaction
 occurs is set assuming a uniform volume distribution.  Starting from this 
position, the subsequent interactions of the produced (secondary) particles 
in their way out of the nucleus are simulated in a later
 stage. Next, the theoretical model for $d\sigma/dE_\ell d\theta_\ell$ 
cross section is used to fix the four-momentum of the
 outgoing lepton. In the case of QE interactions, the momentum of the 
final nucleons are determined from energy-momentum conservation.  
If the first interaction is not QE, additional
 particles (mostly mesons) are also produced in this first step.
 All final-state hadrons interact in the nucleus and it is necessary to take
 into account secondary collisions.  Although the initial position of the particles 
are fixed from the location of the first interaction that originates the cascade, the
 actual starting point of each particle may be
 shifted to take into account the travel length in the nucleus, in the
 case of resonances, or the formation length of  particles. One of
 the simplest ways to estimate the amount of this shift is to use the
 concept of formation zone.  Once the simulation of the secondary
 interactions of hadrons are finished, some of the generators simulate
 also de-excitation gamma ray processes or (multi-)nucleon emission.

The simulation programs are carefully designed
to deal with the different dynamical models, paying special
attention to prevent that some important interactions could be missed
and to avoid unwanted overlaps between different energy regimes.
Some event generators incorporate rather old theoretical models to simulate
neutrino interactions, even though  more sophisticate schemes have
recently become available.  A first reason for this is that the generators need to
simulate several nuclear targets, but sometimes the models are not
applicable for all nuclei. Another reason
is that the models often have limitations in the kinematic ranges
where they can be trustfully used, although this is no less true for 
the old approaches. Because the event generators have to cover 
the entire allowed phase space and  be capable of simulating 
all the relevant exclusive channels, some models cannot be 
accommodated in them. Finally, to be able to produce millions of events, 
the generators have speed limitations that are not complied by all theoretical descriptions.  

As mentioned in Sec.~\ref{subsec:CCQE-Nucleus}, to simulate CCQE and NCQE scattering, several  
generators still use the simple RgFG. This is because it is easy to implement and
allows to change the target nucleus by modifying just a few parameters.
Recently, some of the generators have started to adopt some of the more advanced approaches, with 
spectral functions or the RlFG with RPA corrections, discussed in Sec.~\ref{subsec:CCQE-Nucleus}.

For single meson production via resonances, many generators
implement the relativistic harmonic oscillator model of
Rein-Sehgal~\cite{Rein:1980wg}. Even though this approach underestimates 
significantly pion electroproduction data (see Fig.~\ref{fig:coh0}
in Sec.~\ref{subsec:pion-nucleon} and the related discussion), it is still extensively used. 
Indeed, the code to calculate the helicity amplitudes was provided by the authors, it is easy to implement and includes
higher energy resonances, which allows the extension of the model to simulate the resonant production of  
kaon, eta and multiple pions. Some authors~\cite{Graczyk:2007bc} have improved the vector form factors for the dominant 
$\Delta(1232)$ resonance, taking advantage of the helicity amplitudes obtained from recent pion electroproduction data.
Some modern generators, like GiBUU~\cite{Buss:2011mx} and NuWRO~\cite{Nowak:2006sx},
are using better models, which rely on pion electroproduction data for the vector part 
of the amplitudes, and on the neutrino pion production cross sections from the old ANL and BNL bubble chamber 
neutrino experiments to constrain the resonance axial-vector couplings. 

For the simulation of multi-hadron production, deep inelastic (neutrino-quark) scattering (DIS) is 
assumed. DIS differential cross sections can be expressed in terms of  PDF's. These PDF's are fitted
to  various experimental data sets and provided as a library.
These PDF's are not applicable in the low $q^2 \lesssim 0.8$~GeV$^2$, and 
low resonance mass ($W \lesssim 2$~GeV) regions. 
Therefore, some correction functions have been proposed 
in these kinematic regions~\cite{Bodek:2002ps}.
There could also be some overlaps with the single meson production
contributions depending on the implementation. These overlaps are usually
avoided by limiting $W$ and the multiplicity. The kinematics of hadrons 
produced in DIS, the so-called hadronization, is determined with the 
help of PYTHIA~\cite{Sjostrand:2006za,Sjostrand:2007gs}, the standard library used in the high energy
experiments. PYTHIA is also designed for large  values of 
$W$, at least bigger than 2 GeV. Thus, most 
generators use a scaling function, known as KNO~\cite{Koba:1972ng}, 
whose parameters are also extracted from data.

The particles produced inside the nucleus may interact with the
nucleons on their way out. These are the FSI addressed in different parts of this review. 
Most generators, except GiBUU and NUNDIS/NUNRES,
simulate these rescatterings with a semi-classical cascade. In particular,
the pions are carefully treated because their interaction probability is rather high,
once many of these are produced from resonance decay. Although the basic idea is the same 
in several generators, the actual implementations of   
the particle mean free paths, the determination of the rescattered particle kinematics,
 or the particle production multiplicities are substantially different in the various  generators.
GiBUU takes FSI into account using a semi-classical transport model in coupled channels, which was 
briefly described in Sec.~\ref{sec:pion-inela}. Finally,  NUNDIS/NURES simulates the rescattering in the framework 
of the FLUKA simulation package.

\section{Concluding remarks}

 Neutrino interactions offer unique opportunities for exploring 
fundamental questions in astrophysics, nuclear and particle
 physics. One of these questions regards the neutrino oscillation
 phenomenon, which has been established over the last 15 years.
Neutrino oscillation experiments are currently
 evolving from the discovery to the precision stage. Unavoidably, 
any oscillation experiment faces a major difficulty: the elusive
 nature of the neutrinos. The presence of neutrinos, being chargeless
 particles, can only be inferred by detecting the secondary particles
 created when the neutrinos interact with the nuclear targets used as
 detectors.  A better understanding of the neutrino-nucleus
 interactions is then crucial to minimize systematic uncertainties in
 neutrino oscillation experiments. For nuclear physics this represents
 not only a challenge, but also an opportunity. Indeed, with the recent 
 intense experimental activity a wealth of
 new more precise neutrino-nucleus cross section data have become available,  
and more are awaited in the future.

Most of the relevant event sample falls inside the poorly understood
region of resonance excitation. Experiments like MiniBooNE and
SciBooNE have produced good quality data for quasi-elastic scattering
and pion production at intermediate energies. These data show
interesting deviations from theoretical predictions. Some of them have
found an explanation (CCQE, in terms of 2p2h excitations), 
but other still await for a proper interpretation.  

We have reviewed the recent progress in the physics of neutrino cross
sections, highlighting the open questions revealed 
by the comparison with new experimental data. 
Among others, we have identified both incoherent and coherent pion production on nuclei 
as topical problems, where significant discrepancies 
between data and theoretical predictions exist and have not yet been understood. 
In the first case, state of the art microscopic approaches fail to describe
the CC1$\pi$ MiniBooNE differential and total cross sections: 
theoretical estimates are systematically below the data. Microscopic
models predict a suppression in the pion spectra around the $\Delta(1232)$
resonance region, which is not seen in the experimental distribution of
events. This points out either to an incorrect theoretical description of 
pion FSI effects or to a misunderstanding on the actual (dynamical) origin of the
events measured by MiniBooNE. Regarding coherent pion production, the
SciBooNE measurement of the ratio $\sigma_{{\rm
    CC-COH}\pi^+}/\sigma_{\rm NC-COH\pi^0}$ on carbon is more than
4$\sigma$ away from the bulk of the theoretical results, which are consistent with 
what is expected from the combination of PCAC and isospin invariance.

So far, more precise nuclear cross sections have not allowed to gain insight
into the axial hadron properties because of the intrinsic difficulties
in the treatment of nuclear and FSI effects. The more than 30-year old
ANL and BNL low statistics deuterium pion production data are still
nowadays the best source of information about the $N\Delta$ transition
matrix element. New measurements of neutrino cross sections on
hydrogen and deuterium are thus absolutely necessary, even more when 
there is tension between ANL and BNL $p\pi^+$ data samples.

We should also stress that, to achieve the precision goals
in neutrino oscillation measurements and to reliably extract
information about the axial properties of the nucleon and baryon
resonances, it is crucial that the current theoretical developments are
implemented in the event generators used in the experimental data
analysis. The strategy, adopted by some oscillation experiments, of fitting the new data 
with the parameters available in the MC event generators is  dangerous  
in the long term.  

The MINER$\nu$A experiment at FermiLab, fully dedicated 
to the measurement of neutrino cross sections on nuclear targets, 
has started to take data. This experiment shall
provide valuable information complementary to the findings of
Jefferson Lab with electrons, and stimulate a productive activity 
in the field of neutrino cross sections in the next years.

\begin{acknowledgments}
We thank M. Barbaro, R. Gran, E. Hernandez and M. J. Vicente Vacas for their help with 
some of the figures included in this review.
 Research supported by the Spanish Ministerio de Econom\'\i a y
 Competitividad and European FEDER funds under the contract
 FIS2011-28853-C02-02 and the Spanish
 Consolider-Ingenio 2010 Programme CPAN (CSD2007-00042), by
 Generalitat Valenciana under contract PROMETEO/2009/0090 and by the
 EU HadronPhysics2 project, grant agreement no. 227431.
\end{acknowledgments}

\bibliography{neutrino}

\begin{thebibliography}{276}%
\makeatletter
\providecommand \@ifxundefined [1]{%
 \@ifx{#1\undefined}
}%
\providecommand \@ifnum [1]{%
 \ifnum #1\expandafter \@firstoftwo
 \else \expandafter \@secondoftwo
 \fi
}%
\providecommand \@ifx [1]{%
 \ifx #1\expandafter \@firstoftwo
 \else \expandafter \@secondoftwo
 \fi
}%
\providecommand \natexlab [1]{#1}%
\providecommand \enquote  [1]{``#1''}%
\providecommand \bibnamefont  [1]{#1}%
\providecommand \bibfnamefont [1]{#1}%
\providecommand \citenamefont [1]{#1}%
\providecommand \href@noop [0]{\@secondoftwo}%
\providecommand \href [0]{\begingroup \@sanitize@url \@href}%
\providecommand \@href[1]{\@@startlink{#1}\@@href}%
\providecommand \@@href[1]{\endgroup#1\@@endlink}%
\providecommand \@sanitize@url [0]{\catcode `\\12\catcode `\$12\catcode
  `\&12\catcode `\#12\catcode `\^12\catcode `\_12\catcode `\%12\relax}%
\providecommand \@@startlink[1]{}%
\providecommand \@@endlink[0]{}%
\providecommand \url  [0]{\begingroup\@sanitize@url \@url }%
\providecommand \@url [1]{\endgroup\@href {#1}{\urlprefix }}%
\providecommand \urlprefix  [0]{URL }%
\providecommand \Eprint [0]{\href }%
\providecommand \doibase [0]{http://dx.doi.org/}%
\providecommand \selectlanguage [0]{\@gobble}%
\providecommand \bibinfo  [0]{\@secondoftwo}%
\providecommand \bibfield  [0]{\@secondoftwo}%
\providecommand \translation [1]{[#1]}%
\providecommand \BibitemOpen [0]{}%
\providecommand \bibitemStop [0]{}%
\providecommand \bibitemNoStop [0]{.\EOS\space}%
\providecommand \EOS [0]{\spacefactor3000\relax}%
\providecommand \BibitemShut  [1]{\csname bibitem#1\endcsname}%
\let\auto@bib@innerbib\@empty
\bibitem [{\citenamefont {Bertulani}\ and\ \citenamefont
  {Gade}(2010)}]{Bertulani:2009mf}%
  \BibitemOpen
  \bibfield  {author} {\bibinfo {author} {\bibfnamefont {C.}~\bibnamefont
  {Bertulani}}\ and\ \bibinfo {author} {\bibfnamefont {A.}~\bibnamefont
  {Gade}},\ }\href {\doibase 10.1016/j.physrep.2009.09.002} {\bibfield
  {journal} {\bibinfo  {journal} {Phys.Rept.}\ }\textbf {\bibinfo {volume}
  {485}},\ \bibinfo {pages} {195} (\bibinfo {year} {2010})},\ \Eprint
  {http://arxiv.org/abs/0909.5693} {arXiv:0909.5693 [nucl-th]} \BibitemShut
  {NoStop}%
\bibitem [{\citenamefont {Volpe}(2013)}]{Volpe:2013kxa}%
  \BibitemOpen
  \bibfield  {author} {\bibinfo {author} {\bibfnamefont {C.}~\bibnamefont
  {Volpe}},\ }\href@noop {} {\bibfield  {journal} {\bibinfo  {journal}
  {Ann.Phys.(Berlin)}\ }\textbf {\bibinfo {volume} {525}},\ \bibinfo {pages}
  {588} (\bibinfo {year} {2013})},\ \Eprint {http://arxiv.org/abs/1303.1681}
  {arXiv:1303.1681 [hep-ph]} \BibitemShut {NoStop}%
\bibitem [{\citenamefont {Ohlsson}(2013)}]{Ohlsson:2012kf}%
  \BibitemOpen
  \bibfield  {author} {\bibinfo {author} {\bibfnamefont {T.}~\bibnamefont
  {Ohlsson}},\ }\href {\doibase 10.1088/0034-4885/76/4/044201} {\bibfield
  {journal} {\bibinfo  {journal} {Rept.Prog.Phys.}\ }\textbf {\bibinfo {volume}
  {76}},\ \bibinfo {pages} {044201} (\bibinfo {year} {2013})},\ \Eprint
  {http://arxiv.org/abs/1209.2710} {arXiv:1209.2710 [hep-ph]} \BibitemShut
  {NoStop}%
\bibitem [{\citenamefont {Gallagher}\ \emph {et~al.}(2011)\citenamefont
  {Gallagher}, \citenamefont {Garvey},\ and\ \citenamefont
  {Zeller}}]{Gallagher:2011zza}%
  \BibitemOpen
  \bibfield  {author} {\bibinfo {author} {\bibfnamefont {H.}~\bibnamefont
  {Gallagher}}, \bibinfo {author} {\bibfnamefont {G.}~\bibnamefont {Garvey}}, \
  and\ \bibinfo {author} {\bibfnamefont {G.}~\bibnamefont {Zeller}},\
  }\href@noop {} {\bibfield  {journal} {\bibinfo  {journal}
  {Ann.Rev.Nucl.Part.Sci.}\ }\textbf {\bibinfo {volume} {61}},\ \bibinfo
  {pages} {355} (\bibinfo {year} {2011})}\BibitemShut {NoStop}%
\bibitem [{\citenamefont {Morfin}\ \emph {et~al.}(2012)\citenamefont {Morfin},
  \citenamefont {Nieves},\ and\ \citenamefont {Sobczyk}}]{Morfin:2012kn}%
  \BibitemOpen
  \bibfield  {author} {\bibinfo {author} {\bibfnamefont {J.~G.}\ \bibnamefont
  {Morfin}}, \bibinfo {author} {\bibfnamefont {J.}~\bibnamefont {Nieves}}, \
  and\ \bibinfo {author} {\bibfnamefont {J.~T.}\ \bibnamefont {Sobczyk}},\
  }\href {\doibase 10.1155/2012/934597} {\bibfield  {journal} {\bibinfo
  {journal} {Adv.High Energy Phys.}\ }\textbf {\bibinfo {volume} {2012}},\
  \bibinfo {pages} {934597} (\bibinfo {year} {2012})},\ \Eprint
  {http://arxiv.org/abs/1209.6586} {arXiv:1209.6586 [hep-ex]} \BibitemShut
  {NoStop}%
\bibitem [{\citenamefont {Kopeliovich}\ \emph {et~al.}(2013)\citenamefont
  {Kopeliovich}, \citenamefont {Morfin},\ and\ \citenamefont
  {Schmidt}}]{Kopeliovich:2012kw}%
  \BibitemOpen
  \bibfield  {author} {\bibinfo {author} {\bibfnamefont {B.}~\bibnamefont
  {Kopeliovich}}, \bibinfo {author} {\bibfnamefont {J.}~\bibnamefont {Morfin}},
  \ and\ \bibinfo {author} {\bibfnamefont {I.}~\bibnamefont {Schmidt}},\ }\href
  {\doibase 10.1016/j.ppnp.2012.09.004} {\bibfield  {journal} {\bibinfo
  {journal} {Prog.Part.Nucl.Phys.}\ }\textbf {\bibinfo {volume} {68}},\
  \bibinfo {pages} {314} (\bibinfo {year} {2013})},\ \Eprint
  {http://arxiv.org/abs/1208.6541} {arXiv:1208.6541 [hep-ph]} \BibitemShut
  {NoStop}%
\bibitem [{\citenamefont {Formaggio}\ and\ \citenamefont
  {Zeller}(2012)}]{Formaggio:2013kya}%
  \BibitemOpen
  \bibfield  {author} {\bibinfo {author} {\bibfnamefont {J.}~\bibnamefont
  {Formaggio}}\ and\ \bibinfo {author} {\bibfnamefont {G.}~\bibnamefont
  {Zeller}},\ }\href@noop {} {\bibfield  {journal} {\bibinfo  {journal}
  {Rev.Mod.Phys.}\ }\textbf {\bibinfo {volume} {84}},\ \bibinfo {pages} {1307}
  (\bibinfo {year} {2012})},\ \Eprint {http://arxiv.org/abs/1305.7513}
  {arXiv:1305.7513 [hep-ex]} \BibitemShut {NoStop}%
\bibitem [{\citenamefont {Fukuda}\ \emph {et~al.}(1998)\citenamefont {Fukuda}
  \emph {et~al.}}]{Fukuda:1998mi}%
  \BibitemOpen
  \bibfield  {author} {\bibinfo {author} {\bibfnamefont {Y.}~\bibnamefont
  {Fukuda}} \emph {et~al.} (\bibinfo {collaboration} {Super-Kamiokande}),\
  }\href@noop {} {\bibfield  {journal} {\bibinfo  {journal} {Phys. Rev. Lett.}\
  }\textbf {\bibinfo {volume} {81}},\ \bibinfo {pages} {1562} (\bibinfo {year}
  {1998})},\ \Eprint {http://arxiv.org/abs/hep-ex/9807003}
  {arXiv:hep-ex/9807003} \BibitemShut {NoStop}%
\bibitem [{\citenamefont {Ashie}\ \emph {et~al.}(2005)\citenamefont {Ashie}
  \emph {et~al.}}]{Ashie:2005ik}%
  \BibitemOpen
  \bibfield  {author} {\bibinfo {author} {\bibfnamefont {Y.}~\bibnamefont
  {Ashie}} \emph {et~al.} (\bibinfo {collaboration} {Super-Kamiokande
  Collaboration}),\ }\href {\doibase 10.1103/PhysRevD.71.112005} {\bibfield
  {journal} {\bibinfo  {journal} {Phys.Rev.}\ }\textbf {\bibinfo {volume}
  {D71}},\ \bibinfo {pages} {112005} (\bibinfo {year} {2005})},\ \Eprint
  {http://arxiv.org/abs/hep-ex/0501064} {arXiv:hep-ex/0501064 [hep-ex]}
  \BibitemShut {NoStop}%
\bibitem [{\citenamefont {Abe}\ \emph {et~al.}(2013{\natexlab{a}})\citenamefont
  {Abe} \emph {et~al.}}]{Abe:2012jj}%
  \BibitemOpen
  \bibfield  {author} {\bibinfo {author} {\bibfnamefont {K.}~\bibnamefont
  {Abe}} \emph {et~al.} (\bibinfo {collaboration} {Super-Kamiokande
  Collaboration}),\ }\href {\doibase 10.1103/PhysRevLett.110.181802} {\bibfield
   {journal} {\bibinfo  {journal} {Phys.Rev.Lett.}\ }\textbf {\bibinfo {volume}
  {110}},\ \bibinfo {pages} {181802} (\bibinfo {year} {2013}{\natexlab{a}})},\
  \Eprint {http://arxiv.org/abs/1206.0328} {arXiv:1206.0328 [hep-ex]}
  \BibitemShut {NoStop}%
\bibitem [{\citenamefont {Wendell}\ \emph {et~al.}(2010)\citenamefont {Wendell}
  \emph {et~al.}}]{Wendell:2010md}%
  \BibitemOpen
  \bibfield  {author} {\bibinfo {author} {\bibfnamefont {R.}~\bibnamefont
  {Wendell}} \emph {et~al.} (\bibinfo {collaboration} {Super-Kamiokande
  Collaboration}),\ }\href {\doibase 10.1103/PhysRevD.81.092004} {\bibfield
  {journal} {\bibinfo  {journal} {Phys.Rev.}\ }\textbf {\bibinfo {volume}
  {D81}},\ \bibinfo {pages} {092004} (\bibinfo {year} {2010})},\ \Eprint
  {http://arxiv.org/abs/1002.3471} {arXiv:1002.3471 [hep-ex]} \BibitemShut
  {NoStop}%
\bibitem [{\citenamefont {Ahn}\ \emph {et~al.}(2001)\citenamefont {Ahn} \emph
  {et~al.}}]{Ahn:2001cq}%
  \BibitemOpen
  \bibfield  {author} {\bibinfo {author} {\bibfnamefont {S.~H.}\ \bibnamefont
  {Ahn}} \emph {et~al.} (\bibinfo {collaboration} {K2K}),\ }\href@noop {}
  {\bibfield  {journal} {\bibinfo  {journal} {Phys. Lett.}\ }\textbf {\bibinfo
  {volume} {B511}},\ \bibinfo {pages} {178} (\bibinfo {year} {2001})},\ \Eprint
  {http://arxiv.org/abs/hep-ex/0103001} {arXiv:hep-ex/0103001} \BibitemShut
  {NoStop}%
\bibitem [{\citenamefont {Adamson}\ \emph {et~al.}(2008)\citenamefont {Adamson}
  \emph {et~al.}}]{Adamson:2007gu}%
  \BibitemOpen
  \bibfield  {author} {\bibinfo {author} {\bibfnamefont {P.}~\bibnamefont
  {Adamson}} \emph {et~al.} (\bibinfo {collaboration} {MINOS}),\ }\href
  {\doibase 10.1103/PhysRevD.77.072002} {\bibfield  {journal} {\bibinfo
  {journal} {Phys. Rev.}\ }\textbf {\bibinfo {volume} {D77}},\ \bibinfo {pages}
  {072002} (\bibinfo {year} {2008})},\ \Eprint {http://arxiv.org/abs/0711.0769}
  {arXiv:0711.0769 [hep-ex]} \BibitemShut {NoStop}%
\bibitem [{\citenamefont {Agafonova}\ \emph {et~al.}(2010)\citenamefont
  {Agafonova} \emph {et~al.}}]{Agafonova:2010dc}%
  \BibitemOpen
  \bibfield  {author} {\bibinfo {author} {\bibfnamefont {N.}~\bibnamefont
  {Agafonova}} \emph {et~al.} (\bibinfo {collaboration} {OPERA
  Collaboration}),\ }\href {\doibase 10.1016/j.physletb.2010.06.022} {\bibfield
   {journal} {\bibinfo  {journal} {Phys.Lett.}\ }\textbf {\bibinfo {volume}
  {B691}},\ \bibinfo {pages} {138} (\bibinfo {year} {2010})},\ \Eprint
  {http://arxiv.org/abs/1006.1623} {arXiv:1006.1623 [hep-ex]} \BibitemShut
  {NoStop}%
\bibitem [{\citenamefont {Abgrall}\ \emph {et~al.}(2011)\citenamefont {Abgrall}
  \emph {et~al.}}]{Abe:2011ks}%
  \BibitemOpen
  \bibfield  {author} {\bibinfo {author} {\bibfnamefont {N.}~\bibnamefont
  {Abgrall}} \emph {et~al.} (\bibinfo {collaboration} {T2KK. Abe}),\
  }\href@noop {} {\bibfield  {journal} {\bibinfo  {journal} {Nucl. Instrum.
  Meth.}\ }\textbf {\bibinfo {volume} {A659}},\ \bibinfo {pages} {106}
  (\bibinfo {year} {2011})},\ \Eprint {http://arxiv.org/abs/1106.1238}
  {arXiv:1106.1238 [physics]} \BibitemShut {NoStop}%
\bibitem [{\citenamefont {Gran}\ \emph {et~al.}(2006)\citenamefont {Gran} \emph
  {et~al.}}]{Gran:2006jn}%
  \BibitemOpen
  \bibfield  {author} {\bibinfo {author} {\bibfnamefont {R.}~\bibnamefont
  {Gran}} \emph {et~al.} (\bibinfo {collaboration} {K2K}),\ }\href@noop {}
  {\bibfield  {journal} {\bibinfo  {journal} {Phys. Rev.}\ }\textbf {\bibinfo
  {volume} {D74}},\ \bibinfo {pages} {052002} (\bibinfo {year} {2006})},\
  \Eprint {http://arxiv.org/abs/hep-ex/0603034} {arXiv:hep-ex/0603034}
  \BibitemShut {NoStop}%
\bibitem [{\citenamefont {Bodek}\ and\ \citenamefont
  {Yang}(2003{\natexlab{a}})}]{Bodek:2003wc}%
  \BibitemOpen
  \bibfield  {author} {\bibinfo {author} {\bibfnamefont {A.}~\bibnamefont
  {Bodek}}\ and\ \bibinfo {author} {\bibfnamefont {U.}~\bibnamefont {Yang}},\
  }\href {\doibase 10.1063/1.1594324} {\bibfield  {journal} {\bibinfo
  {journal} {AIP Conf.Proc.}\ }\textbf {\bibinfo {volume} {670}},\ \bibinfo
  {pages} {110} (\bibinfo {year} {2003}{\natexlab{a}})},\ \Eprint
  {http://arxiv.org/abs/hep-ex/0301036} {arXiv:hep-ex/0301036 [hep-ex]}
  \BibitemShut {NoStop}%
\bibitem [{\citenamefont {Nakayama}\ \emph {et~al.}(2005)\citenamefont
  {Nakayama} \emph {et~al.}}]{Nakayama:2004dp}%
  \BibitemOpen
  \bibfield  {author} {\bibinfo {author} {\bibfnamefont {S.}~\bibnamefont
  {Nakayama}} \emph {et~al.} (\bibinfo {collaboration} {K2K Collaboration}),\
  }\href {\doibase 10.1016/j.physletb.2005.05.044} {\bibfield  {journal}
  {\bibinfo  {journal} {Phys.Lett.}\ }\textbf {\bibinfo {volume} {B619}},\
  \bibinfo {pages} {255} (\bibinfo {year} {2005})},\ \Eprint
  {http://arxiv.org/abs/hep-ex/0408134} {arXiv:hep-ex/0408134 [hep-ex]}
  \BibitemShut {NoStop}%
\bibitem [{\citenamefont {Hasegawa}\ \emph {et~al.}(2005)\citenamefont
  {Hasegawa} \emph {et~al.}}]{Hasegawa:2005td}%
  \BibitemOpen
  \bibfield  {author} {\bibinfo {author} {\bibfnamefont {M.}~\bibnamefont
  {Hasegawa}} \emph {et~al.} (\bibinfo {collaboration} {K2K Collaboration}),\
  }\href {\doibase 10.1103/PhysRevLett.95.252301} {\bibfield  {journal}
  {\bibinfo  {journal} {Phys.Rev.Lett.}\ }\textbf {\bibinfo {volume} {95}},\
  \bibinfo {pages} {252301} (\bibinfo {year} {2005})},\ \Eprint
  {http://arxiv.org/abs/hep-ex/0506008} {arXiv:hep-ex/0506008 [hep-ex]}
  \BibitemShut {NoStop}%
\bibitem [{\citenamefont {Adamson}\ \emph {et~al.}(2013)\citenamefont {Adamson}
  \emph {et~al.}}]{Adamson:2013whj}%
  \BibitemOpen
  \bibfield  {author} {\bibinfo {author} {\bibfnamefont {P.}~\bibnamefont
  {Adamson}} \emph {et~al.} (\bibinfo {collaboration} {MINOS Collaboration}),\
  }\href {\doibase 10.1103/PhysRevLett.110.251801} {\bibfield  {journal}
  {\bibinfo  {journal} {Phys.Rev.Lett.}\ }\textbf {\bibinfo {volume} {110}},\
  \bibinfo {pages} {251801} (\bibinfo {year} {2013})},\ \Eprint
  {http://arxiv.org/abs/1304.6335} {arXiv:1304.6335 [hep-ex]} \BibitemShut
  {NoStop}%
\bibitem [{\citenamefont {Dorman}(2009)}]{Dorman:2009zz}%
  \BibitemOpen
  \bibfield  {author} {\bibinfo {author} {\bibfnamefont {M.}~\bibnamefont
  {Dorman}} (\bibinfo {collaboration} {MINOS Collaboration}),\ }\href {\doibase
  10.1063/1.3274143} {\bibfield  {journal} {\bibinfo  {journal} {AIP
  Conf.Proc.}\ }\textbf {\bibinfo {volume} {1189}},\ \bibinfo {pages} {133}
  (\bibinfo {year} {2009})}\BibitemShut {NoStop}%
\bibitem [{\citenamefont {Agafonova}\ \emph {et~al.}(2013)\citenamefont
  {Agafonova} \emph {et~al.}}]{Agafonova:2013dtp}%
  \BibitemOpen
  \bibfield  {author} {\bibinfo {author} {\bibfnamefont {N.}~\bibnamefont
  {Agafonova}} \emph {et~al.} (\bibinfo {collaboration} {OPERA
  Collaboration}),\ }\href {\doibase 10.1007/JHEP11(2013)036} {\bibfield
  {journal} {\bibinfo  {journal} {JHEP}\ }\textbf {\bibinfo {volume} {1311}},\
  \bibinfo {pages} {036} (\bibinfo {year} {2013})},\ \Eprint
  {http://arxiv.org/abs/1308.2553} {arXiv:1308.2553 [hep-ex]} \BibitemShut
  {NoStop}%
\bibitem [{\citenamefont {Aguilar-Arevalo}\ \emph
  {et~al.}(2010{\natexlab{a}})\citenamefont {Aguilar-Arevalo} \emph
  {et~al.}}]{AguilarArevalo:2010zc}%
  \BibitemOpen
  \bibfield  {author} {\bibinfo {author} {\bibfnamefont {A.}~\bibnamefont
  {Aguilar-Arevalo}} \emph {et~al.} (\bibinfo {collaboration} {MiniBooNE
  Collaboration}),\ }\href {\doibase 10.1103/PhysRevD.81.092005} {\bibfield
  {journal} {\bibinfo  {journal} {Phys.Rev.}\ }\textbf {\bibinfo {volume}
  {D81}},\ \bibinfo {pages} {092005} (\bibinfo {year}
  {2010}{\natexlab{a}})}\BibitemShut {NoStop}%
\bibitem [{\citenamefont {Aguilar-Arevalo}\ \emph {et~al.}(2001)\citenamefont
  {Aguilar-Arevalo} \emph {et~al.}}]{Aguilar:2001ty}%
  \BibitemOpen
  \bibfield  {author} {\bibinfo {author} {\bibfnamefont {A.}~\bibnamefont
  {Aguilar-Arevalo}} \emph {et~al.} (\bibinfo {collaboration} {LSND
  Collaboration}),\ }\href {\doibase 10.1103/PhysRevD.64.112007} {\bibfield
  {journal} {\bibinfo  {journal} {Phys.Rev.}\ }\textbf {\bibinfo {volume}
  {D64}},\ \bibinfo {pages} {112007} (\bibinfo {year} {2001})},\ \Eprint
  {http://arxiv.org/abs/hep-ex/0104049} {arXiv:hep-ex/0104049 [hep-ex]}
  \BibitemShut {NoStop}%
\bibitem [{\citenamefont {Aguilar-Arevalo}\ \emph
  {et~al.}(2013{\natexlab{a}})\citenamefont {Aguilar-Arevalo} \emph
  {et~al.}}]{AguilarArevalo:2013hm}%
  \BibitemOpen
  \bibfield  {author} {\bibinfo {author} {\bibfnamefont {A.}~\bibnamefont
  {Aguilar-Arevalo}} \emph {et~al.} (\bibinfo {collaboration} {MiniBooNE
  Collaboration}),\ }\href {\doibase 10.1103/PhysRevD.88.032001} {\bibfield
  {journal} {\bibinfo  {journal} {Phys.Rev.}\ }\textbf {\bibinfo {volume}
  {D88}},\ \bibinfo {pages} {032001} (\bibinfo {year} {2013}{\natexlab{a}})},\
  \Eprint {http://arxiv.org/abs/1301.7067} {arXiv:1301.7067 [hep-ex]}
  \BibitemShut {NoStop}%
\bibitem [{\citenamefont {Katori}(2010)}]{Katori:2009zz}%
  \BibitemOpen
  \bibfield  {author} {\bibinfo {author} {\bibfnamefont {T.}~\bibnamefont
  {Katori}} (\bibinfo {collaboration} {MiniBooNE Collaboration}),\ }\href
  {\doibase 10.1063/1.3399376} {\bibfield  {journal} {\bibinfo  {journal} {AIP
  Conf.Proc.}\ }\textbf {\bibinfo {volume} {1222}},\ \bibinfo {pages} {471}
  (\bibinfo {year} {2010})}\BibitemShut {NoStop}%
\bibitem [{\citenamefont {Aguilar-Arevalo}\ \emph
  {et~al.}(2011{\natexlab{a}})\citenamefont {Aguilar-Arevalo} \emph
  {et~al.}}]{AguilarArevalo:2010xt}%
  \BibitemOpen
  \bibfield  {author} {\bibinfo {author} {\bibfnamefont {A.}~\bibnamefont
  {Aguilar-Arevalo}} \emph {et~al.} (\bibinfo {collaboration} {MiniBooNE
  Collaboration}),\ }\href {\doibase 10.1103/PhysRevD.83.052009} {\bibfield
  {journal} {\bibinfo  {journal} {Phys.Rev.}\ }\textbf {\bibinfo {volume}
  {D83}},\ \bibinfo {pages} {052009} (\bibinfo {year} {2011}{\natexlab{a}})},\
  \Eprint {http://arxiv.org/abs/1010.3264} {arXiv:1010.3264 [hep-ex]}
  \BibitemShut {NoStop}%
\bibitem [{\citenamefont {Aguilar-Arevalo}\ \emph
  {et~al.}(2011{\natexlab{b}})\citenamefont {Aguilar-Arevalo} \emph
  {et~al.}}]{AguilarArevalo:2010bm}%
  \BibitemOpen
  \bibfield  {author} {\bibinfo {author} {\bibfnamefont {A.}~\bibnamefont
  {Aguilar-Arevalo}} \emph {et~al.} (\bibinfo {collaboration} {MiniBooNE
  Collaboration}),\ }\href {\doibase 10.1103/PhysRevD.83.052007} {\bibfield
  {journal} {\bibinfo  {journal} {Phys.Rev.}\ }\textbf {\bibinfo {volume}
  {D83}},\ \bibinfo {pages} {052007} (\bibinfo {year} {2011}{\natexlab{b}})},\
  \Eprint {http://arxiv.org/abs/1011.3572} {arXiv:1011.3572 [hep-ex]}
  \BibitemShut {NoStop}%
\bibitem [{\citenamefont {Aguilar-Arevalo}\ \emph
  {et~al.}(2010{\natexlab{b}})\citenamefont {Aguilar-Arevalo} \emph
  {et~al.}}]{AguilarArevalo:2010cx}%
  \BibitemOpen
  \bibfield  {author} {\bibinfo {author} {\bibfnamefont {A.}~\bibnamefont
  {Aguilar-Arevalo}} \emph {et~al.} (\bibinfo {collaboration} {MiniBooNE
  Collaboration}),\ }\href {\doibase 10.1103/PhysRevD.82.092005} {\bibfield
  {journal} {\bibinfo  {journal} {Phys.Rev.}\ }\textbf {\bibinfo {volume}
  {D82}},\ \bibinfo {pages} {092005} (\bibinfo {year} {2010}{\natexlab{b}})},\
  \Eprint {http://arxiv.org/abs/1007.4730} {arXiv:1007.4730 [hep-ex]}
  \BibitemShut {NoStop}%
\bibitem [{\citenamefont {Aguilar-Arevalo}\ \emph
  {et~al.}(2010{\natexlab{c}})\citenamefont {Aguilar-Arevalo} \emph
  {et~al.}}]{AguilarArevalo:2009ww}%
  \BibitemOpen
  \bibfield  {author} {\bibinfo {author} {\bibfnamefont {A.~A.}\ \bibnamefont
  {Aguilar-Arevalo}} \emph {et~al.} (\bibinfo {collaboration} {MiniBooNE
  Collaboration}),\ }\href {\doibase 10.1103/PhysRevD.81.013005} {\bibfield
  {journal} {\bibinfo  {journal} {Phys.Rev.}\ }\textbf {\bibinfo {volume}
  {D81}},\ \bibinfo {pages} {013005} (\bibinfo {year} {2010}{\natexlab{c}})},\
  \Eprint {http://arxiv.org/abs/0911.2063} {arXiv:0911.2063 [hep-ex]}
  \BibitemShut {NoStop}%
\bibitem [{\citenamefont {Aguilar-Arevalo}\ \emph
  {et~al.}(2009{\natexlab{a}})\citenamefont {Aguilar-Arevalo} \emph
  {et~al.}}]{AguilarArevalo:2008rc}%
  \BibitemOpen
  \bibfield  {author} {\bibinfo {author} {\bibfnamefont {A.}~\bibnamefont
  {Aguilar-Arevalo}} \emph {et~al.} (\bibinfo {collaboration} {MiniBooNE
  Collaboration}),\ }\href {\doibase 10.1103/PhysRevLett.102.101802} {\bibfield
   {journal} {\bibinfo  {journal} {Phys.Rev.Lett.}\ }\textbf {\bibinfo {volume}
  {102}},\ \bibinfo {pages} {101802} (\bibinfo {year} {2009}{\natexlab{a}})},\
  \Eprint {http://arxiv.org/abs/0812.2243} {arXiv:0812.2243 [hep-ex]}
  \BibitemShut {NoStop}%
\bibitem [{\citenamefont {Aguilar-Arevalo}\ \emph
  {et~al.}(2013{\natexlab{b}})\citenamefont {Aguilar-Arevalo} \emph
  {et~al.}}]{Aguilar-Arevalo:2013pmq}%
  \BibitemOpen
  \bibfield  {author} {\bibinfo {author} {\bibfnamefont {A.}~\bibnamefont
  {Aguilar-Arevalo}} \emph {et~al.} (\bibinfo {collaboration} {MiniBooNE
  Collaboration}),\ }\href {\doibase 10.1103/PhysRevLett.110.161801} {\bibfield
   {journal} {\bibinfo  {journal} {Phys.Rev.Lett.}\ }\textbf {\bibinfo {volume}
  {110}},\ \bibinfo {pages} {161801} (\bibinfo {year} {2013}{\natexlab{b}})},\
  \Eprint {http://arxiv.org/abs/1207.4809} {arXiv:1207.4809 [hep-ex]}
  \BibitemShut {NoStop}%
\bibitem [{\citenamefont {Altegoer}\ \emph {et~al.}(1998)\citenamefont
  {Altegoer} \emph {et~al.}}]{Altegoer:1997gv}%
  \BibitemOpen
  \bibfield  {author} {\bibinfo {author} {\bibfnamefont {J.}~\bibnamefont
  {Altegoer}} \emph {et~al.} (\bibinfo {collaboration} {NOMAD Collaboration}),\
  }\href {\doibase 10.1016/S0168-9002(97)01079-6} {\bibfield  {journal}
  {\bibinfo  {journal} {Nucl.Instrum.Meth.}\ }\textbf {\bibinfo {volume}
  {A404}},\ \bibinfo {pages} {96} (\bibinfo {year} {1998})}\BibitemShut
  {NoStop}%
\bibitem [{\citenamefont {Astier}\ \emph {et~al.}(2001)\citenamefont {Astier}
  \emph {et~al.}}]{Astier:2001yj}%
  \BibitemOpen
  \bibfield  {author} {\bibinfo {author} {\bibfnamefont {P.}~\bibnamefont
  {Astier}} \emph {et~al.} (\bibinfo {collaboration} {NOMAD Collaboration}),\
  }\href {\doibase 10.1016/S0550-3213(01)00339-X} {\bibfield  {journal}
  {\bibinfo  {journal} {Nucl.Phys.}\ }\textbf {\bibinfo {volume} {B611}},\
  \bibinfo {pages} {3} (\bibinfo {year} {2001})},\ \Eprint
  {http://arxiv.org/abs/hep-ex/0106102} {arXiv:hep-ex/0106102 [hep-ex]}
  \BibitemShut {NoStop}%
\bibitem [{\citenamefont {Lyubushkin}\ \emph {et~al.}(2009)\citenamefont
  {Lyubushkin} \emph {et~al.}}]{Lyubushkin:2008pe}%
  \BibitemOpen
  \bibfield  {author} {\bibinfo {author} {\bibfnamefont {V.}~\bibnamefont
  {Lyubushkin}} \emph {et~al.} (\bibinfo {collaboration} {NOMAD
  Collaboration}),\ }\href {\doibase 10.1140/epjc/s10052-009-1113-0} {\bibfield
   {journal} {\bibinfo  {journal} {Eur.Phys.J.}\ }\textbf {\bibinfo {volume}
  {C63}},\ \bibinfo {pages} {355} (\bibinfo {year} {2009})},\ \Eprint
  {http://arxiv.org/abs/0812.4543} {arXiv:0812.4543 [hep-ex]} \BibitemShut
  {NoStop}%
\bibitem [{\citenamefont {Kullenberg}\ \emph {et~al.}(2009)\citenamefont
  {Kullenberg} \emph {et~al.}}]{Kullenberg:2009pu}%
  \BibitemOpen
  \bibfield  {author} {\bibinfo {author} {\bibfnamefont {C.}~\bibnamefont
  {Kullenberg}} \emph {et~al.} (\bibinfo {collaboration} {NOMAD
  Collaboration}),\ }\href {\doibase 10.1016/j.physletb.2009.10.083} {\bibfield
   {journal} {\bibinfo  {journal} {Phys.Lett.}\ }\textbf {\bibinfo {volume}
  {B682}},\ \bibinfo {pages} {177} (\bibinfo {year} {2009})},\ \Eprint
  {http://arxiv.org/abs/0910.0062} {arXiv:0910.0062 [hep-ex]} \BibitemShut
  {NoStop}%
\bibitem [{\citenamefont {Tian}(2013)}]{fortheNOMAD:2013gba}%
  \BibitemOpen
  \bibfield  {author} {\bibinfo {author} {\bibfnamefont {X.}~\bibnamefont
  {Tian}} (\bibinfo {collaboration} {NOMAD Collaboration}),\ }\href@noop {} {\
  (\bibinfo {year} {2013})},\ \Eprint {http://arxiv.org/abs/1310.8547}
  {arXiv:1310.8547 [hep-ex]} \BibitemShut {NoStop}%
\bibitem [{\citenamefont {Astier}\ \emph
  {et~al.}(2002{\natexlab{a}})\citenamefont {Astier} \emph
  {et~al.}}]{Astier:2001vi}%
  \BibitemOpen
  \bibfield  {author} {\bibinfo {author} {\bibfnamefont {P.}~\bibnamefont
  {Astier}} \emph {et~al.} (\bibinfo {collaboration} {NOMAD Collaboration}),\
  }\href {\doibase 10.1016/S0550-3213(01)00584-3} {\bibfield  {journal}
  {\bibinfo  {journal} {Nucl.Phys.}\ }\textbf {\bibinfo {volume} {B621}},\
  \bibinfo {pages} {3} (\bibinfo {year} {2002}{\natexlab{a}})},\ \Eprint
  {http://arxiv.org/abs/hep-ex/0111057} {arXiv:hep-ex/0111057 [hep-ex]}
  \BibitemShut {NoStop}%
\bibitem [{\citenamefont {Naumov}\ \emph {et~al.}(2004)\citenamefont {Naumov}
  \emph {et~al.}}]{Naumov:2004wa}%
  \BibitemOpen
  \bibfield  {author} {\bibinfo {author} {\bibfnamefont {D.}~\bibnamefont
  {Naumov}} \emph {et~al.} (\bibinfo {collaboration} {NOMAD Collaboration}),\
  }\href {\doibase 10.1016/j.nuclphysb.2004.09.013} {\bibfield  {journal}
  {\bibinfo  {journal} {Nucl.Phys.}\ }\textbf {\bibinfo {volume} {B700}},\
  \bibinfo {pages} {51} (\bibinfo {year} {2004})},\ \Eprint
  {http://arxiv.org/abs/hep-ex/0409037} {arXiv:hep-ex/0409037 [hep-ex]}
  \BibitemShut {NoStop}%
\bibitem [{\citenamefont {Astier}\ \emph
  {et~al.}(2002{\natexlab{b}})\citenamefont {Astier} \emph
  {et~al.}}]{Astier:2001ri}%
  \BibitemOpen
  \bibfield  {author} {\bibinfo {author} {\bibfnamefont {P.}~\bibnamefont
  {Astier}} \emph {et~al.} (\bibinfo {collaboration} {NOMAD Collaboration}),\
  }\href {\doibase 10.1016/S0370-2693(01)01493-9} {\bibfield  {journal}
  {\bibinfo  {journal} {Phys.Lett.}\ }\textbf {\bibinfo {volume} {B526}},\
  \bibinfo {pages} {278} (\bibinfo {year} {2002}{\natexlab{b}})}\BibitemShut
  {NoStop}%
\bibitem [{\citenamefont {Aunion}(2010)}]{Aunion:2010zz}%
  \BibitemOpen
  \bibfield  {author} {\bibinfo {author} {\bibfnamefont {J.~L.~A.}\
  \bibnamefont {Aunion}},\ }\href@noop {} {\bibfield  {journal} {\bibinfo
  {journal} {PhD Thesis, Barcelona}\ } (\bibinfo {year} {2010})}\BibitemShut
  {NoStop}%
\bibitem [{\citenamefont {Katori}(2013)}]{Katori:2013nca}%
  \BibitemOpen
  \bibfield  {author} {\bibinfo {author} {\bibfnamefont {T.}~\bibnamefont
  {Katori}} (\bibinfo {collaboration} {MiniBooNE Collaboration, SciBooNE
  Collaboration}),\ }\href@noop {} {\  (\bibinfo {year} {2013})},\ \Eprint
  {http://arxiv.org/abs/1304.5325} {arXiv:1304.5325 [hep-ex]} \BibitemShut
  {NoStop}%
\bibitem [{\citenamefont {Hiraide}\ \emph {et~al.}(2008)\citenamefont {Hiraide}
  \emph {et~al.}}]{Hiraide:2008eu}%
  \BibitemOpen
  \bibfield  {author} {\bibinfo {author} {\bibfnamefont {K.}~\bibnamefont
  {Hiraide}} \emph {et~al.} (\bibinfo {collaboration} {SciBooNE
  Collaboration}),\ }\href {\doibase 10.1103/PhysRevD.78.112004} {\bibfield
  {journal} {\bibinfo  {journal} {Phys.Rev.}\ }\textbf {\bibinfo {volume}
  {D78}},\ \bibinfo {pages} {112004} (\bibinfo {year} {2008})},\ \Eprint
  {http://arxiv.org/abs/0811.0369} {arXiv:0811.0369 [hep-ex]} \BibitemShut
  {NoStop}%
\bibitem [{\citenamefont {Kurimoto}\ \emph
  {et~al.}(2010{\natexlab{a}})\citenamefont {Kurimoto} \emph
  {et~al.}}]{Kurimoto:2010rc}%
  \BibitemOpen
  \bibfield  {author} {\bibinfo {author} {\bibfnamefont {Y.}~\bibnamefont
  {Kurimoto}} \emph {et~al.} (\bibinfo {collaboration} {SciBooNE
  Collaboration}),\ }\href {\doibase 10.1103/PhysRevD.81.111102} {\bibfield
  {journal} {\bibinfo  {journal} {Phys.Rev.}\ }\textbf {\bibinfo {volume}
  {D81}},\ \bibinfo {pages} {111102} (\bibinfo {year} {2010}{\natexlab{a}})},\
  \Eprint {http://arxiv.org/abs/1005.0059} {arXiv:1005.0059 [hep-ex]}
  \BibitemShut {NoStop}%
\bibitem [{\citenamefont {Nakajima}\ \emph {et~al.}(2011)\citenamefont
  {Nakajima} \emph {et~al.}}]{Nakajima:2010fp}%
  \BibitemOpen
  \bibfield  {author} {\bibinfo {author} {\bibfnamefont {Y.}~\bibnamefont
  {Nakajima}} \emph {et~al.} (\bibinfo {collaboration} {SciBooNE
  Collaboration}),\ }\href {\doibase 10.1103/PhysRevD.83.012005} {\bibfield
  {journal} {\bibinfo  {journal} {Phys.Rev.}\ }\textbf {\bibinfo {volume}
  {D83}},\ \bibinfo {pages} {012005} (\bibinfo {year} {2011})}\BibitemShut
  {NoStop}%
\bibitem [{\citenamefont {Anderson}\ \emph
  {et~al.}(2012{\natexlab{a}})\citenamefont {Anderson}, \citenamefont
  {Antonello}, \citenamefont {Baller}, \citenamefont {Bolton}, \citenamefont
  {Bromberg} \emph {et~al.}}]{Anderson:2012vc}%
  \BibitemOpen
  \bibfield  {author} {\bibinfo {author} {\bibfnamefont {C.}~\bibnamefont
  {Anderson}}, \bibinfo {author} {\bibfnamefont {M.}~\bibnamefont {Antonello}},
  \bibinfo {author} {\bibfnamefont {B.}~\bibnamefont {Baller}}, \bibinfo
  {author} {\bibfnamefont {T.}~\bibnamefont {Bolton}}, \bibinfo {author}
  {\bibfnamefont {C.}~\bibnamefont {Bromberg}},  \emph {et~al.},\ }\href
  {\doibase 10.1088/1748-0221/7/10/P10019} {\bibfield  {journal} {\bibinfo
  {journal} {JINST}\ }\textbf {\bibinfo {volume} {7}},\ \bibinfo {pages}
  {P10019} (\bibinfo {year} {2012}{\natexlab{a}})},\ \Eprint
  {http://arxiv.org/abs/1205.6747} {arXiv:1205.6747 [physics.ins-det]}
  \BibitemShut {NoStop}%
\bibitem [{\citenamefont {Anderson}\ \emph
  {et~al.}(2012{\natexlab{b}})\citenamefont {Anderson} \emph
  {et~al.}}]{Anderson:2011ce}%
  \BibitemOpen
  \bibfield  {author} {\bibinfo {author} {\bibfnamefont {C.}~\bibnamefont
  {Anderson}} \emph {et~al.} (\bibinfo {collaboration} {ArgoNeuT
  Collaboration}),\ }\href {\doibase 10.1103/PhysRevLett.108.161802} {\bibfield
   {journal} {\bibinfo  {journal} {Phys.Rev.Lett.}\ }\textbf {\bibinfo {volume}
  {108}},\ \bibinfo {pages} {161802} (\bibinfo {year} {2012}{\natexlab{b}})},\
  \Eprint {http://arxiv.org/abs/1111.0103} {arXiv:1111.0103 [hep-ex]}
  \BibitemShut {NoStop}%
\bibitem [{\citenamefont {Fields}\ \emph {et~al.}(2013)\citenamefont {Fields}
  \emph {et~al.}}]{Fields:2013zhk}%
  \BibitemOpen
  \bibfield  {author} {\bibinfo {author} {\bibfnamefont {L.}~\bibnamefont
  {Fields}} \emph {et~al.} (\bibinfo {collaboration} {MINERvA Collaboration}),\
  }\href {\doibase 10.1103/PhysRevLett.111.022501} {\bibfield  {journal}
  {\bibinfo  {journal} {Phys.Rev.Lett.}\ }\textbf {\bibinfo {volume} {111}},\
  \bibinfo {pages} {022501} (\bibinfo {year} {2013})},\ \Eprint
  {http://arxiv.org/abs/1305.2234} {arXiv:1305.2234 [hep-ex]} \BibitemShut
  {NoStop}%
\bibitem [{\citenamefont {Fiorentini}\ \emph {et~al.}(2013)\citenamefont
  {Fiorentini} \emph {et~al.}}]{Fiorentini:2013ezn}%
  \BibitemOpen
  \bibfield  {author} {\bibinfo {author} {\bibfnamefont {G.}~\bibnamefont
  {Fiorentini}} \emph {et~al.} (\bibinfo {collaboration} {MINERvA
  Collaboration}),\ }\href {\doibase 10.1103/PhysRevLett.111.022502} {\bibfield
   {journal} {\bibinfo  {journal} {Phys.Rev.Lett.}\ }\textbf {\bibinfo {volume}
  {111}},\ \bibinfo {pages} {022502} (\bibinfo {year} {2013})},\ \Eprint
  {http://arxiv.org/abs/1305.2243} {arXiv:1305.2243 [hep-ex]} \BibitemShut
  {NoStop}%
\bibitem [{\citenamefont {Abe}\ \emph {et~al.}(2011)\citenamefont {Abe} \emph
  {et~al.}}]{Abe:2011sj}%
  \BibitemOpen
  \bibfield  {author} {\bibinfo {author} {\bibfnamefont {K.}~\bibnamefont
  {Abe}} \emph {et~al.} (\bibinfo {collaboration} {T2K Collaboration}),\ }\href
  {\doibase 10.1103/PhysRevLett.107.041801} {\bibfield  {journal} {\bibinfo
  {journal} {Phys.Rev.Lett.}\ }\textbf {\bibinfo {volume} {107}},\ \bibinfo
  {pages} {041801} (\bibinfo {year} {2011})},\ \Eprint
  {http://arxiv.org/abs/1106.2822} {arXiv:1106.2822 [hep-ex]} \BibitemShut
  {NoStop}%
\bibitem [{\citenamefont {Abe}\ \emph {et~al.}(2014)\citenamefont {Abe} \emph
  {et~al.}}]{Abe:2013hdq}%
  \BibitemOpen
  \bibfield  {author} {\bibinfo {author} {\bibfnamefont {K.}~\bibnamefont
  {Abe}} \emph {et~al.} (\bibinfo {collaboration} {T2K Collaboration}),\ }\href
  {\doibase 10.1103/PhysRevLett.112.061802} {\bibfield  {journal} {\bibinfo
  {journal} {Phys.Rev.Lett.}\ }\textbf {\bibinfo {volume} {112}},\ \bibinfo
  {pages} {061802} (\bibinfo {year} {2014})},\ \Eprint
  {http://arxiv.org/abs/1311.4750} {arXiv:1311.4750 [hep-ex]} \BibitemShut
  {NoStop}%
\bibitem [{\citenamefont {An}\ \emph {et~al.}(2014)\citenamefont {An} \emph
  {et~al.}}]{An:2013zwz}%
  \BibitemOpen
  \bibfield  {author} {\bibinfo {author} {\bibfnamefont {F.}~\bibnamefont {An}}
  \emph {et~al.} (\bibinfo {collaboration} {Daya Bay Collaboration}),\ }\href
  {\doibase 10.1103/PhysRevLett.112.061801} {\bibfield  {journal} {\bibinfo
  {journal} {Phys.Rev.Lett.}\ }\textbf {\bibinfo {volume} {112}},\ \bibinfo
  {pages} {061801} (\bibinfo {year} {2014})},\ \Eprint
  {http://arxiv.org/abs/1310.6732} {arXiv:1310.6732 [hep-ex]} \BibitemShut
  {NoStop}%
\bibitem [{\citenamefont {Ahn}\ \emph {et~al.}(2012)\citenamefont {Ahn} \emph
  {et~al.}}]{Ahn:2012nd}%
  \BibitemOpen
  \bibfield  {author} {\bibinfo {author} {\bibfnamefont {J.}~\bibnamefont
  {Ahn}} \emph {et~al.} (\bibinfo {collaboration} {RENO collaboration}),\
  }\href {\doibase 10.1103/PhysRevLett.108.191802} {\bibfield  {journal}
  {\bibinfo  {journal} {Phys.Rev.Lett.}\ }\textbf {\bibinfo {volume} {108}},\
  \bibinfo {pages} {191802} (\bibinfo {year} {2012})},\ \Eprint
  {http://arxiv.org/abs/1204.0626} {arXiv:1204.0626 [hep-ex]} \BibitemShut
  {NoStop}%
\bibitem [{\citenamefont {Abe}\ \emph {et~al.}(2013{\natexlab{b}})\citenamefont
  {Abe} \emph {et~al.}}]{Abe:2013sxa}%
  \BibitemOpen
  \bibfield  {author} {\bibinfo {author} {\bibfnamefont {Y.}~\bibnamefont
  {Abe}} \emph {et~al.} (\bibinfo {collaboration} {Double Chooz
  Collaboration}),\ }\href {\doibase 10.1016/j.physletb.2013.04.050} {\bibfield
   {journal} {\bibinfo  {journal} {Phys.Lett.}\ }\textbf {\bibinfo {volume}
  {B723}},\ \bibinfo {pages} {66} (\bibinfo {year} {2013}{\natexlab{b}})},\
  \Eprint {http://arxiv.org/abs/1301.2948} {arXiv:1301.2948 [hep-ex]}
  \BibitemShut {NoStop}%
\bibitem [{\citenamefont {Nieves}\ \emph {et~al.}(2004)\citenamefont {Nieves},
  \citenamefont {Amaro},\ and\ \citenamefont {Valverde}}]{Nieves:2004wx}%
  \BibitemOpen
  \bibfield  {author} {\bibinfo {author} {\bibfnamefont {J.}~\bibnamefont
  {Nieves}}, \bibinfo {author} {\bibfnamefont {J.~E.}\ \bibnamefont {Amaro}}, \
  and\ \bibinfo {author} {\bibfnamefont {M.}~\bibnamefont {Valverde}},\ }\href
  {\doibase 10.1103/PhysRevC.70.055503, 10.1103/PhysRevC.72.019902} {\bibfield
  {journal} {\bibinfo  {journal} {Phys.Rev.}\ }\textbf {\bibinfo {volume}
  {C70}},\ \bibinfo {pages} {055503} (\bibinfo {year} {2004})},\ \Eprint
  {http://arxiv.org/abs/nucl-th/0408005} {arXiv:nucl-th/0408005 [nucl-th]}
  \BibitemShut {NoStop}%
\bibitem [{\citenamefont {Nieves}\ \emph {et~al.}(2006)\citenamefont {Nieves},
  \citenamefont {Valverde},\ and\ \citenamefont
  {Vicente~Vacas}}]{Nieves:2005rq}%
  \BibitemOpen
  \bibfield  {author} {\bibinfo {author} {\bibfnamefont {J.}~\bibnamefont
  {Nieves}}, \bibinfo {author} {\bibfnamefont {M.}~\bibnamefont {Valverde}}, \
  and\ \bibinfo {author} {\bibfnamefont {M.}~\bibnamefont {Vicente~Vacas}},\
  }\href {\doibase 10.1103/PhysRevC.73.025504} {\bibfield  {journal} {\bibinfo
  {journal} {Phys.Rev.}\ }\textbf {\bibinfo {volume} {C73}},\ \bibinfo {pages}
  {025504} (\bibinfo {year} {2006})},\ \Eprint
  {http://arxiv.org/abs/hep-ph/0511204} {arXiv:hep-ph/0511204 [hep-ph]}
  \BibitemShut {NoStop}%
\bibitem [{\citenamefont {Llewellyn~Smith}(1972)}]{LlewellynSmith:1971zm}%
  \BibitemOpen
  \bibfield  {author} {\bibinfo {author} {\bibfnamefont {C.}~\bibnamefont
  {Llewellyn~Smith}},\ }\href {\doibase 10.1016/0370-1573(72)90010-5}
  {\bibfield  {journal} {\bibinfo  {journal} {Phys.Rept.}\ }\textbf {\bibinfo
  {volume} {3}},\ \bibinfo {pages} {261} (\bibinfo {year} {1972})}\BibitemShut
  {NoStop}%
\bibitem [{\citenamefont {Bernard}\ \emph {et~al.}(1992)\citenamefont
  {Bernard}, \citenamefont {Kaiser},\ and\ \citenamefont
  {Meissner}}]{Bernard:1992ys}%
  \BibitemOpen
  \bibfield  {author} {\bibinfo {author} {\bibfnamefont {V.}~\bibnamefont
  {Bernard}}, \bibinfo {author} {\bibfnamefont {N.}~\bibnamefont {Kaiser}}, \
  and\ \bibinfo {author} {\bibfnamefont {U.~G.}\ \bibnamefont {Meissner}},\
  }\href {\doibase 10.1103/PhysRevLett.69.1877} {\bibfield  {journal} {\bibinfo
   {journal} {Phys.Rev.Lett.}\ }\textbf {\bibinfo {volume} {69}},\ \bibinfo
  {pages} {1877} (\bibinfo {year} {1992})}\BibitemShut {NoStop}%
\bibitem [{\citenamefont {Bernard}\ \emph {et~al.}(2002)\citenamefont
  {Bernard}, \citenamefont {Elouadrhiri},\ and\ \citenamefont
  {Meissner}}]{Bernard:2001rs}%
  \BibitemOpen
  \bibfield  {author} {\bibinfo {author} {\bibfnamefont {V.}~\bibnamefont
  {Bernard}}, \bibinfo {author} {\bibfnamefont {L.}~\bibnamefont
  {Elouadrhiri}}, \ and\ \bibinfo {author} {\bibfnamefont {U.}~\bibnamefont
  {Meissner}},\ }\href {\doibase 10.1088/0954-3899/28/1/201} {\bibfield
  {journal} {\bibinfo  {journal} {J.Phys.}\ }\textbf {\bibinfo {volume}
  {G28}},\ \bibinfo {pages} {R1} (\bibinfo {year} {2002})},\ \Eprint
  {http://arxiv.org/abs/hep-ph/0107088} {arXiv:hep-ph/0107088 [hep-ph]}
  \BibitemShut {NoStop}%
\bibitem [{\citenamefont {Liesenfeld}\ \emph {et~al.}(1999)\citenamefont
  {Liesenfeld} \emph {et~al.}}]{Liesenfeld:1999mv}%
  \BibitemOpen
  \bibfield  {author} {\bibinfo {author} {\bibfnamefont {A.}~\bibnamefont
  {Liesenfeld}} \emph {et~al.} (\bibinfo {collaboration} {A1 Collaboration}),\
  }\href {\doibase 10.1016/S0370-2693(99)01204-6} {\bibfield  {journal}
  {\bibinfo  {journal} {Phys.Lett.}\ }\textbf {\bibinfo {volume} {B468}},\
  \bibinfo {pages} {20} (\bibinfo {year} {1999})},\ \Eprint
  {http://arxiv.org/abs/nucl-ex/9911003} {arXiv:nucl-ex/9911003 [nucl-ex]}
  \BibitemShut {NoStop}%
\bibitem [{\citenamefont {Gonzalez-Jimenez}\ \emph
  {et~al.}(2013{\natexlab{a}})\citenamefont {Gonzalez-Jimenez}, \citenamefont
  {Caballero},\ and\ \citenamefont {Donnelly}}]{GonzalezJimenez:2011fq}%
  \BibitemOpen
  \bibfield  {author} {\bibinfo {author} {\bibfnamefont {R.}~\bibnamefont
  {Gonzalez-Jimenez}}, \bibinfo {author} {\bibfnamefont {J.}~\bibnamefont
  {Caballero}}, \ and\ \bibinfo {author} {\bibfnamefont {T.}~\bibnamefont
  {Donnelly}},\ }\href {\doibase 10.1016/j.physrep.2012.10.003} {\bibfield
  {journal} {\bibinfo  {journal} {Phys.Rept.}\ }\textbf {\bibinfo {volume}
  {524}},\ \bibinfo {pages} {1} (\bibinfo {year} {2013}{\natexlab{a}})},\
  \Eprint {http://arxiv.org/abs/1111.6918} {arXiv:1111.6918 [nucl-th]}
  \BibitemShut {NoStop}%
\bibitem [{\citenamefont {Bodek}\ \emph {et~al.}(2008)\citenamefont {Bodek},
  \citenamefont {Avvakumov}, \citenamefont {Bradford},\ and\ \citenamefont
  {Budd}}]{Bodek:2007ym}%
  \BibitemOpen
  \bibfield  {author} {\bibinfo {author} {\bibfnamefont {A.}~\bibnamefont
  {Bodek}}, \bibinfo {author} {\bibfnamefont {S.}~\bibnamefont {Avvakumov}},
  \bibinfo {author} {\bibfnamefont {R.}~\bibnamefont {Bradford}}, \ and\
  \bibinfo {author} {\bibfnamefont {H.~S.}\ \bibnamefont {Budd}},\ }\href
  {\doibase 10.1140/epjc/s10052-007-0491-4} {\bibfield  {journal} {\bibinfo
  {journal} {Eur.Phys.J.}\ }\textbf {\bibinfo {volume} {C53}},\ \bibinfo
  {pages} {349} (\bibinfo {year} {2008})},\ \Eprint
  {http://arxiv.org/abs/0708.1946} {arXiv:0708.1946 [hep-ex]} \BibitemShut
  {NoStop}%
\bibitem [{\citenamefont {Crawford}\ \emph {et~al.}(2010)\citenamefont
  {Crawford}, \citenamefont {Akdogan}, \citenamefont {Alarcon}, \citenamefont
  {Bertozzi}, \citenamefont {Booth} \emph {et~al.}}]{Crawford:2010gv}%
  \BibitemOpen
  \bibfield  {author} {\bibinfo {author} {\bibfnamefont {C.}~\bibnamefont
  {Crawford}}, \bibinfo {author} {\bibfnamefont {T.}~\bibnamefont {Akdogan}},
  \bibinfo {author} {\bibfnamefont {R.}~\bibnamefont {Alarcon}}, \bibinfo
  {author} {\bibfnamefont {W.}~\bibnamefont {Bertozzi}}, \bibinfo {author}
  {\bibfnamefont {E.}~\bibnamefont {Booth}},  \emph {et~al.},\ }\href {\doibase
  10.1103/PhysRevC.82.045211} {\bibfield  {journal} {\bibinfo  {journal}
  {Phys.Rev.}\ }\textbf {\bibinfo {volume} {C82}},\ \bibinfo {pages} {045211}
  (\bibinfo {year} {2010})},\ \Eprint {http://arxiv.org/abs/1003.0903}
  {arXiv:1003.0903 [nucl-th]} \BibitemShut {NoStop}%
\bibitem [{\citenamefont {Masjuan}\ \emph {et~al.}(2013)\citenamefont
  {Masjuan}, \citenamefont {Ruiz~Arriola},\ and\ \citenamefont
  {Broniowski}}]{Masjuan:2012sk}%
  \BibitemOpen
  \bibfield  {author} {\bibinfo {author} {\bibfnamefont {P.}~\bibnamefont
  {Masjuan}}, \bibinfo {author} {\bibfnamefont {E.}~\bibnamefont
  {Ruiz~Arriola}}, \ and\ \bibinfo {author} {\bibfnamefont {W.}~\bibnamefont
  {Broniowski}},\ }\href {\doibase 10.1103/PhysRevD.87.014005} {\bibfield
  {journal} {\bibinfo  {journal} {Phys.Rev.}\ }\textbf {\bibinfo {volume}
  {D87}},\ \bibinfo {pages} {014005} (\bibinfo {year} {2013})},\ \Eprint
  {http://arxiv.org/abs/1210.0760} {arXiv:1210.0760 [hep-ph]} \BibitemShut
  {NoStop}%
\bibitem [{\citenamefont {Megias}\ \emph {et~al.}(2013)\citenamefont {Megias},
  \citenamefont {Amaro}, \citenamefont {Barbaro}, \citenamefont {Caballero},\
  and\ \citenamefont {Donnelly}}]{Amaro:2013yna}%
  \BibitemOpen
  \bibfield  {author} {\bibinfo {author} {\bibfnamefont {G.}~\bibnamefont
  {Megias}}, \bibinfo {author} {\bibfnamefont {J.}~\bibnamefont {Amaro}},
  \bibinfo {author} {\bibfnamefont {M.}~\bibnamefont {Barbaro}}, \bibinfo
  {author} {\bibfnamefont {J.}~\bibnamefont {Caballero}}, \ and\ \bibinfo
  {author} {\bibfnamefont {T.}~\bibnamefont {Donnelly}},\ }\href {\doibase
  10.1016/j.physletb.2013.07.004} {\bibfield  {journal} {\bibinfo  {journal}
  {Phys.Lett.}\ }\textbf {\bibinfo {volume} {B725}},\ \bibinfo {pages} {170}
  (\bibinfo {year} {2013})},\ \Eprint {http://arxiv.org/abs/1305.6884}
  {arXiv:1305.6884 [nucl-th]} \BibitemShut {NoStop}%
\bibitem [{\citenamefont {Pate}\ and\ \citenamefont
  {Trujillo}(2013)}]{Pate:2013wra}%
  \BibitemOpen
  \bibfield  {author} {\bibinfo {author} {\bibfnamefont {S.}~\bibnamefont
  {Pate}}\ and\ \bibinfo {author} {\bibfnamefont {D.}~\bibnamefont
  {Trujillo}},\ }\href@noop {} {\  (\bibinfo {year} {2013})},\ \Eprint
  {http://arxiv.org/abs/1308.5694} {arXiv:1308.5694 [hep-ph]} \BibitemShut
  {NoStop}%
\bibitem [{\citenamefont {Gonzalez-Jimenez}\ \emph
  {et~al.}(2013{\natexlab{b}})\citenamefont {Gonzalez-Jimenez}, \citenamefont
  {Ivanov}, \citenamefont {Barbaro}, \citenamefont {Caballero},\ and\
  \citenamefont {Udias}}]{GonzalezJimenez:2012bz}%
  \BibitemOpen
  \bibfield  {author} {\bibinfo {author} {\bibfnamefont {R.}~\bibnamefont
  {Gonzalez-Jimenez}}, \bibinfo {author} {\bibfnamefont {M.}~\bibnamefont
  {Ivanov}}, \bibinfo {author} {\bibfnamefont {M.}~\bibnamefont {Barbaro}},
  \bibinfo {author} {\bibfnamefont {J.}~\bibnamefont {Caballero}}, \ and\
  \bibinfo {author} {\bibfnamefont {J.}~\bibnamefont {Udias}},\ }\href
  {\doibase 10.1016/j.physletb.2012.11.065} {\bibfield  {journal} {\bibinfo
  {journal} {Phys.Lett.}\ }\textbf {\bibinfo {volume} {B718}},\ \bibinfo
  {pages} {1471} (\bibinfo {year} {2013}{\natexlab{b}})},\ \Eprint
  {http://arxiv.org/abs/1210.6344} {arXiv:1210.6344 [nucl-th]} \BibitemShut
  {NoStop}%
\bibitem [{\citenamefont {Dharmapalan}\ \emph {et~al.}(2013)\citenamefont
  {Dharmapalan} \emph {et~al.}}]{Dharmapalan:2013zcy}%
  \BibitemOpen
  \bibfield  {author} {\bibinfo {author} {\bibfnamefont {R.}~\bibnamefont
  {Dharmapalan}} \emph {et~al.} (\bibinfo {collaboration} {MiniBooNE+
  Collaboration}),\ }\href@noop {} {\  (\bibinfo {year} {2013})},\ \Eprint
  {http://arxiv.org/abs/1310.0076} {arXiv:1310.0076 [hep-ex]} \BibitemShut
  {NoStop}%
\bibitem [{\citenamefont {Chen}\ \emph {et~al.}(2007)\citenamefont {Chen} \emph
  {et~al.}}]{Chen:2007ae}%
  \BibitemOpen
  \bibfield  {author} {\bibinfo {author} {\bibfnamefont {H.}~\bibnamefont
  {Chen}} \emph {et~al.} (\bibinfo {collaboration} {MicroBooNE
  Collaboration}),\ }\href@noop {} {\bibfield  {journal} {\bibinfo  {journal}
  {Fermilab Proposal}\ }\textbf {\bibinfo {volume} {0974}} (\bibinfo {year}
  {2007})}\BibitemShut {NoStop}%
\bibitem [{\citenamefont {Leitner}\ \emph
  {et~al.}(2006{\natexlab{a}})\citenamefont {Leitner}, \citenamefont
  {Alvarez-Ruso},\ and\ \citenamefont {Mosel}}]{Leitner:2006sp}%
  \BibitemOpen
  \bibfield  {author} {\bibinfo {author} {\bibfnamefont {T.}~\bibnamefont
  {Leitner}}, \bibinfo {author} {\bibfnamefont {L.}~\bibnamefont
  {Alvarez-Ruso}}, \ and\ \bibinfo {author} {\bibfnamefont {U.}~\bibnamefont
  {Mosel}},\ }\href {\doibase 10.1103/PhysRevC.74.065502} {\bibfield  {journal}
  {\bibinfo  {journal} {Phys.Rev.}\ }\textbf {\bibinfo {volume} {C74}},\
  \bibinfo {pages} {065502} (\bibinfo {year} {2006}{\natexlab{a}})},\ \Eprint
  {http://arxiv.org/abs/nucl-th/0606058} {arXiv:nucl-th/0606058 [nucl-th]}
  \BibitemShut {NoStop}%
\bibitem [{\citenamefont {Fetter}\ and\ \citenamefont
  {Walecka}(2003)}]{fetterwalecka}%
  \BibitemOpen
  \bibfield  {author} {\bibinfo {author} {\bibfnamefont {A.~L.}\ \bibnamefont
  {Fetter}}\ and\ \bibinfo {author} {\bibfnamefont {J.~D.}\ \bibnamefont
  {Walecka}},\ }\href@noop {} {\emph {\bibinfo {title} {Quantum Theory of
  Many-particle Systems}}}\ (\bibinfo  {publisher} {Dover},\ \bibinfo {year}
  {2003})\BibitemShut {NoStop}%
\bibitem [{\citenamefont {Smith}\ and\ \citenamefont
  {Moniz}(1972)}]{Smith:1972xh}%
  \BibitemOpen
  \bibfield  {author} {\bibinfo {author} {\bibfnamefont {R.}~\bibnamefont
  {Smith}}\ and\ \bibinfo {author} {\bibfnamefont {E.}~\bibnamefont {Moniz}},\
  }\href {\doibase 10.1016/0550-3213(72)90040-5} {\bibfield  {journal}
  {\bibinfo  {journal} {Nucl.Phys.}\ }\textbf {\bibinfo {volume} {B43}},\
  \bibinfo {pages} {605} (\bibinfo {year} {1972})}\BibitemShut {NoStop}%
\bibitem [{\citenamefont {Benhar}\ \emph {et~al.}(1994)\citenamefont {Benhar},
  \citenamefont {Fabrocini}, \citenamefont {Fantoni},\ and\ \citenamefont
  {Sick}}]{Benhar:1994hw}%
  \BibitemOpen
  \bibfield  {author} {\bibinfo {author} {\bibfnamefont {O.}~\bibnamefont
  {Benhar}}, \bibinfo {author} {\bibfnamefont {A.}~\bibnamefont {Fabrocini}},
  \bibinfo {author} {\bibfnamefont {S.}~\bibnamefont {Fantoni}}, \ and\
  \bibinfo {author} {\bibfnamefont {I.}~\bibnamefont {Sick}},\ }\href {\doibase
  10.1016/0375-9474(94)90920-2} {\bibfield  {journal} {\bibinfo  {journal}
  {Nucl.Phys.}\ }\textbf {\bibinfo {volume} {A579}},\ \bibinfo {pages} {493}
  (\bibinfo {year} {1994})}\BibitemShut {NoStop}%
\bibitem [{\citenamefont {Ciofi~degli Atti}\ and\ \citenamefont
  {Simula}(1996)}]{CiofidegliAtti:1995qe}%
  \BibitemOpen
  \bibfield  {author} {\bibinfo {author} {\bibfnamefont {C.}~\bibnamefont
  {Ciofi~degli Atti}}\ and\ \bibinfo {author} {\bibfnamefont {S.}~\bibnamefont
  {Simula}},\ }\href {\doibase 10.1103/PhysRevC.53.1689} {\bibfield  {journal}
  {\bibinfo  {journal} {Phys.Rev.}\ }\textbf {\bibinfo {volume} {C53}},\
  \bibinfo {pages} {1689} (\bibinfo {year} {1996})},\ \Eprint
  {http://arxiv.org/abs/nucl-th/9507024} {arXiv:nucl-th/9507024 [nucl-th]}
  \BibitemShut {NoStop}%
\bibitem [{\citenamefont {Fernandez~de Cordoba}\ and\ \citenamefont
  {Oset}(1992)}]{FernandezdeCordoba:1991wf}%
  \BibitemOpen
  \bibfield  {author} {\bibinfo {author} {\bibfnamefont {P.}~\bibnamefont
  {Fernandez~de Cordoba}}\ and\ \bibinfo {author} {\bibfnamefont
  {E.}~\bibnamefont {Oset}},\ }\href {\doibase 10.1103/PhysRevC.46.1697}
  {\bibfield  {journal} {\bibinfo  {journal} {Phys.Rev.}\ }\textbf {\bibinfo
  {volume} {C46}},\ \bibinfo {pages} {1697} (\bibinfo {year}
  {1992})}\BibitemShut {NoStop}%
\bibitem [{\citenamefont {Kulagin}\ and\ \citenamefont
  {Petti}(2006)}]{Kulagin:2004ie}%
  \BibitemOpen
  \bibfield  {author} {\bibinfo {author} {\bibfnamefont {S.~A.}\ \bibnamefont
  {Kulagin}}\ and\ \bibinfo {author} {\bibfnamefont {R.}~\bibnamefont
  {Petti}},\ }\href {\doibase 10.1016/j.nuclphysa.2005.10.011} {\bibfield
  {journal} {\bibinfo  {journal} {Nucl.Phys.}\ }\textbf {\bibinfo {volume}
  {A765}},\ \bibinfo {pages} {126} (\bibinfo {year} {2006})},\ \Eprint
  {http://arxiv.org/abs/hep-ph/0412425} {arXiv:hep-ph/0412425 [hep-ph]}
  \BibitemShut {NoStop}%
\bibitem [{\citenamefont {Ciofi~degli Atti}\ \emph {et~al.}(1990)\citenamefont
  {Ciofi~degli Atti}, \citenamefont {Liuti},\ and\ \citenamefont
  {Simula}}]{CiofidegliAtti:1990vn}%
  \BibitemOpen
  \bibfield  {author} {\bibinfo {author} {\bibfnamefont {C.}~\bibnamefont
  {Ciofi~degli Atti}}, \bibinfo {author} {\bibfnamefont {S.}~\bibnamefont
  {Liuti}}, \ and\ \bibinfo {author} {\bibfnamefont {S.}~\bibnamefont
  {Simula}},\ }\href {\doibase 10.1103/PhysRevC.41.R2474} {\bibfield  {journal}
  {\bibinfo  {journal} {Phys.Rev.}\ }\textbf {\bibinfo {volume} {C41}},\
  \bibinfo {pages} {R2474} (\bibinfo {year} {1990})}\BibitemShut {NoStop}%
\bibitem [{\citenamefont {Gil}\ \emph {et~al.}(1997{\natexlab{a}})\citenamefont
  {Gil}, \citenamefont {Nieves},\ and\ \citenamefont {Oset}}]{Gil:1997bm}%
  \BibitemOpen
  \bibfield  {author} {\bibinfo {author} {\bibfnamefont {A.}~\bibnamefont
  {Gil}}, \bibinfo {author} {\bibfnamefont {J.}~\bibnamefont {Nieves}}, \ and\
  \bibinfo {author} {\bibfnamefont {E.}~\bibnamefont {Oset}},\ }\href {\doibase
  10.1016/S0375-9474(97)00513-7} {\bibfield  {journal} {\bibinfo  {journal}
  {Nucl.Phys.}\ }\textbf {\bibinfo {volume} {A627}},\ \bibinfo {pages} {543}
  (\bibinfo {year} {1997}{\natexlab{a}})},\ \Eprint
  {http://arxiv.org/abs/nucl-th/9711009} {arXiv:nucl-th/9711009 [nucl-th]}
  \BibitemShut {NoStop}%
\bibitem [{\citenamefont {Benhar}\ \emph {et~al.}(2005)\citenamefont {Benhar},
  \citenamefont {Farina}, \citenamefont {Nakamura}, \citenamefont {Sakuda},\
  and\ \citenamefont {Seki}}]{Benhar:2005dj}%
  \BibitemOpen
  \bibfield  {author} {\bibinfo {author} {\bibfnamefont {O.}~\bibnamefont
  {Benhar}}, \bibinfo {author} {\bibfnamefont {N.}~\bibnamefont {Farina}},
  \bibinfo {author} {\bibfnamefont {H.}~\bibnamefont {Nakamura}}, \bibinfo
  {author} {\bibfnamefont {M.}~\bibnamefont {Sakuda}}, \ and\ \bibinfo {author}
  {\bibfnamefont {R.}~\bibnamefont {Seki}},\ }\href {\doibase
  10.1103/PhysRevD.72.053005} {\bibfield  {journal} {\bibinfo  {journal}
  {Phys.Rev.}\ }\textbf {\bibinfo {volume} {D72}},\ \bibinfo {pages} {053005}
  (\bibinfo {year} {2005})},\ \Eprint {http://arxiv.org/abs/hep-ph/0506116}
  {arXiv:hep-ph/0506116 [hep-ph]} \BibitemShut {NoStop}%
\bibitem [{\citenamefont {Benhar}\ and\ \citenamefont
  {Meloni}(2007)}]{Benhar:2006nr}%
  \BibitemOpen
  \bibfield  {author} {\bibinfo {author} {\bibfnamefont {O.}~\bibnamefont
  {Benhar}}\ and\ \bibinfo {author} {\bibfnamefont {D.}~\bibnamefont
  {Meloni}},\ }\href {\doibase 10.1016/j.nuclphysa.2007.02.015} {\bibfield
  {journal} {\bibinfo  {journal} {Nucl.Phys.}\ }\textbf {\bibinfo {volume}
  {A789}},\ \bibinfo {pages} {379} (\bibinfo {year} {2007})},\ \Eprint
  {http://arxiv.org/abs/hep-ph/0610403} {arXiv:hep-ph/0610403 [hep-ph]}
  \BibitemShut {NoStop}%
\bibitem [{\citenamefont {Ankowski}\ and\ \citenamefont
  {Sobczyk}(2008)}]{Ankowski:2007uy}%
  \BibitemOpen
  \bibfield  {author} {\bibinfo {author} {\bibfnamefont {A.~M.}\ \bibnamefont
  {Ankowski}}\ and\ \bibinfo {author} {\bibfnamefont {J.~T.}\ \bibnamefont
  {Sobczyk}},\ }\href {\doibase 10.1103/PhysRevC.77.044311} {\bibfield
  {journal} {\bibinfo  {journal} {Phys.Rev.}\ }\textbf {\bibinfo {volume}
  {C77}},\ \bibinfo {pages} {044311} (\bibinfo {year} {2008})},\ \Eprint
  {http://arxiv.org/abs/0711.2031} {arXiv:0711.2031 [nucl-th]} \BibitemShut
  {NoStop}%
\bibitem [{\citenamefont {Butkevich}(2012)}]{Butkevich:2012zr}%
  \BibitemOpen
  \bibfield  {author} {\bibinfo {author} {\bibfnamefont {A.}~\bibnamefont
  {Butkevich}},\ }\href {\doibase 10.1103/PhysRevC.85.065501} {\bibfield
  {journal} {\bibinfo  {journal} {Phys.Rev.}\ }\textbf {\bibinfo {volume}
  {C85}},\ \bibinfo {pages} {065501} (\bibinfo {year} {2012})},\ \Eprint
  {http://arxiv.org/abs/1204.3160} {arXiv:1204.3160 [nucl-th]} \BibitemShut
  {NoStop}%
\bibitem [{\citenamefont {Ankowski}\ and\ \citenamefont
  {Sobczyk}(2006)}]{Ankowski:2005wi}%
  \BibitemOpen
  \bibfield  {author} {\bibinfo {author} {\bibfnamefont {A.~M.}\ \bibnamefont
  {Ankowski}}\ and\ \bibinfo {author} {\bibfnamefont {J.~T.}\ \bibnamefont
  {Sobczyk}},\ }\href {\doibase 10.1103/PhysRevC.74.054316} {\bibfield
  {journal} {\bibinfo  {journal} {Phys.Rev.}\ }\textbf {\bibinfo {volume}
  {C74}},\ \bibinfo {pages} {054316} (\bibinfo {year} {2006})},\ \Eprint
  {http://arxiv.org/abs/nucl-th/0512004} {arXiv:nucl-th/0512004 [nucl-th]}
  \BibitemShut {NoStop}%
\bibitem [{\citenamefont {Nieves}\ \emph {et~al.}(1993)\citenamefont {Nieves},
  \citenamefont {Oset},\ and\ \citenamefont {Garcia-Recio}}]{Nieves:1993ev}%
  \BibitemOpen
  \bibfield  {author} {\bibinfo {author} {\bibfnamefont {J.}~\bibnamefont
  {Nieves}}, \bibinfo {author} {\bibfnamefont {E.}~\bibnamefont {Oset}}, \ and\
  \bibinfo {author} {\bibfnamefont {C.}~\bibnamefont {Garcia-Recio}},\ }\href
  {\doibase 10.1016/0375-9474(93)90245-S} {\bibfield  {journal} {\bibinfo
  {journal} {Nucl.Phys.}\ }\textbf {\bibinfo {volume} {A554}},\ \bibinfo
  {pages} {509} (\bibinfo {year} {1993})}\BibitemShut {NoStop}%
\bibitem [{\citenamefont {Shneor}\ \emph {et~al.}(2007)\citenamefont {Shneor}
  \emph {et~al.}}]{Shneor:2007tu}%
  \BibitemOpen
  \bibfield  {author} {\bibinfo {author} {\bibfnamefont {R.}~\bibnamefont
  {Shneor}} \emph {et~al.} (\bibinfo {collaboration} {Jefferson Lab Hall A
  Collaboration}),\ }\href {\doibase 10.1103/PhysRevLett.99.072501} {\bibfield
  {journal} {\bibinfo  {journal} {Phys.Rev.Lett.}\ }\textbf {\bibinfo {volume}
  {99}},\ \bibinfo {pages} {072501} (\bibinfo {year} {2007})},\ \Eprint
  {http://arxiv.org/abs/nucl-ex/0703023} {arXiv:nucl-ex/0703023 [nucl-ex]}
  \BibitemShut {NoStop}%
\bibitem [{\citenamefont {Buss}\ \emph {et~al.}(2012)\citenamefont {Buss},
  \citenamefont {Gaitanos}, \citenamefont {Gallmeister}, \citenamefont {van
  Hees}, \citenamefont {Kaskulov} \emph {et~al.}}]{Buss:2011mx}%
  \BibitemOpen
  \bibfield  {author} {\bibinfo {author} {\bibfnamefont {O.}~\bibnamefont
  {Buss}}, \bibinfo {author} {\bibfnamefont {T.}~\bibnamefont {Gaitanos}},
  \bibinfo {author} {\bibfnamefont {K.}~\bibnamefont {Gallmeister}}, \bibinfo
  {author} {\bibfnamefont {H.}~\bibnamefont {van Hees}}, \bibinfo {author}
  {\bibfnamefont {M.}~\bibnamefont {Kaskulov}},  \emph {et~al.},\ }\href
  {\doibase 10.1016/j.physrep.2011.12.001} {\bibfield  {journal} {\bibinfo
  {journal} {Phys.Rept.}\ }\textbf {\bibinfo {volume} {512}},\ \bibinfo {pages}
  {1} (\bibinfo {year} {2012})},\ \Eprint {http://arxiv.org/abs/1106.1344}
  {arXiv:1106.1344 [hep-ph]} \BibitemShut {NoStop}%
\bibitem [{\citenamefont {Alberico}\ \emph {et~al.}(1997)\citenamefont
  {Alberico}, \citenamefont {Barbaro}, \citenamefont {Bilenky}, \citenamefont
  {Caballero}, \citenamefont {Giunti} \emph {et~al.}}]{Alberico:1997vh}%
  \BibitemOpen
  \bibfield  {author} {\bibinfo {author} {\bibfnamefont {W.}~\bibnamefont
  {Alberico}}, \bibinfo {author} {\bibfnamefont {M.}~\bibnamefont {Barbaro}},
  \bibinfo {author} {\bibfnamefont {S.~M.}\ \bibnamefont {Bilenky}}, \bibinfo
  {author} {\bibfnamefont {J.}~\bibnamefont {Caballero}}, \bibinfo {author}
  {\bibfnamefont {C.}~\bibnamefont {Giunti}},  \emph {et~al.},\ }\href
  {\doibase 10.1016/S0375-9474(97)00416-8} {\bibfield  {journal} {\bibinfo
  {journal} {Nucl.Phys.}\ }\textbf {\bibinfo {volume} {A623}},\ \bibinfo
  {pages} {471} (\bibinfo {year} {1997})},\ \Eprint
  {http://arxiv.org/abs/hep-ph/9703415} {arXiv:hep-ph/9703415 [hep-ph]}
  \BibitemShut {NoStop}%
\bibitem [{\citenamefont {Maieron}\ \emph {et~al.}(2003)\citenamefont
  {Maieron}, \citenamefont {Martinez}, \citenamefont {Caballero},\ and\
  \citenamefont {Udias}}]{Maieron:2003df}%
  \BibitemOpen
  \bibfield  {author} {\bibinfo {author} {\bibfnamefont {C.}~\bibnamefont
  {Maieron}}, \bibinfo {author} {\bibfnamefont {M.}~\bibnamefont {Martinez}},
  \bibinfo {author} {\bibfnamefont {J.}~\bibnamefont {Caballero}}, \ and\
  \bibinfo {author} {\bibfnamefont {J.}~\bibnamefont {Udias}},\ }\href
  {\doibase 10.1103/PhysRevC.68.048501} {\bibfield  {journal} {\bibinfo
  {journal} {Phys.Rev.}\ }\textbf {\bibinfo {volume} {C68}},\ \bibinfo {pages}
  {048501} (\bibinfo {year} {2003})},\ \Eprint
  {http://arxiv.org/abs/nucl-th/0303075} {arXiv:nucl-th/0303075 [nucl-th]}
  \BibitemShut {NoStop}%
\bibitem [{\citenamefont {Butkevich}\ and\ \citenamefont
  {Kulagin}(2007)}]{Butkevich:2007gm}%
  \BibitemOpen
  \bibfield  {author} {\bibinfo {author} {\bibfnamefont {A.}~\bibnamefont
  {Butkevich}}\ and\ \bibinfo {author} {\bibfnamefont {S.~A.}\ \bibnamefont
  {Kulagin}},\ }\href {\doibase 10.1103/PhysRevC.76.045502} {\bibfield
  {journal} {\bibinfo  {journal} {Phys.Rev.}\ }\textbf {\bibinfo {volume}
  {C76}},\ \bibinfo {pages} {045502} (\bibinfo {year} {2007})},\ \Eprint
  {http://arxiv.org/abs/0705.1051} {arXiv:0705.1051 [nucl-th]} \BibitemShut
  {NoStop}%
\bibitem [{\citenamefont {Meucci}\ \emph
  {et~al.}(2004{\natexlab{a}})\citenamefont {Meucci}, \citenamefont {Giusti},\
  and\ \citenamefont {Pacati}}]{Meucci:2004ip}%
  \BibitemOpen
  \bibfield  {author} {\bibinfo {author} {\bibfnamefont {A.}~\bibnamefont
  {Meucci}}, \bibinfo {author} {\bibfnamefont {C.}~\bibnamefont {Giusti}}, \
  and\ \bibinfo {author} {\bibfnamefont {F.~D.}\ \bibnamefont {Pacati}},\
  }\href {\doibase 10.1016/j.nuclphysa.2004.08.023} {\bibfield  {journal}
  {\bibinfo  {journal} {Nucl.Phys.}\ }\textbf {\bibinfo {volume} {A744}},\
  \bibinfo {pages} {307} (\bibinfo {year} {2004}{\natexlab{a}})},\ \Eprint
  {http://arxiv.org/abs/nucl-th/0405004} {arXiv:nucl-th/0405004 [nucl-th]}
  \BibitemShut {NoStop}%
\bibitem [{\citenamefont {Benhar}\ \emph {et~al.}(1991)\citenamefont {Benhar},
  \citenamefont {Fabrocini}, \citenamefont {Fantoni}, \citenamefont {Miller},
  \citenamefont {Pandharipande} \emph {et~al.}}]{Benhar:1991af}%
  \BibitemOpen
  \bibfield  {author} {\bibinfo {author} {\bibfnamefont {O.}~\bibnamefont
  {Benhar}}, \bibinfo {author} {\bibfnamefont {A.}~\bibnamefont {Fabrocini}},
  \bibinfo {author} {\bibfnamefont {S.}~\bibnamefont {Fantoni}}, \bibinfo
  {author} {\bibfnamefont {G.}~\bibnamefont {Miller}}, \bibinfo {author}
  {\bibfnamefont {V.}~\bibnamefont {Pandharipande}},  \emph {et~al.},\ }\href
  {\doibase 10.1103/PhysRevC.44.2328} {\bibfield  {journal} {\bibinfo
  {journal} {Phys.Rev.}\ }\textbf {\bibinfo {volume} {C44}},\ \bibinfo {pages}
  {2328} (\bibinfo {year} {1991})}\BibitemShut {NoStop}%
\bibitem [{\citenamefont {Leitner}\ \emph
  {et~al.}(2009{\natexlab{a}})\citenamefont {Leitner}, \citenamefont {Buss},
  \citenamefont {Alvarez-Ruso},\ and\ \citenamefont {Mosel}}]{Leitner:2008ue}%
  \BibitemOpen
  \bibfield  {author} {\bibinfo {author} {\bibfnamefont {T.}~\bibnamefont
  {Leitner}}, \bibinfo {author} {\bibfnamefont {O.}~\bibnamefont {Buss}},
  \bibinfo {author} {\bibfnamefont {L.}~\bibnamefont {Alvarez-Ruso}}, \ and\
  \bibinfo {author} {\bibfnamefont {U.}~\bibnamefont {Mosel}},\ }\href
  {\doibase 10.1103/PhysRevC.79.034601} {\bibfield  {journal} {\bibinfo
  {journal} {Phys.Rev.}\ }\textbf {\bibinfo {volume} {C79}},\ \bibinfo {pages}
  {034601} (\bibinfo {year} {2009}{\natexlab{a}})},\ \Eprint
  {http://arxiv.org/abs/0812.0587} {arXiv:0812.0587 [nucl-th]} \BibitemShut
  {NoStop}%
\bibitem [{\citenamefont {Martinez}\ \emph {et~al.}(2006)\citenamefont
  {Martinez}, \citenamefont {Lava}, \citenamefont {Jachowicz}, \citenamefont
  {Ryckebusch}, \citenamefont {Vantournhout} \emph {et~al.}}]{Martinez:2005xe}%
  \BibitemOpen
  \bibfield  {author} {\bibinfo {author} {\bibfnamefont {M.}~\bibnamefont
  {Martinez}}, \bibinfo {author} {\bibfnamefont {P.}~\bibnamefont {Lava}},
  \bibinfo {author} {\bibfnamefont {N.}~\bibnamefont {Jachowicz}}, \bibinfo
  {author} {\bibfnamefont {J.}~\bibnamefont {Ryckebusch}}, \bibinfo {author}
  {\bibfnamefont {K.}~\bibnamefont {Vantournhout}},  \emph {et~al.},\ }\href
  {\doibase 10.1103/PhysRevC.73.024607} {\bibfield  {journal} {\bibinfo
  {journal} {Phys.Rev.}\ }\textbf {\bibinfo {volume} {C73}},\ \bibinfo {pages}
  {024607} (\bibinfo {year} {2006})},\ \Eprint
  {http://arxiv.org/abs/nucl-th/0505008} {arXiv:nucl-th/0505008 [nucl-th]}
  \BibitemShut {NoStop}%
\bibitem [{\citenamefont {Meucci}\ \emph
  {et~al.}(2011{\natexlab{a}})\citenamefont {Meucci}, \citenamefont {Giusti},\
  and\ \citenamefont {Pacati}}]{Meucci:2011ce}%
  \BibitemOpen
  \bibfield  {author} {\bibinfo {author} {\bibfnamefont {A.}~\bibnamefont
  {Meucci}}, \bibinfo {author} {\bibfnamefont {C.}~\bibnamefont {Giusti}}, \
  and\ \bibinfo {author} {\bibfnamefont {F.~D.}\ \bibnamefont {Pacati}},\
  }\href {\doibase 10.1103/PhysRevD.84.113003} {\bibfield  {journal} {\bibinfo
  {journal} {Phys.Rev.}\ }\textbf {\bibinfo {volume} {D84}},\ \bibinfo {pages}
  {113003} (\bibinfo {year} {2011}{\natexlab{a}})},\ \Eprint
  {http://arxiv.org/abs/1110.3928} {arXiv:1110.3928 [nucl-th]} \BibitemShut
  {NoStop}%
\bibitem [{\citenamefont {Boffi}\ \emph {et~al.}(1993)\citenamefont {Boffi},
  \citenamefont {Giusti},\ and\ \citenamefont {Pacati}}]{Boffi:1993gs}%
  \BibitemOpen
  \bibfield  {author} {\bibinfo {author} {\bibfnamefont {S.}~\bibnamefont
  {Boffi}}, \bibinfo {author} {\bibfnamefont {C.}~\bibnamefont {Giusti}}, \
  and\ \bibinfo {author} {\bibfnamefont {F.}~\bibnamefont {Pacati}},\ }\href
  {\doibase 10.1016/0370-1573(93)90132-W} {\bibfield  {journal} {\bibinfo
  {journal} {Phys.Rept.}\ }\textbf {\bibinfo {volume} {226}},\ \bibinfo {pages}
  {1} (\bibinfo {year} {1993})}\BibitemShut {NoStop}%
\bibitem [{\citenamefont {Udias}\ \emph {et~al.}(1993)\citenamefont {Udias},
  \citenamefont {Sarriguren}, \citenamefont {Moya~de Guerra}, \citenamefont
  {Garrido},\ and\ \citenamefont {Caballero}}]{Udias:1993xy}%
  \BibitemOpen
  \bibfield  {author} {\bibinfo {author} {\bibfnamefont {J.}~\bibnamefont
  {Udias}}, \bibinfo {author} {\bibfnamefont {P.}~\bibnamefont {Sarriguren}},
  \bibinfo {author} {\bibfnamefont {E.}~\bibnamefont {Moya~de Guerra}},
  \bibinfo {author} {\bibfnamefont {E.}~\bibnamefont {Garrido}}, \ and\
  \bibinfo {author} {\bibfnamefont {J.}~\bibnamefont {Caballero}},\ }\href
  {\doibase 10.1103/PhysRevC.48.2731} {\bibfield  {journal} {\bibinfo
  {journal} {Phys.Rev.}\ }\textbf {\bibinfo {volume} {C48}},\ \bibinfo {pages}
  {2731} (\bibinfo {year} {1993})},\ \Eprint
  {http://arxiv.org/abs/nucl-th/9310004} {arXiv:nucl-th/9310004 [nucl-th]}
  \BibitemShut {NoStop}%
\bibitem [{\citenamefont {Giusti}\ \emph {et~al.}(2011)\citenamefont {Giusti},
  \citenamefont {Meucci}, \citenamefont {Pacati}, \citenamefont {Co'},\ and\
  \citenamefont {De~Donno}}]{Giusti:2011it}%
  \BibitemOpen
  \bibfield  {author} {\bibinfo {author} {\bibfnamefont {C.}~\bibnamefont
  {Giusti}}, \bibinfo {author} {\bibfnamefont {A.}~\bibnamefont {Meucci}},
  \bibinfo {author} {\bibfnamefont {F.}~\bibnamefont {Pacati}}, \bibinfo
  {author} {\bibfnamefont {G.}~\bibnamefont {Co'}}, \ and\ \bibinfo {author}
  {\bibfnamefont {V.}~\bibnamefont {De~Donno}},\ }\href {\doibase
  10.1103/PhysRevC.84.024615} {\bibfield  {journal} {\bibinfo  {journal}
  {Phys.Rev.}\ }\textbf {\bibinfo {volume} {C84}},\ \bibinfo {pages} {024615}
  (\bibinfo {year} {2011})},\ \Eprint {http://arxiv.org/abs/1105.1295}
  {arXiv:1105.1295 [nucl-th]} \BibitemShut {NoStop}%
\bibitem [{\citenamefont {Giusti}\ \emph {et~al.}(2009)\citenamefont {Giusti},
  \citenamefont {Meucci}, \citenamefont {Pacati}, \citenamefont {Caballero},\
  and\ \citenamefont {Udias}}]{Giusti:2009ym}%
  \BibitemOpen
  \bibfield  {author} {\bibinfo {author} {\bibfnamefont {C.}~\bibnamefont
  {Giusti}}, \bibinfo {author} {\bibfnamefont {A.}~\bibnamefont {Meucci}},
  \bibinfo {author} {\bibfnamefont {F.}~\bibnamefont {Pacati}}, \bibinfo
  {author} {\bibfnamefont {J.}~\bibnamefont {Caballero}}, \ and\ \bibinfo
  {author} {\bibfnamefont {J.}~\bibnamefont {Udias}},\ }\href {\doibase
  10.1063/1.3274139} {\bibfield  {journal} {\bibinfo  {journal} {AIP
  Conf.Proc.}\ }\textbf {\bibinfo {volume} {1189}},\ \bibinfo {pages} {107}
  (\bibinfo {year} {2009})},\ \Eprint {http://arxiv.org/abs/0909.0620}
  {arXiv:0909.0620 [nucl-th]} \BibitemShut {NoStop}%
\bibitem [{\citenamefont {Meucci}\ \emph
  {et~al.}(2004{\natexlab{b}})\citenamefont {Meucci}, \citenamefont {Giusti},\
  and\ \citenamefont {Pacati}}]{Meucci:2003cv}%
  \BibitemOpen
  \bibfield  {author} {\bibinfo {author} {\bibfnamefont {A.}~\bibnamefont
  {Meucci}}, \bibinfo {author} {\bibfnamefont {C.}~\bibnamefont {Giusti}}, \
  and\ \bibinfo {author} {\bibfnamefont {F.~D.}\ \bibnamefont {Pacati}},\
  }\href {\doibase 10.1016/j.nuclphysa.2004.04.108} {\bibfield  {journal}
  {\bibinfo  {journal} {Nucl.Phys.}\ }\textbf {\bibinfo {volume} {A739}},\
  \bibinfo {pages} {277} (\bibinfo {year} {2004}{\natexlab{b}})},\ \Eprint
  {http://arxiv.org/abs/nucl-th/0311081} {arXiv:nucl-th/0311081 [nucl-th]}
  \BibitemShut {NoStop}%
\bibitem [{\citenamefont {Meucci}\ \emph
  {et~al.}(2011{\natexlab{b}})\citenamefont {Meucci}, \citenamefont {Barbaro},
  \citenamefont {Caballero}, \citenamefont {Giusti},\ and\ \citenamefont
  {Udias}}]{Meucci:2011vd}%
  \BibitemOpen
  \bibfield  {author} {\bibinfo {author} {\bibfnamefont {A.}~\bibnamefont
  {Meucci}}, \bibinfo {author} {\bibfnamefont {M.}~\bibnamefont {Barbaro}},
  \bibinfo {author} {\bibfnamefont {J.}~\bibnamefont {Caballero}}, \bibinfo
  {author} {\bibfnamefont {C.}~\bibnamefont {Giusti}}, \ and\ \bibinfo {author}
  {\bibfnamefont {J.}~\bibnamefont {Udias}},\ }\href {\doibase
  10.1103/PhysRevLett.107.172501} {\bibfield  {journal} {\bibinfo  {journal}
  {Phys.Rev.Lett.}\ }\textbf {\bibinfo {volume} {107}},\ \bibinfo {pages}
  {172501} (\bibinfo {year} {2011}{\natexlab{b}})},\ \Eprint
  {http://arxiv.org/abs/1107.5145} {arXiv:1107.5145 [nucl-th]} \BibitemShut
  {NoStop}%
\bibitem [{\citenamefont {Meucci}\ and\ \citenamefont
  {Giusti}(2012)}]{Meucci:2012yq}%
  \BibitemOpen
  \bibfield  {author} {\bibinfo {author} {\bibfnamefont {A.}~\bibnamefont
  {Meucci}}\ and\ \bibinfo {author} {\bibfnamefont {C.}~\bibnamefont
  {Giusti}},\ }\href {\doibase 10.1103/PhysRevD.85.093002} {\bibfield
  {journal} {\bibinfo  {journal} {Phys.Rev.}\ }\textbf {\bibinfo {volume}
  {D85}},\ \bibinfo {pages} {093002} (\bibinfo {year} {2012})},\ \Eprint
  {http://arxiv.org/abs/1202.4312} {arXiv:1202.4312 [nucl-th]} \BibitemShut
  {NoStop}%
\bibitem [{\citenamefont {Meucci}\ and\ \citenamefont
  {Giusti}(2014)}]{Meucci:2014pka}%
  \BibitemOpen
  \bibfield  {author} {\bibinfo {author} {\bibfnamefont {A.}~\bibnamefont
  {Meucci}}\ and\ \bibinfo {author} {\bibfnamefont {C.}~\bibnamefont
  {Giusti}},\ }\href {\doibase 10.1103/PhysRevD.89.057302} {\bibfield
  {journal} {\bibinfo  {journal} {Phys.Rev.}\ }\textbf {\bibinfo {volume}
  {D89}},\ \bibinfo {pages} {057302} (\bibinfo {year} {2014})},\ \Eprint
  {http://arxiv.org/abs/1401.3650} {arXiv:1401.3650 [nucl-th]} \BibitemShut
  {NoStop}%
\bibitem [{\citenamefont {Gonz\'alez-Jim\'enez}\ \emph
  {et~al.}(2013)\citenamefont {Gonz\'alez-Jim\'enez}, \citenamefont
  {Caballero}, \citenamefont {Meucci}, \citenamefont {Giusti}, \citenamefont
  {Barbaro} \emph {et~al.}}]{Gonzalez-Jimenez:2013xpa}%
  \BibitemOpen
  \bibfield  {author} {\bibinfo {author} {\bibfnamefont {R.}~\bibnamefont
  {Gonz\'alez-Jim\'enez}}, \bibinfo {author} {\bibfnamefont {J.}~\bibnamefont
  {Caballero}}, \bibinfo {author} {\bibfnamefont {A.}~\bibnamefont {Meucci}},
  \bibinfo {author} {\bibfnamefont {C.}~\bibnamefont {Giusti}}, \bibinfo
  {author} {\bibfnamefont {M.}~\bibnamefont {Barbaro}},  \emph {et~al.},\
  }\href {\doibase 10.1103/PhysRevC.88.025502} {\bibfield  {journal} {\bibinfo
  {journal} {Phys.Rev.}\ }\textbf {\bibinfo {volume} {C88}},\ \bibinfo {pages}
  {025502} (\bibinfo {year} {2013})},\ \Eprint {http://arxiv.org/abs/1307.4309}
  {arXiv:1307.4309 [nucl-th]} \BibitemShut {NoStop}%
\bibitem [{\citenamefont {Leitner}\ \emph
  {et~al.}(2006{\natexlab{b}})\citenamefont {Leitner}, \citenamefont
  {Alvarez-Ruso},\ and\ \citenamefont {Mosel}}]{Leitner:2006ww}%
  \BibitemOpen
  \bibfield  {author} {\bibinfo {author} {\bibfnamefont {T.}~\bibnamefont
  {Leitner}}, \bibinfo {author} {\bibfnamefont {L.}~\bibnamefont
  {Alvarez-Ruso}}, \ and\ \bibinfo {author} {\bibfnamefont {U.}~\bibnamefont
  {Mosel}},\ }\href {\doibase 10.1103/PhysRevC.73.065502} {\bibfield  {journal}
  {\bibinfo  {journal} {Phys.Rev.}\ }\textbf {\bibinfo {volume} {C73}},\
  \bibinfo {pages} {065502} (\bibinfo {year} {2006}{\natexlab{b}})},\ \Eprint
  {http://arxiv.org/abs/nucl-th/0601103} {arXiv:nucl-th/0601103 [nucl-th]}
  \BibitemShut {NoStop}%
\bibitem [{\citenamefont {Meziani}\ \emph {et~al.}(1984)\citenamefont
  {Meziani}, \citenamefont {Barreau}, \citenamefont {Bernheim}, \citenamefont
  {Morgenstern}, \citenamefont {Turck-Chieze} \emph {et~al.}}]{Meziani:1984is}%
  \BibitemOpen
  \bibfield  {author} {\bibinfo {author} {\bibfnamefont {Z.}~\bibnamefont
  {Meziani}}, \bibinfo {author} {\bibfnamefont {P.}~\bibnamefont {Barreau}},
  \bibinfo {author} {\bibfnamefont {M.}~\bibnamefont {Bernheim}}, \bibinfo
  {author} {\bibfnamefont {J.}~\bibnamefont {Morgenstern}}, \bibinfo {author}
  {\bibfnamefont {S.}~\bibnamefont {Turck-Chieze}},  \emph {et~al.},\ }\href
  {\doibase 10.1103/PhysRevLett.52.2130} {\bibfield  {journal} {\bibinfo
  {journal} {Phys.Rev.Lett.}\ }\textbf {\bibinfo {volume} {52}},\ \bibinfo
  {pages} {2130} (\bibinfo {year} {1984})}\BibitemShut {NoStop}%
\bibitem [{\citenamefont {Meucci}\ \emph {et~al.}(2003)\citenamefont {Meucci},
  \citenamefont {Capuzzi}, \citenamefont {Giusti},\ and\ \citenamefont
  {Pacati}}]{Meucci:2003uy}%
  \BibitemOpen
  \bibfield  {author} {\bibinfo {author} {\bibfnamefont {A.}~\bibnamefont
  {Meucci}}, \bibinfo {author} {\bibfnamefont {F.}~\bibnamefont {Capuzzi}},
  \bibinfo {author} {\bibfnamefont {C.}~\bibnamefont {Giusti}}, \ and\ \bibinfo
  {author} {\bibfnamefont {F.~D.}\ \bibnamefont {Pacati}},\ }\href {\doibase
  10.1103/PhysRevC.67.054601} {\bibfield  {journal} {\bibinfo  {journal}
  {Phys.Rev.}\ }\textbf {\bibinfo {volume} {C67}},\ \bibinfo {pages} {054601}
  (\bibinfo {year} {2003})},\ \Eprint {http://arxiv.org/abs/nucl-th/0301084}
  {arXiv:nucl-th/0301084 [nucl-th]} \BibitemShut {NoStop}%
\bibitem [{\citenamefont {Benhar}\ \emph {et~al.}(2008)\citenamefont {Benhar},
  \citenamefont {Day},\ and\ \citenamefont {Sick}}]{Benhar:2006wy}%
  \BibitemOpen
  \bibfield  {author} {\bibinfo {author} {\bibfnamefont {O.}~\bibnamefont
  {Benhar}}, \bibinfo {author} {\bibfnamefont {D.}~\bibnamefont {Day}}, \ and\
  \bibinfo {author} {\bibfnamefont {I.}~\bibnamefont {Sick}},\ }\href {\doibase
  10.1103/RevModPhys.80.189} {\bibfield  {journal} {\bibinfo  {journal}
  {Rev.Mod.Phys.}\ }\textbf {\bibinfo {volume} {80}},\ \bibinfo {pages} {189}
  (\bibinfo {year} {2008})},\ \Eprint {http://arxiv.org/abs/nucl-ex/0603029}
  {arXiv:nucl-ex/0603029 [nucl-ex]} \BibitemShut {NoStop}%
\bibitem [{\citenamefont {Fabrocini}\ and\ \citenamefont
  {Fantoni}(1989)}]{Fabrocini:1989nw}%
  \BibitemOpen
  \bibfield  {author} {\bibinfo {author} {\bibfnamefont {A.}~\bibnamefont
  {Fabrocini}}\ and\ \bibinfo {author} {\bibfnamefont {S.}~\bibnamefont
  {Fantoni}},\ }\href@noop {} {\bibfield  {journal} {\bibinfo  {journal}
  {Nucl.Phys.}\ }\textbf {\bibinfo {volume} {A503}},\ \bibinfo {pages} {375}
  (\bibinfo {year} {1989})}\BibitemShut {NoStop}%
\bibitem [{\citenamefont {Donnelly}\ and\ \citenamefont
  {Sick}(1999{\natexlab{a}})}]{Donnelly:1998xg}%
  \BibitemOpen
  \bibfield  {author} {\bibinfo {author} {\bibfnamefont {T.}~\bibnamefont
  {Donnelly}}\ and\ \bibinfo {author} {\bibfnamefont {I.}~\bibnamefont
  {Sick}},\ }\href {\doibase 10.1103/PhysRevLett.82.3212} {\bibfield  {journal}
  {\bibinfo  {journal} {Phys.Rev.Lett.}\ }\textbf {\bibinfo {volume} {82}},\
  \bibinfo {pages} {3212} (\bibinfo {year} {1999}{\natexlab{a}})},\ \Eprint
  {http://arxiv.org/abs/nucl-th/9809063} {arXiv:nucl-th/9809063 [nucl-th]}
  \BibitemShut {NoStop}%
\bibitem [{\citenamefont {Donnelly}\ and\ \citenamefont
  {Sick}(1999{\natexlab{b}})}]{Donnelly:1999sw}%
  \BibitemOpen
  \bibfield  {author} {\bibinfo {author} {\bibfnamefont {T.}~\bibnamefont
  {Donnelly}}\ and\ \bibinfo {author} {\bibfnamefont {I.}~\bibnamefont
  {Sick}},\ }\href {\doibase 10.1103/PhysRevC.60.065502} {\bibfield  {journal}
  {\bibinfo  {journal} {Phys.Rev.}\ }\textbf {\bibinfo {volume} {C60}},\
  \bibinfo {pages} {065502} (\bibinfo {year} {1999}{\natexlab{b}})},\ \Eprint
  {http://arxiv.org/abs/nucl-th/9905060} {arXiv:nucl-th/9905060 [nucl-th]}
  \BibitemShut {NoStop}%
\bibitem [{\citenamefont {Amaro}\ \emph
  {et~al.}(2005{\natexlab{a}})\citenamefont {Amaro}, \citenamefont {Barbaro},
  \citenamefont {Caballero}, \citenamefont {Donnelly}, \citenamefont {Molinari}
  \emph {et~al.}}]{Amaro:2004bs}%
  \BibitemOpen
  \bibfield  {author} {\bibinfo {author} {\bibfnamefont {J.~E.}\ \bibnamefont
  {Amaro}}, \bibinfo {author} {\bibfnamefont {M.}~\bibnamefont {Barbaro}},
  \bibinfo {author} {\bibfnamefont {J.}~\bibnamefont {Caballero}}, \bibinfo
  {author} {\bibfnamefont {T.}~\bibnamefont {Donnelly}}, \bibinfo {author}
  {\bibfnamefont {A.}~\bibnamefont {Molinari}},  \emph {et~al.},\ }\href
  {\doibase 10.1103/PhysRevC.71.015501} {\bibfield  {journal} {\bibinfo
  {journal} {Phys.Rev.}\ }\textbf {\bibinfo {volume} {C71}},\ \bibinfo {pages}
  {015501} (\bibinfo {year} {2005}{\natexlab{a}})},\ \Eprint
  {http://arxiv.org/abs/nucl-th/0409078} {arXiv:nucl-th/0409078 [nucl-th]}
  \BibitemShut {NoStop}%
\bibitem [{\citenamefont {Jourdan}(1996)}]{Jourdan:1996np}%
  \BibitemOpen
  \bibfield  {author} {\bibinfo {author} {\bibfnamefont {J.}~\bibnamefont
  {Jourdan}},\ }\href {\doibase 10.1016/0375-9474(96)00143-1} {\bibfield
  {journal} {\bibinfo  {journal} {Nucl.Phys.}\ }\textbf {\bibinfo {volume}
  {A603}},\ \bibinfo {pages} {117} (\bibinfo {year} {1996})}\BibitemShut
  {NoStop}%
\bibitem [{\citenamefont {Caballero}\ \emph {et~al.}(2005)\citenamefont
  {Caballero}, \citenamefont {Amaro}, \citenamefont {Barbaro}, \citenamefont
  {Donnelly}, \citenamefont {Maieron} \emph {et~al.}}]{Caballero:2005sj}%
  \BibitemOpen
  \bibfield  {author} {\bibinfo {author} {\bibfnamefont {J.}~\bibnamefont
  {Caballero}}, \bibinfo {author} {\bibfnamefont {J.~E.}\ \bibnamefont
  {Amaro}}, \bibinfo {author} {\bibfnamefont {M.}~\bibnamefont {Barbaro}},
  \bibinfo {author} {\bibfnamefont {T.}~\bibnamefont {Donnelly}}, \bibinfo
  {author} {\bibfnamefont {C.}~\bibnamefont {Maieron}},  \emph {et~al.},\
  }\href {\doibase 10.1103/PhysRevLett.95.252502} {\bibfield  {journal}
  {\bibinfo  {journal} {Phys.Rev.Lett.}\ }\textbf {\bibinfo {volume} {95}},\
  \bibinfo {pages} {252502} (\bibinfo {year} {2005})},\ \Eprint
  {http://arxiv.org/abs/nucl-th/0504040} {arXiv:nucl-th/0504040 [nucl-th]}
  \BibitemShut {NoStop}%
\bibitem [{\citenamefont {Caballero}\ \emph {et~al.}(2007)\citenamefont
  {Caballero}, \citenamefont {Amaro}, \citenamefont {Barbaro}, \citenamefont
  {Donnelly},\ and\ \citenamefont {Udias}}]{Caballero:2007tz}%
  \BibitemOpen
  \bibfield  {author} {\bibinfo {author} {\bibfnamefont {J.}~\bibnamefont
  {Caballero}}, \bibinfo {author} {\bibfnamefont {J.~E.}\ \bibnamefont
  {Amaro}}, \bibinfo {author} {\bibfnamefont {M.}~\bibnamefont {Barbaro}},
  \bibinfo {author} {\bibfnamefont {T.}~\bibnamefont {Donnelly}}, \ and\
  \bibinfo {author} {\bibfnamefont {J.}~\bibnamefont {Udias}},\ }\href
  {\doibase 10.1016/j.physletb.2007.08.018} {\bibfield  {journal} {\bibinfo
  {journal} {Phys.Lett.}\ }\textbf {\bibinfo {volume} {B653}},\ \bibinfo
  {pages} {366} (\bibinfo {year} {2007})},\ \Eprint
  {http://arxiv.org/abs/0705.1429} {arXiv:0705.1429 [nucl-th]} \BibitemShut
  {NoStop}%
\bibitem [{\citenamefont {Antonov}\ \emph {et~al.}(2011)\citenamefont
  {Antonov}, \citenamefont {Ivanov}, \citenamefont {Caballero}, \citenamefont
  {Barbaro}, \citenamefont {Udias} \emph {et~al.}}]{Antonov:2011bi}%
  \BibitemOpen
  \bibfield  {author} {\bibinfo {author} {\bibfnamefont {A.}~\bibnamefont
  {Antonov}}, \bibinfo {author} {\bibfnamefont {M.}~\bibnamefont {Ivanov}},
  \bibinfo {author} {\bibfnamefont {J.}~\bibnamefont {Caballero}}, \bibinfo
  {author} {\bibfnamefont {M.}~\bibnamefont {Barbaro}}, \bibinfo {author}
  {\bibfnamefont {J.}~\bibnamefont {Udias}},  \emph {et~al.},\ }\href {\doibase
  10.1103/PhysRevC.83.045504} {\bibfield  {journal} {\bibinfo  {journal}
  {Phys.Rev.}\ }\textbf {\bibinfo {volume} {C83}},\ \bibinfo {pages} {045504}
  (\bibinfo {year} {2011})},\ \Eprint {http://arxiv.org/abs/1104.0125}
  {arXiv:1104.0125 [nucl-th]} \BibitemShut {NoStop}%
\bibitem [{\citenamefont {Amaro}\ \emph {et~al.}(2007)\citenamefont {Amaro},
  \citenamefont {Barbaro}, \citenamefont {Caballero},\ and\ \citenamefont
  {Donnelly}}]{Amaro:2006tf}%
  \BibitemOpen
  \bibfield  {author} {\bibinfo {author} {\bibfnamefont {J.~E.}\ \bibnamefont
  {Amaro}}, \bibinfo {author} {\bibfnamefont {M.}~\bibnamefont {Barbaro}},
  \bibinfo {author} {\bibfnamefont {J.}~\bibnamefont {Caballero}}, \ and\
  \bibinfo {author} {\bibfnamefont {T.}~\bibnamefont {Donnelly}},\ }\href
  {\doibase 10.1103/PhysRevLett.98.242501} {\bibfield  {journal} {\bibinfo
  {journal} {Phys.Rev.Lett.}\ }\textbf {\bibinfo {volume} {98}},\ \bibinfo
  {pages} {242501} (\bibinfo {year} {2007})},\ \Eprint
  {http://arxiv.org/abs/nucl-th/0612046} {arXiv:nucl-th/0612046 [nucl-th]}
  \BibitemShut {NoStop}%
\bibitem [{\citenamefont {Singh}\ and\ \citenamefont
  {Oset}(1992)}]{Singh:1992dc}%
  \BibitemOpen
  \bibfield  {author} {\bibinfo {author} {\bibfnamefont {S.}~\bibnamefont
  {Singh}}\ and\ \bibinfo {author} {\bibfnamefont {E.}~\bibnamefont {Oset}},\
  }\href {\doibase 10.1016/0375-9474(92)90259-M} {\bibfield  {journal}
  {\bibinfo  {journal} {Nucl.Phys.}\ }\textbf {\bibinfo {volume} {A542}},\
  \bibinfo {pages} {587} (\bibinfo {year} {1992})}\BibitemShut {NoStop}%
\bibitem [{\citenamefont {Kosmas}\ and\ \citenamefont
  {Oset}(1996)}]{Kosmas:1996fh}%
  \BibitemOpen
  \bibfield  {author} {\bibinfo {author} {\bibfnamefont {T.}~\bibnamefont
  {Kosmas}}\ and\ \bibinfo {author} {\bibfnamefont {E.}~\bibnamefont {Oset}},\
  }\href {\doibase 10.1103/PhysRevC.53.1409} {\bibfield  {journal} {\bibinfo
  {journal} {Phys.Rev.}\ }\textbf {\bibinfo {volume} {C53}},\ \bibinfo {pages}
  {1409} (\bibinfo {year} {1996})}\BibitemShut {NoStop}%
\bibitem [{\citenamefont {Singh}\ \emph
  {et~al.}(1998{\natexlab{a}})\citenamefont {Singh}, \citenamefont
  {Mukhopadhyay},\ and\ \citenamefont {Oset}}]{Singh:1998md}%
  \BibitemOpen
  \bibfield  {author} {\bibinfo {author} {\bibfnamefont {S.}~\bibnamefont
  {Singh}}, \bibinfo {author} {\bibfnamefont {N.~C.}\ \bibnamefont
  {Mukhopadhyay}}, \ and\ \bibinfo {author} {\bibfnamefont {E.}~\bibnamefont
  {Oset}},\ }\href {\doibase 10.1103/PhysRevC.57.2687} {\bibfield  {journal}
  {\bibinfo  {journal} {Phys.Rev.}\ }\textbf {\bibinfo {volume} {C57}},\
  \bibinfo {pages} {2687} (\bibinfo {year} {1998}{\natexlab{a}})},\ \Eprint
  {http://arxiv.org/abs/nucl-th/9802059} {arXiv:nucl-th/9802059 [nucl-th]}
  \BibitemShut {NoStop}%
\bibitem [{\citenamefont {Sajjad~Athar}\ \emph {et~al.}(2006)\citenamefont
  {Sajjad~Athar}, \citenamefont {Ahmad},\ and\ \citenamefont
  {Singh}}]{SajjadAthar:2005ke}%
  \BibitemOpen
  \bibfield  {author} {\bibinfo {author} {\bibfnamefont {M.}~\bibnamefont
  {Sajjad~Athar}}, \bibinfo {author} {\bibfnamefont {S.}~\bibnamefont {Ahmad}},
  \ and\ \bibinfo {author} {\bibfnamefont {S.}~\bibnamefont {Singh}},\ }\href
  {\doibase 10.1016/j.nuclphysa.2005.09.017} {\bibfield  {journal} {\bibinfo
  {journal} {Nucl.Phys.}\ }\textbf {\bibinfo {volume} {A764}},\ \bibinfo
  {pages} {551} (\bibinfo {year} {2006})},\ \Eprint
  {http://arxiv.org/abs/nucl-th/0506046} {arXiv:nucl-th/0506046 [nucl-th]}
  \BibitemShut {NoStop}%
\bibitem [{\citenamefont {Athar}\ \emph {et~al.}(2005)\citenamefont {Athar},
  \citenamefont {Ahmad},\ and\ \citenamefont {Singh}}]{Athar:2005hu}%
  \BibitemOpen
  \bibfield  {author} {\bibinfo {author} {\bibfnamefont {M.~S.}\ \bibnamefont
  {Athar}}, \bibinfo {author} {\bibfnamefont {S.}~\bibnamefont {Ahmad}}, \ and\
  \bibinfo {author} {\bibfnamefont {S.}~\bibnamefont {Singh}},\ }\href
  {\doibase 10.1140/epja/i2004-10234-2} {\bibfield  {journal} {\bibinfo
  {journal} {Eur.Phys.J.}\ }\textbf {\bibinfo {volume} {A24}},\ \bibinfo
  {pages} {459} (\bibinfo {year} {2005})},\ \Eprint
  {http://arxiv.org/abs/nucl-th/0506057} {arXiv:nucl-th/0506057 [nucl-th]}
  \BibitemShut {NoStop}%
\bibitem [{\citenamefont {Marteau}(1999)}]{Marteau:1999kt}%
  \BibitemOpen
  \bibfield  {author} {\bibinfo {author} {\bibfnamefont {J.}~\bibnamefont
  {Marteau}},\ }\href {\doibase 10.1007/s100500050274} {\bibfield  {journal}
  {\bibinfo  {journal} {Eur.Phys.J.}\ }\textbf {\bibinfo {volume} {A5}},\
  \bibinfo {pages} {183} (\bibinfo {year} {1999})},\ \Eprint
  {http://arxiv.org/abs/hep-ph/9902210} {arXiv:hep-ph/9902210 [hep-ph]}
  \BibitemShut {NoStop}%
\bibitem [{\citenamefont {Marteau}\ \emph {et~al.}(2000)\citenamefont
  {Marteau}, \citenamefont {Delorme},\ and\ \citenamefont
  {Ericson}}]{Marteau:1999jp}%
  \BibitemOpen
  \bibfield  {author} {\bibinfo {author} {\bibfnamefont {J.}~\bibnamefont
  {Marteau}}, \bibinfo {author} {\bibfnamefont {J.}~\bibnamefont {Delorme}}, \
  and\ \bibinfo {author} {\bibfnamefont {M.}~\bibnamefont {Ericson}},\ }\href
  {\doibase 10.1016/S0168-9002(00)00375-2} {\bibfield  {journal} {\bibinfo
  {journal} {Nucl.Instrum.Meth.}\ }\textbf {\bibinfo {volume} {A451}},\
  \bibinfo {pages} {76} (\bibinfo {year} {2000})}\BibitemShut {NoStop}%
\bibitem [{\citenamefont {Martini}\ \emph {et~al.}(2009)\citenamefont
  {Martini}, \citenamefont {Ericson}, \citenamefont {Chanfray},\ and\
  \citenamefont {Marteau}}]{Martini:2009uj}%
  \BibitemOpen
  \bibfield  {author} {\bibinfo {author} {\bibfnamefont {M.}~\bibnamefont
  {Martini}}, \bibinfo {author} {\bibfnamefont {M.}~\bibnamefont {Ericson}},
  \bibinfo {author} {\bibfnamefont {G.}~\bibnamefont {Chanfray}}, \ and\
  \bibinfo {author} {\bibfnamefont {J.}~\bibnamefont {Marteau}},\ }\href
  {\doibase 10.1103/PhysRevC.80.065501} {\bibfield  {journal} {\bibinfo
  {journal} {Phys.Rev.}\ }\textbf {\bibinfo {volume} {C80}},\ \bibinfo {pages}
  {065501} (\bibinfo {year} {2009})},\ \Eprint {http://arxiv.org/abs/0910.2622}
  {arXiv:0910.2622 [nucl-th]} \BibitemShut {NoStop}%
\bibitem [{\citenamefont {Kim}\ \emph {et~al.}(1995)\citenamefont {Kim},
  \citenamefont {Piekarewicz},\ and\ \citenamefont {Horowitz}}]{Kim:1994zea}%
  \BibitemOpen
  \bibfield  {author} {\bibinfo {author} {\bibfnamefont {H.-c.}\ \bibnamefont
  {Kim}}, \bibinfo {author} {\bibfnamefont {J.}~\bibnamefont {Piekarewicz}}, \
  and\ \bibinfo {author} {\bibfnamefont {C.}~\bibnamefont {Horowitz}},\ }\href
  {\doibase 10.1103/PhysRevC.51.2739} {\bibfield  {journal} {\bibinfo
  {journal} {Phys.Rev.}\ }\textbf {\bibinfo {volume} {C51}},\ \bibinfo {pages}
  {2739} (\bibinfo {year} {1995})},\ \Eprint
  {http://arxiv.org/abs/nucl-th/9412017} {arXiv:nucl-th/9412017 [nucl-th]}
  \BibitemShut {NoStop}%
\bibitem [{\citenamefont {Graczyk}\ and\ \citenamefont
  {Sobczyk}(2003)}]{Graczyk:2003ru}%
  \BibitemOpen
  \bibfield  {author} {\bibinfo {author} {\bibfnamefont {K.~M.}\ \bibnamefont
  {Graczyk}}\ and\ \bibinfo {author} {\bibfnamefont {J.~T.}\ \bibnamefont
  {Sobczyk}},\ }\href {\doibase 10.1140/epjc/s2003-01338-6} {\bibfield
  {journal} {\bibinfo  {journal} {Eur.Phys.J.}\ }\textbf {\bibinfo {volume}
  {C31}},\ \bibinfo {pages} {177} (\bibinfo {year} {2003})},\ \Eprint
  {http://arxiv.org/abs/nucl-th/0303054} {arXiv:nucl-th/0303054 [nucl-th]}
  \BibitemShut {NoStop}%
\bibitem [{\citenamefont {Graczyk}(2005)}]{Graczyk:2004uy}%
  \BibitemOpen
  \bibfield  {author} {\bibinfo {author} {\bibfnamefont {K.~M.}\ \bibnamefont
  {Graczyk}},\ }\href {\doibase 10.1016/j.nuclphysa.2004.10.029} {\bibfield
  {journal} {\bibinfo  {journal} {Nucl.Phys.}\ }\textbf {\bibinfo {volume}
  {A748}},\ \bibinfo {pages} {313} (\bibinfo {year} {2005})},\ \Eprint
  {http://arxiv.org/abs/hep-ph/0407275} {arXiv:hep-ph/0407275 [hep-ph]}
  \BibitemShut {NoStop}%
\bibitem [{\citenamefont {Aguilar-Arevalo}\ \emph
  {et~al.}(2009{\natexlab{b}})\citenamefont {Aguilar-Arevalo} \emph
  {et~al.}}]{AguilarArevalo:2008yp}%
  \BibitemOpen
  \bibfield  {author} {\bibinfo {author} {\bibfnamefont {A.}~\bibnamefont
  {Aguilar-Arevalo}} \emph {et~al.} (\bibinfo {collaboration} {MiniBooNE
  Collaboration}),\ }\href {\doibase 10.1103/PhysRevD.79.072002} {\bibfield
  {journal} {\bibinfo  {journal} {Phys.Rev.}\ }\textbf {\bibinfo {volume}
  {D79}},\ \bibinfo {pages} {072002} (\bibinfo {year} {2009}{\natexlab{b}})},\
  \Eprint {http://arxiv.org/abs/0806.1449} {arXiv:0806.1449 [hep-ex]}
  \BibitemShut {NoStop}%
\bibitem [{\citenamefont {Alvarez-Ruso}\ \emph {et~al.}(2009)\citenamefont
  {Alvarez-Ruso}, \citenamefont {Buss}, \citenamefont {Leitner},\ and\
  \citenamefont {Mosel}}]{AlvarezRuso:2009ad}%
  \BibitemOpen
  \bibfield  {author} {\bibinfo {author} {\bibfnamefont {L.}~\bibnamefont
  {Alvarez-Ruso}}, \bibinfo {author} {\bibfnamefont {O.}~\bibnamefont {Buss}},
  \bibinfo {author} {\bibfnamefont {T.}~\bibnamefont {Leitner}}, \ and\
  \bibinfo {author} {\bibfnamefont {U.}~\bibnamefont {Mosel}},\ }\href
  {\doibase 10.1063/1.3274146} {\bibfield  {journal} {\bibinfo  {journal} {AIP
  Conf.Proc.}\ }\textbf {\bibinfo {volume} {1189}},\ \bibinfo {pages} {151}
  (\bibinfo {year} {2009})},\ \Eprint {http://arxiv.org/abs/0909.5123}
  {arXiv:0909.5123 [nucl-th]} \BibitemShut {NoStop}%
\bibitem [{\citenamefont {Speth}\ \emph {et~al.}(1980)\citenamefont {Speth},
  \citenamefont {Klemt}, \citenamefont {Wambach},\ and\ \citenamefont
  {Brown}}]{Speth:1980kw}%
  \BibitemOpen
  \bibfield  {author} {\bibinfo {author} {\bibfnamefont {J.}~\bibnamefont
  {Speth}}, \bibinfo {author} {\bibfnamefont {V.}~\bibnamefont {Klemt}},
  \bibinfo {author} {\bibfnamefont {J.}~\bibnamefont {Wambach}}, \ and\
  \bibinfo {author} {\bibfnamefont {G.}~\bibnamefont {Brown}},\ }\href
  {\doibase 10.1016/0375-9474(80)90660-0} {\bibfield  {journal} {\bibinfo
  {journal} {Nucl.Phys.}\ }\textbf {\bibinfo {volume} {A343}},\ \bibinfo
  {pages} {382} (\bibinfo {year} {1980})}\BibitemShut {NoStop}%
\bibitem [{\citenamefont {Kolbe}\ \emph {et~al.}(1992)\citenamefont {Kolbe},
  \citenamefont {Langanke}, \citenamefont {Krewald},\ and\ \citenamefont
  {Thielemann}}]{Kolbe:1992xu}%
  \BibitemOpen
  \bibfield  {author} {\bibinfo {author} {\bibfnamefont {E.}~\bibnamefont
  {Kolbe}}, \bibinfo {author} {\bibfnamefont {K.}~\bibnamefont {Langanke}},
  \bibinfo {author} {\bibfnamefont {S.}~\bibnamefont {Krewald}}, \ and\
  \bibinfo {author} {\bibfnamefont {F.}~\bibnamefont {Thielemann}},\ }\href
  {\doibase 10.1016/0375-9474(92)90175-J} {\bibfield  {journal} {\bibinfo
  {journal} {Nucl.Phys.}\ }\textbf {\bibinfo {volume} {A540}},\ \bibinfo
  {pages} {599} (\bibinfo {year} {1992})}\BibitemShut {NoStop}%
\bibitem [{\citenamefont {Volpe}\ \emph {et~al.}(2000)\citenamefont {Volpe},
  \citenamefont {Auerbach}, \citenamefont {Colo}, \citenamefont {Suzuki},\ and\
  \citenamefont {Van~Giai}}]{Volpe:2000zn}%
  \BibitemOpen
  \bibfield  {author} {\bibinfo {author} {\bibfnamefont {C.}~\bibnamefont
  {Volpe}}, \bibinfo {author} {\bibfnamefont {N.}~\bibnamefont {Auerbach}},
  \bibinfo {author} {\bibfnamefont {G.}~\bibnamefont {Colo}}, \bibinfo {author}
  {\bibfnamefont {T.}~\bibnamefont {Suzuki}}, \ and\ \bibinfo {author}
  {\bibfnamefont {N.}~\bibnamefont {Van~Giai}},\ }\href {\doibase
  10.1103/PhysRevC.62.015501} {\bibfield  {journal} {\bibinfo  {journal}
  {Phys.Rev.}\ }\textbf {\bibinfo {volume} {C62}},\ \bibinfo {pages} {015501}
  (\bibinfo {year} {2000})},\ \Eprint {http://arxiv.org/abs/nucl-th/0001050}
  {arXiv:nucl-th/0001050 [nucl-th]} \BibitemShut {NoStop}%
\bibitem [{\citenamefont {Jachowicz}\ \emph {et~al.}(2002)\citenamefont
  {Jachowicz}, \citenamefont {Heyde}, \citenamefont {Ryckebusch},\ and\
  \citenamefont {Rombouts}}]{Jachowicz:2002rr}%
  \BibitemOpen
  \bibfield  {author} {\bibinfo {author} {\bibfnamefont {N.}~\bibnamefont
  {Jachowicz}}, \bibinfo {author} {\bibfnamefont {K.}~\bibnamefont {Heyde}},
  \bibinfo {author} {\bibfnamefont {J.}~\bibnamefont {Ryckebusch}}, \ and\
  \bibinfo {author} {\bibfnamefont {S.}~\bibnamefont {Rombouts}},\ }\href
  {\doibase 10.1103/PhysRevC.65.025501} {\bibfield  {journal} {\bibinfo
  {journal} {Phys.Rev.}\ }\textbf {\bibinfo {volume} {C65}},\ \bibinfo {pages}
  {025501} (\bibinfo {year} {2002})}\BibitemShut {NoStop}%
\bibitem [{\citenamefont {Pandey}\ \emph {et~al.}(2014)\citenamefont {Pandey},
  \citenamefont {Jachowicz}, \citenamefont {Ryckebusch}, \citenamefont
  {Van~Cuyck},\ and\ \citenamefont {Cosyn}}]{Pandey:2013cca}%
  \BibitemOpen
  \bibfield  {author} {\bibinfo {author} {\bibfnamefont {V.}~\bibnamefont
  {Pandey}}, \bibinfo {author} {\bibfnamefont {N.}~\bibnamefont {Jachowicz}},
  \bibinfo {author} {\bibfnamefont {J.}~\bibnamefont {Ryckebusch}}, \bibinfo
  {author} {\bibfnamefont {T.}~\bibnamefont {Van~Cuyck}}, \ and\ \bibinfo
  {author} {\bibfnamefont {W.}~\bibnamefont {Cosyn}},\ }\href {\doibase
  10.1103/PhysRevC.89.024601} {\bibfield  {journal} {\bibinfo  {journal}
  {Phys.Rev.}\ }\textbf {\bibinfo {volume} {C89}},\ \bibinfo {pages} {024601}
  (\bibinfo {year} {2014})},\ \Eprint {http://arxiv.org/abs/1310.6885}
  {arXiv:1310.6885 [nucl-th]} \BibitemShut {NoStop}%
\bibitem [{\citenamefont {Amaro}\ \emph
  {et~al.}(2005{\natexlab{b}})\citenamefont {Amaro}, \citenamefont {Maieron},
  \citenamefont {Nieves},\ and\ \citenamefont {Valverde}}]{Amaro:2004cm}%
  \BibitemOpen
  \bibfield  {author} {\bibinfo {author} {\bibfnamefont {J.~E.}\ \bibnamefont
  {Amaro}}, \bibinfo {author} {\bibfnamefont {C.}~\bibnamefont {Maieron}},
  \bibinfo {author} {\bibfnamefont {J.}~\bibnamefont {Nieves}}, \ and\ \bibinfo
  {author} {\bibfnamefont {M.}~\bibnamefont {Valverde}},\ }\href {\doibase
  10.1140/epja/i2005-10034-2} {\bibfield  {journal} {\bibinfo  {journal}
  {Eur.Phys.J.}\ }\textbf {\bibinfo {volume} {A24}},\ \bibinfo {pages} {343}
  (\bibinfo {year} {2005}{\natexlab{b}})},\ \Eprint
  {http://arxiv.org/abs/nucl-th/0409017} {arXiv:nucl-th/0409017 [nucl-th]}
  \BibitemShut {NoStop}%
\bibitem [{\citenamefont {Katori}(2009)}]{Katori:2009du}%
  \BibitemOpen
  \bibfield  {author} {\bibinfo {author} {\bibfnamefont {T.}~\bibnamefont
  {Katori}} (\bibinfo {collaboration} {MiniBooNE Collaboration}),\ }\href
  {\doibase 10.1063/1.3274144} {\bibfield  {journal} {\bibinfo  {journal} {AIP
  Conf.Proc.}\ }\textbf {\bibinfo {volume} {1189}},\ \bibinfo {pages} {139}
  (\bibinfo {year} {2009})},\ \Eprint {http://arxiv.org/abs/0909.1996}
  {arXiv:0909.1996 [hep-ex]} \BibitemShut {NoStop}%
\bibitem [{\citenamefont {Boyd}\ \emph {et~al.}(2009)\citenamefont {Boyd},
  \citenamefont {Dytman}, \citenamefont {Hernandez}, \citenamefont {Sobczyk},\
  and\ \citenamefont {Tacik}}]{Boyd:2009zz}%
  \BibitemOpen
  \bibfield  {author} {\bibinfo {author} {\bibfnamefont {S.}~\bibnamefont
  {Boyd}}, \bibinfo {author} {\bibfnamefont {S.}~\bibnamefont {Dytman}},
  \bibinfo {author} {\bibfnamefont {E.}~\bibnamefont {Hernandez}}, \bibinfo
  {author} {\bibfnamefont {J.}~\bibnamefont {Sobczyk}}, \ and\ \bibinfo
  {author} {\bibfnamefont {R.}~\bibnamefont {Tacik}},\ }\href {\doibase
  10.1063/1.3274191} {\bibfield  {journal} {\bibinfo  {journal} {AIP
  Conf.Proc.}\ }\textbf {\bibinfo {volume} {1189}},\ \bibinfo {pages} {60}
  (\bibinfo {year} {2009})}\BibitemShut {NoStop}%
\bibitem [{\citenamefont {Sajjad~Athar}\ \emph
  {et~al.}(2010{\natexlab{a}})\citenamefont {Sajjad~Athar}, \citenamefont
  {Chauhan},\ and\ \citenamefont {Singh}}]{SajjadAthar:2009rd}%
  \BibitemOpen
  \bibfield  {author} {\bibinfo {author} {\bibfnamefont {M.}~\bibnamefont
  {Sajjad~Athar}}, \bibinfo {author} {\bibfnamefont {S.}~\bibnamefont
  {Chauhan}}, \ and\ \bibinfo {author} {\bibfnamefont {S.}~\bibnamefont
  {Singh}},\ }\href {\doibase 10.1140/epja/i2010-10908-0} {\bibfield  {journal}
  {\bibinfo  {journal} {Eur.Phys.J.}\ }\textbf {\bibinfo {volume} {A43}},\
  \bibinfo {pages} {209} (\bibinfo {year} {2010}{\natexlab{a}})},\ \Eprint
  {http://arxiv.org/abs/0908.1443} {arXiv:0908.1443 [nucl-th]} \BibitemShut
  {NoStop}%
\bibitem [{\citenamefont {Nieves}\ \emph
  {et~al.}(2012{\natexlab{a}})\citenamefont {Nieves}, \citenamefont
  {Ruiz~Simo},\ and\ \citenamefont {Vicente~Vacas}}]{Nieves:2011yp}%
  \BibitemOpen
  \bibfield  {author} {\bibinfo {author} {\bibfnamefont {J.}~\bibnamefont
  {Nieves}}, \bibinfo {author} {\bibfnamefont {I.}~\bibnamefont {Ruiz~Simo}}, \
  and\ \bibinfo {author} {\bibfnamefont {M.}~\bibnamefont {Vicente~Vacas}},\
  }\href {\doibase 10.1016/j.physletb.2011.11.061} {\bibfield  {journal}
  {\bibinfo  {journal} {Phys.Lett.}\ }\textbf {\bibinfo {volume} {B707}},\
  \bibinfo {pages} {72} (\bibinfo {year} {2012}{\natexlab{a}})}\BibitemShut
  {NoStop}%
\bibitem [{\citenamefont {Shimizu}\ and\ \citenamefont
  {Faessler}(1980)}]{Shimizu:1980kb}%
  \BibitemOpen
  \bibfield  {author} {\bibinfo {author} {\bibfnamefont {K.}~\bibnamefont
  {Shimizu}}\ and\ \bibinfo {author} {\bibfnamefont {A.}~\bibnamefont
  {Faessler}},\ }\href {\doibase 10.1016/0375-9474(80)90112-8} {\bibfield
  {journal} {\bibinfo  {journal} {Nucl.Phys.}\ }\textbf {\bibinfo {volume}
  {A333}},\ \bibinfo {pages} {495} (\bibinfo {year} {1980})}\BibitemShut
  {NoStop}%
\bibitem [{\citenamefont {Alberico}\ \emph {et~al.}(1984)\citenamefont
  {Alberico}, \citenamefont {Ericson},\ and\ \citenamefont
  {Molinari}}]{Alberico:1983zg}%
  \BibitemOpen
  \bibfield  {author} {\bibinfo {author} {\bibfnamefont {W.}~\bibnamefont
  {Alberico}}, \bibinfo {author} {\bibfnamefont {M.}~\bibnamefont {Ericson}}, \
  and\ \bibinfo {author} {\bibfnamefont {A.}~\bibnamefont {Molinari}},\ }\href
  {\doibase 10.1016/0003-4916(84)90155-6} {\bibfield  {journal} {\bibinfo
  {journal} {Annals Phys.}\ }\textbf {\bibinfo {volume} {154}},\ \bibinfo
  {pages} {356} (\bibinfo {year} {1984})}\BibitemShut {NoStop}%
\bibitem [{\citenamefont {Nieves}\ \emph {et~al.}(2011)\citenamefont {Nieves},
  \citenamefont {Ruiz~Simo},\ and\ \citenamefont
  {Vicente~Vacas}}]{Nieves:2011pp}%
  \BibitemOpen
  \bibfield  {author} {\bibinfo {author} {\bibfnamefont {J.}~\bibnamefont
  {Nieves}}, \bibinfo {author} {\bibfnamefont {I.}~\bibnamefont {Ruiz~Simo}}, \
  and\ \bibinfo {author} {\bibfnamefont {M.}~\bibnamefont {Vicente~Vacas}},\
  }\href {\doibase 10.1103/PhysRevC.83.045501} {\bibfield  {journal} {\bibinfo
  {journal} {Phys.Rev.}\ }\textbf {\bibinfo {volume} {C83}},\ \bibinfo {pages}
  {045501} (\bibinfo {year} {2011})},\ \Eprint {http://arxiv.org/abs/1102.2777}
  {arXiv:1102.2777 [hep-ph]} \BibitemShut {NoStop}%
\bibitem [{\citenamefont {Lovato}\ \emph {et~al.}(2014)\citenamefont {Lovato},
  \citenamefont {Gandolfi}, \citenamefont {Carlson}, \citenamefont {Pieper},\
  and\ \citenamefont {Schiavilla}}]{Lovato:2014eva}%
  \BibitemOpen
  \bibfield  {author} {\bibinfo {author} {\bibfnamefont {A.}~\bibnamefont
  {Lovato}}, \bibinfo {author} {\bibfnamefont {S.}~\bibnamefont {Gandolfi}},
  \bibinfo {author} {\bibfnamefont {J.}~\bibnamefont {Carlson}}, \bibinfo
  {author} {\bibfnamefont {S.~C.}\ \bibnamefont {Pieper}}, \ and\ \bibinfo
  {author} {\bibfnamefont {R.}~\bibnamefont {Schiavilla}},\ }\href {\doibase
  10.1103/PhysRevLett.112.182502} {\bibfield  {journal} {\bibinfo  {journal}
  {Phys.Rev.Lett.}\ }\textbf {\bibinfo {volume} {112}},\ \bibinfo {pages}
  {182502} (\bibinfo {year} {2014})},\ \Eprint {http://arxiv.org/abs/1401.2605}
  {arXiv:1401.2605 [nucl-th]} \BibitemShut {NoStop}%
\bibitem [{\citenamefont {Lovato}\ \emph {et~al.}(2013)\citenamefont {Lovato},
  \citenamefont {Gandolfi}, \citenamefont {Butler}, \citenamefont {Carlson},
  \citenamefont {Lusk} \emph {et~al.}}]{Lovato:2013cua}%
  \BibitemOpen
  \bibfield  {author} {\bibinfo {author} {\bibfnamefont {A.}~\bibnamefont
  {Lovato}}, \bibinfo {author} {\bibfnamefont {S.}~\bibnamefont {Gandolfi}},
  \bibinfo {author} {\bibfnamefont {R.}~\bibnamefont {Butler}}, \bibinfo
  {author} {\bibfnamefont {J.}~\bibnamefont {Carlson}}, \bibinfo {author}
  {\bibfnamefont {E.}~\bibnamefont {Lusk}},  \emph {et~al.},\ }\href {\doibase
  10.1103/PhysRevLett.111.092501} {\bibfield  {journal} {\bibinfo  {journal}
  {Phys.Rev.Lett.}\ }\textbf {\bibinfo {volume} {111}},\ \bibinfo {pages}
  {092501} (\bibinfo {year} {2013})},\ \Eprint {http://arxiv.org/abs/1305.6959}
  {arXiv:1305.6959 [nucl-th]} \BibitemShut {NoStop}%
\bibitem [{\citenamefont {Benhar}\ \emph {et~al.}(2013)\citenamefont {Benhar},
  \citenamefont {Lovato},\ and\ \citenamefont {Rocco}}]{Benhar:2013bba}%
  \BibitemOpen
  \bibfield  {author} {\bibinfo {author} {\bibfnamefont {O.}~\bibnamefont
  {Benhar}}, \bibinfo {author} {\bibfnamefont {A.}~\bibnamefont {Lovato}}, \
  and\ \bibinfo {author} {\bibfnamefont {N.}~\bibnamefont {Rocco}},\
  }\href@noop {} {\  (\bibinfo {year} {2013})},\ \Eprint
  {http://arxiv.org/abs/1312.1210} {arXiv:1312.1210 [nucl-th]} \BibitemShut
  {NoStop}%
\bibitem [{\citenamefont {Amaro}\ \emph {et~al.}(2011)\citenamefont {Amaro},
  \citenamefont {Barbaro}, \citenamefont {Caballero}, \citenamefont
  {Donnelly},\ and\ \citenamefont {Williamson}}]{Amaro:2010sd}%
  \BibitemOpen
  \bibfield  {author} {\bibinfo {author} {\bibfnamefont {J.}~\bibnamefont
  {Amaro}}, \bibinfo {author} {\bibfnamefont {M.}~\bibnamefont {Barbaro}},
  \bibinfo {author} {\bibfnamefont {J.}~\bibnamefont {Caballero}}, \bibinfo
  {author} {\bibfnamefont {T.}~\bibnamefont {Donnelly}}, \ and\ \bibinfo
  {author} {\bibfnamefont {C.}~\bibnamefont {Williamson}},\ }\href {\doibase
  10.1016/j.physletb.2010.12.007} {\bibfield  {journal} {\bibinfo  {journal}
  {Phys.Lett.}\ }\textbf {\bibinfo {volume} {B696}},\ \bibinfo {pages} {151}
  (\bibinfo {year} {2011})}\BibitemShut {NoStop}%
\bibitem [{\citenamefont {Martini}\ \emph {et~al.}(2011)\citenamefont
  {Martini}, \citenamefont {Ericson},\ and\ \citenamefont
  {Chanfray}}]{Martini:2011wp}%
  \BibitemOpen
  \bibfield  {author} {\bibinfo {author} {\bibfnamefont {M.}~\bibnamefont
  {Martini}}, \bibinfo {author} {\bibfnamefont {M.}~\bibnamefont {Ericson}}, \
  and\ \bibinfo {author} {\bibfnamefont {G.}~\bibnamefont {Chanfray}},\ }\href
  {\doibase 10.1103/PhysRevC.84.055502} {\bibfield  {journal} {\bibinfo
  {journal} {Phys.Rev.}\ }\textbf {\bibinfo {volume} {C84}},\ \bibinfo {pages}
  {055502} (\bibinfo {year} {2011})}\BibitemShut {NoStop}%
\bibitem [{\citenamefont {Amaro}\ \emph {et~al.}(2012)\citenamefont {Amaro},
  \citenamefont {Barbaro}, \citenamefont {Caballero},\ and\ \citenamefont
  {Donnelly}}]{Amaro:2011aa}%
  \BibitemOpen
  \bibfield  {author} {\bibinfo {author} {\bibfnamefont {J.}~\bibnamefont
  {Amaro}}, \bibinfo {author} {\bibfnamefont {M.}~\bibnamefont {Barbaro}},
  \bibinfo {author} {\bibfnamefont {J.}~\bibnamefont {Caballero}}, \ and\
  \bibinfo {author} {\bibfnamefont {T.}~\bibnamefont {Donnelly}},\ }\href
  {\doibase 10.1103/PhysRevLett.108.152501} {\bibfield  {journal} {\bibinfo
  {journal} {Phys.Rev.Lett.}\ }\textbf {\bibinfo {volume} {108}},\ \bibinfo
  {pages} {152501} (\bibinfo {year} {2012})}\BibitemShut {NoStop}%
\bibitem [{\citenamefont {Nieves}\ \emph {et~al.}(2013)\citenamefont {Nieves},
  \citenamefont {Ruiz~Simo},\ and\ \citenamefont
  {Vicente~Vacas}}]{Nieves:2013fr}%
  \BibitemOpen
  \bibfield  {author} {\bibinfo {author} {\bibfnamefont {J.}~\bibnamefont
  {Nieves}}, \bibinfo {author} {\bibfnamefont {I.}~\bibnamefont {Ruiz~Simo}}, \
  and\ \bibinfo {author} {\bibfnamefont {M.}~\bibnamefont {Vicente~Vacas}},\
  }\href {\doibase 10.1016/j.physletb.2013.03.002} {\bibfield  {journal}
  {\bibinfo  {journal} {Phys.Lett.}\ }\textbf {\bibinfo {volume} {B721}},\
  \bibinfo {pages} {90} (\bibinfo {year} {2013})},\ \Eprint
  {http://arxiv.org/abs/1302.0703} {arXiv:1302.0703 [hep-ph]} \BibitemShut
  {NoStop}%
\bibitem [{\citenamefont {Martini}\ and\ \citenamefont
  {Ericson}(2013)}]{Martini:2013sha}%
  \BibitemOpen
  \bibfield  {author} {\bibinfo {author} {\bibfnamefont {M.}~\bibnamefont
  {Martini}}\ and\ \bibinfo {author} {\bibfnamefont {M.}~\bibnamefont
  {Ericson}},\ }\href {\doibase 10.1103/PhysRevC.87.065501} {\bibfield
  {journal} {\bibinfo  {journal} {Phys.Rev.}\ }\textbf {\bibinfo {volume}
  {C87}},\ \bibinfo {pages} {065501} (\bibinfo {year} {2013})},\ \Eprint
  {http://arxiv.org/abs/1303.7199} {arXiv:1303.7199 [nucl-th]} \BibitemShut
  {NoStop}%
\bibitem [{\citenamefont {Lalakulich}\ \emph
  {et~al.}(2012{\natexlab{a}})\citenamefont {Lalakulich}, \citenamefont
  {Gallmeister},\ and\ \citenamefont {Mosel}}]{Lalakulich:2012ac}%
  \BibitemOpen
  \bibfield  {author} {\bibinfo {author} {\bibfnamefont {O.}~\bibnamefont
  {Lalakulich}}, \bibinfo {author} {\bibfnamefont {K.}~\bibnamefont
  {Gallmeister}}, \ and\ \bibinfo {author} {\bibfnamefont {U.}~\bibnamefont
  {Mosel}},\ }\href {\doibase 10.1103/PhysRevC.86.014614} {\bibfield  {journal}
  {\bibinfo  {journal} {Phys.Rev.}\ }\textbf {\bibinfo {volume} {C86}},\
  \bibinfo {pages} {014614} (\bibinfo {year} {2012}{\natexlab{a}})}\BibitemShut
  {NoStop}%
\bibitem [{\citenamefont {Sobczyk}(2012)}]{Sobczyk:2012ms}%
  \BibitemOpen
  \bibfield  {author} {\bibinfo {author} {\bibfnamefont {J.~T.}\ \bibnamefont
  {Sobczyk}},\ }\href {\doibase 10.1103/PhysRevC.86.015504} {\bibfield
  {journal} {\bibinfo  {journal} {Phys.Rev.}\ }\textbf {\bibinfo {volume}
  {C86}},\ \bibinfo {pages} {015504} (\bibinfo {year} {2012})},\ \Eprint
  {http://arxiv.org/abs/1201.3673} {arXiv:1201.3673 [hep-ph]} \BibitemShut
  {NoStop}%
\bibitem [{\citenamefont {Kordosky}(2006)}]{Kordosky:2006gt}%
  \BibitemOpen
  \bibfield  {author} {\bibinfo {author} {\bibfnamefont {M.}~\bibnamefont
  {Kordosky}},\ }\href {\doibase 10.1016/j.nuclphysbps.2006.08.040} {\bibfield
  {journal} {\bibinfo  {journal} {Nucl.Phys.Proc.Suppl.}\ }\textbf {\bibinfo
  {volume} {159}},\ \bibinfo {pages} {223} (\bibinfo {year} {2006})},\ \Eprint
  {http://arxiv.org/abs/hep-ex/0602029} {arXiv:hep-ex/0602029 [hep-ex]}
  \BibitemShut {NoStop}%
\bibitem [{\citenamefont {Martini}\ \emph {et~al.}(2012)\citenamefont
  {Martini}, \citenamefont {Ericson},\ and\ \citenamefont
  {Chanfray}}]{Martini:2012fa}%
  \BibitemOpen
  \bibfield  {author} {\bibinfo {author} {\bibfnamefont {M.}~\bibnamefont
  {Martini}}, \bibinfo {author} {\bibfnamefont {M.}~\bibnamefont {Ericson}}, \
  and\ \bibinfo {author} {\bibfnamefont {G.}~\bibnamefont {Chanfray}},\ }\href
  {\doibase 10.1103/PhysRevD.85.093012} {\bibfield  {journal} {\bibinfo
  {journal} {Phys.Rev.}\ }\textbf {\bibinfo {volume} {D85}},\ \bibinfo {pages}
  {093012} (\bibinfo {year} {2012})}\BibitemShut {NoStop}%
\bibitem [{\citenamefont {Nieves}\ \emph
  {et~al.}(2012{\natexlab{b}})\citenamefont {Nieves}, \citenamefont {Sanchez},
  \citenamefont {Ruiz~Simo},\ and\ \citenamefont
  {Vicente~Vacas}}]{Nieves:2012yz}%
  \BibitemOpen
  \bibfield  {author} {\bibinfo {author} {\bibfnamefont {J.}~\bibnamefont
  {Nieves}}, \bibinfo {author} {\bibfnamefont {F.}~\bibnamefont {Sanchez}},
  \bibinfo {author} {\bibfnamefont {I.}~\bibnamefont {Ruiz~Simo}}, \ and\
  \bibinfo {author} {\bibfnamefont {M.}~\bibnamefont {Vicente~Vacas}},\ }\href
  {\doibase 10.1103/PhysRevD.85.113008} {\bibfield  {journal} {\bibinfo
  {journal} {Phys.Rev.}\ }\textbf {\bibinfo {volume} {D85}},\ \bibinfo {pages}
  {113008} (\bibinfo {year} {2012}{\natexlab{b}})},\ \Eprint
  {http://arxiv.org/abs/1204.5404} {arXiv:1204.5404 [hep-ph]} \BibitemShut
  {NoStop}%
\bibitem [{\citenamefont {Martini}\ \emph {et~al.}(2013)\citenamefont
  {Martini}, \citenamefont {Ericson},\ and\ \citenamefont
  {Chanfray}}]{Martini:2012uc}%
  \BibitemOpen
  \bibfield  {author} {\bibinfo {author} {\bibfnamefont {M.}~\bibnamefont
  {Martini}}, \bibinfo {author} {\bibfnamefont {M.}~\bibnamefont {Ericson}}, \
  and\ \bibinfo {author} {\bibfnamefont {G.}~\bibnamefont {Chanfray}},\ }\href
  {\doibase 10.1103/PhysRevD.87.013009} {\bibfield  {journal} {\bibinfo
  {journal} {Phys.Rev.}\ }\textbf {\bibinfo {volume} {D87}},\ \bibinfo {pages}
  {013009} (\bibinfo {year} {2013})},\ \Eprint {http://arxiv.org/abs/1211.1523}
  {arXiv:1211.1523 [hep-ph]} \BibitemShut {NoStop}%
\bibitem [{\citenamefont {Lalakulich}\ and\ \citenamefont
  {Mosel}(2012)}]{Lalakulich:2012hs}%
  \BibitemOpen
  \bibfield  {author} {\bibinfo {author} {\bibfnamefont {O.}~\bibnamefont
  {Lalakulich}}\ and\ \bibinfo {author} {\bibfnamefont {U.}~\bibnamefont
  {Mosel}},\ }\href {\doibase 10.1103/PhysRevC.86.054606} {\bibfield  {journal}
  {\bibinfo  {journal} {Phys.Rev.}\ }\textbf {\bibinfo {volume} {C86}},\
  \bibinfo {pages} {054606} (\bibinfo {year} {2012})}\BibitemShut {NoStop}%
\bibitem [{\citenamefont {Mosel}\ \emph
  {et~al.}(2014{\natexlab{a}})\citenamefont {Mosel}, \citenamefont
  {Lalakulich},\ and\ \citenamefont {Gallmeister}}]{Mosel:2013fxa}%
  \BibitemOpen
  \bibfield  {author} {\bibinfo {author} {\bibfnamefont {U.}~\bibnamefont
  {Mosel}}, \bibinfo {author} {\bibfnamefont {O.}~\bibnamefont {Lalakulich}}, \
  and\ \bibinfo {author} {\bibfnamefont {K.}~\bibnamefont {Gallmeister}},\
  }\href {\doibase 10.1103/PhysRevLett.112.151802} {\bibfield  {journal}
  {\bibinfo  {journal} {Phys.Rev.Lett.}\ }\textbf {\bibinfo {volume} {112}},\
  \bibinfo {pages} {151802} (\bibinfo {year} {2014}{\natexlab{a}})},\ \Eprint
  {http://arxiv.org/abs/1311.7288} {arXiv:1311.7288 [nucl-th]} \BibitemShut
  {NoStop}%
\bibitem [{\citenamefont {Coloma}\ and\ \citenamefont
  {Huber}(2013)}]{Coloma:2013rqa}%
  \BibitemOpen
  \bibfield  {author} {\bibinfo {author} {\bibfnamefont {P.}~\bibnamefont
  {Coloma}}\ and\ \bibinfo {author} {\bibfnamefont {P.}~\bibnamefont {Huber}},\
  }\href {\doibase 10.1103/PhysRevLett.111.221802} {\bibfield  {journal}
  {\bibinfo  {journal} {Phys.Rev.Lett.}\ }\textbf {\bibinfo {volume} {111}},\
  \bibinfo {pages} {221802} (\bibinfo {year} {2013})},\ \Eprint
  {http://arxiv.org/abs/1307.1243} {arXiv:1307.1243 [hep-ph]} \BibitemShut
  {NoStop}%
\bibitem [{\citenamefont {Coloma}\ \emph {et~al.}(2014)\citenamefont {Coloma},
  \citenamefont {Huber}, \citenamefont {Jen},\ and\ \citenamefont
  {Mariani}}]{Coloma:2013tba}%
  \BibitemOpen
  \bibfield  {author} {\bibinfo {author} {\bibfnamefont {P.}~\bibnamefont
  {Coloma}}, \bibinfo {author} {\bibfnamefont {P.}~\bibnamefont {Huber}},
  \bibinfo {author} {\bibfnamefont {C.-M.}\ \bibnamefont {Jen}}, \ and\
  \bibinfo {author} {\bibfnamefont {C.}~\bibnamefont {Mariani}},\ }\href
  {\doibase 10.1103/PhysRevD.89.073015} {\bibfield  {journal} {\bibinfo
  {journal} {Phys.Rev.}\ }\textbf {\bibinfo {volume} {D89}},\ \bibinfo {pages}
  {073015} (\bibinfo {year} {2014})},\ \Eprint {http://arxiv.org/abs/1311.4506}
  {arXiv:1311.4506 [hep-ph]} \BibitemShut {NoStop}%
\bibitem [{\citenamefont {Gran}\ \emph {et~al.}(2013)\citenamefont {Gran},
  \citenamefont {Nieves}, \citenamefont {Sanchez},\ and\ \citenamefont
  {Vacas}}]{Gran:2013kda}%
  \BibitemOpen
  \bibfield  {author} {\bibinfo {author} {\bibfnamefont {R.}~\bibnamefont
  {Gran}}, \bibinfo {author} {\bibfnamefont {J.}~\bibnamefont {Nieves}},
  \bibinfo {author} {\bibfnamefont {F.}~\bibnamefont {Sanchez}}, \ and\
  \bibinfo {author} {\bibfnamefont {M.~J.~V.}\ \bibnamefont {Vacas}},\ }\href
  {\doibase 10.1103/PhysRevD.88.113007} {\bibfield  {journal} {\bibinfo
  {journal} {Phys.Rev.}\ }\textbf {\bibinfo {volume} {D88}},\ \bibinfo {pages}
  {113007} (\bibinfo {year} {2013})},\ \Eprint {http://arxiv.org/abs/1307.8105}
  {arXiv:1307.8105 [hep-ph]} \BibitemShut {NoStop}%
\bibitem [{\citenamefont {Cabibbo}\ \emph {et~al.}(2003)\citenamefont
  {Cabibbo}, \citenamefont {Swallow},\ and\ \citenamefont
  {Winston}}]{Cabibbo:2003cu}%
  \BibitemOpen
  \bibfield  {author} {\bibinfo {author} {\bibfnamefont {N.}~\bibnamefont
  {Cabibbo}}, \bibinfo {author} {\bibfnamefont {E.~C.}\ \bibnamefont
  {Swallow}}, \ and\ \bibinfo {author} {\bibfnamefont {R.}~\bibnamefont
  {Winston}},\ }\href {\doibase 10.1146/annurev.nucl.53.013103.155258}
  {\bibfield  {journal} {\bibinfo  {journal} {Ann.Rev.Nucl.Part.Sci.}\ }\textbf
  {\bibinfo {volume} {53}},\ \bibinfo {pages} {39} (\bibinfo {year} {2003})},\
  \Eprint {http://arxiv.org/abs/hep-ph/0307298} {arXiv:hep-ph/0307298 [hep-ph]}
  \BibitemShut {NoStop}%
\bibitem [{\citenamefont {Singh}\ and\ \citenamefont
  {Vicente~Vacas}(2006)}]{Singh:2006xp}%
  \BibitemOpen
  \bibfield  {author} {\bibinfo {author} {\bibfnamefont {S.}~\bibnamefont
  {Singh}}\ and\ \bibinfo {author} {\bibfnamefont {M.}~\bibnamefont
  {Vicente~Vacas}},\ }\href {\doibase 10.1103/PhysRevD.74.053009} {\bibfield
  {journal} {\bibinfo  {journal} {Phys.Rev.}\ }\textbf {\bibinfo {volume}
  {D74}},\ \bibinfo {pages} {053009} (\bibinfo {year} {2006})},\ \Eprint
  {http://arxiv.org/abs/hep-ph/0606235} {arXiv:hep-ph/0606235 [hep-ph]}
  \BibitemShut {NoStop}%
\bibitem [{\citenamefont {Mintz}\ and\ \citenamefont
  {Wen}(2007)}]{Mintz:2007zz}%
  \BibitemOpen
  \bibfield  {author} {\bibinfo {author} {\bibfnamefont {S.}~\bibnamefont
  {Mintz}}\ and\ \bibinfo {author} {\bibfnamefont {L.}~\bibnamefont {Wen}},\
  }\href {\doibase 10.1140/epja/i2007-10472-8} {\bibfield  {journal} {\bibinfo
  {journal} {Eur.Phys.J.}\ }\textbf {\bibinfo {volume} {A33}},\ \bibinfo
  {pages} {299} (\bibinfo {year} {2007})}\BibitemShut {NoStop}%
\bibitem [{\citenamefont {Kuzmin}\ and\ \citenamefont
  {Naumov}(2009)}]{Kuzmin:2008zz}%
  \BibitemOpen
  \bibfield  {author} {\bibinfo {author} {\bibfnamefont {K.}~\bibnamefont
  {Kuzmin}}\ and\ \bibinfo {author} {\bibfnamefont {V.}~\bibnamefont
  {Naumov}},\ }\href {\doibase 10.1134/S1063778809090105} {\bibfield  {journal}
  {\bibinfo  {journal} {Phys.Atom.Nucl.}\ }\textbf {\bibinfo {volume} {72}},\
  \bibinfo {pages} {1501} (\bibinfo {year} {2009})}\BibitemShut {NoStop}%
\bibitem [{\citenamefont {Zhu}\ \emph {et~al.}(2001)\citenamefont {Zhu},
  \citenamefont {Puglia},\ and\ \citenamefont {Ramsey-Musolf}}]{Zhu:2000zf}%
  \BibitemOpen
  \bibfield  {author} {\bibinfo {author} {\bibfnamefont {S.-L.}\ \bibnamefont
  {Zhu}}, \bibinfo {author} {\bibfnamefont {S.}~\bibnamefont {Puglia}}, \ and\
  \bibinfo {author} {\bibfnamefont {M.}~\bibnamefont {Ramsey-Musolf}},\ }\href
  {\doibase 10.1103/PhysRevD.63.034002} {\bibfield  {journal} {\bibinfo
  {journal} {Phys.Rev.}\ }\textbf {\bibinfo {volume} {D63}},\ \bibinfo {pages}
  {034002} (\bibinfo {year} {2001})},\ \Eprint
  {http://arxiv.org/abs/hep-ph/0009159} {arXiv:hep-ph/0009159 [hep-ph]}
  \BibitemShut {NoStop}%
\bibitem [{\citenamefont {Alam}\ \emph
  {et~al.}(2013{\natexlab{a}})\citenamefont {Alam}, \citenamefont {Chauhan},
  \citenamefont {Athar},\ and\ \citenamefont {Singh}}]{Alam:2013cra}%
  \BibitemOpen
  \bibfield  {author} {\bibinfo {author} {\bibfnamefont {M.~R.}\ \bibnamefont
  {Alam}}, \bibinfo {author} {\bibfnamefont {S.}~\bibnamefont {Chauhan}},
  \bibinfo {author} {\bibfnamefont {M.~S.}\ \bibnamefont {Athar}}, \ and\
  \bibinfo {author} {\bibfnamefont {S.}~\bibnamefont {Singh}},\ }\href
  {\doibase 10.1103/PhysRevD.88.077301} {\bibfield  {journal} {\bibinfo
  {journal} {Phys.Rev.}\ }\textbf {\bibinfo {volume} {D88}},\ \bibinfo {pages}
  {077301} (\bibinfo {year} {2013}{\natexlab{a}})},\ \Eprint
  {http://arxiv.org/abs/1310.7704} {arXiv:1310.7704 [nucl-th]} \BibitemShut
  {NoStop}%
\bibitem [{\citenamefont {Kurimoto}\ \emph
  {et~al.}(2010{\natexlab{b}})\citenamefont {Kurimoto} \emph
  {et~al.}}]{Kurimoto:2009wq}%
  \BibitemOpen
  \bibfield  {author} {\bibinfo {author} {\bibfnamefont {Y.}~\bibnamefont
  {Kurimoto}} \emph {et~al.} (\bibinfo {collaboration} {SciBooNE
  Collaboration}),\ }\href {\doibase 10.1103/PhysRevD.81.033004} {\bibfield
  {journal} {\bibinfo  {journal} {Phys.Rev.}\ }\textbf {\bibinfo {volume}
  {D81}},\ \bibinfo {pages} {033004} (\bibinfo {year} {2010}{\natexlab{b}})},\
  \Eprint {http://arxiv.org/abs/0910.5768} {arXiv:0910.5768 [hep-ex]}
  \BibitemShut {NoStop}%
\bibitem [{\citenamefont {Campbell}\ \emph {et~al.}(1973)\citenamefont
  {Campbell}, \citenamefont {Charlton}, \citenamefont {Cho}, \citenamefont
  {Derrick}, \citenamefont {Engelmann} \emph {et~al.}}]{Campbell:1973wg}%
  \BibitemOpen
  \bibfield  {author} {\bibinfo {author} {\bibfnamefont {J.}~\bibnamefont
  {Campbell}}, \bibinfo {author} {\bibfnamefont {G.}~\bibnamefont {Charlton}},
  \bibinfo {author} {\bibfnamefont {Y.}~\bibnamefont {Cho}}, \bibinfo {author}
  {\bibfnamefont {M.}~\bibnamefont {Derrick}}, \bibinfo {author} {\bibfnamefont
  {R.}~\bibnamefont {Engelmann}},  \emph {et~al.},\ }\href {\doibase
  10.1103/PhysRevLett.30.335} {\bibfield  {journal} {\bibinfo  {journal}
  {Phys.Rev.Lett.}\ }\textbf {\bibinfo {volume} {30}},\ \bibinfo {pages} {335}
  (\bibinfo {year} {1973})}\BibitemShut {NoStop}%
\bibitem [{\citenamefont {Radecky}\ \emph {et~al.}(1982)\citenamefont
  {Radecky}, \citenamefont {Barnes}, \citenamefont {Carmony}, \citenamefont
  {Garfinkel}, \citenamefont {Derrick} \emph {et~al.}}]{Radecky:1981fn}%
  \BibitemOpen
  \bibfield  {author} {\bibinfo {author} {\bibfnamefont {G.}~\bibnamefont
  {Radecky}}, \bibinfo {author} {\bibfnamefont {V.}~\bibnamefont {Barnes}},
  \bibinfo {author} {\bibfnamefont {D.}~\bibnamefont {Carmony}}, \bibinfo
  {author} {\bibfnamefont {A.}~\bibnamefont {Garfinkel}}, \bibinfo {author}
  {\bibfnamefont {M.}~\bibnamefont {Derrick}},  \emph {et~al.},\ }\href
  {\doibase 10.1103/PhysRevD.25.1161, 10.1103/PhysRevD.26.3297} {\bibfield
  {journal} {\bibinfo  {journal} {Phys.Rev.}\ }\textbf {\bibinfo {volume}
  {D25}},\ \bibinfo {pages} {1161} (\bibinfo {year} {1982})}\BibitemShut
  {NoStop}%
\bibitem [{\citenamefont {Kitagaki}\ \emph {et~al.}(1986)\citenamefont
  {Kitagaki}, \citenamefont {Yuta}, \citenamefont {Tanaka}, \citenamefont
  {Yamaguchi}, \citenamefont {Abe} \emph {et~al.}}]{Kitagaki:1986ct}%
  \BibitemOpen
  \bibfield  {author} {\bibinfo {author} {\bibfnamefont {T.}~\bibnamefont
  {Kitagaki}}, \bibinfo {author} {\bibfnamefont {H.}~\bibnamefont {Yuta}},
  \bibinfo {author} {\bibfnamefont {S.}~\bibnamefont {Tanaka}}, \bibinfo
  {author} {\bibfnamefont {A.}~\bibnamefont {Yamaguchi}}, \bibinfo {author}
  {\bibfnamefont {K.}~\bibnamefont {Abe}},  \emph {et~al.},\ }\href {\doibase
  10.1103/PhysRevD.34.2554} {\bibfield  {journal} {\bibinfo  {journal}
  {Phys.Rev.}\ }\textbf {\bibinfo {volume} {D34}},\ \bibinfo {pages} {2554}
  (\bibinfo {year} {1986})}\BibitemShut {NoStop}%
\bibitem [{\citenamefont {Schreiner}\ and\ \citenamefont
  {Von~Hippel}(1973)}]{Schreiner:1973ka}%
  \BibitemOpen
  \bibfield  {author} {\bibinfo {author} {\bibfnamefont {P.}~\bibnamefont
  {Schreiner}}\ and\ \bibinfo {author} {\bibfnamefont {F.}~\bibnamefont
  {Von~Hippel}},\ }\href {\doibase 10.1103/PhysRevLett.30.339} {\bibfield
  {journal} {\bibinfo  {journal} {Phys.Rev.Lett.}\ }\textbf {\bibinfo {volume}
  {30}},\ \bibinfo {pages} {339} (\bibinfo {year} {1973})}\BibitemShut
  {NoStop}%
\bibitem [{\citenamefont {Fogli}\ and\ \citenamefont
  {Nardulli}(1979)}]{Fogli:1979cz}%
  \BibitemOpen
  \bibfield  {author} {\bibinfo {author} {\bibfnamefont {G.~L.}\ \bibnamefont
  {Fogli}}\ and\ \bibinfo {author} {\bibfnamefont {G.}~\bibnamefont
  {Nardulli}},\ }\href {\doibase 10.1016/0550-3213(79)90233-5} {\bibfield
  {journal} {\bibinfo  {journal} {Nucl.Phys.}\ }\textbf {\bibinfo {volume}
  {B160}},\ \bibinfo {pages} {116} (\bibinfo {year} {1979})}\BibitemShut
  {NoStop}%
\bibitem [{\citenamefont {Fogli}\ and\ \citenamefont
  {Nardulli}(1980)}]{Fogli:1979qj}%
  \BibitemOpen
  \bibfield  {author} {\bibinfo {author} {\bibfnamefont {G.~L.}\ \bibnamefont
  {Fogli}}\ and\ \bibinfo {author} {\bibfnamefont {G.}~\bibnamefont
  {Nardulli}},\ }\href {\doibase 10.1016/0550-3213(80)90312-0} {\bibfield
  {journal} {\bibinfo  {journal} {Nucl.Phys.}\ }\textbf {\bibinfo {volume}
  {B165}},\ \bibinfo {pages} {162} (\bibinfo {year} {1980})}\BibitemShut
  {NoStop}%
\bibitem [{\citenamefont {Rein}\ and\ \citenamefont
  {Sehgal}(1981{\natexlab{a}})}]{Rein:1980wg}%
  \BibitemOpen
  \bibfield  {author} {\bibinfo {author} {\bibfnamefont {D.}~\bibnamefont
  {Rein}}\ and\ \bibinfo {author} {\bibfnamefont {L.~M.}\ \bibnamefont
  {Sehgal}},\ }\href {\doibase 10.1016/0003-4916(81)90242-6} {\bibfield
  {journal} {\bibinfo  {journal} {Annals Phys.}\ }\textbf {\bibinfo {volume}
  {133}},\ \bibinfo {pages} {79} (\bibinfo {year}
  {1981}{\natexlab{a}})}\BibitemShut {NoStop}%
\bibitem [{\citenamefont {Alvarez-Ruso}\ \emph {et~al.}(1998)\citenamefont
  {Alvarez-Ruso}, \citenamefont {Singh},\ and\ \citenamefont
  {Vicente~Vacas}}]{AlvarezRuso:1997jr}%
  \BibitemOpen
  \bibfield  {author} {\bibinfo {author} {\bibfnamefont {L.}~\bibnamefont
  {Alvarez-Ruso}}, \bibinfo {author} {\bibfnamefont {S.}~\bibnamefont {Singh}},
  \ and\ \bibinfo {author} {\bibfnamefont {M.}~\bibnamefont {Vicente~Vacas}},\
  }\href {\doibase 10.1103/PhysRevC.57.2693} {\bibfield  {journal} {\bibinfo
  {journal} {Phys.Rev.}\ }\textbf {\bibinfo {volume} {C57}},\ \bibinfo {pages}
  {2693} (\bibinfo {year} {1998})},\ \Eprint
  {http://arxiv.org/abs/nucl-th/9712058} {arXiv:nucl-th/9712058 [nucl-th]}
  \BibitemShut {NoStop}%
\bibitem [{\citenamefont {Alvarez-Ruso}\ \emph {et~al.}(1999)\citenamefont
  {Alvarez-Ruso}, \citenamefont {Singh},\ and\ \citenamefont
  {Vicente~Vacas}}]{AlvarezRuso:1998hi}%
  \BibitemOpen
  \bibfield  {author} {\bibinfo {author} {\bibfnamefont {L.}~\bibnamefont
  {Alvarez-Ruso}}, \bibinfo {author} {\bibfnamefont {S.}~\bibnamefont {Singh}},
  \ and\ \bibinfo {author} {\bibfnamefont {M.}~\bibnamefont {Vicente~Vacas}},\
  }\href {\doibase 10.1103/PhysRevC.59.3386} {\bibfield  {journal} {\bibinfo
  {journal} {Phys.Rev.}\ }\textbf {\bibinfo {volume} {C59}},\ \bibinfo {pages}
  {3386} (\bibinfo {year} {1999})},\ \Eprint
  {http://arxiv.org/abs/nucl-th/9804007} {arXiv:nucl-th/9804007 [nucl-th]}
  \BibitemShut {NoStop}%
\bibitem [{\citenamefont {Sato}\ \emph {et~al.}(2003)\citenamefont {Sato},
  \citenamefont {Uno},\ and\ \citenamefont {Lee}}]{Sato:2003rq}%
  \BibitemOpen
  \bibfield  {author} {\bibinfo {author} {\bibfnamefont {T.}~\bibnamefont
  {Sato}}, \bibinfo {author} {\bibfnamefont {D.}~\bibnamefont {Uno}}, \ and\
  \bibinfo {author} {\bibfnamefont {T.}~\bibnamefont {Lee}},\ }\href {\doibase
  10.1103/PhysRevC.67.065201} {\bibfield  {journal} {\bibinfo  {journal}
  {Phys.Rev.}\ }\textbf {\bibinfo {volume} {C67}},\ \bibinfo {pages} {065201}
  (\bibinfo {year} {2003})},\ \Eprint {http://arxiv.org/abs/nucl-th/0303050}
  {arXiv:nucl-th/0303050 [nucl-th]} \BibitemShut {NoStop}%
\bibitem [{\citenamefont {Paschos}\ \emph {et~al.}(2004)\citenamefont
  {Paschos}, \citenamefont {Yu},\ and\ \citenamefont
  {Sakuda}}]{Paschos:2003qr}%
  \BibitemOpen
  \bibfield  {author} {\bibinfo {author} {\bibfnamefont {E.~A.}\ \bibnamefont
  {Paschos}}, \bibinfo {author} {\bibfnamefont {J.-Y.}\ \bibnamefont {Yu}}, \
  and\ \bibinfo {author} {\bibfnamefont {M.}~\bibnamefont {Sakuda}},\ }\href
  {\doibase 10.1103/PhysRevD.69.014013} {\bibfield  {journal} {\bibinfo
  {journal} {Phys.Rev.}\ }\textbf {\bibinfo {volume} {D69}},\ \bibinfo {pages}
  {014013} (\bibinfo {year} {2004})},\ \Eprint
  {http://arxiv.org/abs/hep-ph/0308130} {arXiv:hep-ph/0308130 [hep-ph]}
  \BibitemShut {NoStop}%
\bibitem [{\citenamefont {Lalakulich}\ and\ \citenamefont
  {Paschos}(2005)}]{Lalakulich:2005cs}%
  \BibitemOpen
  \bibfield  {author} {\bibinfo {author} {\bibfnamefont {O.}~\bibnamefont
  {Lalakulich}}\ and\ \bibinfo {author} {\bibfnamefont {E.~A.}\ \bibnamefont
  {Paschos}},\ }\href {\doibase 10.1103/PhysRevD.71.074003} {\bibfield
  {journal} {\bibinfo  {journal} {Phys.Rev.}\ }\textbf {\bibinfo {volume}
  {D71}},\ \bibinfo {pages} {074003} (\bibinfo {year} {2005})},\ \Eprint
  {http://arxiv.org/abs/hep-ph/0501109} {arXiv:hep-ph/0501109 [hep-ph]}
  \BibitemShut {NoStop}%
\bibitem [{\citenamefont {Lalakulich}\ \emph {et~al.}(2006)\citenamefont
  {Lalakulich}, \citenamefont {Paschos},\ and\ \citenamefont
  {Piranishvili}}]{Lalakulich:2006sw}%
  \BibitemOpen
  \bibfield  {author} {\bibinfo {author} {\bibfnamefont {O.}~\bibnamefont
  {Lalakulich}}, \bibinfo {author} {\bibfnamefont {E.~A.}\ \bibnamefont
  {Paschos}}, \ and\ \bibinfo {author} {\bibfnamefont {G.}~\bibnamefont
  {Piranishvili}},\ }\href {\doibase 10.1103/PhysRevD.74.014009} {\bibfield
  {journal} {\bibinfo  {journal} {Phys.Rev.}\ }\textbf {\bibinfo {volume}
  {D74}},\ \bibinfo {pages} {014009} (\bibinfo {year} {2006})},\ \Eprint
  {http://arxiv.org/abs/hep-ph/0602210} {arXiv:hep-ph/0602210 [hep-ph]}
  \BibitemShut {NoStop}%
\bibitem [{\citenamefont {Hernandez}\ \emph {et~al.}(2007)\citenamefont
  {Hernandez}, \citenamefont {Nieves},\ and\ \citenamefont
  {Valverde}}]{Hernandez:2007qq}%
  \BibitemOpen
  \bibfield  {author} {\bibinfo {author} {\bibfnamefont {E.}~\bibnamefont
  {Hernandez}}, \bibinfo {author} {\bibfnamefont {J.}~\bibnamefont {Nieves}}, \
  and\ \bibinfo {author} {\bibfnamefont {M.}~\bibnamefont {Valverde}},\ }\href
  {\doibase 10.1103/PhysRevD.76.033005} {\bibfield  {journal} {\bibinfo
  {journal} {Phys.Rev.}\ }\textbf {\bibinfo {volume} {D76}},\ \bibinfo {pages}
  {033005} (\bibinfo {year} {2007})},\ \Eprint
  {http://arxiv.org/abs/hep-ph/0701149} {arXiv:hep-ph/0701149 [hep-ph]}
  \BibitemShut {NoStop}%
\bibitem [{\citenamefont {Hernandez}\ \emph
  {et~al.}(2010{\natexlab{a}})\citenamefont {Hernandez}, \citenamefont
  {Nieves}, \citenamefont {Valverde},\ and\ \citenamefont
  {Vicente~Vacas}}]{Hernandez:2010bx}%
  \BibitemOpen
  \bibfield  {author} {\bibinfo {author} {\bibfnamefont {E.}~\bibnamefont
  {Hernandez}}, \bibinfo {author} {\bibfnamefont {J.}~\bibnamefont {Nieves}},
  \bibinfo {author} {\bibfnamefont {M.}~\bibnamefont {Valverde}}, \ and\
  \bibinfo {author} {\bibfnamefont {M.}~\bibnamefont {Vicente~Vacas}},\ }\href
  {\doibase 10.1103/PhysRevD.81.085046} {\bibfield  {journal} {\bibinfo
  {journal} {Phys.Rev.}\ }\textbf {\bibinfo {volume} {D81}},\ \bibinfo {pages}
  {085046} (\bibinfo {year} {2010}{\natexlab{a}})},\ \Eprint
  {http://arxiv.org/abs/1001.4416} {arXiv:1001.4416 [hep-ph]} \BibitemShut
  {NoStop}%
\bibitem [{\citenamefont {Hernández}\ \emph {et~al.}(2013)\citenamefont
  {Hernández}, \citenamefont {Nieves},\ and\ \citenamefont
  {Vacas}}]{Hernandez:2013jka}%
  \BibitemOpen
  \bibfield  {author} {\bibinfo {author} {\bibfnamefont {E.}~\bibnamefont
  {Hernández}}, \bibinfo {author} {\bibfnamefont {J.}~\bibnamefont {Nieves}},
  \ and\ \bibinfo {author} {\bibfnamefont {M.~J.~V.}\ \bibnamefont {Vacas}},\
  }\href {\doibase 10.1103/PhysRevD.87.113009} {\bibfield  {journal} {\bibinfo
  {journal} {Phys.Rev.}\ }\textbf {\bibinfo {volume} {D87}},\ \bibinfo {pages}
  {113009} (\bibinfo {year} {2013})},\ \Eprint {http://arxiv.org/abs/1304.1320}
  {arXiv:1304.1320 [hep-ph]} \BibitemShut {NoStop}%
\bibitem [{\citenamefont {Graczyk}\ and\ \citenamefont
  {Sobczyk}(2008{\natexlab{a}})}]{Graczyk:2007bc}%
  \BibitemOpen
  \bibfield  {author} {\bibinfo {author} {\bibfnamefont {K.~M.}\ \bibnamefont
  {Graczyk}}\ and\ \bibinfo {author} {\bibfnamefont {J.~T.}\ \bibnamefont
  {Sobczyk}},\ }\href {\doibase 10.1103/PhysRevD.79.079903,
  10.1103/PhysRevD.77.053001} {\bibfield  {journal} {\bibinfo  {journal}
  {Phys.Rev.}\ }\textbf {\bibinfo {volume} {D77}},\ \bibinfo {pages} {053001}
  (\bibinfo {year} {2008}{\natexlab{a}})},\ \Eprint
  {http://arxiv.org/abs/0707.3561} {arXiv:0707.3561 [hep-ph]} \BibitemShut
  {NoStop}%
\bibitem [{\citenamefont {Graczyk}\ \emph {et~al.}(2009)\citenamefont
  {Graczyk}, \citenamefont {Kielczewska}, \citenamefont {Przewlocki},\ and\
  \citenamefont {Sobczyk}}]{Graczyk:2009qm}%
  \BibitemOpen
  \bibfield  {author} {\bibinfo {author} {\bibfnamefont {K.}~\bibnamefont
  {Graczyk}}, \bibinfo {author} {\bibfnamefont {D.}~\bibnamefont
  {Kielczewska}}, \bibinfo {author} {\bibfnamefont {P.}~\bibnamefont
  {Przewlocki}}, \ and\ \bibinfo {author} {\bibfnamefont {J.}~\bibnamefont
  {Sobczyk}},\ }\href {\doibase 10.1103/PhysRevD.80.093001} {\bibfield
  {journal} {\bibinfo  {journal} {Phys.Rev.}\ }\textbf {\bibinfo {volume}
  {D80}},\ \bibinfo {pages} {093001} (\bibinfo {year} {2009})},\ \Eprint
  {http://arxiv.org/abs/0908.2175} {arXiv:0908.2175 [hep-ph]} \BibitemShut
  {NoStop}%
\bibitem [{\citenamefont {Barbero}\ \emph {et~al.}(2008)\citenamefont
  {Barbero}, \citenamefont {Lopez~Castro},\ and\ \citenamefont
  {Mariano}}]{Barbero:2008zza}%
  \BibitemOpen
  \bibfield  {author} {\bibinfo {author} {\bibfnamefont {C.}~\bibnamefont
  {Barbero}}, \bibinfo {author} {\bibfnamefont {G.}~\bibnamefont
  {Lopez~Castro}}, \ and\ \bibinfo {author} {\bibfnamefont {A.}~\bibnamefont
  {Mariano}},\ }\href {\doibase 10.1016/j.physletb.2008.05.011} {\bibfield
  {journal} {\bibinfo  {journal} {Phys.Lett.}\ }\textbf {\bibinfo {volume}
  {B664}},\ \bibinfo {pages} {70} (\bibinfo {year} {2008})}\BibitemShut
  {NoStop}%
\bibitem [{\citenamefont {Barbero}\ \emph {et~al.}(2014)\citenamefont
  {Barbero}, \citenamefont {López~Castro},\ and\ \citenamefont
  {Mariano}}]{Barbero:2013eqa}%
  \BibitemOpen
  \bibfield  {author} {\bibinfo {author} {\bibfnamefont {C.}~\bibnamefont
  {Barbero}}, \bibinfo {author} {\bibfnamefont {G.}~\bibnamefont
  {López~Castro}}, \ and\ \bibinfo {author} {\bibfnamefont {A.}~\bibnamefont
  {Mariano}},\ }\href {\doibase 10.1016/j.physletb.2013.12.006} {\bibfield
  {journal} {\bibinfo  {journal} {Phys.Lett.}\ }\textbf {\bibinfo {volume}
  {B728}},\ \bibinfo {pages} {282} (\bibinfo {year} {2014})},\ \Eprint
  {http://arxiv.org/abs/1311.3542} {arXiv:1311.3542 [nucl-th]} \BibitemShut
  {NoStop}%
\bibitem [{\citenamefont {Serot}\ and\ \citenamefont
  {Zhang}(2012)}]{Serot:2012rd}%
  \BibitemOpen
  \bibfield  {author} {\bibinfo {author} {\bibfnamefont {B.~D.}\ \bibnamefont
  {Serot}}\ and\ \bibinfo {author} {\bibfnamefont {X.}~\bibnamefont {Zhang}},\
  }\href {\doibase 10.1103/PhysRevC.86.015501} {\bibfield  {journal} {\bibinfo
  {journal} {Phys.Rev.}\ }\textbf {\bibinfo {volume} {C86}},\ \bibinfo {pages}
  {015501} (\bibinfo {year} {2012})},\ \Eprint {http://arxiv.org/abs/1206.3812}
  {arXiv:1206.3812 [nucl-th]} \BibitemShut {NoStop}%
\bibitem [{\citenamefont {MAID}()}]{MAID}%
  \BibitemOpen
  \bibfield  {author} {\bibinfo {author} {\bibnamefont {MAID}},\ }\href@noop {}
  {}\bibinfo {howpublished}
  {\url{http://www.kph.uni-mainz.de/MAID}}\BibitemShut {NoStop}%
\bibitem [{\citenamefont {Tiator}\ \emph {et~al.}(2011)\citenamefont {Tiator},
  \citenamefont {Drechsel}, \citenamefont {Kamalov},\ and\ \citenamefont
  {Vanderhaeghen}}]{Tiator:2011pw}%
  \BibitemOpen
  \bibfield  {author} {\bibinfo {author} {\bibfnamefont {L.}~\bibnamefont
  {Tiator}}, \bibinfo {author} {\bibfnamefont {D.}~\bibnamefont {Drechsel}},
  \bibinfo {author} {\bibfnamefont {S.}~\bibnamefont {Kamalov}}, \ and\
  \bibinfo {author} {\bibfnamefont {M.}~\bibnamefont {Vanderhaeghen}},\ }\href
  {\doibase 10.1140/epjst/e2011-01488-9} {\bibfield  {journal} {\bibinfo
  {journal} {Eur.Phys.J.ST}\ }\textbf {\bibinfo {volume} {198}},\ \bibinfo
  {pages} {141} (\bibinfo {year} {2011})},\ \Eprint
  {http://arxiv.org/abs/1109.6745} {arXiv:1109.6745 [nucl-th]} \BibitemShut
  {NoStop}%
\bibitem [{\citenamefont {Feynman}\ \emph {et~al.}(1971)\citenamefont
  {Feynman}, \citenamefont {Kislinger},\ and\ \citenamefont
  {Ravndal}}]{Feynman:1971wr}%
  \BibitemOpen
  \bibfield  {author} {\bibinfo {author} {\bibfnamefont {R.}~\bibnamefont
  {Feynman}}, \bibinfo {author} {\bibfnamefont {M.}~\bibnamefont {Kislinger}},
  \ and\ \bibinfo {author} {\bibfnamefont {F.}~\bibnamefont {Ravndal}},\ }\href
  {\doibase 10.1103/PhysRevD.3.2706} {\bibfield  {journal} {\bibinfo  {journal}
  {Phys.Rev.}\ }\textbf {\bibinfo {volume} {D3}},\ \bibinfo {pages} {2706}
  (\bibinfo {year} {1971})}\BibitemShut {NoStop}%
\bibitem [{\citenamefont {Berger}\ and\ \citenamefont
  {Sehgal}(2007)}]{Berger:2007rq}%
  \BibitemOpen
  \bibfield  {author} {\bibinfo {author} {\bibfnamefont {C.}~\bibnamefont
  {Berger}}\ and\ \bibinfo {author} {\bibfnamefont {L.}~\bibnamefont
  {Sehgal}},\ }\href {\doibase 10.1103/PhysRevD.76.113004} {\bibfield
  {journal} {\bibinfo  {journal} {Phys.Rev.}\ }\textbf {\bibinfo {volume}
  {D76}},\ \bibinfo {pages} {113004} (\bibinfo {year} {2007})},\ \Eprint
  {http://arxiv.org/abs/0709.4378} {arXiv:0709.4378 [hep-ph]} \BibitemShut
  {NoStop}%
\bibitem [{\citenamefont {Graczyk}\ and\ \citenamefont
  {Sobczyk}(2008{\natexlab{b}})}]{Graczyk:2007xk}%
  \BibitemOpen
  \bibfield  {author} {\bibinfo {author} {\bibfnamefont {K.~M.}\ \bibnamefont
  {Graczyk}}\ and\ \bibinfo {author} {\bibfnamefont {J.~T.}\ \bibnamefont
  {Sobczyk}},\ }\href {\doibase 10.1103/PhysRevD.77.053003} {\bibfield
  {journal} {\bibinfo  {journal} {Phys.Rev.}\ }\textbf {\bibinfo {volume}
  {D77}},\ \bibinfo {pages} {053003} (\bibinfo {year} {2008}{\natexlab{b}})},\
  \Eprint {http://arxiv.org/abs/0709.4634} {arXiv:0709.4634 [hep-ph]}
  \BibitemShut {NoStop}%
\bibitem [{\citenamefont {Leitner}\ \emph {et~al.}(2008)\citenamefont
  {Leitner}, \citenamefont {Buss}, \citenamefont {Mosel},\ and\ \citenamefont
  {Alvarez-Ruso}}]{Leitner:2008fg}%
  \BibitemOpen
  \bibfield  {author} {\bibinfo {author} {\bibfnamefont {T.}~\bibnamefont
  {Leitner}}, \bibinfo {author} {\bibfnamefont {O.}~\bibnamefont {Buss}},
  \bibinfo {author} {\bibfnamefont {U.}~\bibnamefont {Mosel}}, \ and\ \bibinfo
  {author} {\bibfnamefont {L.}~\bibnamefont {Alvarez-Ruso}},\ }\href@noop {}
  {\bibfield  {journal} {\bibinfo  {journal} {PoS}\ }\textbf {\bibinfo {volume}
  {NUFACT08}},\ \bibinfo {pages} {009} (\bibinfo {year} {2008})},\ \Eprint
  {http://arxiv.org/abs/0809.3986} {arXiv:0809.3986 [nucl-th]} \BibitemShut
  {NoStop}%
\bibitem [{\citenamefont {Paschos}\ and\ \citenamefont
  {Schalla}(2011)}]{Paschos:2011ye}%
  \BibitemOpen
  \bibfield  {author} {\bibinfo {author} {\bibfnamefont {E.}~\bibnamefont
  {Paschos}}\ and\ \bibinfo {author} {\bibfnamefont {D.}~\bibnamefont
  {Schalla}},\ }\href {\doibase 10.1103/PhysRevD.84.013004} {\bibfield
  {journal} {\bibinfo  {journal} {Phys.Rev.}\ }\textbf {\bibinfo {volume}
  {D84}},\ \bibinfo {pages} {013004} (\bibinfo {year} {2011})},\ \Eprint
  {http://arxiv.org/abs/1102.4466} {arXiv:1102.4466 [hep-ph]} \BibitemShut
  {NoStop}%
\bibitem [{\citenamefont {Kamano}\ \emph {et~al.}(2012)\citenamefont {Kamano},
  \citenamefont {Nakamura}, \citenamefont {Lee},\ and\ \citenamefont
  {Sato}}]{Kamano:2012id}%
  \BibitemOpen
  \bibfield  {author} {\bibinfo {author} {\bibfnamefont {H.}~\bibnamefont
  {Kamano}}, \bibinfo {author} {\bibfnamefont {S.}~\bibnamefont {Nakamura}},
  \bibinfo {author} {\bibfnamefont {T.-S.}\ \bibnamefont {Lee}}, \ and\
  \bibinfo {author} {\bibfnamefont {T.}~\bibnamefont {Sato}},\ }\href {\doibase
  10.1103/PhysRevD.86.097503} {\bibfield  {journal} {\bibinfo  {journal}
  {Phys.Rev.}\ }\textbf {\bibinfo {volume} {D86}},\ \bibinfo {pages} {097503}
  (\bibinfo {year} {2012})},\ \Eprint {http://arxiv.org/abs/1207.5724}
  {arXiv:1207.5724 [nucl-th]} \BibitemShut {NoStop}%
\bibitem [{\citenamefont {Nakamura}\ \emph {et~al.}(2013)\citenamefont
  {Nakamura}, \citenamefont {Kamano}, \citenamefont {Lee},\ and\ \citenamefont
  {Sato}}]{Nakamura:2013zaa}%
  \BibitemOpen
  \bibfield  {author} {\bibinfo {author} {\bibfnamefont {S.}~\bibnamefont
  {Nakamura}}, \bibinfo {author} {\bibfnamefont {H.}~\bibnamefont {Kamano}},
  \bibinfo {author} {\bibfnamefont {T.~S.~H.}\ \bibnamefont {Lee}}, \ and\
  \bibinfo {author} {\bibfnamefont {T.}~\bibnamefont {Sato}},\ }\href@noop {}
  {\  (\bibinfo {year} {2013})},\ \Eprint {http://arxiv.org/abs/1303.4152}
  {arXiv:1303.4152 [hep-ph]} \BibitemShut {NoStop}%
\bibitem [{\citenamefont {JLab}()}]{JLab}%
  \BibitemOpen
  \bibfield  {author} {\bibinfo {author} {\bibnamefont {JLab}},\ }\href@noop {}
  {}\bibinfo {howpublished}
  {\url{https://hallcweb.jlab.org/resdata/database/jlabh2.txt}}\BibitemShut
  {NoStop}%
\bibitem [{\citenamefont {Adler}(1968)}]{Adler:1968tw}%
  \BibitemOpen
  \bibfield  {author} {\bibinfo {author} {\bibfnamefont {S.~L.}\ \bibnamefont
  {Adler}},\ }\href {\doibase 10.1016/0003-4916(68)90278-9} {\bibfield
  {journal} {\bibinfo  {journal} {Annals Phys.}\ }\textbf {\bibinfo {volume}
  {50}},\ \bibinfo {pages} {189} (\bibinfo {year} {1968})}\BibitemShut
  {NoStop}%
\bibitem [{\citenamefont {Bijtebier}(1970)}]{Bijtebier:1970ku}%
  \BibitemOpen
  \bibfield  {author} {\bibinfo {author} {\bibfnamefont {J.}~\bibnamefont
  {Bijtebier}},\ }\href@noop {} {\bibfield  {journal} {\bibinfo  {journal}
  {Nucl.Phys.}\ }\textbf {\bibinfo {volume} {B21}},\ \bibinfo {pages} {158}
  (\bibinfo {year} {1970})}\BibitemShut {NoStop}%
\bibitem [{\citenamefont {Lalakulich}\ \emph {et~al.}(2010)\citenamefont
  {Lalakulich}, \citenamefont {Leitner}, \citenamefont {Buss},\ and\
  \citenamefont {Mosel}}]{Lalakulich:2010ss}%
  \BibitemOpen
  \bibfield  {author} {\bibinfo {author} {\bibfnamefont {O.}~\bibnamefont
  {Lalakulich}}, \bibinfo {author} {\bibfnamefont {T.}~\bibnamefont {Leitner}},
  \bibinfo {author} {\bibfnamefont {O.}~\bibnamefont {Buss}}, \ and\ \bibinfo
  {author} {\bibfnamefont {U.}~\bibnamefont {Mosel}},\ }\href {\doibase
  10.1103/PhysRevD.82.093001} {\bibfield  {journal} {\bibinfo  {journal}
  {Phys.Rev.}\ }\textbf {\bibinfo {volume} {D82}},\ \bibinfo {pages} {093001}
  (\bibinfo {year} {2010})},\ \Eprint {http://arxiv.org/abs/1007.0925}
  {arXiv:1007.0925 [hep-ph]} \BibitemShut {NoStop}%
\bibitem [{\citenamefont {Kim}\ \emph {et~al.}(1996)\citenamefont {Kim},
  \citenamefont {Schramm},\ and\ \citenamefont {Horowitz}}]{Kim:1996bt}%
  \BibitemOpen
  \bibfield  {author} {\bibinfo {author} {\bibfnamefont {H.-C.}\ \bibnamefont
  {Kim}}, \bibinfo {author} {\bibfnamefont {S.}~\bibnamefont {Schramm}}, \ and\
  \bibinfo {author} {\bibfnamefont {C.}~\bibnamefont {Horowitz}},\ }\href
  {\doibase 10.1103/PhysRevC.53.2468} {\bibfield  {journal} {\bibinfo
  {journal} {Phys.Rev.}\ }\textbf {\bibinfo {volume} {C53}},\ \bibinfo {pages}
  {2468} (\bibinfo {year} {1996})},\ \Eprint
  {http://arxiv.org/abs/nucl-th/9507006} {arXiv:nucl-th/9507006 [nucl-th]}
  \BibitemShut {NoStop}%
\bibitem [{\citenamefont {Singh}\ \emph
  {et~al.}(1998{\natexlab{b}})\citenamefont {Singh}, \citenamefont
  {Vicente-Vacas},\ and\ \citenamefont {Oset}}]{Singh:1998ha}%
  \BibitemOpen
  \bibfield  {author} {\bibinfo {author} {\bibfnamefont {S.}~\bibnamefont
  {Singh}}, \bibinfo {author} {\bibfnamefont {M.}~\bibnamefont
  {Vicente-Vacas}}, \ and\ \bibinfo {author} {\bibfnamefont {E.}~\bibnamefont
  {Oset}},\ }\href {\doibase 10.1016/S0370-2693(97)01325-7} {\bibfield
  {journal} {\bibinfo  {journal} {Phys.Lett.}\ }\textbf {\bibinfo {volume}
  {B416}},\ \bibinfo {pages} {23} (\bibinfo {year}
  {1998}{\natexlab{b}})}\BibitemShut {NoStop}%
\bibitem [{\citenamefont {Praet}\ \emph {et~al.}(2009)\citenamefont {Praet},
  \citenamefont {Lalakulich}, \citenamefont {Jachowicz},\ and\ \citenamefont
  {Ryckebusch}}]{Praet:2008yn}%
  \BibitemOpen
  \bibfield  {author} {\bibinfo {author} {\bibfnamefont {C.}~\bibnamefont
  {Praet}}, \bibinfo {author} {\bibfnamefont {O.}~\bibnamefont {Lalakulich}},
  \bibinfo {author} {\bibfnamefont {N.}~\bibnamefont {Jachowicz}}, \ and\
  \bibinfo {author} {\bibfnamefont {J.}~\bibnamefont {Ryckebusch}},\ }\href
  {\doibase 10.1103/PhysRevC.79.044603} {\bibfield  {journal} {\bibinfo
  {journal} {Phys.Rev.}\ }\textbf {\bibinfo {volume} {C79}},\ \bibinfo {pages}
  {044603} (\bibinfo {year} {2009})},\ \Eprint {http://arxiv.org/abs/0804.2750}
  {arXiv:0804.2750 [nucl-th]} \BibitemShut {NoStop}%
\bibitem [{\citenamefont {Casper}(2002)}]{Casper:2002sd}%
  \BibitemOpen
  \bibfield  {author} {\bibinfo {author} {\bibfnamefont {D.}~\bibnamefont
  {Casper}},\ }\href {\doibase 10.1016/S0920-5632(02)01756-5} {\bibfield
  {journal} {\bibinfo  {journal} {Nucl.Phys.Proc.Suppl.}\ }\textbf {\bibinfo
  {volume} {112}},\ \bibinfo {pages} {161} (\bibinfo {year} {2002})},\ \Eprint
  {http://arxiv.org/abs/hep-ph/0208030} {arXiv:hep-ph/0208030 [hep-ph]}
  \BibitemShut {NoStop}%
\bibitem [{\citenamefont {Lalakulich}\ and\ \citenamefont
  {Mosel}(2013{\natexlab{a}})}]{Lalakulich:2012cj}%
  \BibitemOpen
  \bibfield  {author} {\bibinfo {author} {\bibfnamefont {O.}~\bibnamefont
  {Lalakulich}}\ and\ \bibinfo {author} {\bibfnamefont {U.}~\bibnamefont
  {Mosel}},\ }\href {\doibase 10.1103/PhysRevC.87.014602} {\bibfield  {journal}
  {\bibinfo  {journal} {Phys.Rev.}\ }\textbf {\bibinfo {volume} {C87}},\
  \bibinfo {pages} {014602} (\bibinfo {year} {2013}{\natexlab{a}})},\ \Eprint
  {http://arxiv.org/abs/1210.4717} {arXiv:1210.4717 [nucl-th]} \BibitemShut
  {NoStop}%
\bibitem [{\citenamefont {Leitner}\ \emph {et~al.}(2010)\citenamefont
  {Leitner}, \citenamefont {Lalakulich}, \citenamefont {Buss}, \citenamefont
  {Mosel},\ and\ \citenamefont {Alvarez-Ruso}}]{Leitner:2009de}%
  \BibitemOpen
  \bibfield  {author} {\bibinfo {author} {\bibfnamefont {T.}~\bibnamefont
  {Leitner}}, \bibinfo {author} {\bibfnamefont {O.}~\bibnamefont {Lalakulich}},
  \bibinfo {author} {\bibfnamefont {O.}~\bibnamefont {Buss}}, \bibinfo {author}
  {\bibfnamefont {U.}~\bibnamefont {Mosel}}, \ and\ \bibinfo {author}
  {\bibfnamefont {L.}~\bibnamefont {Alvarez-Ruso}},\ }\href {\doibase
  10.1063/1.3399298} {\bibfield  {journal} {\bibinfo  {journal} {AIP
  Conf.Proc.}\ }\textbf {\bibinfo {volume} {1222}},\ \bibinfo {pages} {212}
  (\bibinfo {year} {2010})},\ \Eprint {http://arxiv.org/abs/0910.2835}
  {arXiv:0910.2835 [nucl-th]} \BibitemShut {NoStop}%
\bibitem [{\citenamefont {Leitner}\ \emph
  {et~al.}(2009{\natexlab{b}})\citenamefont {Leitner}, \citenamefont {Buss},
  \citenamefont {Mosel},\ and\ \citenamefont {Alvarez-Ruso}}]{Leitner:2009ec}%
  \BibitemOpen
  \bibfield  {author} {\bibinfo {author} {\bibfnamefont {T.}~\bibnamefont
  {Leitner}}, \bibinfo {author} {\bibfnamefont {O.}~\bibnamefont {Buss}},
  \bibinfo {author} {\bibfnamefont {U.}~\bibnamefont {Mosel}}, \ and\ \bibinfo
  {author} {\bibfnamefont {L.}~\bibnamefont {Alvarez-Ruso}},\ }\href {\doibase
  10.1063/1.3274157} {\bibfield  {journal} {\bibinfo  {journal} {AIP
  Conf.Proc.}\ }\textbf {\bibinfo {volume} {1189}},\ \bibinfo {pages} {207}
  (\bibinfo {year} {2009}{\natexlab{b}})},\ \Eprint
  {http://arxiv.org/abs/0909.0838} {arXiv:0909.0838 [nucl-th]} \BibitemShut
  {NoStop}%
\bibitem [{\citenamefont {Leitner}\ \emph
  {et~al.}(2009{\natexlab{c}})\citenamefont {Leitner}, \citenamefont {Buss},
  \citenamefont {Mosel},\ and\ \citenamefont {Alvarez-Ruso}}]{Leitner:2008wx}%
  \BibitemOpen
  \bibfield  {author} {\bibinfo {author} {\bibfnamefont {T.}~\bibnamefont
  {Leitner}}, \bibinfo {author} {\bibfnamefont {O.}~\bibnamefont {Buss}},
  \bibinfo {author} {\bibfnamefont {U.}~\bibnamefont {Mosel}}, \ and\ \bibinfo
  {author} {\bibfnamefont {L.}~\bibnamefont {Alvarez-Ruso}},\ }\href {\doibase
  10.1103/PhysRevC.79.038501} {\bibfield  {journal} {\bibinfo  {journal}
  {Phys.Rev.}\ }\textbf {\bibinfo {volume} {C79}},\ \bibinfo {pages} {038501}
  (\bibinfo {year} {2009}{\natexlab{c}})},\ \Eprint
  {http://arxiv.org/abs/0812.1787} {arXiv:0812.1787 [nucl-th]} \BibitemShut
  {NoStop}%
\bibitem [{\citenamefont {Amaro}\ \emph {et~al.}(2009)\citenamefont {Amaro},
  \citenamefont {Hernandez}, \citenamefont {Nieves},\ and\ \citenamefont
  {Valverde}}]{Amaro:2008hd}%
  \BibitemOpen
  \bibfield  {author} {\bibinfo {author} {\bibfnamefont {J.}~\bibnamefont
  {Amaro}}, \bibinfo {author} {\bibfnamefont {E.}~\bibnamefont {Hernandez}},
  \bibinfo {author} {\bibfnamefont {J.}~\bibnamefont {Nieves}}, \ and\ \bibinfo
  {author} {\bibfnamefont {M.}~\bibnamefont {Valverde}},\ }\href {\doibase
  10.1103/PhysRevD.79.013002} {\bibfield  {journal} {\bibinfo  {journal}
  {Phys.Rev.}\ }\textbf {\bibinfo {volume} {D79}},\ \bibinfo {pages} {013002}
  (\bibinfo {year} {2009})},\ \Eprint {http://arxiv.org/abs/0811.1421}
  {arXiv:0811.1421 [hep-ph]} \BibitemShut {NoStop}%
\bibitem [{\citenamefont {Hernandez}\ \emph
  {et~al.}(2010{\natexlab{b}})\citenamefont {Hernandez}, \citenamefont
  {Nieves},\ and\ \citenamefont {Valverde}}]{Hernandez:2010jf}%
  \BibitemOpen
  \bibfield  {author} {\bibinfo {author} {\bibfnamefont {E.}~\bibnamefont
  {Hernandez}}, \bibinfo {author} {\bibfnamefont {J.}~\bibnamefont {Nieves}}, \
  and\ \bibinfo {author} {\bibfnamefont {M.}~\bibnamefont {Valverde}},\ }\href
  {\doibase 10.1103/PhysRevD.82.077303} {\bibfield  {journal} {\bibinfo
  {journal} {Phys.Rev.}\ }\textbf {\bibinfo {volume} {D82}},\ \bibinfo {pages}
  {077303} (\bibinfo {year} {2010}{\natexlab{b}})},\ \Eprint
  {http://arxiv.org/abs/1007.3685} {arXiv:1007.3685 [hep-ph]} \BibitemShut
  {NoStop}%
\bibitem [{\citenamefont {Oset}\ and\ \citenamefont
  {Salcedo}(1987)}]{Oset:1987re}%
  \BibitemOpen
  \bibfield  {author} {\bibinfo {author} {\bibfnamefont {E.}~\bibnamefont
  {Oset}}\ and\ \bibinfo {author} {\bibfnamefont {L.}~\bibnamefont {Salcedo}},\
  }\href {\doibase 10.1016/0375-9474(87)90185-0} {\bibfield  {journal}
  {\bibinfo  {journal} {Nucl.Phys.}\ }\textbf {\bibinfo {volume} {A468}},\
  \bibinfo {pages} {631} (\bibinfo {year} {1987})}\BibitemShut {NoStop}%
\bibitem [{\citenamefont {Salcedo}\ \emph {et~al.}(1988)\citenamefont
  {Salcedo}, \citenamefont {Oset}, \citenamefont {Vicente-Vacas},\ and\
  \citenamefont {Garcia-Recio}}]{Salcedo:1987md}%
  \BibitemOpen
  \bibfield  {author} {\bibinfo {author} {\bibfnamefont {L.}~\bibnamefont
  {Salcedo}}, \bibinfo {author} {\bibfnamefont {E.}~\bibnamefont {Oset}},
  \bibinfo {author} {\bibfnamefont {M.}~\bibnamefont {Vicente-Vacas}}, \ and\
  \bibinfo {author} {\bibfnamefont {C.}~\bibnamefont {Garcia-Recio}},\ }\href
  {\doibase 10.1016/0375-9474(88)90310-7} {\bibfield  {journal} {\bibinfo
  {journal} {Nucl.Phys.}\ }\textbf {\bibinfo {volume} {A484}},\ \bibinfo
  {pages} {557} (\bibinfo {year} {1988})}\BibitemShut {NoStop}%
\bibitem [{\citenamefont {Gil}\ \emph {et~al.}(1997{\natexlab{b}})\citenamefont
  {Gil}, \citenamefont {Nieves},\ and\ \citenamefont {Oset}}]{Gil:1997jg}%
  \BibitemOpen
  \bibfield  {author} {\bibinfo {author} {\bibfnamefont {A.}~\bibnamefont
  {Gil}}, \bibinfo {author} {\bibfnamefont {J.}~\bibnamefont {Nieves}}, \ and\
  \bibinfo {author} {\bibfnamefont {E.}~\bibnamefont {Oset}},\ }\href {\doibase
  10.1016/S0375-9474(97)00515-0} {\bibfield  {journal} {\bibinfo  {journal}
  {Nucl.Phys.}\ }\textbf {\bibinfo {volume} {A627}},\ \bibinfo {pages} {599}
  (\bibinfo {year} {1997}{\natexlab{b}})},\ \Eprint
  {http://arxiv.org/abs/nucl-th/9710070} {arXiv:nucl-th/9710070 [nucl-th]}
  \BibitemShut {NoStop}%
\bibitem [{\citenamefont {Juszczak}\ \emph {et~al.}(2006)\citenamefont
  {Juszczak}, \citenamefont {Nowak},\ and\ \citenamefont
  {Sobczyk}}]{Juszczak:2005zs}%
  \BibitemOpen
  \bibfield  {author} {\bibinfo {author} {\bibfnamefont {C.}~\bibnamefont
  {Juszczak}}, \bibinfo {author} {\bibfnamefont {J.~A.}\ \bibnamefont {Nowak}},
  \ and\ \bibinfo {author} {\bibfnamefont {J.~T.}\ \bibnamefont {Sobczyk}},\
  }\href {\doibase 10.1016/j.nuclphysbps.2006.08.069} {\bibfield  {journal}
  {\bibinfo  {journal} {Nucl.Phys.Proc.Suppl.}\ }\textbf {\bibinfo {volume}
  {159}},\ \bibinfo {pages} {211} (\bibinfo {year} {2006})},\ \Eprint
  {http://arxiv.org/abs/hep-ph/0512365} {arXiv:hep-ph/0512365 [hep-ph]}
  \BibitemShut {NoStop}%
\bibitem [{\citenamefont {Juszczak}(2009)}]{Juszczak:2009qa}%
  \BibitemOpen
  \bibfield  {author} {\bibinfo {author} {\bibfnamefont {C.}~\bibnamefont
  {Juszczak}},\ }\href@noop {} {\bibfield  {journal} {\bibinfo  {journal} {Acta
  Phys.Polon.}\ }\textbf {\bibinfo {volume} {B40}},\ \bibinfo {pages} {2507}
  (\bibinfo {year} {2009})},\ \Eprint {http://arxiv.org/abs/0909.1492}
  {arXiv:0909.1492 [hep-ex]} \BibitemShut {NoStop}%
\bibitem [{\citenamefont {Golan}\ \emph {et~al.}(2012)\citenamefont {Golan},
  \citenamefont {Juszczak},\ and\ \citenamefont {Sobczyk}}]{Golan:2012wx}%
  \BibitemOpen
  \bibfield  {author} {\bibinfo {author} {\bibfnamefont {T.}~\bibnamefont
  {Golan}}, \bibinfo {author} {\bibfnamefont {C.}~\bibnamefont {Juszczak}}, \
  and\ \bibinfo {author} {\bibfnamefont {J.~T.}\ \bibnamefont {Sobczyk}},\
  }\href {\doibase 10.1103/PhysRevC.86.015505} {\bibfield  {journal} {\bibinfo
  {journal} {Phys.Rev.}\ }\textbf {\bibinfo {volume} {C86}},\ \bibinfo {pages}
  {015505} (\bibinfo {year} {2012})},\ \Eprint {http://arxiv.org/abs/1202.4197}
  {arXiv:1202.4197 [nucl-th]} \BibitemShut {NoStop}%
\bibitem [{\citenamefont {Rein}\ and\ \citenamefont
  {Sehgal}(1983)}]{Rein:1982pf}%
  \BibitemOpen
  \bibfield  {author} {\bibinfo {author} {\bibfnamefont {D.}~\bibnamefont
  {Rein}}\ and\ \bibinfo {author} {\bibfnamefont {L.~M.}\ \bibnamefont
  {Sehgal}},\ }\href {\doibase 10.1016/0550-3213(83)90090-1} {\bibfield
  {journal} {\bibinfo  {journal} {Nucl.Phys.}\ }\textbf {\bibinfo {volume}
  {B223}},\ \bibinfo {pages} {29} (\bibinfo {year} {1983})}\BibitemShut
  {NoStop}%
\bibitem [{\citenamefont {Pinzon}(2013)}]{piano1}%
  \BibitemOpen
  \bibfield  {author} {\bibinfo {author} {\bibfnamefont {E.}~\bibnamefont
  {Pinzon}},\ }\href@noop {} {\bibfield  {journal} {\bibinfo  {journal} {{\it
  talk at NuFacT 13, Beijing, China }}\ } (\bibinfo {year} {2013})}\BibitemShut
  {NoStop}%
\bibitem [{\citenamefont {Aguilar-Arevalo}\ \emph
  {et~al.}(2009{\natexlab{c}})\citenamefont {Aguilar-Arevalo} \emph
  {et~al.}}]{AguilarArevalo:2009eb}%
  \BibitemOpen
  \bibfield  {author} {\bibinfo {author} {\bibfnamefont {A.}~\bibnamefont
  {Aguilar-Arevalo}} \emph {et~al.} (\bibinfo {collaboration} {MiniBooNE
  Collaboration}),\ }\href {\doibase 10.1103/PhysRevLett.103.081801} {\bibfield
   {journal} {\bibinfo  {journal} {Phys.Rev.Lett.}\ }\textbf {\bibinfo {volume}
  {103}},\ \bibinfo {pages} {081801} (\bibinfo {year} {2009}{\natexlab{c}})},\
  \Eprint {http://arxiv.org/abs/0904.3159} {arXiv:0904.3159 [hep-ex]}
  \BibitemShut {NoStop}%
\bibitem [{\citenamefont {Rodriguez}\ \emph {et~al.}(2008)\citenamefont
  {Rodriguez} \emph {et~al.}}]{Rodriguez:2008aa}%
  \BibitemOpen
  \bibfield  {author} {\bibinfo {author} {\bibfnamefont {A.}~\bibnamefont
  {Rodriguez}} \emph {et~al.} (\bibinfo {collaboration} {K2K Collaboration}),\
  }\href {\doibase 10.1103/PhysRevD.78.032003} {\bibfield  {journal} {\bibinfo
  {journal} {Phys.Rev.}\ }\textbf {\bibinfo {volume} {D78}},\ \bibinfo {pages}
  {032003} (\bibinfo {year} {2008})},\ \Eprint {http://arxiv.org/abs/0805.0186}
  {arXiv:0805.0186 [hep-ex]} \BibitemShut {NoStop}%
\bibitem [{\citenamefont {Sajjad~Athar}\ \emph
  {et~al.}(2010{\natexlab{b}})\citenamefont {Sajjad~Athar}, \citenamefont
  {Chauhan},\ and\ \citenamefont {Singh}}]{SajjadAthar:2009rc}%
  \BibitemOpen
  \bibfield  {author} {\bibinfo {author} {\bibfnamefont {M.}~\bibnamefont
  {Sajjad~Athar}}, \bibinfo {author} {\bibfnamefont {S.}~\bibnamefont
  {Chauhan}}, \ and\ \bibinfo {author} {\bibfnamefont {S.}~\bibnamefont
  {Singh}},\ }\href {\doibase 10.1088/0954-3899/37/1/015005} {\bibfield
  {journal} {\bibinfo  {journal} {J.Phys.}\ }\textbf {\bibinfo {volume}
  {G37}},\ \bibinfo {pages} {015005} (\bibinfo {year} {2010}{\natexlab{b}})},\
  \Eprint {http://arxiv.org/abs/0908.1442} {arXiv:0908.1442 [nucl-th]}
  \BibitemShut {NoStop}%
\bibitem [{\citenamefont {Lalakulich}\ and\ \citenamefont
  {Mosel}(2013{\natexlab{b}})}]{Lalakulich:2013iaa}%
  \BibitemOpen
  \bibfield  {author} {\bibinfo {author} {\bibfnamefont {O.}~\bibnamefont
  {Lalakulich}}\ and\ \bibinfo {author} {\bibfnamefont {U.}~\bibnamefont
  {Mosel}},\ }\href {\doibase 10.1103/PhysRevC.88.017601} {\bibfield  {journal}
  {\bibinfo  {journal} {Phys.Rev.}\ }\textbf {\bibinfo {volume} {C88}},\
  \bibinfo {pages} {017601} (\bibinfo {year} {2013}{\natexlab{b}})},\ \Eprint
  {http://arxiv.org/abs/1305.3861} {arXiv:1305.3861 [nucl-th]} \BibitemShut
  {NoStop}%
\bibitem [{\citenamefont {Mosel}\ \emph
  {et~al.}(2014{\natexlab{b}})\citenamefont {Mosel}, \citenamefont
  {Lalakulich},\ and\ \citenamefont {Gallmeister}}]{Mosel:2014lja}%
  \BibitemOpen
  \bibfield  {author} {\bibinfo {author} {\bibfnamefont {U.}~\bibnamefont
  {Mosel}}, \bibinfo {author} {\bibfnamefont {O.}~\bibnamefont {Lalakulich}}, \
  and\ \bibinfo {author} {\bibfnamefont {K.}~\bibnamefont {Gallmeister}},\
  }\href@noop {} {\  (\bibinfo {year} {2014}{\natexlab{b}})},\ \Eprint
  {http://arxiv.org/abs/1402.0297} {arXiv:1402.0297 [nucl-th]} \BibitemShut
  {NoStop}%
\bibitem [{\citenamefont {Dewan}(1981)}]{Dewan:1981ab}%
  \BibitemOpen
  \bibfield  {author} {\bibinfo {author} {\bibfnamefont {H.}~\bibnamefont
  {Dewan}},\ }\href {\doibase 10.1103/PhysRevD.24.2369} {\bibfield  {journal}
  {\bibinfo  {journal} {Phys.Rev.}\ }\textbf {\bibinfo {volume} {D24}},\
  \bibinfo {pages} {2369} (\bibinfo {year} {1981})}\BibitemShut {NoStop}%
\bibitem [{\citenamefont {Rafi~Alam}\ \emph {et~al.}(2010)\citenamefont
  {Rafi~Alam}, \citenamefont {Ruiz~Simo}, \citenamefont {Sajjad~Athar},\ and\
  \citenamefont {Vicente~Vacas}}]{RafiAlam:2010kf}%
  \BibitemOpen
  \bibfield  {author} {\bibinfo {author} {\bibfnamefont {M.}~\bibnamefont
  {Rafi~Alam}}, \bibinfo {author} {\bibfnamefont {I.}~\bibnamefont
  {Ruiz~Simo}}, \bibinfo {author} {\bibfnamefont {M.}~\bibnamefont
  {Sajjad~Athar}}, \ and\ \bibinfo {author} {\bibfnamefont {M.}~\bibnamefont
  {Vicente~Vacas}},\ }\href {\doibase 10.1103/PhysRevD.82.033001} {\bibfield
  {journal} {\bibinfo  {journal} {Phys.Rev.}\ }\textbf {\bibinfo {volume}
  {D82}},\ \bibinfo {pages} {033001} (\bibinfo {year} {2010})},\ \Eprint
  {http://arxiv.org/abs/1004.5484} {arXiv:1004.5484 [hep-ph]} \BibitemShut
  {NoStop}%
\bibitem [{\citenamefont {Alam}\ \emph
  {et~al.}(2012{\natexlab{a}})\citenamefont {Alam}, \citenamefont {Simo},
  \citenamefont {Athar},\ and\ \citenamefont {Vicente~Vacas}}]{Alam:2011xq}%
  \BibitemOpen
  \bibfield  {author} {\bibinfo {author} {\bibfnamefont {M.~R.}\ \bibnamefont
  {Alam}}, \bibinfo {author} {\bibfnamefont {I.~R.}\ \bibnamefont {Simo}},
  \bibinfo {author} {\bibfnamefont {M.~S.}\ \bibnamefont {Athar}}, \ and\
  \bibinfo {author} {\bibfnamefont {M.}~\bibnamefont {Vicente~Vacas}},\ }\href
  {\doibase 10.1103/PhysRevD.85.013014} {\bibfield  {journal} {\bibinfo
  {journal} {Phys.Rev.}\ }\textbf {\bibinfo {volume} {D85}},\ \bibinfo {pages}
  {013014} (\bibinfo {year} {2012}{\natexlab{a}})},\ \Eprint
  {http://arxiv.org/abs/1111.0863} {arXiv:1111.0863 [hep-ph]} \BibitemShut
  {NoStop}%
\bibitem [{\citenamefont {Shrock}(1975)}]{Shrock:1975an}%
  \BibitemOpen
  \bibfield  {author} {\bibinfo {author} {\bibfnamefont {R.~E.}\ \bibnamefont
  {Shrock}},\ }\href {\doibase 10.1103/PhysRevD.12.2049} {\bibfield  {journal}
  {\bibinfo  {journal} {Phys.Rev.}\ }\textbf {\bibinfo {volume} {D12}},\
  \bibinfo {pages} {2049} (\bibinfo {year} {1975})}\BibitemShut {NoStop}%
\bibitem [{\citenamefont {Adera}\ \emph {et~al.}(2010)\citenamefont {Adera},
  \citenamefont {Van Der~Ventel}, \citenamefont {van Niekerk},\ and\
  \citenamefont {Mart}}]{Adera:2010zz}%
  \BibitemOpen
  \bibfield  {author} {\bibinfo {author} {\bibfnamefont {G.}~\bibnamefont
  {Adera}}, \bibinfo {author} {\bibfnamefont {B.}~\bibnamefont {Van
  Der~Ventel}}, \bibinfo {author} {\bibfnamefont {D.}~\bibnamefont {van
  Niekerk}}, \ and\ \bibinfo {author} {\bibfnamefont {T.}~\bibnamefont
  {Mart}},\ }\href {\doibase 10.1103/PhysRevC.82.025501} {\bibfield  {journal}
  {\bibinfo  {journal} {Phys.Rev.}\ }\textbf {\bibinfo {volume} {C82}},\
  \bibinfo {pages} {025501} (\bibinfo {year} {2010})},\ \Eprint
  {http://arxiv.org/abs/1112.5748} {arXiv:1112.5748 [nucl-th]} \BibitemShut
  {NoStop}%
\bibitem [{\citenamefont {Rafi~Alam}\ \emph {et~al.}(2013)\citenamefont
  {Rafi~Alam}, \citenamefont {Ruiz~Simo}, \citenamefont {Athar}, \citenamefont
  {Alvarez-Ruso},\ and\ \citenamefont {Vicente~Vacas}}]{Alam:2013woa}%
  \BibitemOpen
  \bibfield  {author} {\bibinfo {author} {\bibfnamefont {M.}~\bibnamefont
  {Rafi~Alam}}, \bibinfo {author} {\bibfnamefont {I.}~\bibnamefont
  {Ruiz~Simo}}, \bibinfo {author} {\bibfnamefont {M.~S.}\ \bibnamefont
  {Athar}}, \bibinfo {author} {\bibfnamefont {L.}~\bibnamefont {Alvarez-Ruso}},
  \ and\ \bibinfo {author} {\bibfnamefont {M.~J.}\ \bibnamefont
  {Vicente~Vacas}},\ }\href@noop {} {\  (\bibinfo {year} {2013})},\ \Eprint
  {http://arxiv.org/abs/1303.5924} {arXiv:1303.5924 [hep-ph]} \BibitemShut
  {NoStop}%
\bibitem [{\citenamefont {Alam}\ \emph
  {et~al.}(2013{\natexlab{b}})\citenamefont {Alam}, \citenamefont {Athar},
  \citenamefont {Alvarez-Ruso}, \citenamefont {Simo}, \citenamefont {Vacas}
  \emph {et~al.}}]{Alam:2013vwa}%
  \BibitemOpen
  \bibfield  {author} {\bibinfo {author} {\bibfnamefont {M.~R.}\ \bibnamefont
  {Alam}}, \bibinfo {author} {\bibfnamefont {M.~S.}\ \bibnamefont {Athar}},
  \bibinfo {author} {\bibfnamefont {L.}~\bibnamefont {Alvarez-Ruso}}, \bibinfo
  {author} {\bibfnamefont {I.~R.}\ \bibnamefont {Simo}}, \bibinfo {author}
  {\bibfnamefont {M.~J.~V.}\ \bibnamefont {Vacas}},  \emph {et~al.},\
  }\href@noop {} {\  (\bibinfo {year} {2013}{\natexlab{b}})},\ \Eprint
  {http://arxiv.org/abs/1311.2293} {arXiv:1311.2293 [hep-ph]} \BibitemShut
  {NoStop}%
\bibitem [{\citenamefont {Lalakulich}\ \emph
  {et~al.}(2012{\natexlab{b}})\citenamefont {Lalakulich}, \citenamefont
  {Gallmeister},\ and\ \citenamefont {Mosel}}]{Lalakulich:2012gm}%
  \BibitemOpen
  \bibfield  {author} {\bibinfo {author} {\bibfnamefont {O.}~\bibnamefont
  {Lalakulich}}, \bibinfo {author} {\bibfnamefont {K.}~\bibnamefont
  {Gallmeister}}, \ and\ \bibinfo {author} {\bibfnamefont {U.}~\bibnamefont
  {Mosel}},\ }\href {\doibase 10.1103/PhysRevC.86.014607} {\bibfield  {journal}
  {\bibinfo  {journal} {Phys.Rev.}\ }\textbf {\bibinfo {volume} {C86}},\
  \bibinfo {pages} {014607} (\bibinfo {year} {2012}{\natexlab{b}})},\ \Eprint
  {http://arxiv.org/abs/1205.1061} {arXiv:1205.1061 [nucl-th]} \BibitemShut
  {NoStop}%
\bibitem [{\citenamefont {Ankowski}\ \emph {et~al.}(2012)\citenamefont
  {Ankowski}, \citenamefont {Benhar}, \citenamefont {Mori}, \citenamefont
  {Yamaguchi},\ and\ \citenamefont {Sakuda}}]{Ankowski:2011ei}%
  \BibitemOpen
  \bibfield  {author} {\bibinfo {author} {\bibfnamefont {A.~M.}\ \bibnamefont
  {Ankowski}}, \bibinfo {author} {\bibfnamefont {O.}~\bibnamefont {Benhar}},
  \bibinfo {author} {\bibfnamefont {T.}~\bibnamefont {Mori}}, \bibinfo {author}
  {\bibfnamefont {R.}~\bibnamefont {Yamaguchi}}, \ and\ \bibinfo {author}
  {\bibfnamefont {M.}~\bibnamefont {Sakuda}},\ }\href {\doibase
  10.1103/PhysRevLett.108.052505} {\bibfield  {journal} {\bibinfo  {journal}
  {Phys.Rev.Lett.}\ }\textbf {\bibinfo {volume} {108}},\ \bibinfo {pages}
  {052505} (\bibinfo {year} {2012})},\ \Eprint {http://arxiv.org/abs/1110.0679}
  {arXiv:1110.0679 [nucl-th]} \BibitemShut {NoStop}%
\bibitem [{\citenamefont {Aguilar-Arevalo}\ \emph {et~al.}(2007)\citenamefont
  {Aguilar-Arevalo} \emph {et~al.}}]{AguilarArevalo:2007it}%
  \BibitemOpen
  \bibfield  {author} {\bibinfo {author} {\bibfnamefont {A.}~\bibnamefont
  {Aguilar-Arevalo}} \emph {et~al.} (\bibinfo {collaboration} {MiniBooNE
  Collaboration}),\ }\href {\doibase 10.1103/PhysRevLett.98.231801} {\bibfield
  {journal} {\bibinfo  {journal} {Phys.Rev.Lett.}\ }\textbf {\bibinfo {volume}
  {98}},\ \bibinfo {pages} {231801} (\bibinfo {year} {2007})},\ \Eprint
  {http://arxiv.org/abs/0704.1500} {arXiv:0704.1500 [hep-ex]} \BibitemShut
  {NoStop}%
\bibitem [{\citenamefont {Hill}(2010)}]{Hill:2009ek}%
  \BibitemOpen
  \bibfield  {author} {\bibinfo {author} {\bibfnamefont {R.~J.}\ \bibnamefont
  {Hill}},\ }\href {\doibase 10.1103/PhysRevD.81.013008} {\bibfield  {journal}
  {\bibinfo  {journal} {Phys.Rev.}\ }\textbf {\bibinfo {volume} {D81}},\
  \bibinfo {pages} {013008} (\bibinfo {year} {2010})},\ \Eprint
  {http://arxiv.org/abs/0905.0291} {arXiv:0905.0291 [hep-ph]} \BibitemShut
  {NoStop}%
\bibitem [{\citenamefont {Zhang}\ and\ \citenamefont
  {Serot}(2013)}]{Zhang:2012xn}%
  \BibitemOpen
  \bibfield  {author} {\bibinfo {author} {\bibfnamefont {X.}~\bibnamefont
  {Zhang}}\ and\ \bibinfo {author} {\bibfnamefont {B.~D.}\ \bibnamefont
  {Serot}},\ }\href {\doibase 10.1016/j.physletb.2013.01.057} {\bibfield
  {journal} {\bibinfo  {journal} {Phys.Lett.}\ }\textbf {\bibinfo {volume}
  {B719}},\ \bibinfo {pages} {409} (\bibinfo {year} {2013})},\ \Eprint
  {http://arxiv.org/abs/1210.3610} {arXiv:1210.3610 [nucl-th]} \BibitemShut
  {NoStop}%
\bibitem [{\citenamefont {Wang}\ \emph
  {et~al.}(2014{\natexlab{a}})\citenamefont {Wang}, \citenamefont
  {Alvarez-Ruso},\ and\ \citenamefont {Nieves}}]{Wang:2013wva}%
  \BibitemOpen
  \bibfield  {author} {\bibinfo {author} {\bibfnamefont {E.}~\bibnamefont
  {Wang}}, \bibinfo {author} {\bibfnamefont {L.}~\bibnamefont {Alvarez-Ruso}},
  \ and\ \bibinfo {author} {\bibfnamefont {J.}~\bibnamefont {Nieves}},\ }\href
  {\doibase 10.1103/PhysRevC.89.015503} {\bibfield  {journal} {\bibinfo
  {journal} {Phys.Rev.}\ }\textbf {\bibinfo {volume} {C89}},\ \bibinfo {pages}
  {015503} (\bibinfo {year} {2014}{\natexlab{a}})},\ \Eprint
  {http://arxiv.org/abs/1311.2151} {arXiv:1311.2151 [nucl-th]} \BibitemShut
  {NoStop}%
\bibitem [{\citenamefont {Harvey}\ \emph {et~al.}(2007)\citenamefont {Harvey},
  \citenamefont {Hill},\ and\ \citenamefont {Hill}}]{Harvey:2007rd}%
  \BibitemOpen
  \bibfield  {author} {\bibinfo {author} {\bibfnamefont {J.~A.}\ \bibnamefont
  {Harvey}}, \bibinfo {author} {\bibfnamefont {C.~T.}\ \bibnamefont {Hill}}, \
  and\ \bibinfo {author} {\bibfnamefont {R.~J.}\ \bibnamefont {Hill}},\ }\href
  {\doibase 10.1103/PhysRevLett.99.261601} {\bibfield  {journal} {\bibinfo
  {journal} {Phys.Rev.Lett.}\ }\textbf {\bibinfo {volume} {99}},\ \bibinfo
  {pages} {261601} (\bibinfo {year} {2007})},\ \Eprint
  {http://arxiv.org/abs/0708.1281} {arXiv:0708.1281 [hep-ph]} \BibitemShut
  {NoStop}%
\bibitem [{\citenamefont {Harada}\ \emph {et~al.}(2011)\citenamefont {Harada},
  \citenamefont {Matsuzaki},\ and\ \citenamefont {Yamawaki}}]{Harada:2011xx}%
  \BibitemOpen
  \bibfield  {author} {\bibinfo {author} {\bibfnamefont {M.}~\bibnamefont
  {Harada}}, \bibinfo {author} {\bibfnamefont {S.}~\bibnamefont {Matsuzaki}}, \
  and\ \bibinfo {author} {\bibfnamefont {K.}~\bibnamefont {Yamawaki}},\ }\href
  {\doibase 10.1103/PhysRevD.84.036010} {\bibfield  {journal} {\bibinfo
  {journal} {Phys.Rev.}\ }\textbf {\bibinfo {volume} {D84}},\ \bibinfo {pages}
  {036010} (\bibinfo {year} {2011})},\ \Eprint {http://arxiv.org/abs/1104.3286}
  {arXiv:1104.3286 [hep-ph]} \BibitemShut {NoStop}%
\bibitem [{\citenamefont {Zhang}\ and\ \citenamefont
  {Serot}(2012{\natexlab{a}})}]{Zhang:2012aka}%
  \BibitemOpen
  \bibfield  {author} {\bibinfo {author} {\bibfnamefont {X.}~\bibnamefont
  {Zhang}}\ and\ \bibinfo {author} {\bibfnamefont {B.~D.}\ \bibnamefont
  {Serot}},\ }\href {\doibase 10.1103/PhysRevC.86.035502} {\bibfield  {journal}
  {\bibinfo  {journal} {Phys.Rev.}\ }\textbf {\bibinfo {volume} {C86}},\
  \bibinfo {pages} {035502} (\bibinfo {year} {2012}{\natexlab{a}})},\ \Eprint
  {http://arxiv.org/abs/1206.6324} {arXiv:1206.6324 [nucl-th]} \BibitemShut
  {NoStop}%
\bibitem [{\citenamefont {Hill}(2011)}]{Hill:2010zy}%
  \BibitemOpen
  \bibfield  {author} {\bibinfo {author} {\bibfnamefont {R.~J.}\ \bibnamefont
  {Hill}},\ }\href {\doibase 10.1103/PhysRevD.84.017501} {\bibfield  {journal}
  {\bibinfo  {journal} {Phys.Rev.}\ }\textbf {\bibinfo {volume} {D84}},\
  \bibinfo {pages} {017501} (\bibinfo {year} {2011})},\ \Eprint
  {http://arxiv.org/abs/1002.4215} {arXiv:1002.4215 [hep-ph]} \BibitemShut
  {NoStop}%
\bibitem [{\citenamefont {MiniBooNE}(2012)}]{MiniBooNEweb}%
  \BibitemOpen
  \bibfield  {author} {\bibinfo {author} {\bibnamefont {MiniBooNE}},\
  }\href@noop {} {}\bibinfo {howpublished}
  {\url{http://www-boone.fnal.gov/for_physicists/data_release/nue_nuebar_2012/efficiency/MB_nu_nubar_combined_release.html}}
  (\bibinfo {year} {2012})\BibitemShut {NoStop}%
\bibitem [{\citenamefont {Alvarez-Ruso}\ \emph
  {et~al.}(2013{\natexlab{a}})\citenamefont {Alvarez-Ruso}, \citenamefont
  {Nieves},\ and\ \citenamefont {Wang}}]{Alvarez-Ruso:2013ica}%
  \BibitemOpen
  \bibfield  {author} {\bibinfo {author} {\bibfnamefont {L.}~\bibnamefont
  {Alvarez-Ruso}}, \bibinfo {author} {\bibfnamefont {J.}~\bibnamefont
  {Nieves}}, \ and\ \bibinfo {author} {\bibfnamefont {E.}~\bibnamefont
  {Wang}},\ }\href@noop {} {\  (\bibinfo {year} {2013}{\natexlab{a}})},\
  \Eprint {http://arxiv.org/abs/1304.2702} {arXiv:1304.2702 [nucl-th]}
  \BibitemShut {NoStop}%
\bibitem [{\citenamefont {Wang}\ \emph
  {et~al.}(2014{\natexlab{b}})\citenamefont {Wang}, \citenamefont
  {Alvarez-Ruso},\ and\ \citenamefont {Nieves}}]{WangMB}%
  \BibitemOpen
  \bibfield  {author} {\bibinfo {author} {\bibfnamefont {E.}~\bibnamefont
  {Wang}}, \bibinfo {author} {\bibfnamefont {L.}~\bibnamefont {Alvarez-Ruso}},
  \ and\ \bibinfo {author} {\bibfnamefont {J.}~\bibnamefont {Nieves}},\
  }\href@noop {} {} (\bibinfo {year} {2014}{\natexlab{b}}),\ \bibinfo {note}
  {in preparation}\BibitemShut {NoStop}%
\bibitem [{\citenamefont {Hiraide}(2006)}]{Hiraide:2006zq}%
  \BibitemOpen
  \bibfield  {author} {\bibinfo {author} {\bibfnamefont {K.}~\bibnamefont
  {Hiraide}} (\bibinfo {collaboration} {SciBooNE Collaboration}),\ }\href
  {\doibase 10.1016/j.nuclphysbps.2006.08.052} {\bibfield  {journal} {\bibinfo
  {journal} {Nucl.Phys.Proc.Suppl.}\ }\textbf {\bibinfo {volume} {159}},\
  \bibinfo {pages} {85} (\bibinfo {year} {2006})}\BibitemShut {NoStop}%
\bibitem [{\citenamefont {Carrasco}\ \emph {et~al.}(1993)\citenamefont
  {Carrasco}, \citenamefont {Nieves},\ and\ \citenamefont
  {Oset}}]{Carrasco:1991we}%
  \BibitemOpen
  \bibfield  {author} {\bibinfo {author} {\bibfnamefont {R.}~\bibnamefont
  {Carrasco}}, \bibinfo {author} {\bibfnamefont {J.}~\bibnamefont {Nieves}}, \
  and\ \bibinfo {author} {\bibfnamefont {E.}~\bibnamefont {Oset}},\ }\href
  {\doibase 10.1016/0375-9474(93)90005-I} {\bibfield  {journal} {\bibinfo
  {journal} {Nucl.Phys.}\ }\textbf {\bibinfo {volume} {A565}},\ \bibinfo
  {pages} {797} (\bibinfo {year} {1993})}\BibitemShut {NoStop}%
\bibitem [{\citenamefont {Hirenzaki}\ \emph {et~al.}(1993)\citenamefont
  {Hirenzaki}, \citenamefont {Nieves}, \citenamefont {Oset},\ and\
  \citenamefont {Vicente-Vacas}}]{Hirenzaki:1993jc}%
  \BibitemOpen
  \bibfield  {author} {\bibinfo {author} {\bibfnamefont {S.}~\bibnamefont
  {Hirenzaki}}, \bibinfo {author} {\bibfnamefont {J.}~\bibnamefont {Nieves}},
  \bibinfo {author} {\bibfnamefont {E.}~\bibnamefont {Oset}}, \ and\ \bibinfo
  {author} {\bibfnamefont {M.}~\bibnamefont {Vicente-Vacas}},\ }\href {\doibase
  10.1016/0370-2693(93)90282-M} {\bibfield  {journal} {\bibinfo  {journal}
  {Phys.Lett.}\ }\textbf {\bibinfo {volume} {B304}},\ \bibinfo {pages} {198}
  (\bibinfo {year} {1993})}\BibitemShut {NoStop}%
\bibitem [{\citenamefont {Fernandez~de Cordoba}\ \emph
  {et~al.}(1993)\citenamefont {Fernandez~de Cordoba}, \citenamefont {Nieves},
  \citenamefont {Oset},\ and\ \citenamefont
  {Vicente-Vacas}}]{FernandezdeCordoba:1992ky}%
  \BibitemOpen
  \bibfield  {author} {\bibinfo {author} {\bibfnamefont {P.}~\bibnamefont
  {Fernandez~de Cordoba}}, \bibinfo {author} {\bibfnamefont {J.}~\bibnamefont
  {Nieves}}, \bibinfo {author} {\bibfnamefont {E.}~\bibnamefont {Oset}}, \ and\
  \bibinfo {author} {\bibfnamefont {M.}~\bibnamefont {Vicente-Vacas}},\ }\href
  {\doibase 10.1016/0370-2693(93)91744-8} {\bibfield  {journal} {\bibinfo
  {journal} {Phys.Lett.}\ }\textbf {\bibinfo {volume} {B319}},\ \bibinfo
  {pages} {416} (\bibinfo {year} {1993})}\BibitemShut {NoStop}%
\bibitem [{\citenamefont {Leitner}\ \emph
  {et~al.}(2009{\natexlab{d}})\citenamefont {Leitner}, \citenamefont {Mosel},\
  and\ \citenamefont {Winkelmann}}]{Leitner:2009ph}%
  \BibitemOpen
  \bibfield  {author} {\bibinfo {author} {\bibfnamefont {T.}~\bibnamefont
  {Leitner}}, \bibinfo {author} {\bibfnamefont {U.}~\bibnamefont {Mosel}}, \
  and\ \bibinfo {author} {\bibfnamefont {S.}~\bibnamefont {Winkelmann}},\
  }\href {\doibase 10.1103/PhysRevC.79.057601} {\bibfield  {journal} {\bibinfo
  {journal} {Phys.Rev.}\ }\textbf {\bibinfo {volume} {C79}},\ \bibinfo {pages}
  {057601} (\bibinfo {year} {2009}{\natexlab{d}})},\ \Eprint
  {http://arxiv.org/abs/0901.2837} {arXiv:0901.2837 [nucl-th]} \BibitemShut
  {NoStop}%
\bibitem [{\citenamefont {Nakamura}\ \emph {et~al.}(2010)\citenamefont
  {Nakamura}, \citenamefont {Sato}, \citenamefont {Lee}, \citenamefont
  {Szczerbinska},\ and\ \citenamefont {Kubodera}}]{Nakamura:2009iq}%
  \BibitemOpen
  \bibfield  {author} {\bibinfo {author} {\bibfnamefont {S.}~\bibnamefont
  {Nakamura}}, \bibinfo {author} {\bibfnamefont {T.}~\bibnamefont {Sato}},
  \bibinfo {author} {\bibfnamefont {T.-S.}\ \bibnamefont {Lee}}, \bibinfo
  {author} {\bibfnamefont {B.}~\bibnamefont {Szczerbinska}}, \ and\ \bibinfo
  {author} {\bibfnamefont {K.}~\bibnamefont {Kubodera}},\ }\href {\doibase
  10.1103/PhysRevC.81.035502} {\bibfield  {journal} {\bibinfo  {journal}
  {Phys.Rev.}\ }\textbf {\bibinfo {volume} {C81}},\ \bibinfo {pages} {035502}
  (\bibinfo {year} {2010})},\ \Eprint {http://arxiv.org/abs/0910.1057}
  {arXiv:0910.1057 [nucl-th]} \BibitemShut {NoStop}%
\bibitem [{\citenamefont {Kopeliovich}\ and\ \citenamefont
  {Marage}(1993)}]{Kopeliovich:1992ym}%
  \BibitemOpen
  \bibfield  {author} {\bibinfo {author} {\bibfnamefont {B.}~\bibnamefont
  {Kopeliovich}}\ and\ \bibinfo {author} {\bibfnamefont {P.}~\bibnamefont
  {Marage}},\ }\href {\doibase 10.1142/S0217751X93000631} {\bibfield  {journal}
  {\bibinfo  {journal} {Int.J.Mod.Phys.}\ }\textbf {\bibinfo {volume} {A8}},\
  \bibinfo {pages} {1513} (\bibinfo {year} {1993})}\BibitemShut {NoStop}%
\bibitem [{\citenamefont {Rein}\ and\ \citenamefont
  {Sehgal}(2007)}]{Rein:2006di}%
  \BibitemOpen
  \bibfield  {author} {\bibinfo {author} {\bibfnamefont {D.}~\bibnamefont
  {Rein}}\ and\ \bibinfo {author} {\bibfnamefont {L.}~\bibnamefont {Sehgal}},\
  }\href {\doibase 10.1016/j.physletb.2007.10.025} {\bibfield  {journal}
  {\bibinfo  {journal} {Phys.Lett.}\ }\textbf {\bibinfo {volume} {B657}},\
  \bibinfo {pages} {207} (\bibinfo {year} {2007})},\ \Eprint
  {http://arxiv.org/abs/hep-ph/0606185} {arXiv:hep-ph/0606185 [hep-ph]}
  \BibitemShut {NoStop}%
\bibitem [{\citenamefont {Paschos}\ \emph {et~al.}(2006)\citenamefont
  {Paschos}, \citenamefont {Kartavtsev},\ and\ \citenamefont
  {Gounaris}}]{Paschos:2005km}%
  \BibitemOpen
  \bibfield  {author} {\bibinfo {author} {\bibfnamefont {E.}~\bibnamefont
  {Paschos}}, \bibinfo {author} {\bibfnamefont {A.}~\bibnamefont {Kartavtsev}},
  \ and\ \bibinfo {author} {\bibfnamefont {G.}~\bibnamefont {Gounaris}},\
  }\href {\doibase 10.1103/PhysRevD.74.054007} {\bibfield  {journal} {\bibinfo
  {journal} {Phys.Rev.}\ }\textbf {\bibinfo {volume} {D74}},\ \bibinfo {pages}
  {054007} (\bibinfo {year} {2006})},\ \Eprint
  {http://arxiv.org/abs/hep-ph/0512139} {arXiv:hep-ph/0512139 [hep-ph]}
  \BibitemShut {NoStop}%
\bibitem [{\citenamefont {Singh}\ \emph {et~al.}(2006)\citenamefont {Singh},
  \citenamefont {Sajjad~Athar},\ and\ \citenamefont {Ahmad}}]{Singh:2006bm}%
  \BibitemOpen
  \bibfield  {author} {\bibinfo {author} {\bibfnamefont {S.}~\bibnamefont
  {Singh}}, \bibinfo {author} {\bibfnamefont {M.}~\bibnamefont {Sajjad~Athar}},
  \ and\ \bibinfo {author} {\bibfnamefont {S.}~\bibnamefont {Ahmad}},\ }\href
  {\doibase 10.1103/PhysRevLett.96.241801} {\bibfield  {journal} {\bibinfo
  {journal} {Phys.Rev.Lett.}\ }\textbf {\bibinfo {volume} {96}},\ \bibinfo
  {pages} {241801} (\bibinfo {year} {2006})}\BibitemShut {NoStop}%
\bibitem [{\citenamefont {Alvarez-Ruso}\ \emph
  {et~al.}(2007{\natexlab{a}})\citenamefont {Alvarez-Ruso}, \citenamefont
  {Geng}, \citenamefont {Hirenzaki},\ and\ \citenamefont
  {Vicente~Vacas}}]{AlvarezRuso:2007tt}%
  \BibitemOpen
  \bibfield  {author} {\bibinfo {author} {\bibfnamefont {L.}~\bibnamefont
  {Alvarez-Ruso}}, \bibinfo {author} {\bibfnamefont {L.}~\bibnamefont {Geng}},
  \bibinfo {author} {\bibfnamefont {S.}~\bibnamefont {Hirenzaki}}, \ and\
  \bibinfo {author} {\bibfnamefont {M.}~\bibnamefont {Vicente~Vacas}},\ }\href
  {\doibase 10.1103/PhysRevC.75.055501, 10.1103/PhysRevC.80.019906} {\bibfield
  {journal} {\bibinfo  {journal} {Phys.Rev.}\ }\textbf {\bibinfo {volume}
  {C75}},\ \bibinfo {pages} {055501} (\bibinfo {year} {2007}{\natexlab{a}})},\
  \Eprint {http://arxiv.org/abs/nucl-th/0701098} {arXiv:nucl-th/0701098
  [nucl-th]} \BibitemShut {NoStop}%
\bibitem [{\citenamefont {Alvarez-Ruso}\ \emph
  {et~al.}(2007{\natexlab{b}})\citenamefont {Alvarez-Ruso}, \citenamefont
  {Geng},\ and\ \citenamefont {Vicente~Vacas}}]{AlvarezRuso:2007it}%
  \BibitemOpen
  \bibfield  {author} {\bibinfo {author} {\bibfnamefont {L.}~\bibnamefont
  {Alvarez-Ruso}}, \bibinfo {author} {\bibfnamefont {L.}~\bibnamefont {Geng}},
  \ and\ \bibinfo {author} {\bibfnamefont {M.}~\bibnamefont {Vicente~Vacas}},\
  }\href {\doibase 10.1103/PhysRevC.76.068501, 10.1103/PhysRevC.80.029904}
  {\bibfield  {journal} {\bibinfo  {journal} {Phys.Rev.}\ }\textbf {\bibinfo
  {volume} {C76}},\ \bibinfo {pages} {068501} (\bibinfo {year}
  {2007}{\natexlab{b}})},\ \Eprint {http://arxiv.org/abs/0707.2172}
  {arXiv:0707.2172 [nucl-th]} \BibitemShut {NoStop}%
\bibitem [{\citenamefont {Zhang}\ and\ \citenamefont
  {Serot}(2012{\natexlab{b}})}]{Zhang:2012xi}%
  \BibitemOpen
  \bibfield  {author} {\bibinfo {author} {\bibfnamefont {X.}~\bibnamefont
  {Zhang}}\ and\ \bibinfo {author} {\bibfnamefont {B.~D.}\ \bibnamefont
  {Serot}},\ }\href {\doibase 10.1103/PhysRevC.86.035504} {\bibfield  {journal}
  {\bibinfo  {journal} {Phys.Rev.}\ }\textbf {\bibinfo {volume} {C86}},\
  \bibinfo {pages} {035504} (\bibinfo {year} {2012}{\natexlab{b}})},\ \Eprint
  {http://arxiv.org/abs/1208.1553} {arXiv:1208.1553 [nucl-th]} \BibitemShut
  {NoStop}%
\bibitem [{\citenamefont {Hernandez}\ \emph {et~al.}(2009)\citenamefont
  {Hernandez}, \citenamefont {Nieves},\ and\ \citenamefont
  {Vicente-Vacas}}]{Hernandez:2009vm}%
  \BibitemOpen
  \bibfield  {author} {\bibinfo {author} {\bibfnamefont {E.}~\bibnamefont
  {Hernandez}}, \bibinfo {author} {\bibfnamefont {J.}~\bibnamefont {Nieves}}, \
  and\ \bibinfo {author} {\bibfnamefont {M.}~\bibnamefont {Vicente-Vacas}},\
  }\href {\doibase 10.1103/PhysRevD.80.013003} {\bibfield  {journal} {\bibinfo
  {journal} {Phys.Rev.}\ }\textbf {\bibinfo {volume} {D80}},\ \bibinfo {pages}
  {013003} (\bibinfo {year} {2009})},\ \Eprint {http://arxiv.org/abs/0903.5285}
  {arXiv:0903.5285 [hep-ph]} \BibitemShut {NoStop}%
\bibitem [{\citenamefont {Berger}\ and\ \citenamefont
  {Sehgal}(2009)}]{Berger:2008xs}%
  \BibitemOpen
  \bibfield  {author} {\bibinfo {author} {\bibfnamefont {C.}~\bibnamefont
  {Berger}}\ and\ \bibinfo {author} {\bibfnamefont {L.}~\bibnamefont
  {Sehgal}},\ }\href {\doibase 10.1103/PhysRevD.79.053003} {\bibfield
  {journal} {\bibinfo  {journal} {Phys.Rev.}\ }\textbf {\bibinfo {volume}
  {D79}},\ \bibinfo {pages} {053003} (\bibinfo {year} {2009})},\ \Eprint
  {http://arxiv.org/abs/0812.2653} {arXiv:0812.2653 [hep-ph]} \BibitemShut
  {NoStop}%
\bibitem [{\citenamefont {Paschos}\ and\ \citenamefont
  {Schalla}(2009)}]{Paschos:2009ag}%
  \BibitemOpen
  \bibfield  {author} {\bibinfo {author} {\bibfnamefont {E.}~\bibnamefont
  {Paschos}}\ and\ \bibinfo {author} {\bibfnamefont {D.}~\bibnamefont
  {Schalla}},\ }\href {\doibase 10.1103/PhysRevD.80.033005} {\bibfield
  {journal} {\bibinfo  {journal} {Phys.Rev.}\ }\textbf {\bibinfo {volume}
  {D80}},\ \bibinfo {pages} {033005} (\bibinfo {year} {2009})},\ \Eprint
  {http://arxiv.org/abs/0903.0451} {arXiv:0903.0451 [hep-ph]} \BibitemShut
  {NoStop}%
\bibitem [{\citenamefont {Adler}(1964)}]{Adler:1964yx}%
  \BibitemOpen
  \bibfield  {author} {\bibinfo {author} {\bibfnamefont {S.~L.}\ \bibnamefont
  {Adler}},\ }\href {\doibase 10.1103/PhysRev.135.B963} {\bibfield  {journal}
  {\bibinfo  {journal} {Phys.Rev.}\ }\textbf {\bibinfo {volume} {135}},\
  \bibinfo {pages} {B963} (\bibinfo {year} {1964})}\BibitemShut {NoStop}%
\bibitem [{\citenamefont {Alvarez-Ruso}\ \emph
  {et~al.}(2013{\natexlab{b}})\citenamefont {Alvarez-Ruso}, \citenamefont
  {Nieves}, \citenamefont {Simo}, \citenamefont {Valverde},\ and\ \citenamefont
  {Vicente~Vacas}}]{AlvarezRuso:2012fc}%
  \BibitemOpen
  \bibfield  {author} {\bibinfo {author} {\bibfnamefont {L.}~\bibnamefont
  {Alvarez-Ruso}}, \bibinfo {author} {\bibfnamefont {J.}~\bibnamefont
  {Nieves}}, \bibinfo {author} {\bibfnamefont {I.~R.}\ \bibnamefont {Simo}},
  \bibinfo {author} {\bibfnamefont {M.}~\bibnamefont {Valverde}}, \ and\
  \bibinfo {author} {\bibfnamefont {M.}~\bibnamefont {Vicente~Vacas}},\ }\href
  {\doibase 10.1103/PhysRevC.87.015503} {\bibfield  {journal} {\bibinfo
  {journal} {Phys.Rev.}\ }\textbf {\bibinfo {volume} {C87}},\ \bibinfo {pages}
  {015503} (\bibinfo {year} {2013}{\natexlab{b}})},\ \Eprint
  {http://arxiv.org/abs/1205.4863} {arXiv:1205.4863 [nucl-th]} \BibitemShut
  {NoStop}%
\bibitem [{\citenamefont {Alam}\ \emph
  {et~al.}(2012{\natexlab{b}})\citenamefont {Alam}, \citenamefont {Simo},
  \citenamefont {Athar},\ and\ \citenamefont {Vicente~Vacas}}]{Alam:2012zz}%
  \BibitemOpen
  \bibfield  {author} {\bibinfo {author} {\bibfnamefont {M.~R.}\ \bibnamefont
  {Alam}}, \bibinfo {author} {\bibfnamefont {I.~R.}\ \bibnamefont {Simo}},
  \bibinfo {author} {\bibfnamefont {M.~S.}\ \bibnamefont {Athar}}, \ and\
  \bibinfo {author} {\bibfnamefont {M.}~\bibnamefont {Vicente~Vacas}},\ }\href
  {\doibase 10.1103/PhysRevD.85.013014} {\bibfield  {journal} {\bibinfo
  {journal} {Phys.Rev.}\ }\textbf {\bibinfo {volume} {D85}},\ \bibinfo {pages}
  {013014} (\bibinfo {year} {2012}{\natexlab{b}})},\ \Eprint
  {http://arxiv.org/abs/1111.0863} {arXiv:1111.0863 [hep-ph]} \BibitemShut
  {NoStop}%
\bibitem [{\citenamefont {Gershtein}\ \emph {et~al.}(1981)\citenamefont
  {Gershtein}, \citenamefont {Komachenko},\ and\ \citenamefont
  {Khlopov}}]{Gershtein:1980wu}%
  \BibitemOpen
  \bibfield  {author} {\bibinfo {author} {\bibfnamefont {S.}~\bibnamefont
  {Gershtein}}, \bibinfo {author} {\bibfnamefont {Y.~Y.}\ \bibnamefont
  {Komachenko}}, \ and\ \bibinfo {author} {\bibfnamefont {M.~Y.~a.}\
  \bibnamefont {Khlopov}},\ }\href@noop {} {\bibfield  {journal} {\bibinfo
  {journal} {Sov.J.Nucl.Phys.}\ }\textbf {\bibinfo {volume} {33}},\ \bibinfo
  {pages} {860} (\bibinfo {year} {1981})}\BibitemShut {NoStop}%
\bibitem [{\citenamefont {Rein}\ and\ \citenamefont
  {Sehgal}(1981{\natexlab{b}})}]{Rein:1981ys}%
  \BibitemOpen
  \bibfield  {author} {\bibinfo {author} {\bibfnamefont {D.}~\bibnamefont
  {Rein}}\ and\ \bibinfo {author} {\bibfnamefont {L.}~\bibnamefont {Sehgal}},\
  }\href {\doibase 10.1016/0370-2693(81)90706-1} {\bibfield  {journal}
  {\bibinfo  {journal} {Phys.Lett.}\ }\textbf {\bibinfo {volume} {B104}},\
  \bibinfo {pages} {394} (\bibinfo {year} {1981}{\natexlab{b}})}\BibitemShut
  {NoStop}%
\bibitem [{\citenamefont {Gazizov}\ and\ \citenamefont
  {Kowalski}(2005)}]{Gazizov:2004va}%
  \BibitemOpen
  \bibfield  {author} {\bibinfo {author} {\bibfnamefont {A.}~\bibnamefont
  {Gazizov}}\ and\ \bibinfo {author} {\bibfnamefont {M.~P.}\ \bibnamefont
  {Kowalski}},\ }\href {\doibase 10.1016/j.cpc.2005.03.113} {\bibfield
  {journal} {\bibinfo  {journal} {Comput.Phys.Commun.}\ }\textbf {\bibinfo
  {volume} {172}},\ \bibinfo {pages} {203} (\bibinfo {year} {2005})},\ \Eprint
  {http://arxiv.org/abs/astro-ph/0406439} {arXiv:astro-ph/0406439 [astro-ph]}
  \BibitemShut {NoStop}%
\bibitem [{\citenamefont {Andreopoulos}\ \emph {et~al.}(2010)\citenamefont
  {Andreopoulos}, \citenamefont {Bell}, \citenamefont {Bhattacharya},
  \citenamefont {Cavanna}, \citenamefont {Dobson} \emph
  {et~al.}}]{Andreopoulos:2009rq}%
  \BibitemOpen
  \bibfield  {author} {\bibinfo {author} {\bibfnamefont {C.}~\bibnamefont
  {Andreopoulos}}, \bibinfo {author} {\bibfnamefont {A.}~\bibnamefont {Bell}},
  \bibinfo {author} {\bibfnamefont {D.}~\bibnamefont {Bhattacharya}}, \bibinfo
  {author} {\bibfnamefont {F.}~\bibnamefont {Cavanna}}, \bibinfo {author}
  {\bibfnamefont {J.}~\bibnamefont {Dobson}},  \emph {et~al.},\ }\href
  {\doibase 10.1016/j.nima.2009.12.009} {\bibfield  {journal} {\bibinfo
  {journal} {Nucl.Instrum.Meth.}\ }\textbf {\bibinfo {volume} {A614}},\
  \bibinfo {pages} {87} (\bibinfo {year} {2010})},\ \Eprint
  {http://arxiv.org/abs/0905.2517} {arXiv:0905.2517 [hep-ph]} \BibitemShut
  {NoStop}%
\bibitem [{\citenamefont {Autiero}(2005)}]{Autiero:2005ve}%
  \BibitemOpen
  \bibfield  {author} {\bibinfo {author} {\bibfnamefont {D.}~\bibnamefont
  {Autiero}},\ }\href {\doibase 10.1016/j.nuclphysbps.2004.11.168} {\bibfield
  {journal} {\bibinfo  {journal} {Nucl.Phys.Proc.Suppl.}\ }\textbf {\bibinfo
  {volume} {139}},\ \bibinfo {pages} {253} (\bibinfo {year}
  {2005})}\BibitemShut {NoStop}%
\bibitem [{\citenamefont {Hayato}(2009)}]{Hayato:2009zz}%
  \BibitemOpen
  \bibfield  {author} {\bibinfo {author} {\bibfnamefont {Y.}~\bibnamefont
  {Hayato}},\ }\href@noop {} {\bibfield  {journal} {\bibinfo  {journal} {Acta
  Phys.Polon.}\ }\textbf {\bibinfo {volume} {B40}},\ \bibinfo {pages} {2477}
  (\bibinfo {year} {2009})}\BibitemShut {NoStop}%
\bibitem [{\citenamefont {Battistoni}\ \emph {et~al.}(2009)\citenamefont
  {Battistoni}, \citenamefont {Sala}, \citenamefont {Lantz}, \citenamefont
  {Ferrari},\ and\ \citenamefont {Smirnov}}]{Battistoni:2009zzb}%
  \BibitemOpen
  \bibfield  {author} {\bibinfo {author} {\bibfnamefont {G.}~\bibnamefont
  {Battistoni}}, \bibinfo {author} {\bibfnamefont {P.}~\bibnamefont {Sala}},
  \bibinfo {author} {\bibfnamefont {M.}~\bibnamefont {Lantz}}, \bibinfo
  {author} {\bibfnamefont {A.}~\bibnamefont {Ferrari}}, \ and\ \bibinfo
  {author} {\bibfnamefont {G.}~\bibnamefont {Smirnov}},\ }\href@noop {}
  {\bibfield  {journal} {\bibinfo  {journal} {Acta Phys.Polon.}\ }\textbf
  {\bibinfo {volume} {B40}},\ \bibinfo {pages} {2491} (\bibinfo {year}
  {2009})}\BibitemShut {NoStop}%
\bibitem [{\citenamefont {Nowak}(2006)}]{Nowak:2006sx}%
  \BibitemOpen
  \bibfield  {author} {\bibinfo {author} {\bibfnamefont {J.~A.}\ \bibnamefont
  {Nowak}},\ }\href {\doibase 10.1088/0031-8949/2006/T127/025} {\bibfield
  {journal} {\bibinfo  {journal} {Phys.Scripta}\ }\textbf {\bibinfo {volume}
  {T127}},\ \bibinfo {pages} {70} (\bibinfo {year} {2006})},\ \Eprint
  {http://arxiv.org/abs/hep-ph/0607081} {arXiv:hep-ph/0607081 [hep-ph]}
  \BibitemShut {NoStop}%
\bibitem [{\citenamefont {Bodek}\ and\ \citenamefont
  {Yang}(2003{\natexlab{b}})}]{Bodek:2002ps}%
  \BibitemOpen
  \bibfield  {author} {\bibinfo {author} {\bibfnamefont {A.}~\bibnamefont
  {Bodek}}\ and\ \bibinfo {author} {\bibfnamefont {U.}~\bibnamefont {Yang}},\
  }\href {\doibase 10.1088/0954-3899/29/8/369} {\bibfield  {journal} {\bibinfo
  {journal} {J.Phys.}\ }\textbf {\bibinfo {volume} {G29}},\ \bibinfo {pages}
  {1899} (\bibinfo {year} {2003}{\natexlab{b}})},\ \Eprint
  {http://arxiv.org/abs/hep-ex/0210024} {arXiv:hep-ex/0210024 [hep-ex]}
  \BibitemShut {NoStop}%
\bibitem [{\citenamefont {Sjostrand}\ \emph {et~al.}(2006)\citenamefont
  {Sjostrand}, \citenamefont {Mrenna},\ and\ \citenamefont
  {Skands}}]{Sjostrand:2006za}%
  \BibitemOpen
  \bibfield  {author} {\bibinfo {author} {\bibfnamefont {T.}~\bibnamefont
  {Sjostrand}}, \bibinfo {author} {\bibfnamefont {S.}~\bibnamefont {Mrenna}}, \
  and\ \bibinfo {author} {\bibfnamefont {P.~Z.}\ \bibnamefont {Skands}},\
  }\href {\doibase 10.1088/1126-6708/2006/05/026} {\bibfield  {journal}
  {\bibinfo  {journal} {JHEP}\ }\textbf {\bibinfo {volume} {0605}},\ \bibinfo
  {pages} {026} (\bibinfo {year} {2006})},\ \Eprint
  {http://arxiv.org/abs/hep-ph/0603175} {arXiv:hep-ph/0603175 [hep-ph]}
  \BibitemShut {NoStop}%
\bibitem [{\citenamefont {Sjostrand}\ \emph {et~al.}(2008)\citenamefont
  {Sjostrand}, \citenamefont {Mrenna},\ and\ \citenamefont
  {Skands}}]{Sjostrand:2007gs}%
  \BibitemOpen
  \bibfield  {author} {\bibinfo {author} {\bibfnamefont {T.}~\bibnamefont
  {Sjostrand}}, \bibinfo {author} {\bibfnamefont {S.}~\bibnamefont {Mrenna}}, \
  and\ \bibinfo {author} {\bibfnamefont {P.~Z.}\ \bibnamefont {Skands}},\
  }\href {\doibase 10.1016/j.cpc.2008.01.036} {\bibfield  {journal} {\bibinfo
  {journal} {Comput.Phys.Commun.}\ }\textbf {\bibinfo {volume} {178}},\
  \bibinfo {pages} {852} (\bibinfo {year} {2008})},\ \Eprint
  {http://arxiv.org/abs/0710.3820} {arXiv:0710.3820 [hep-ph]} \BibitemShut
  {NoStop}%
\bibitem [{\citenamefont {Koba}\ \emph {et~al.}(1972)\citenamefont {Koba},
  \citenamefont {Nielsen},\ and\ \citenamefont {Olesen}}]{Koba:1972ng}%
  \BibitemOpen
  \bibfield  {author} {\bibinfo {author} {\bibfnamefont {Z.}~\bibnamefont
  {Koba}}, \bibinfo {author} {\bibfnamefont {H.~B.}\ \bibnamefont {Nielsen}}, \
  and\ \bibinfo {author} {\bibfnamefont {P.}~\bibnamefont {Olesen}},\ }\href
  {\doibase 10.1016/0550-3213(72)90551-2} {\bibfield  {journal} {\bibinfo
  {journal} {Nucl.Phys.}\ }\textbf {\bibinfo {volume} {B40}},\ \bibinfo {pages}
  {317} (\bibinfo {year} {1972})}\BibitemShut {NoStop}%
\end{thebibliography}%
\end{document}